\documentclass{mn2e}

\usepackage{amsmath}
\usepackage{amsbsy}

\input epsf
\usepackage{amsmath,amssymb,subfigure}
\usepackage{graphicx}
\usepackage{epsfig}
\usepackage{color}

\newcommand{\be}{\begin{equation}}
\newcommand{\ee}{\end{equation}}

\newcommand{\beq}{\begin{equation}}
\newcommand{\eeq}{\end{equation}}
\newcommand{\beqa}{\begin{eqnarray}}
\newcommand{\eeqa}{\end{eqnarray}}

\newcommand{\ls}{\mathrel{\raise0.27ex\hbox{$<$}\kern-0.70em \lower0.71ex\hbox{{
$\scriptstyle \sim$}}}}

\newcommand{\APS}{APS }

\title
{Determining Frequentist Confidence Limits Using a Directed Parameter Space
Search} 
\author[Daniel, Connolly, and Schneider]
{Scott F.\ Daniel$^1$, Andrew J.\ Connolly$^1$, and Jeff Schneider$^2$\\
$^1$Department of Astronomy, University of Washington, Seattle, WA\\ 
$^2$Machine Learning Department, Carnegie Mellon University, Pittsburgh, PA
}
\begin{document} 

\pagerange{\pageref{firstpage}--\pageref{lastpage}}

\label{firstpage}

\date{\today}

\begin{abstract} 
We consider the problem of inferring constraints on a 
high-dimensional parameter
space with a computationally expensive likelihood function.
We propose a machine learning algorithm that maps out the 
Frequentist confidence limit on parameter space by intelligently targeting
likelihood evaluations so as to quickly and accurately characterize the likelihood
surface in both low- and high-likelihood regions.
We compare our algorithm to Bayesian credible limits derived by 
the well-tested Markov Chain Monte Carlo (MCMC)
algorithm using both multi-modal toy likelihood functions and the 
7-year WMAP cosmic microwave background likelihood function.  We find that our
algorithm correctly identifies the location, general size, and general shape of
high-likelihood regions in parameter space while being more robust against
multi-modality than MCMC.
\end{abstract} 

\maketitle

\section{Introduction}
\label{sec:intro}

We consider the following problem.  A researcher wants to constrain a theory
described by some number $N_p$ of tunable parameters, which we will represent by
the $N_p$-dimensional vector $\vec{\theta}$.  The available data measure some
function $f$ which is an implicit function of $\vec{\theta}$.  The researcher
is able to calculate a $\chi^2$ value quantifying the fit of a 
model $f(\vec{\theta})$ to the data and wishes to find a confidence limit
defined by some rule, e.g $\chi^2\le\chi^2_\text{lim}$.  This is a general
phrasing of a Frequentist confidence limit as discussed in Appendix
\ref{sec:frequentism}.  We discuss alternatives for determining
$\chi^2_\text{lim}$ in Section \ref{sec:chi} below.  We will also consider
the case that the reseracher wishes to find Bayesian credible limt
to some probability $(1-\alpha)\%$.

If the calculation going from $\vec{\theta}\rightarrow\chi^2$ is rapid, this
problem is trivial.  One generates a fine grid of points in the 
$N_p$-dimensional parameter space, evaluates $\chi^2$ at each of these points, 
and defines the confidence limit as only those points which satisfy the rule.  
Often, however, the calcuation $\vec{\theta}\rightarrow\chi^2$ is not rapid, 
or $N_p$
is large enough that such a grid-based method will take a prohibitively long
time.  In that case, one must make intelligent choices regarding which points in
parameter space are actually interesting enough to warrant
 $\vec{\theta}\rightarrow\chi^2$ evaluation.

Markov Chain Monte Carlo (MCMC) methods propose to find the desired
limits by drawing random samples from the parameter space which are drawn
according to a probability density (the Bayesian posterior) which can be
interpreted as the probability that some value of $\vec{\theta}$ represents the
true value of $\vec{\theta}$ realized by the physical process underlying
the data.  
With enough of these samples, one can integrate this
probability density and find regions which contain some set amount of the total
probability.  In this case, ``credible limits'' are phrased as target values
for the probability integral $(1-\alpha)\%$, i.e. 
68\% for ``1-$\sigma$'' limits, 95\% for ``2-$\sigma$'' limits
and so on \cite{mcmc}.

We propose an alternative solution, using methods from machine learning
(to wit, Gaussian processes, though the 
algorithm admits many possible drivers; see
Section \ref{sec:gp})
to find the desired confidence limit by directly searching the
parameter space of $\vec{\theta}$ for points which lie on the boundary of the
desired confidence limit $\chi^2\le\chi^2_\text{lim}$.  
We refer to this algorithm as Active Parameter Searching (APS).
We present the algorithm in detail in Sections \ref{sec:algorithm} and
\ref{sec:gp}.
As it searches, APS uses
the knowledge it has already learned about $\vec{\theta}\rightarrow\chi^2$ 
to improve its search according to a metric that rewards the
algorithm both for finding points on the confidence limit and for sampling
points from previously unexplored regions of parameter space.  We find that this
behavior makes APS more robust against multi-modal $\chi^2$ functions (see
Section \ref{sec:toy}) without
sacrificing the speed of convergence 
associated with traditional MCMC methods (see Section \ref{sec:wmap}).
We show that, while APS is designed specifically to
yield Frequentist confidence
limits (which we define in Appendix \ref{sec:frequentism}),
it is also capable of roughly approximating
Bayesian credible limits on the space of
$\vec{\theta}$ (see Sections \ref{sec:bayes},\ref{sec:toy} and \ref{sec:wmap}).

Readers should be aware that, while MCMC can be analytically shown to
asymptotically sample the Bayesian posterior, APS is designed to discover
Frequentist confidence limits purely through exploration, i.e. there is no {\it a
priori} motivation behind the APS algorithm.  Our only demonstration of APS's
efficacy is empirical.  It has done well on the problems we have tested it on,
though we have tried to choose those problems such that they are representative of
problems likely to be encountered by the community.  Readers who require
theoretically-motivated results will likely need to resort to a sampling routine
like MCMC to derive their final constraints.  
IN the case of multi-modal likelihood functions, the MultiNest algorithm (Feroz
and Hobson 2008, Feroz {\it et al}. 2009) has been shown to be effective.
We hope to show in Section
\ref{sec:toy} that, even in these use cases, APS can provide a valuable service by
identifying the general regions of parameter space over which the Bayesian
posterior needs to be integrated, i.e. by guiding users in the process of setting
priors and proposal densities for a sampling algorithm, 
a problem which has posed significant problems
for MCMC in the case of multi-modal likelihood functions.

\section{The APS Algorithm}
\label{sec:algorithm}

The APS algorithm was originally presented by Bryan (2007) as

\begin{itemize}
\item(1A) Generate some initial number $N_s$ of randomly-distributed points in 
parameter space.  Find the values of $\chi^2$ corresponding to these points.
\\
\item(2A) Generate some number $N_c$ of candidate points.  
For each of these points, find the $N_g$ nearest points already sampled
by APS (we use a k-d tree as described in Bentley 1975). 
Use these nearest neighbor points and their corresponding $\chi^2$ values
to predict the value of $\chi^2$ at each of these points.
 We will signify this prediction by $\mu$.  Calculate some value $\sigma$
 quantifying confidence in $\mu$ as a prediction of $\chi^2$.  Low values of
 $\sigma$ should correspond to confident predictions.
 \\
\item(3A) Choose the candidate point which maximizes the statistic
\begin{equation}
\label{eqn:sstat}
S=\sigma-|\mu-\chi^2_\text{lim}|
\end{equation}
and evaluate $\vec{\theta}\rightarrow\chi^2$ at that point.  Add this point and
its $\chi^2$ value to the list of sampled points begun in (1A). 
Return to (2A) and repeat until convergence.
\\
\end{itemize}
The confidence limit reported by APS is the list of all points found
for which $\chi^2\le\chi^2_\text{lim}$.  
These points, when plotted in one- or two-dimensional slices of parameter
space, ought to sketch out a region which contains the true value of $\vec{\theta}$
with the desired confidence (as defined in Appendix \ref{sec:frequentism}).
This is in contrast to the credible
limits of MCMC, which represent an integral over the sampled points containing some
fraction of the total posterior probability.  We present an alternative
interpretation for yielding credible limits from APS in Section \ref{sec:bayes}.

The $|\mu-\chi^2_\text{lim}|$ term in equation (\ref{eqn:sstat}) provides an
incentive for APS to choose points that it believes will lie on the boundary of
the confidence limit.  
The $\sigma$ term provides an incentive for APS to explore unknown regions of
parameter space, making the algorithm robust against multimodal $\chi^2$
functions.  

In the present work, we introduce modifications and expansions to this basic algorithm as
described below.

\subsection{Determining $\chi^2_\text{lim}$}
\label{sec:chi}

As discussed above, APS draws confidence limits by mapping
iso-$\chi^2$ contours of $\chi^2=\chi^2_\text{lim}$ in parameter space.
Thus, the user is presented with the problem of determining an
appropriate $\chi^2_\text{lim}$.  APS admits two possible solutions.
The user can fiat a definite value of $\chi^2_\text{lim}$
based on the characteristics of the data at hand,
or $\chi^2_\text{lim}$ can be set adaptively, based on APS'
knowledge of $\vec{\theta}\rightarrow\chi^2$.

The first possibility, a user-defined $\chi^2_\text{lim}$, is based
on the theory of Frequentist confidence intervals, which we briefly discuss
here and consider in much greater detail in Appendix \ref{sec:frequentism}.
If a data set is comprised of $N_d$ Gaussian-distributed data points
with a known covariance structure, then 
\begin{equation}
\chi^2\equiv\sum_{i,j}^{N_d}(d_i-f_i(\vec{\theta}))C^{-1}_{i,j}
(d_j-f_j(\vec{\theta}))
\end{equation}
will be distributed according to a $\chi^2$-distribution with $N_d$ degrees
of freedom.  This distribution can be integrated to determine the value
of $\chi^2$ containing $(1-\alpha)\%$ of the total probability.  This limiting
value will be the user-defined $\chi^2_\text{lim}$.

If the user does not feel comfortable settig $\chi^2_\text{lim}$
a priori, APS $\chi^2_\text{lim}$ can learn $\chi^2_\text{lim}$ by
defining it as $\chi^2_\text{min}+\Delta\chi^2$.  Here, $\chi^2_\text{min}$
is the minimum value of $\chi^2$ discovered by APS.  
$\Delta \chi^2$ is set
based on the properties of the $N_p$-dimensional parameter space being explored.
To wit, Wilks (1938) proves that $\Delta\chi^2$ for an $N_p$-dimensional
parameter space will be distributed according to a $\chi^2$-distribution
with $N_p$ degrees of freedom.  One can perform the same integral discussed
above and set $\Delta\chi^2$ to that value which encloses the desired confidence
limit.  This is the well-known Likelihood Ratio Test \cite{np}.
In this case, the user is relying on APS to eventually find the true
$\chi^2_\text{min}$ of the likelihood surface being explored.  
Section \ref{sec:simplex} below outlines an additional means of search
by which our expanded APS is capable of efficiently locating $\chi^2_\text{lim}$.

\subsection{Modified search algorithms}

Bryan {\it et al.} (2007) show that the APS search algorithm as presented in steps (1A-3A)
is robust against multi-modal
likelihood functions, using it to identify a disjoint region of high likelihood in
cosmological parameter space as constrained by the 1-year WMAP CMB anisotropy power
spectrum.  This is the principal advantage to selecting points for evaluation according
to the statistic (\ref{eqn:sstat}).  Unfortunately, it is possible that the focus on 
exploring
unknown regions of parameter space might slow down convergence in the case of uni-modal
likelihood functions.  Therefore, we augment steps (1A-3A) with
the following modified search algorithms
to ensure that APS adequately characterizes known regions of high likelihood while
simultaneously searching for new regions to explore.

These modified searches will require APS to keep track of information it learns about the
$\vec{\theta}\rightarrow\chi^2$ fuction as it searches.  
To facilitate this discussion, we introduce the following notation.
Symbols denoted with $\vec{x}$ represent points in parameter space.  Variables denoted with
$\{\hat{X}\}$ denote sets of points in parameter space.  Variables denoted $\{X\}$ are sets
of numbers stored by APS.

We are greatly indebted to Eric Linder of the Lawrence Berkeley National
Laboratory who first suggested to us the modifications described below
in sections \ref{sec:focus} and \ref{sec:simplex}.

\subsubsection{Characterization of known low-$\chi^2$ regions}
\label{sec:focus}

As it searches, APS will keep track of a list of all of the discovered points
which are believed to be local minima of $\chi^2$ 
(how it finds these will be described below
in Section \ref{sec:simplex}).  These points will be stored in the set
$\{\hat{\Gamma}\}$.  After each cycle of the main algorithm (steps 1A-3A above),
our version of APS attempts ``focused searches'' about the points in $\{\hat{\Gamma}\}$.  

For each point $\vec{M}$ in $\{\hat{\Gamma}\}$, APS performs the following:
\begin{itemize}
\item (1B) If fewer than $N_p$ focused searches have been proposed about $\vec{M}$, 
randomly choose a point from a small sphere surrounding $\vec{M}$ in parameter space and 
evaluate $\chi^2$ at that point.  
\\
\item(2B) If the point does not satisfy $\chi^2=\chi^2_\text{lim}$ (which is highly probable), 
use bisection
along the line connecting the proposed point to $\vec{M}$ to find a point which does
satisfy $\chi^2=\chi^2_\text{lim}$, i.e. find a point on that line for which
$\chi^2<\chi^2_\text{lim}$ -- often $\vec{M}$ itself -- and a point along that line
for which $\chi^2>\chi^2_\text{lim}$ and iteratively step half the distance between them until
arriving at the desired $\chi^2_\text{lim}$ \cite{minuit}.  The points discovered in this way \
with $\chi^2=\chi^2_\text{lim}$ are stored in the set $\{\hat{B}(\vec{M})\}$ of 
``boundary points'' about $\vec{M}$ (each $\vec{M}$ in $\{\hat{\Gamma}\}$ will have its own
set $\{\hat{B}(\vec{M})\}$).  
\\
\item(3B) If $N_p$ boundary points have been found for $\vec{M}$, skip steps (1B) and (2B).
Instead, examine the points in $\{\hat{B}(\vec{M})\}$ and identify the points
corresponding to both the maximum and minimum values of each parameter $\theta_i$.  These
points will be referred to as $\{\hat{B}_\text{MaxMin}(\vec{M})\}$.  
There will be $2\times N_p$ points in $\{\hat{B}_\text{MaxMin}(\vec{M})\}$, one for
the maximum and one for the minimum in each of our $N_p$ parameters.  
\\
\item(4B) For each point $\vec{b}$ in $\{\hat{B}_\text{MaxMin}(\vec{M})\}$, 
propose some number $N_v$ of new points (we have chosen
$N_v=20$ for the sake of speed) $\vec{v} = \vec{b}+\vec{\epsilon}$ where $\vec{\epsilon}$ is a
random, small vector in parameter space constrained to be perpendicular to
$\vec{b}-\vec{M}$.  There will now be $2\times N_p \times N_v$ points $\vec{v}$
($N_v$ such points for each of the $2\times N_p$ extremal boundary points in
$\{\hat{B}_\text{MaxMin}(\vec{M})\}$).
\\
\item(5B) At each point $\vec{v}$ above, find the $N_g$ nearest neighbor points as in step (2A) above
and evaluate equation (\ref{eqn:sstat}).  
Find the point $\vec{v}$ which maximizes $S$ and evaluate $\vec{\theta}\rightarrow\chi^2$ at
that point.  If
$\chi^2 \neq \chi^2_\text{lim}$, use bisection to find a new boundary point.
Store the resulting $\chi^2=\chi^2_\text{lim}$ point in $\{\hat{B}(\vec{M})\}$
\end{itemize}

The modification suggested in section \ref{sec:focus} is designed to explore 
$\chi^2=\chi^2_\text{lim}$ surface by starting from its extreme corners and stepping
perpendicularly to its ``radii'' (the vector $\vec{b}-\vec{M}$ would be a 
radius if
$\chi^2=\chi^2_\text{lim}$ described a sphere), with the hope of being able
to extend the bounds of $\{\hat{B}_\text{MaxMin}(\vec{M})\}$.  In this way,
our extended APS attempts to plumb the full Frequentist confidence limit $\chi^2\le\chi^2_\text{lim}$
as rapidly as possible.  However, we do not want this to come at the expense of the wide-ranging
exploration of steps (1A-3A).  Our version of APS alternates between performing steps (1A-3A) and steps
(1B-5B).  If there are $N_m$ distinct known local minima in $\{\hat{\Gamma}\}$, 
our algoritm will perform
$N_m$ iterations of steps (1A-3A), and then perform steps (1B-5B) once for each
$\vec{M}$ in $\{\hat{\Gamma}\}$.

\subsubsection{Extension of the $\chi^2=\chi^2_\text{lim}$ contour}
\label{sec:unitSphere}

It is, however, also possible
that the $S$-maximization in steps (1A-3A) could learn something about the
shape of the $\chi^2=\chi^2_\text{lim}$ contour, even if it does not place
points directly on that contour.  To explore that possibility, our
extended APS performs the
following modified search after each iteration of steps (1A-3A).

\begin{itemize}
\item(1C) For each point $\vec{M}$ in $\{\hat{\Gamma}\}$, store, 
in addition to the boundary
points $\{\hat{B}(\vec{M})\}$, the projection of those boundary points
onto a unit sphere in parameter space, i.e. for each point $\vec{b}$ in $\{\hat{B}(\vec{M})\}$, draw a line
connecting $\vec{b}$ with $\vec{M}$ and keep track of
where a unit sphere centered on $\vec{M}$ intersects that line.  Call this set
of projections $\{\hat{U}\}$.
\\
\item(2C) Call the point for which $\chi^2$ has been evaluated in step (3A)
$\vec{P}$.  Find the
point $\vec{M}$ in $\{\hat{\Gamma}\}$ that is nearest to $\vec{P}$ in parameter space.
\\
\item(3C) Find the
projection of $\vec{P}$ onto the unit sphere surrounding $\vec{M}$.  
Call this projected point $\vec{p}$.  Find the point $\vec{u}$ in
$\{\hat{U}\}$ that is nearest to $\vec{p}$.  Record the parameter space 
distance between $\vec{u}$ and $\vec{p}$.
Add this distance to the set $\{D\}$ ($\{D\}$ will be a running list
of the $\vec{u}-\vec{p}$ distance for all points $\vec{P}$ found by steps
1A-3A, regardless of which point in $\vec{M}$ they are nearest).
\\
\item(4C) If the $\vec{p}-\vec{u}$ distance is greater than the 2/3 quantile of $\{D\}$,
then perform bisection between $\vec{M}$ and the point evaluated in step (3A).
Add the resulting $\chi^2=\chi^2_\text{lim}$ point to $\{\hat{B}(\vec{M})\}$.
\end{itemize}

Step (4C) ensures that, if the point $\vec{P}$ is in a novel direction from
$\vec{M}$, APS will find a point on the $\chi^2=\chi^2_\text{lim}$ contour
in that direction.  Because $\vec{P}$ was
chosen by maximizing equation (\ref{eqn:sstat}), we assume that there was something
``interesting'' about that direction from $\vec{M}$ (perhaps a sharp corner or distended
extension of the $\chi^2=\chi^2_\text{lim}$ contour).  In this way, our extended APS
will attempt to cover the surface of $\chi^2=\chi^2_\text{lim}$ in the most informative
way possible.

Because we expect the unit spheres surrounding the points $\vec{M}$ to become more densely
populated as the algorithm runs, we periodically delete the list $\{D\}$ so that
new directions away from $\vec{M}$ can still be explored.

\subsubsection{Function minimization}
\label{sec:simplex}

The penultimate modification to the original APS algorithm which we propose in this
work is to utilize function minimization to identify local minima in $\chi^2$.
One of the principal drawbacks of the algorithm presented in steps (1A-3A) is
that it makes no account of ``near misses.''  If step (3A) finds that $\chi^2$
is lower than average but still greater than $\chi^2_\text{lim}$, the original
APS algorithm does nothing with that information.  We have tried to improve upon
that with our incorporation of bisection in sections \ref{sec:focus} and
\ref{sec:unitSphere} above.  An additional innovation is to allow APS to use points
with 
\begin{eqnarray}
\chi^2_\text{lim}&\le&\chi^2\le\chi^2_\text{threshold}\label{eqn:simplexCriterion}\\
\chi_\text{threshold}&\equiv&\chi^2_\text{min}+G\times(\chi^2_{1/10}-\chi^2_\text{min})
\label{eqn:threshold}
\end{eqnarray}
to seed a simplex function minimizer as developed by Nelder and Mead (1965).
Here, $G$ is a user-set parameter, $\chi^2_\text{min}$ is the minimum discovered
$\chi^2$ value and $\chi^2_{1/10}$ is the 1/10 quantile of the $\chi^2$ values
discovered by steps (1A-3A) (not those discovered by any of the modifications
discussed in sections \ref{sec:focus}, \ref{sec:unitSphere} or
\ref{sec:simplex}).

Because the Nelder-Mead simplex minimizer can be very time intensive, we balance our
algorithm by keeping track of how many times $\vec{\theta}\rightarrow\chi^2$
is called by the different searches and
demanding that the number of $\chi^2$ evaluations devoted to all
other variants of search (i.e. steps (1A-3A), (1B-5B), and (1C-4C)) equal the number of
$\chi^2$ evaluations devoted to the simplex minimizer before a new simplex
minimization can occur.

If the simplex minimizer converges to a point such that
$\chi^2<\chi^2_\text{lim}$, our extended APS determines whether or not it
has found a new local minimum to be added to $\{\hat{\Gamma}\}$ by comparing the end
point $\vec{y}$ of the minimization to each of the points already in
$\{\hat{\Gamma}\}$.  For each $\vec{M}$ in $\{\hat{\Gamma}\}$, APS calculates the mid-point
\begin{equation}
\label{eqn:midpt}
\vec{m}(\vec{M})=0.5\times\left(\vec{y}+\vec{M}\right)
\end{equation}
If, for all $\vec{M}$, 
$\chi^2$ at $\vec{m}(\vec{M})$ is greater
than $\chi^2_\text{lim}$, then it is deemed that $\vec{y}$ is a new local
minimum and is added to $\{\hat{\Gamma}\}$.  If not, $\vec{y}$ is assumed to be a
part of the one of the previously identified loci of low $\chi^2$ and it is not
added to $\{\hat{\Gamma}\}$.  Regardless of whether or not $\vec{y}$ passes the test in
equation (\ref{eqn:midpt}), $\vec{y}$ is added to the set $\{\hat{Y}\}$ of all
simplex end points discovered by APS.

As a further guard against excessive simplex searching, all points $\vec{w}$ 
that pass the test in equation (\ref{eqn:simplexCriterion}) are further required
to satisfy the condition
\begin{eqnarray}
\chi^2(\vec{z})&>&\chi^2(\vec{w})-0.25\times\left(\chi^2(\vec{w})-\chi^2(\vec{y})\right)\\
\vec{z}&\equiv&0.5\times\left(\vec{w}+\vec{y}\right)\\
\vec{y}&\in&\{\hat{Y}\}
\end{eqnarray}
before they are allowed to seed a new simplex search.  In this way, our 
extended APS avoids seeding simplex
searches with points that are inside of already known regions of low $\chi^2$.

\subsubsection{Simplex-driven APS search}
\label{sec:simplex_strad}

Having introduced the concept of the Nelder-Mead (1965) simplex search in
Section \ref{sec:simplex} above, we additionally use this tool to improve upon
the original APS search algorith, steps (1A)-(3A).  Broadly speaking, the
purpose of the original APS algorithm is to identify the point which maximizes
the $S$ statistic in equation (\ref{eqn:sstat}) and evaluate
$\vec{\theta}\rightarrow\chi^2$ at that point.  As more points are sampled, $S$
will, presumably, come to be maximized at points that are near the
$\chi^2=\chi^2_\text{lim}$ contour, but far from other sampled points, so APS
will efficiently explore the parameter space.  Steps (1A)-(3A) approximately
achieve this end by proposing a random sample of candidate points and evaluating
$\vec{\theta}\rightarrow\chi^2$ at the point which maximizes $S$ within that
sample.  We use the Nelder-Mead (1965) simplex to replace this process as follows:
\begin{itemize}
\item(1\~A) Perform step (1A) as above.
\\
\item(2\~A) Generate $N_p+1$ randomly chosen points $\{\hat{R}\}$ in parameter space.  
For each point $\vec{r}$ in $\{\hat{R}\}$, use equation (\ref{eqn:sstat}) as in steps 
(2A-3A) above to calculate $S$.  This set of $S$ values will be called
$\{S\}$.
\\
\item(3\~A) Use $\{\hat{R}\}$ and $\{S\}$ as the seed for a
Nelder-Mead (1965) simplex search.  At each subsequent point sampled by this
search, calculate $S$ as above.  Run this simplex until $S$ converges to a local maximum.
\\
\item(4\~A) Evaluate $\vec{\theta}\rightarrow\chi^2$ at the point which
maximized $S$.  Add this point to the list of points sampled by APS.  Return to
step (2\~A).  Repeat until convergence.
\end{itemize}
By choosing the maximum $S$ point based on a simplex search rather than a
finite random sample, we hope to find a true local maximum in $S$, rather than
just the maximum relative to the limited number of sample points chosen.  Steps
(1\~A)-(4\~A) replace steps (1A)-(3A) in our expanded APS algorithm.

\subsection{Code implementing APS}
\label{sec:user}

We make code implementing our extended APS for 
an arbitrary $\vec{\theta}\rightarrow\chi^2$
function available at \verb|https://github.com/uwssg/APS/|.
The code is written in C++.  It is designed for users to use
with any $\vec{\theta}\rightarrow\chi^2$ function.  Users need only
to write a sub-class of the \verb|chisquared| class defined in
\verb|chisq.h| and \verb|chisq.cpp| which evaluates their desired
$\vec{\theta}\rightarrow\chi^2$ and pass an instantiation of this
sub-class to an instantiation of the \verb|aps| class defined in
\verb|aps.h| and \verb|aps.cpp|.  It is this \verb|aps| class which
does the work of performing the searches described above.
The user should not need to modify this class (though she is 
certainly welcome to).

Our \verb|aps| class provides the option for several
user-specified parameters.  We list them below along with the C++ 
methods (belonging to the class \verb|aps|) used to set them.

\begin{itemize}
\item The parameter $G$ from equation (\ref{eqn:threshold}) 
above can be set using the method
\verb|aps.set_grat|
\\
\item $\chi^2_\text{lim}$ can be manually set using the 
method \verb|aps.set_target|
\\
\item $\Delta \chi^2$ can be set in the constructor for the class \verb|aps|.
\\
\item The number $N_g$ of nearest neighbors used in 
steps (2\~A)-(3\~A) can be set in the constructor
for the class \verb|aps|
\\
\item The minimum and maximum bounds in parameter space can be set using the 
method \verb|aps.set_max| and \verb|aps.set_min|.  These bounds will constrain the range of
points sampled in step (2\~A)-(3\~A).  
They will not contrain searches performed with bisection or the 
simplex minimizer.
\\
\item If there is a characteristic length scale associated with a parameter, this can be set
using the method \verb|aps.set_characteristic_length|.  This length scale will be used to
normalize all distances in parameter space.  If it is not set, it will default to the
difference between the maximum and minimum values set by \verb|aps.set_max| and
\verb|aps.set_min|.  Setting a characteristic length scale will principally
affect what points are considered nearest neighbors for the purposes of steps
(2\~A)-(3\~A).
If the characteristic length scale of a parameter is large, that parameter will
be down-weighted when calculating the parameter-space distance between two points.
If the characteristic length scale of a parameter is small, that parameter will
be up-weighted in length calculations.
\end{itemize}

More complete documentation is made available with the source code.

\subsection{Interpreting APS outputs}
\label{sec:bayes}

Originally, APS was conceived as a way to draw Frequentist confidence limits $\chi^2\le\chi^2_\text{lim}$. 
This is by far the most direct way to interpret APS outputs: a list of points is given with
corresponding values of $\chi^2$.  All of the points with $\chi^2\le\chi^2_\text{lim}$ are
taken to cover the desired confidence limit.  There is, however, also a way to use the
outputs from APS to draw an approximate Bayesian credible limit.

As discussed in the introduction, Bayesian parameter constraints assume that there is some
probability distribution (``the posterior'') over $\vec{\theta}$.  
$(1-\alpha)\%$ credible limits are drawn by
integrating over this probability distribution and selecting that region which contains
$(1-\alpha)\%$ of the total probability.  This integral can be very time-consuming if the
dimensionality of $\vec{\theta}$ is large.  Markov Chain Monte Carlo methods overcome this
difficulty by randomly drawing samples from the posterior and assuming that, if enough samples
are drawn, the distribution of those samples will be sufficient to reconstruct the attributes
of the full posterior probability density.

APS outputs cannot be used this way, since APS does not select its points
probabilisticaly, but in an attempt to recreate the behavior of the 
$\vec{\theta}\rightarrow\chi^2$ function in
interesting regions of parameter space.  We can still, however, use these points to approximate
the integral over the posterior probability density.  The algorithm we propose is as follows.
The total set of points sampled by APS will be referred to below as $\{\hat{A}\}$.

For each $\vec{a}$ in $\{\hat{A}\}$:
\begin{itemize}
\item(1D) Initialize two vectors $\vec{a}_\text{max}$ and
$\vec{a}_\text{min}$.  Set all of the values in these vectors to some nonsense number.

\item(2D) Select the $3N_p+1$ nearest neighbor points to $\vec{a}$ from
$\{\hat{A}\}$ (this will obviously include $\vec{a}$ itself, which can be discarded).  Call
this set of nearest neighbors $\{\hat{N}\}$.
\\
\item(3D) For each point $\vec{n}$ in $\{\hat{N}\}$, calculate the vector
$\vec{\delta}$ such that $\delta_i=|n_i-a_i|$ (here the vertical bars denote the absolute value).  
For each component
$\delta_i$ of $\vec{\delta}$, starting with the largest and working towards the smallest:
\\
\begin{itemize}
\item(3Da) If $n_i>a_i$ and $a_{\text{max},i}$ has not been set, set $a_{\text{max},i}=n_i$.
\\
\item(3Db) If $n_i<a_i$ and $a_{\text{min},i}$ has not been set, set $a_{\text{min},i}=n_i$.
\\
\item(3Dc) If both $a_\text{max,i}$ and $a_\text{min,i}$ have already been set, move on to the
next largest $\delta_i$ and repeat at step (3Da).
\\
\end{itemize}
\item(4D) $\vec{a}_\text{max}$ and $\vec{a}_\text{min}$ should now describe an asymmetric,
rectangular hyperbox about $\vec{a}$.  To symmetrize it, step through all $N_p$ dimensions
and reset either $a_{\text{max},i}$ or $a_{\text{min},i}$ so that 
$d_\text{min}=a_i-a_{\text{min},i}$ and
$d_\text{max}=a_{\text{max},i}-a_i$ are equal to the minimum of $\{d_\text{max},d_\text{min}\}$
from the asymmetric hyperbox.  Return to step (1D) and repeat for the next $\vec{a}$.
\end{itemize}

At this point, each $\vec{a}$ in $\{\hat{A}\}$ should have a corresponding symmetric hyperbox
surrounding it described by the corresponding 
$\vec{a}_\text{max}(\vec{a})$ and $\vec{a}_\text{min}(\vec{a})$.  
To integrate the posterior probability density, we will assume that 
all of the parameter space points in each hyperbox correspond to
$\chi^2=\chi^2(\vec{a})$, i.e. the $\chi^2$ discovered by APS for the point at
the center of the hyperbox is the $\chi^2$ value for all of the points inside
the hyperbox.
The posterior probability associated with each box is then
\begin{eqnarray}
P(\vec{a})&=&\frac{V(\vec{a})}{P_\text{tot}}\times\exp\left[-\chi^2(\vec{a})/2\right]
\label{eqn:posterior}\\
P_\text{tot}&\equiv&\sum_{\vec{a}} V(\vec{a})\times\exp\left[-\chi^2(\vec{a})/2\right]
\end{eqnarray}
where $V(\vec{a})$ is the hypervolume of the hyperbox surrounding $\vec{a}$.
To plot the $(1-\alpha)\%$ credible limit, one need only plot the points in hyperspace that
enclose the lowest-$\chi^2$ hyperboxes containing $(1-\alpha)\%$ of the total probability.
We test this algorithm on a function with an exactly known credible limit
in Section \ref{sec:toy} (see Figures \ref{fig:toyBayes2}-\ref{fig:toyBayes4}) 
and on a real, physical dataset in Section \ref{sec:wmap} (see Figure \ref{fig:contours_bayes}).  
In  Section \ref{sec:toy}, we find that our results agree with the known credible limit.
In Section \ref{sec:wmap} we find that APS correctly identifies the region and general
shape of the Bayesian credible limits, however, with significant noise.  The Bayesian
credible limits found by APS are systematically larger than the Bayesian credible
limits found by MCMC.  This motivates our claim in the introduction that APS may be
principally useful for identifying regions of high posterior density on largely unknown
likelihood functions, allowing users to more shrewdly set priors for sampling
algorithms like MCMC.

The functionality described by this section is provided by the classes defined in the source
code file \verb|aps_extractor.cpp| in our software package.

\section{Gaussian processes}
\label{sec:gp}

Steps (2\~A)-(3\~A) of \APS requires that a prediction $\mu$ be made regarding the value
of $\chi^2$ at points as yet unsampled by the algorithm.  An uncertainty $\sigma$
must also be assigned to this prediction.  APS writ large is agnostic regarding
the method used to find $\mu$ and $\sigma$.  We choose to use a Gaussian process.

Gaussian processes take noisy, sparse, measurements of an unknown function (in our
case, $\vec{\theta}\rightarrow\chi^2$ at
the points in parameter space already sampled by APS) and use them to
make inferences about unmeasured values of the function by assuming that the
function is a random process.  Gaussian processes have already found use in physics
and astronomy constructing non-parametric models of the cosmic expansion
\cite{ericgp}, interpolating the point spread function across 
astronomical images
\cite{psf}, and interpolating models of the non-linear matter
power spectrum from N-body simulations \cite{habib}.
They are useful in the current context because, not only do they
give robust predictions $\mu$ for the unmeasured values of the
function being modeled, but they also
provide well-motivated prediction uncertainties $\sigma$, a requirement of
equation (\ref{eqn:sstat}).  We present below an introduction to the
formalism of Gaussian processes drawn heavily
from Chapter 2 and Appendix A 
of \cite{gp}.

Gaussian processes use the sampled data $\{\vec{\theta}^{(i)},\chi^{2,(i)}\}$, 
where $i$ indexes over all of the sampled points, to predict
$\mu=\chi^{2,(q)}$ at the unknown point 
$\vec{\theta}^{(q)}$ by assuming that the function
$\vec{\theta}\rightarrow\chi^2$ represents a sample drawn from a random process 
distributed
across parameter space.  At each point in parameter space, 
$\chi^2$ is assumed to be
distributed according to a normal distribution with mean 
$\bar{\chi}^2$
and variance dictated by the ``covariogram''
$C_{ij}(\vec{\theta}^{(i)},\vec{\theta}^{(j)})$.
If we have $N_g$ measurements of $\chi^2$ 
(recall from steps (2\~A-3\~A) above that, for
each point where we must predict the value of $\chi^2$, we are only using the $N_g$
nearest neighbor sampling points as data),
 then we assume
they are distributed according to
\begin{equation}
\label{eqn:likelihood}
P(\vec{\chi}^2)=
\frac{\exp\bigg[-\frac{1}{2}K^{-1}\sum_{i,j}^{N_g}
(\chi^{2,(i)}-\bar{\chi}^2)C^{-1}_{ij}
(\chi^{2,(j)}-\bar{\chi}^2)\bigg]}{\sqrt{2\pi}^{N_g}\text{det}|KC|^{N_g/2}}
\end{equation}
$C_{ij}$ is a function assumed by the user encoding how variations in
$\chi^2$ at one point in parameter space affect variations in 
$\chi^2$ at other
points in parameter space.  $K$ is a parameter controlling the
normalization of $C_{ij}$.
The diagonal elements $C_{ii}=1+\sigma^2_{ii}$ where 
$\sigma^2_{ii}$ is the intrinsic variance in the value
of $\chi^2$ at a single point $\vec{\theta}^{(i)}$.  
Rasmussen and Williams (2006) treat 
the special case
$\bar{\chi}^2=0$ and find (see their equations 2.19, 2.25 and 2.26)
\begin{eqnarray}
\mu&=&\sum_{i,j}^{N_g} C_{qi}C^{-1}_{ij}\chi^{2,(j)}\nonumber\\
\sigma^2&=&K\times\left(C_{qq}-\sum_{i,j}^{N_g}C_{qi}C^{-1}_{ij}C_{jq}\right)\label{sig}
\end{eqnarray}
where the sums over $i$ and $j$ are sums over the sampled points 
in parameter space.  $C_{iq}$ relates the sampled point $i$ to the candidate
point $q$.
We do not wish to assume that the mean value of $\chi^2$ is zero everywhere.
Therefore, we modify the equation for $\mu$ to give
\begin{equation}
\label{mu}
\mu=\bar{\chi}^2+\sum_{i,j}^{N_g} C_{qi}C^{-1}_{ij}
(\chi^{2,(j)}-\bar{\chi}^2)
\end{equation}
where $\bar{\chi}^2$ is the algebraic mean of the 
sampled $\chi^{2,(i)}$.
Note the similarity to a multi-dimensional Taylor series expansion with the
covariance matrix playing the role of the derivatives.
Equation (\ref{sig}) differs from equation (6) in \cite{bryan} because
they used the semivariance
$$\gamma_{ij}=\text{var}[\chi^2(\vec{\theta}^{(i)})
-\chi^2(\vec{\theta}^{(j)})]$$
in place of the covariance $C_{ij}$.  In practice, the two assumptions result in
equivalently valid $\mu$ and $\sigma$.

The form of the covariogram
$C_{ij}(\vec{\theta}^{(i)},\vec{\theta}^{(j)})$ must be
assumed.  One possibility, 
taken from equation 4.9 of Rasmussen and Williams (2006) is
\begin{equation}
\label{eqn:GaussianCovar}
C_{ij}=\exp\left[-\frac{1}{2}D^2_{ij}\right]
\end{equation}
where $D_{ij}$ is the normalized distance in parameter space between the points 
$\vec{\theta}^{(i)}$
and $\vec{\theta}^{(j)}$.
\begin{equation}
D^2_{ij}\equiv\sum_n^{N_p}
\left(\frac{\theta^{(i)}_n-\theta^{(j)}_n}{\ell_n\times(\text{max}_n-\text{min}_n)}\right)^2
\end{equation}
where $\text{max}_n-\text{min}_n$ is the difference between the maximum and minimum values of
the $n$th parameter taken from the $N_g$ points used to seed the Gaussian process and $\ell_n$
is a ``hyper-parameter'' associated with the $n$th parameter.  We discuss the normalization
paramter $K$ below in section \ref{sec:kriging}.  We discuss the setting of
hyper-parameters in section \ref{sec:hyperparams}.
The exponential form of $C_{ij}$ quantifies the assumption 
that distant points should
not be very correlated.

Another possibility (equation 4.29 of Rasmussen and Williams) is
\begin{eqnarray}
C_{ij}=\phantom{\sin
\bigg[\frac{2(\sigma_0+\tilde{\vec{\theta}}^{(i)}\cdot\tilde{\vec{\theta}}^{(j)})}
{\sqrt{(1+2(\sigma_1+\tilde{\vec{\theta}}^{(i)}\cdot\tilde{\vec{\theta}}^{(i)}))
(1+2(\sigma_1+\tilde{\vec{\theta}}^{(j)}\cdot\tilde{\vec{\theta}}^{(j)}))}}\bigg]}
\label{eqn:NNCovar}\\
\frac{2}{\pi}\sin^{-1}
\bigg[\frac{2(\sigma_0+\tilde{\vec{\theta}}^{(i)}\cdot\tilde{\vec{\theta}}^{(j)})}
{\sqrt{(1+2(\sigma_1+\tilde{\vec{\theta}}^{(i)}\cdot\tilde{\vec{\theta}}^{(i)}))
(1+2(\sigma_1+\tilde{\vec{\theta}}^{(j)}\cdot\tilde{\vec{\theta}}^{(j)}))}}\bigg]
\nonumber
\end{eqnarray}
corresponding to the covariance function of a neural network with a single
hidden layer.  In this function $\tilde{\vec{\theta}}$ is the vector of
parameters $\vec{\theta}$ recentered and renormalized relative to the span
$(\text{max}_n-\text{min}_n)$ in each dimension.  $\sigma_0$ and $\sigma_1$
are hyper-parameters.  
In Section \ref{sec:wmap}, we test APS using both covarigorams (\ref{eqn:GaussianCovar})
and (\ref{eqn:NNCovar}) and find no appreciable difference
in the resulting parameter constraints.

\subsection{Normalizing the covariogram}
\label{sec:kriging}

The normalization constant $K$ in equation (\ref{eqn:likelihood}) -- known as the
``Kriging parameter'' for the geophysicist who pioneered the overall method -- 
also
must be assumed.  Determining the value of $K$ 
is somewhat problematic because, examining equation
(\ref{mu}), one sees that the factors of $K$ and $K^{-1}$
cancel out of
the prediction $\mu$, so that the assumed value of $K$ has no effect on the accuracy
of the prediction.  
If the opposite had been true, one could heuristically set $K$ to
maximize the accuracy of $\mu$.  
Instead, we set $K$ to the value that maximizes the likelihood
of the input data, i.e. the $N_g$ nearest neighbors in parameter space
used to perform the Gaussian process inference.

Recall from equation (\ref{eqn:likelihood}) that we are treating our
input data as though they were samples drawn from a probability distribution.
To maximize the probability of the observed data (i.e. maximizing the value
of $P(\vec{\chi}^2)$ in equation \ref{eqn:likelihood}) we must set $K$ equal to
\begin{equation}
\label{eqn:kriging}
K=\frac{\sum_{i,j}^{N_g}(\chi^{2,(i)}-\bar{\chi}^2)
C^{-1}_{ij}(\chi^{2,(j)}-\bar{\chi}^2)}
{N_g}
\end{equation}
We adopt this assumption throughout this work.

\begin{figure}
\includegraphics[scale=0.3]{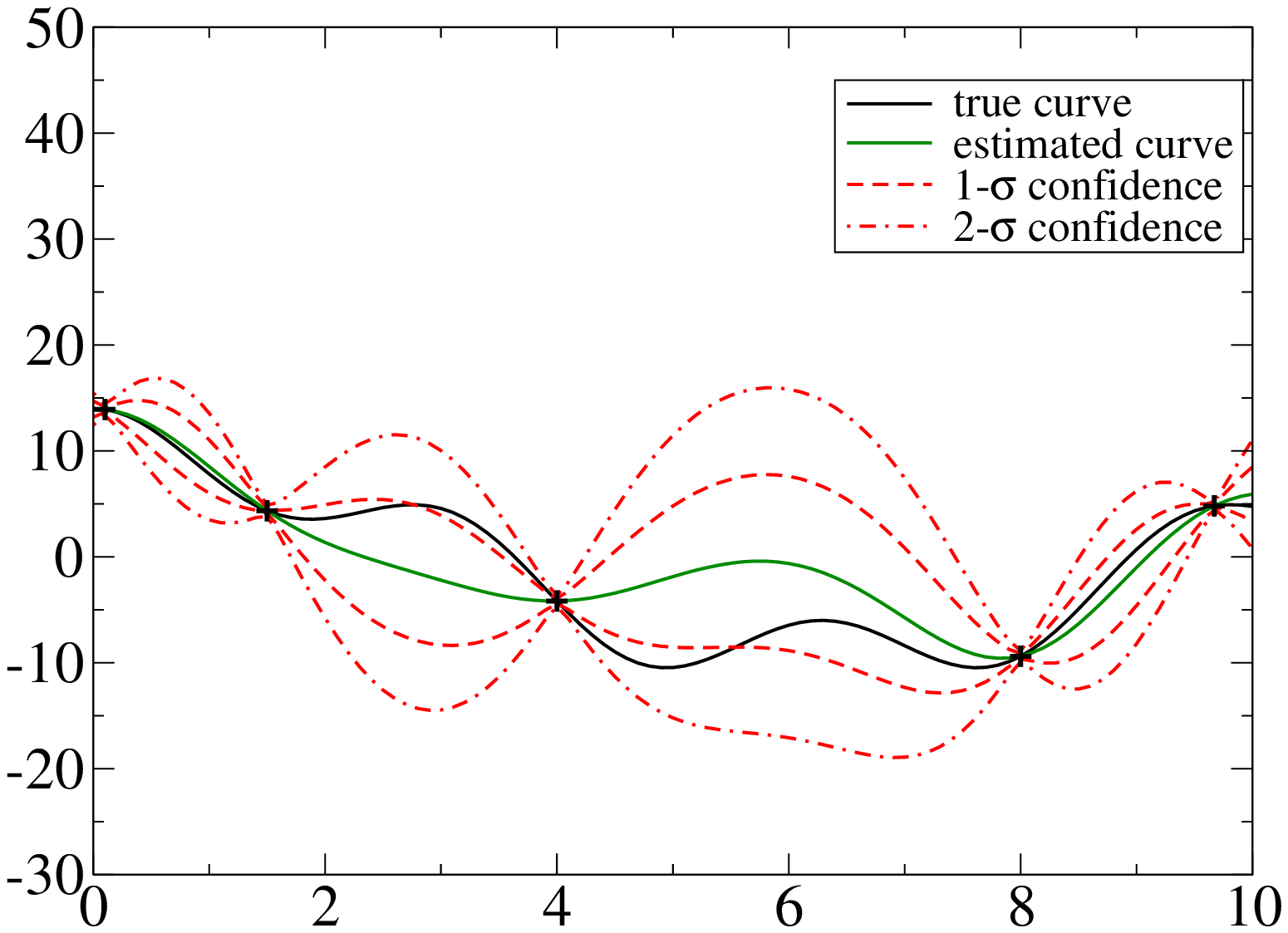}
\caption{
A one-dimensional example of prediction using Gaussian processes.  
The black curve is the function
being considered.  The crosses are the points at which it has been sampled.  The
green curve is the resulting prediction and the red curves represent the 
1- and
2-$\sigma$ uncertainty bounds.  We set $K$ according to 
equation (\ref{eqn:kriging}) wth $N_g=5$ for the 5 sampled points.
}
\label{fig:gp}
\end{figure}

Figure \ref{fig:gp} applies the Gaussian process of equations (\ref{sig}), 
(\ref{mu}), and (\ref{eqn:kriging}) with covariogram (\ref{eqn:GaussianCovar})
to a toy one-dimensional function.  Inspection shows many desirable
behaviors in $\mu$ and $\sigma$.  
As $\theta^{(q)}$ approaches the sampled points $\theta^{(i)}$, $\mu$
approaches $\chi^{2,(i)}$ and $\sigma$ approaches zero.  
Closer to a sampled point,
the Gaussian process knows more about the true behavior of the function. 
Far from the $\theta^{(i)}$, $\sigma$ is larger, and the $S$ statistic in
equation (\ref{eqn:sstat}) will induce the \APS
algorithm to examine the true value of $\chi^2$.

\subsection{Setting hyper-parameters}
\label{sec:hyperparams}

The covariograms $C_{ij}$ presented above in equations (\ref{eqn:GaussianCovar})
and (\ref{eqn:NNCovar}) (indeed, any covariogram)
each contain arbitrary hyper-parameters whose values can be set to fine-tune
the Gaussian process model ($\ell_n$ in equation \ref{eqn:GaussianCovar}
and $\{\sigma_0,\sigma_1\}$ in equation \ref{eqn:NNCovar}).  
In order to set the hyper-parameters used in the APS Gaussian process, we randomly
select a set $\{\hat{V}\}$ of up to 3000 points from those already sampled by APS 
and use these to construct
the error function $E(\vec{h})$ where $\vec{h}$ is a vector specifying the hyper-parameters
required by the chosen $C_{ij}$.  $E$ is defined as
\begin{equation}
\label{eqn:hyperparamError}
E(\vec{h})=\sum_{\vec{v}\in\{\hat{V}\}}\left(\mu(\vec{v},\vec{h})-\chi^2(\vec{v})\right)^2
\end{equation}
where $\mu(\vec{v},\vec{h})$ is calculated according to equation \ref{mu} using
hyper-parameters $\vec{h}$ (and ignoring $\vec{v}$ as its own nearest neighbor) and
$\chi^2(\vec{v})$ is already known, since $\vec{v}$ has been chosen from the points previously
sampled by APS.  If $\vec{h}$ has more than two dimensions, we minimize $E$ on $\vec{h}$-space
using the simplex
minimizer of Nelder and Mead (1965).  Otherwise, we do a simple grid search for the value of
$\vec{h}$ which minimizes $E$.  This optimal value of $\vec{h}$ is used to set
the hyper-parameters for our Gaussian process model.

As APS samples more points in parameter space, we expect it to learn more about the
$\vec{\theta}\rightarrow\chi^2$ function and thus be capable of a more accurate Gaussian
process, so we allow it to periodically reset the hyper-parameters $\vec{h}$.  However,
minimizing $E$ can be a very time-consuming process, so we have APS keep track of the amount of
time elapsed since the algorithm began.  
We divide this value by the number of calls made to the 
$\vec{\theta} \rightarrow \chi^2$ function to give the average total time spent per 
call to $\vec{\theta} \rightarrow \chi^2$ with all of the overhead imposed by APS 
included.  We compare this average with the average amount of time spent on 
just $\vec{\theta} \rightarrow \chi^2$, without any of the overhead imposed by APS.
APS is only allowed to
reset its hyper-parameters if the difference between these two times is less than 0.1 
second per call to $\vec{\theta}\rightarrow\chi^2$
(of course, we do require that APS optimize its hyper parameters at least once, so that
we can have confidence in our Gaussian process model).

\section{A Toy Example}
\label{sec:toy}

We will now test the ability of APS to learn the confidence limits of an
artificial $\chi^2$ function and compare it to that of MCMC driven by the
traditional Metropolis-Hastings algorithm \cite{mcmc,cosmomc}.

The algorithm (1A)-(3A) in Section \ref{sec:algorithm} 
was originally presented and tested against the 1-year data
release from the WMAP satellite in Bryan {\it et al}. (2007).  They found that
the algorithm detected a second locus of low $\chi^2$
that had gone undetected by previous, MCMC-based analyses of the data.  
This second fit to the data has disappeared as the signal-to-noise
ratio of the data has improved (as we will see in Section \ref{sec:wmap}),
therefore, we will use for our test a toy $\chi^2$ function
with multiple locii of low $\chi^2$.  

We construct our toy $\chi^2$ as a series of low-$\chi^2$ ellipses in a 5-dimensional
parameter space (we choose a 5-dimensional parameter space because it allows us to
reasonably show all of the 2-dimensional sub-spaces in the plots which follow).
We consider functions with 2, 3, and 4 discrete minima in $\chi^2$.
The location of the low-$\chi^2$ regions are generated using a pseudo random number
generator taken from Marsaglia (2003) with parameters taken from Press {\it et al.} (2007).
$\chi^2(\vec{\theta})$ for this toy function is calculated as 
\begin{equation}
\label{eqn:toychi}
\chi^2(\vec{\theta})=
\sum_i^{N_p=5}\left(\frac{\theta_i-c_{\text{nearest},i}}{w_{\text{nearest},i}}\right)^2
\end{equation}
where $c_{\text{nearest},i}$ is the $i$th coordinate of the center of the nearest low-$\chi^2$
region and $w_{\text{nearest},i}$ is the (pseudo-randomly generated) width of that ellipse in
the $i$th coordinate.  We refrain from introducing any interesting parameter degeneracies so
that the $\chi^2(\vec{\theta})$ function remains integrable in 2-dimensional sub-spaces.
This gives us a hard-truth ``control'' against which to evaluate our test APS runs below.
Readers wishing to re-create this $\chi^2$ function can find it in our software
package as the class \verb|ellipses_integrable| in the source code files \verb|chisq.h| and
\verb|chisq.cpp|.

\begin{figure*}
\subfigure[]{
\includegraphics[scale=0.15]{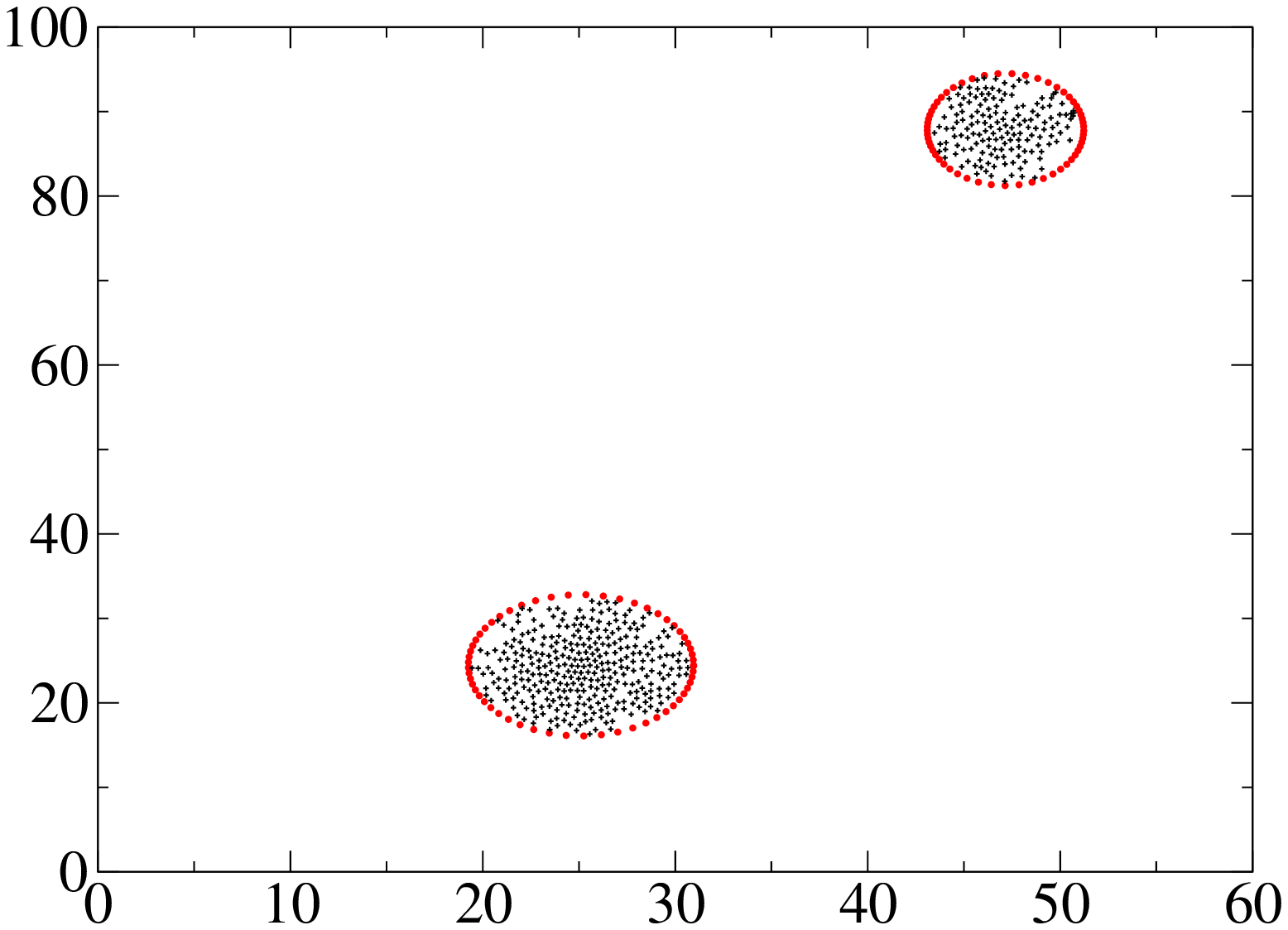}
}
\subfigure[]{
\includegraphics[scale=0.15]{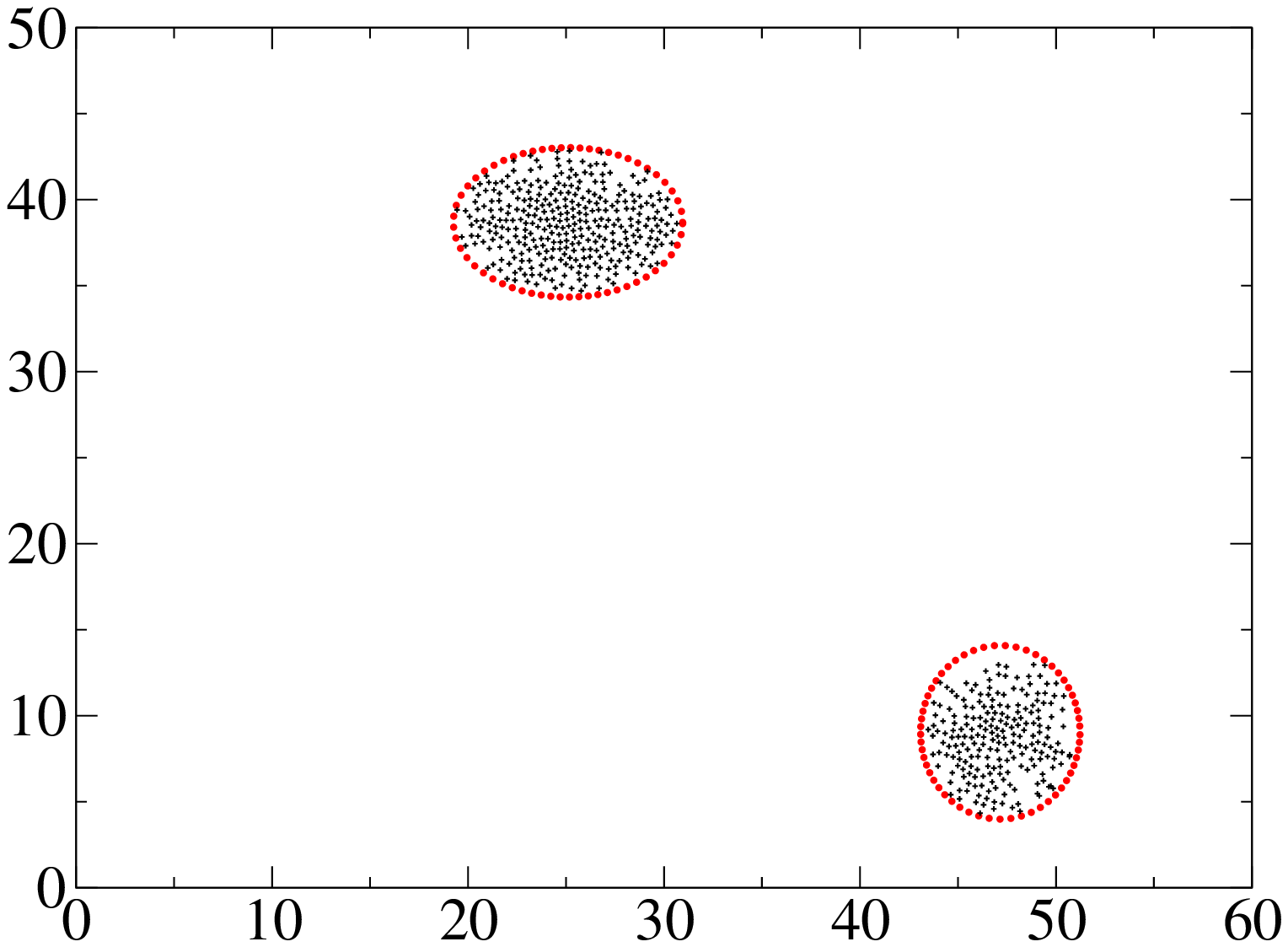}
}
\subfigure[]{
\includegraphics[scale=0.15]{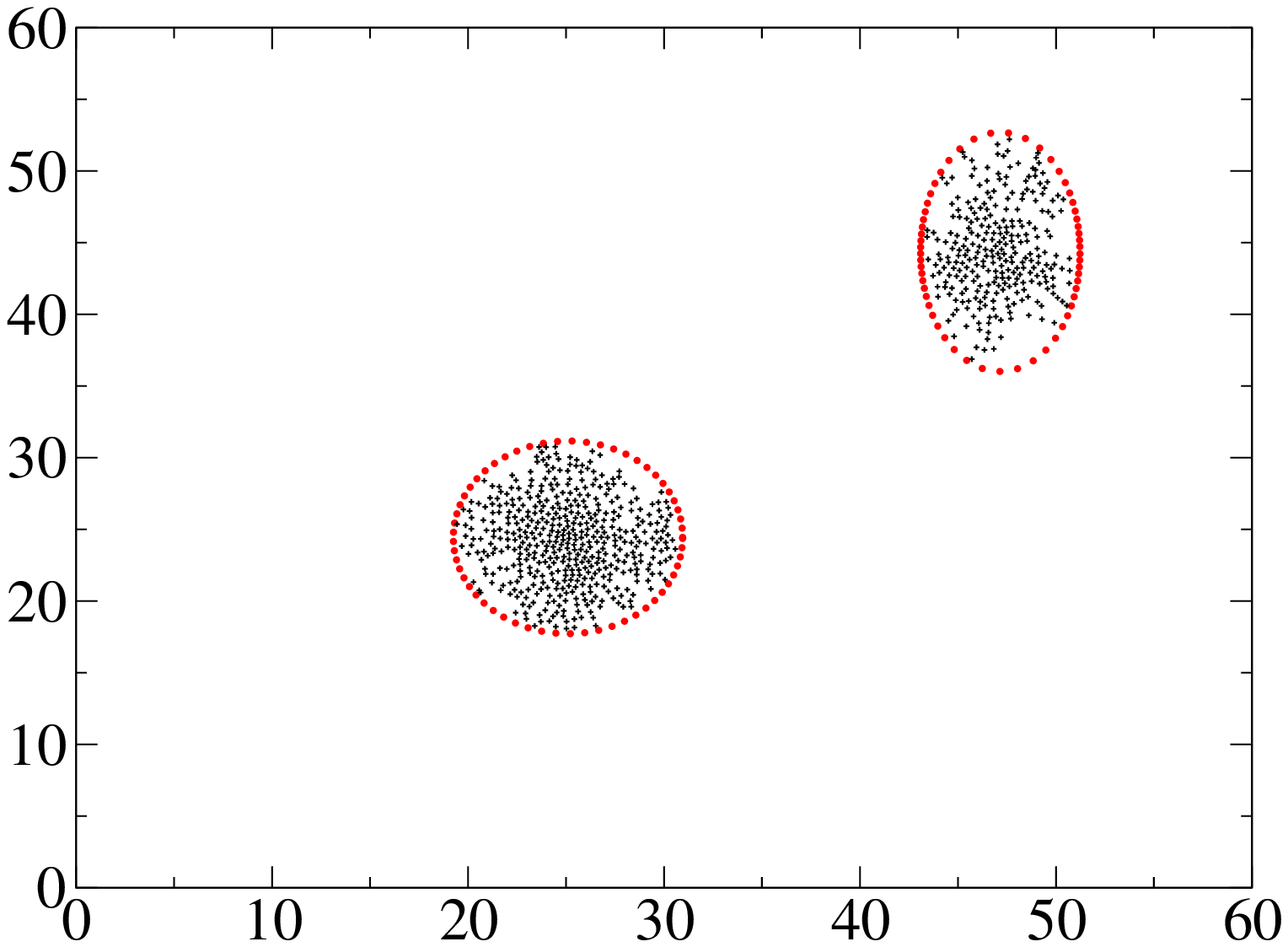}
}
\subfigure[]{
\includegraphics[scale=0.15]{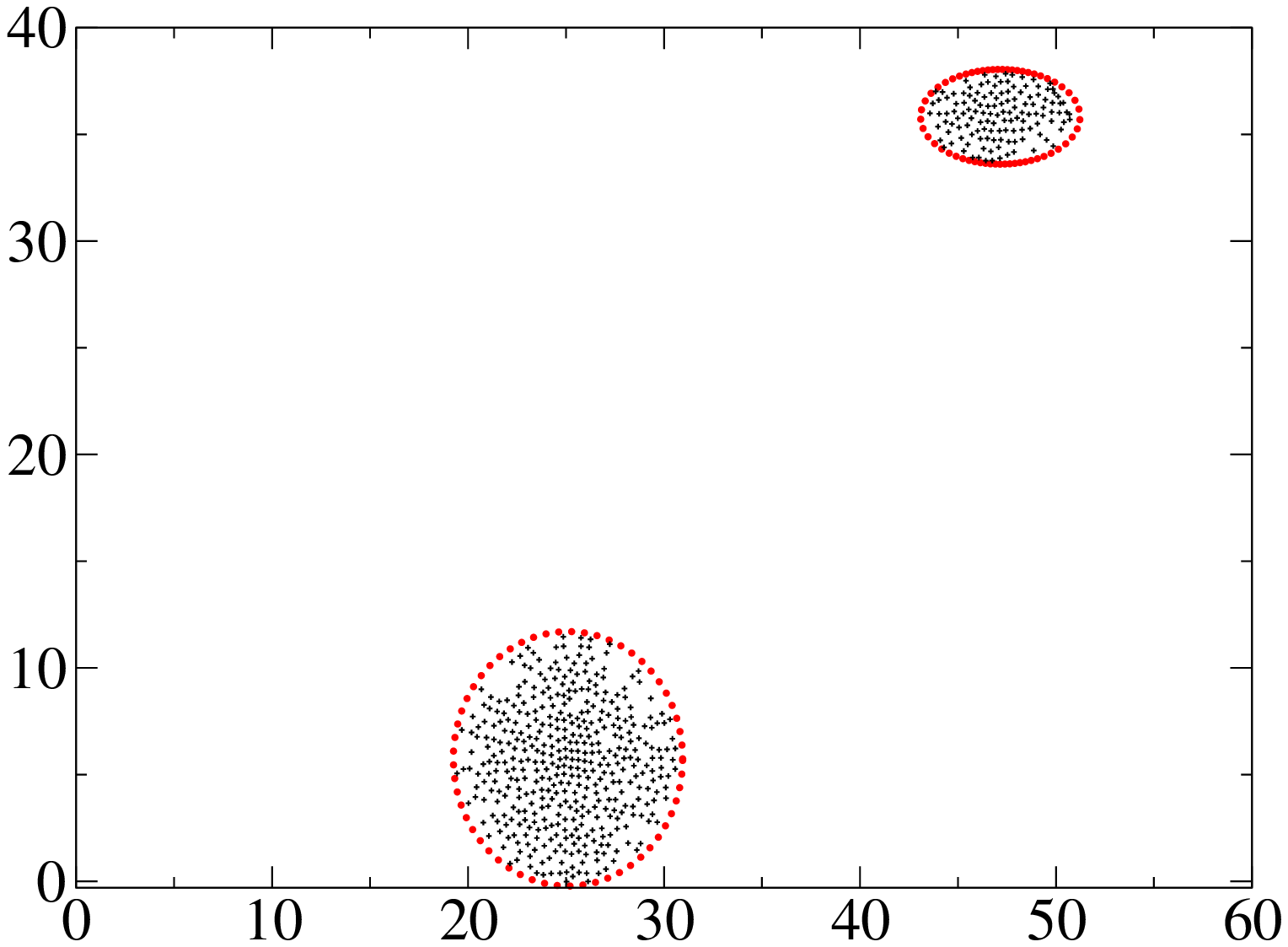}
}
\subfigure[]{
\includegraphics[scale=0.15]{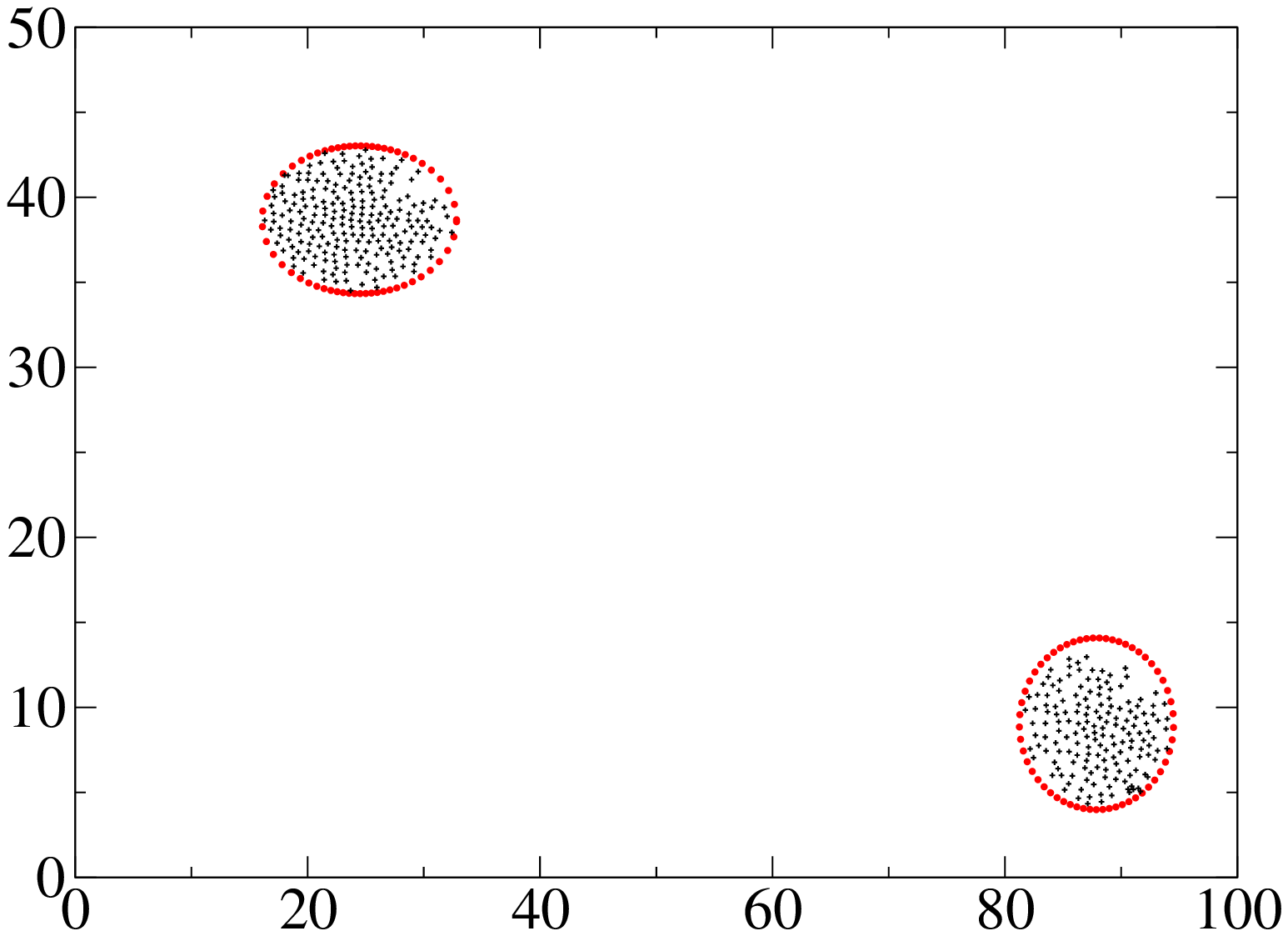}
}
\subfigure[]{
\includegraphics[scale=0.15]{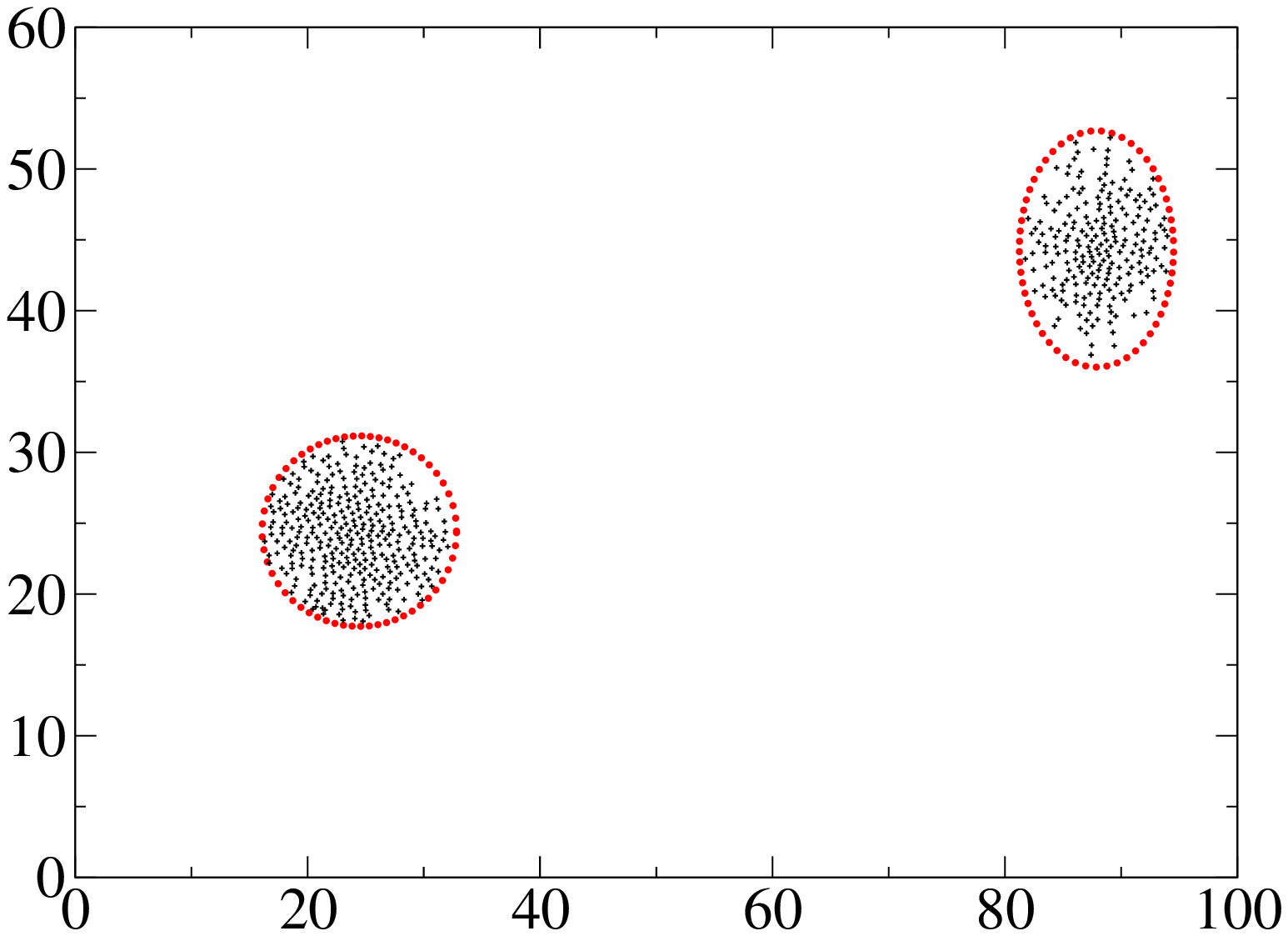}
}
\subfigure[]{
\includegraphics[scale=0.15]{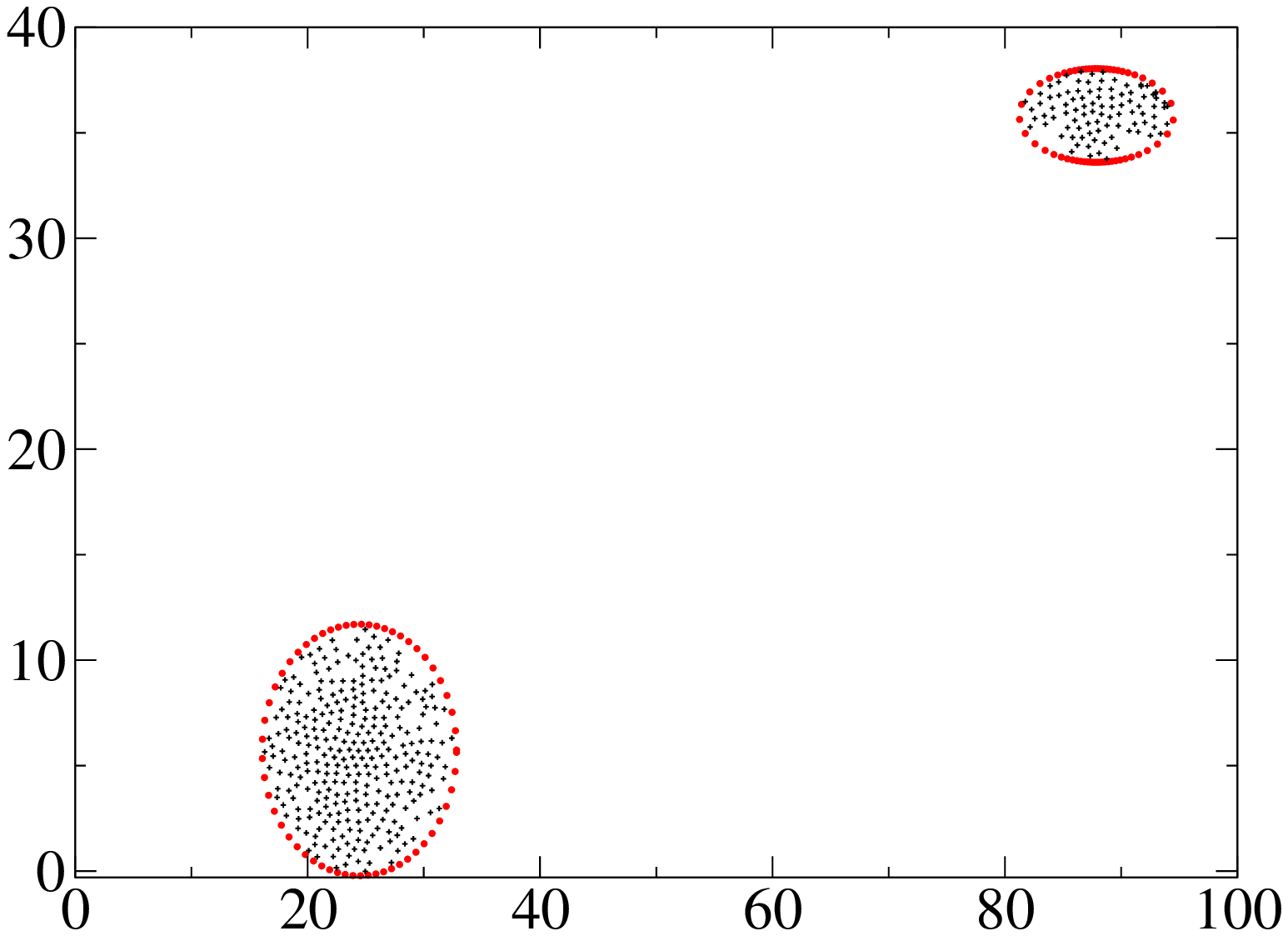}
}
\subfigure[]{
\includegraphics[scale=0.15]{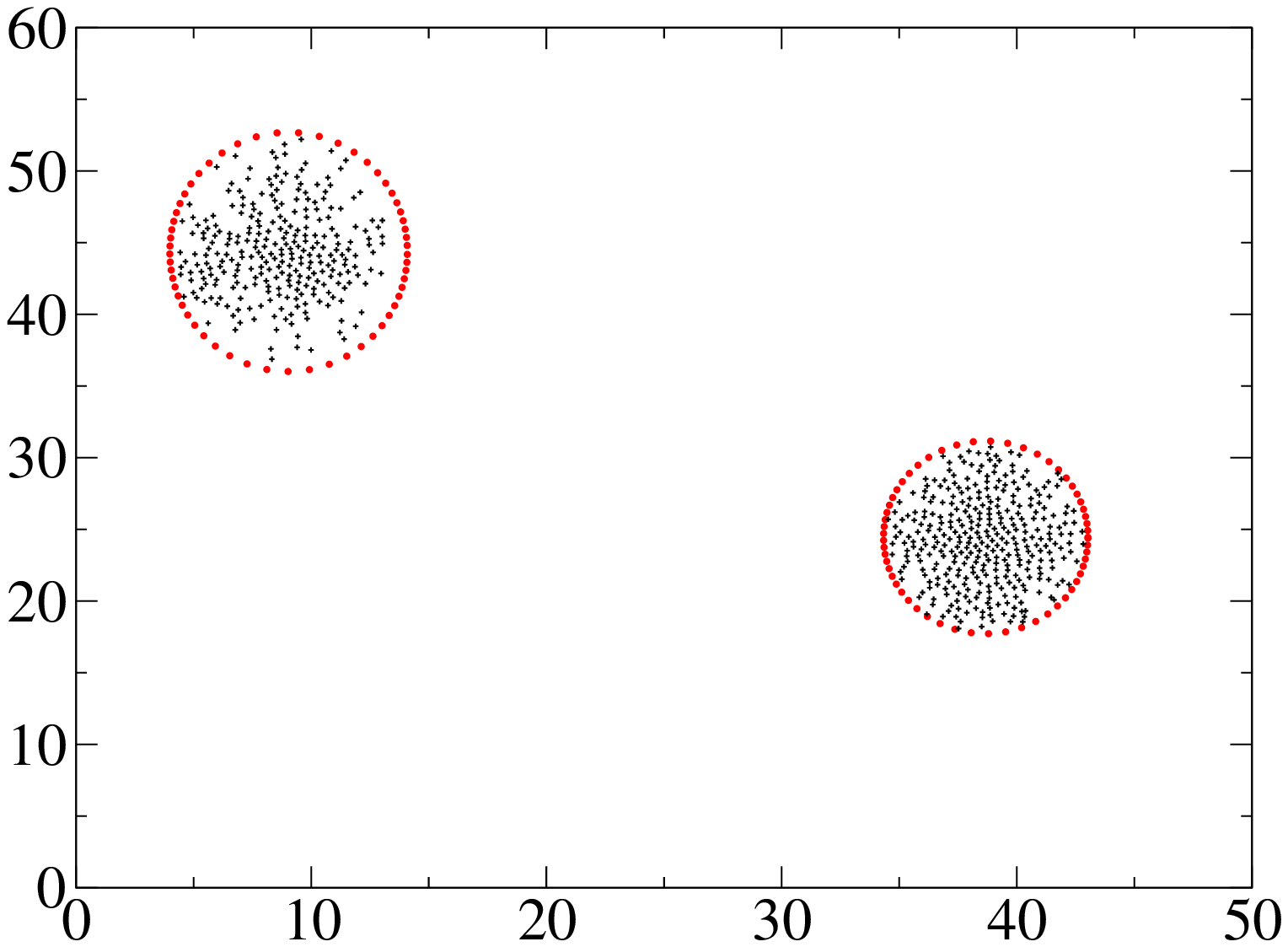}
}
\subfigure[]{
\includegraphics[scale=0.15]{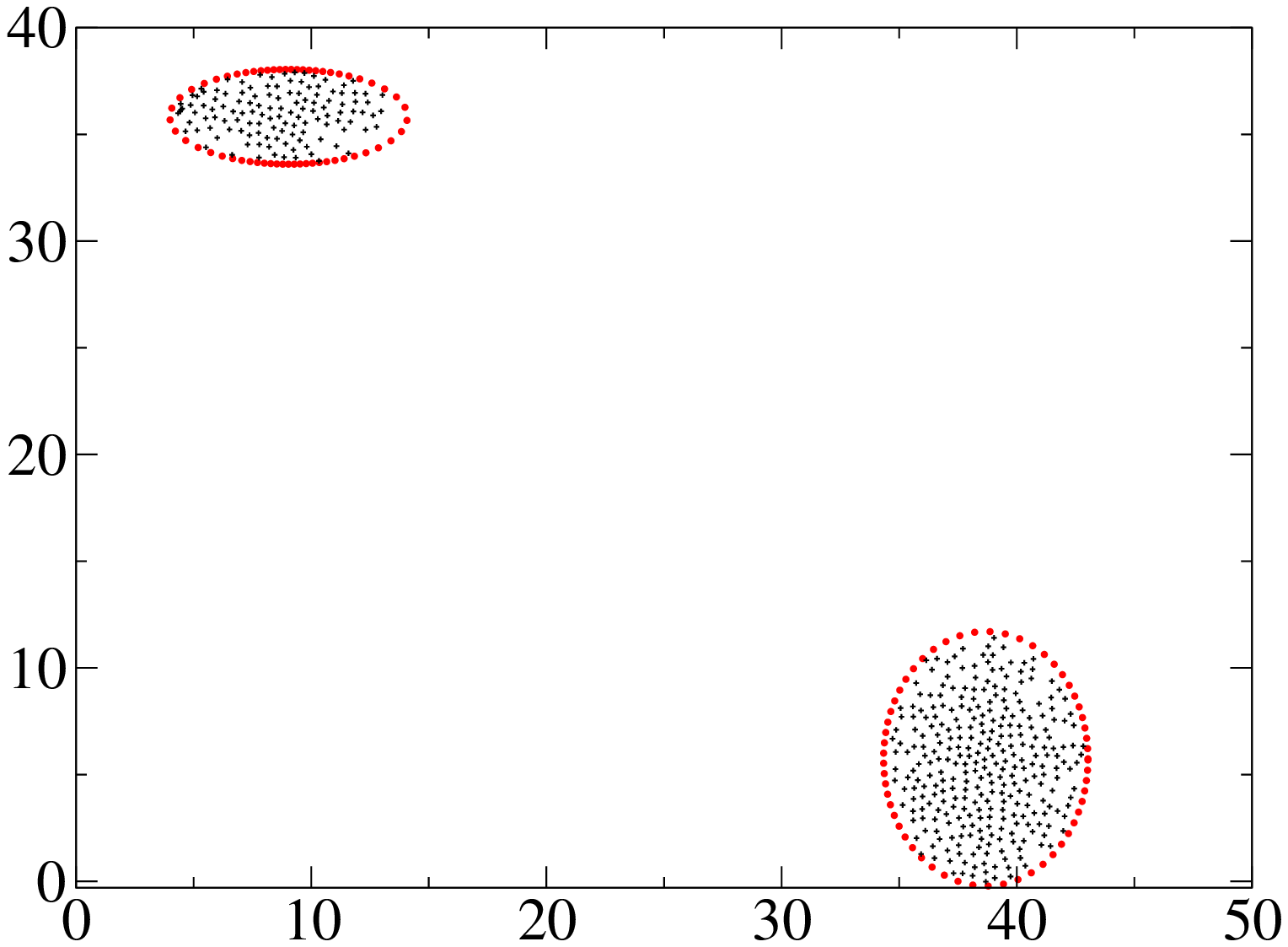}
}
\subfigure[]{
\includegraphics[scale=0.15]{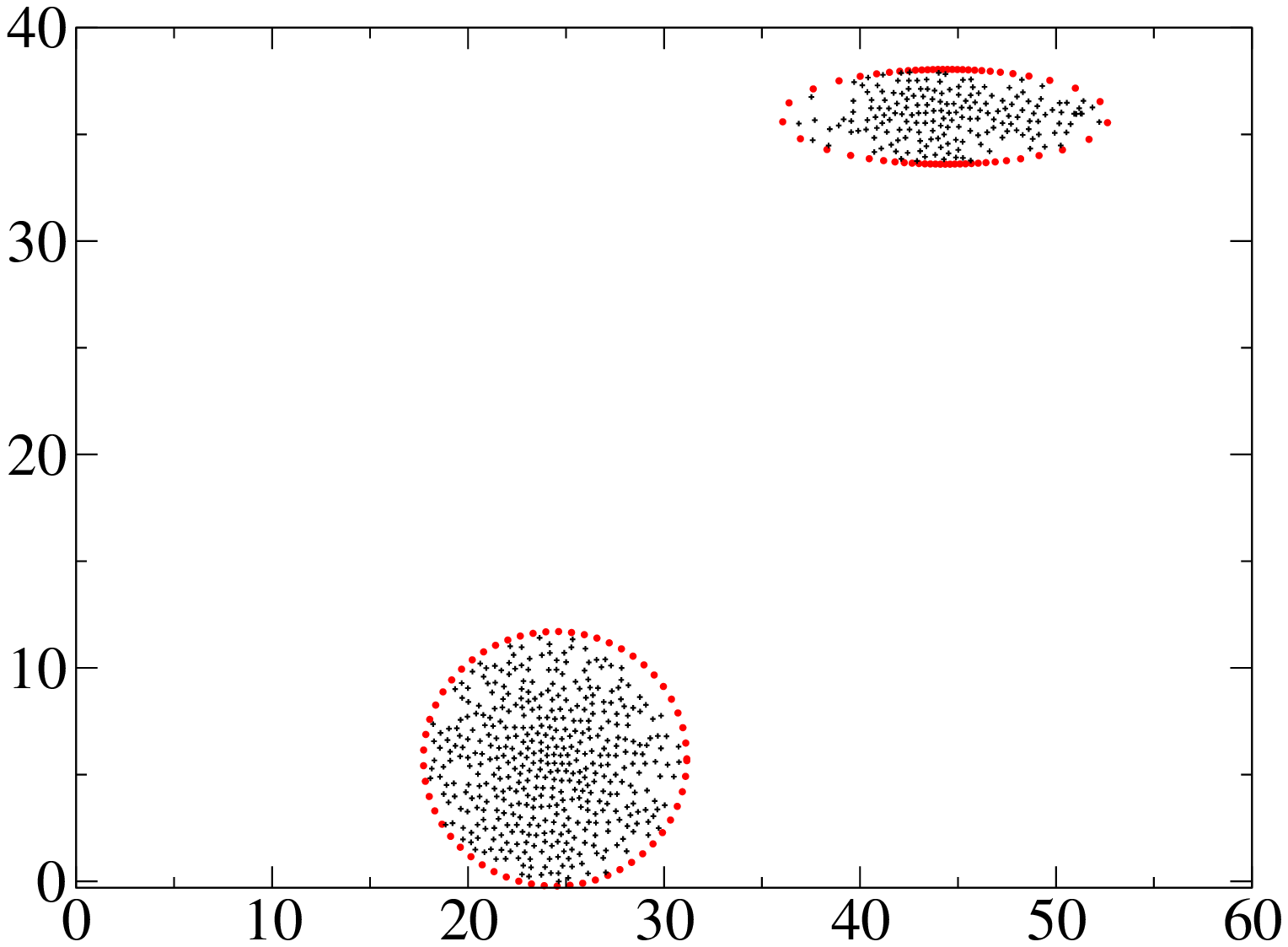}
}
\caption{
This plot shows all 10 2-dimensional sub-spaces of our 5-dimensional toy likelihood
function with 2 minima in $\chi^2$.  The black points are $\chi^2\le\chi^2_\text{lim}$
found by APS after 10,000 samplings.  The red contours are the known $\chi^2=\chi^2_\text{lim}$
contours of the function.
}
\label{fig:toyFreq2}
\end{figure*}

\begin{figure*}
\subfigure[]{
\includegraphics[scale=0.15]{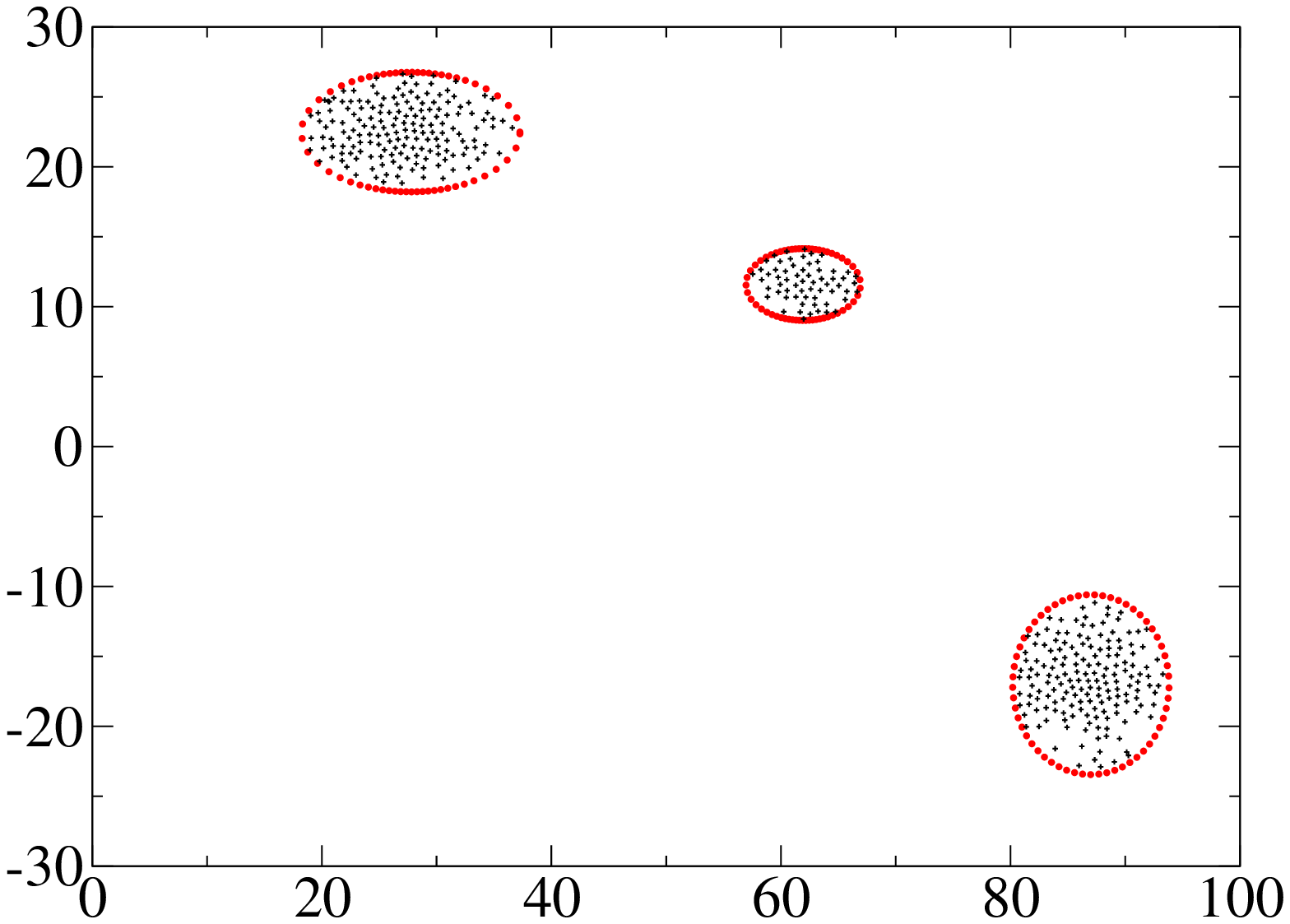}
}
\subfigure[]{
\includegraphics[scale=0.15]{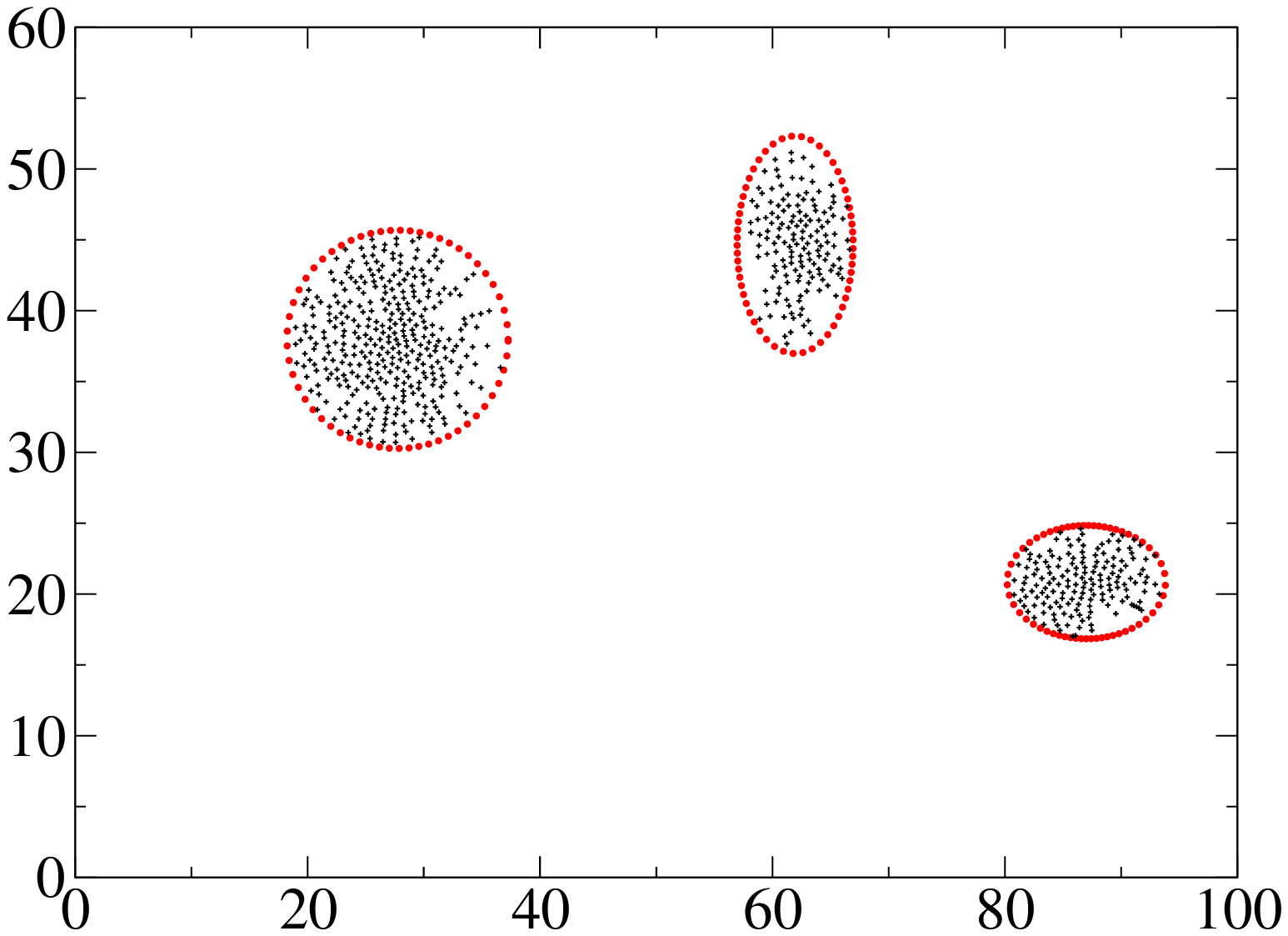}
}
\subfigure[]{
\includegraphics[scale=0.15]{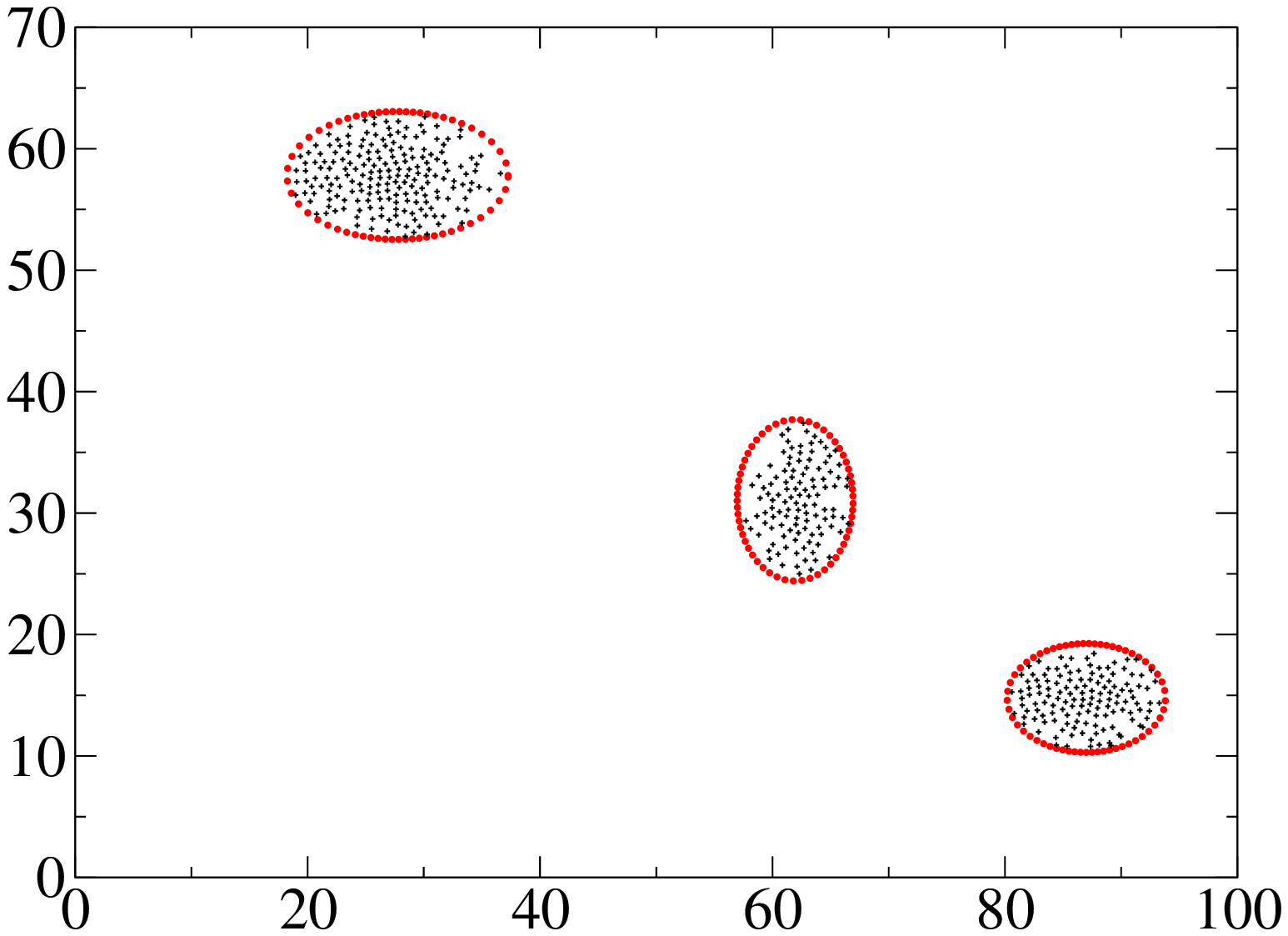}
}
\subfigure[]{
\includegraphics[scale=0.15]{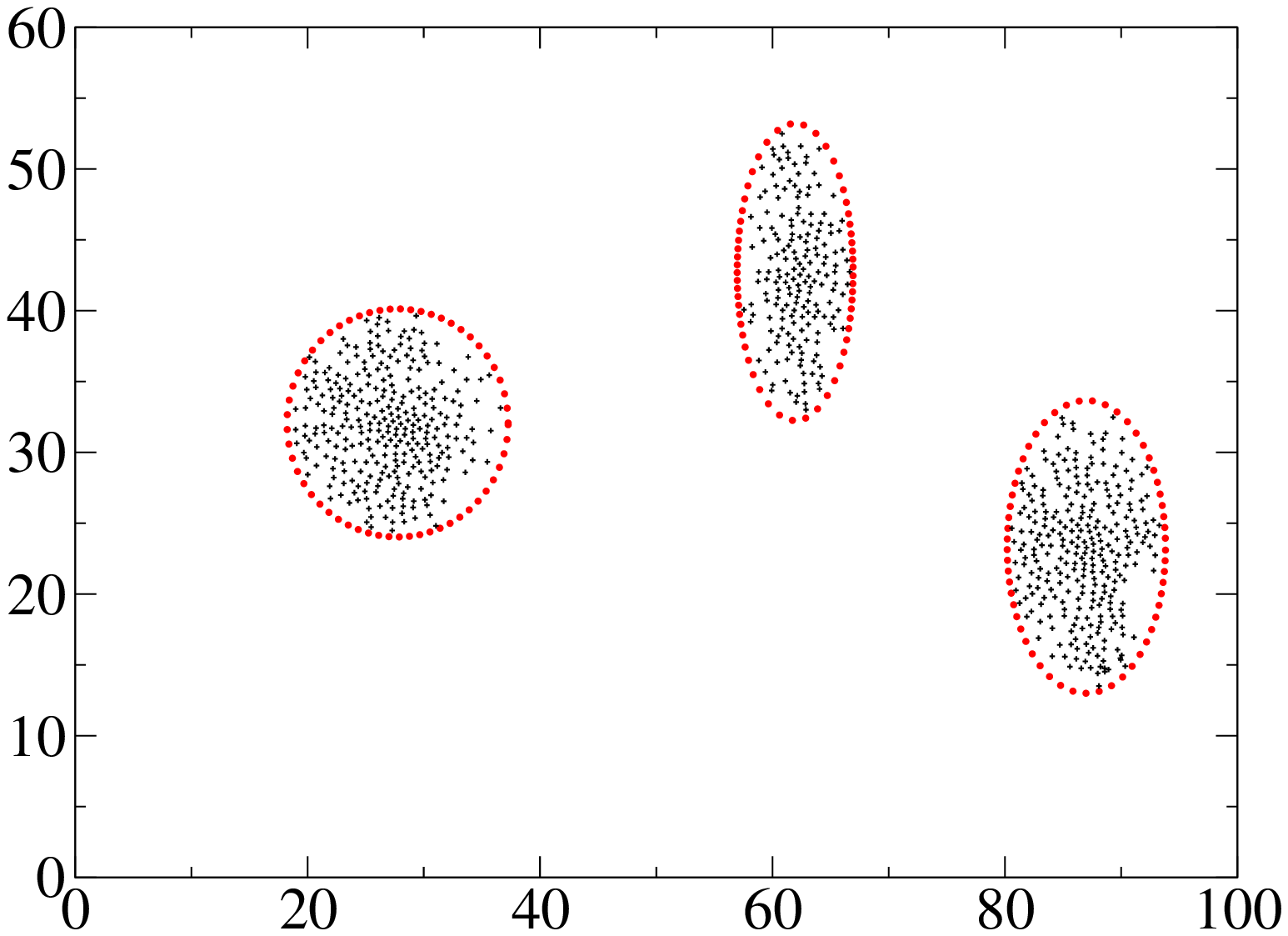}
}
\subfigure[]{
\includegraphics[scale=0.15]{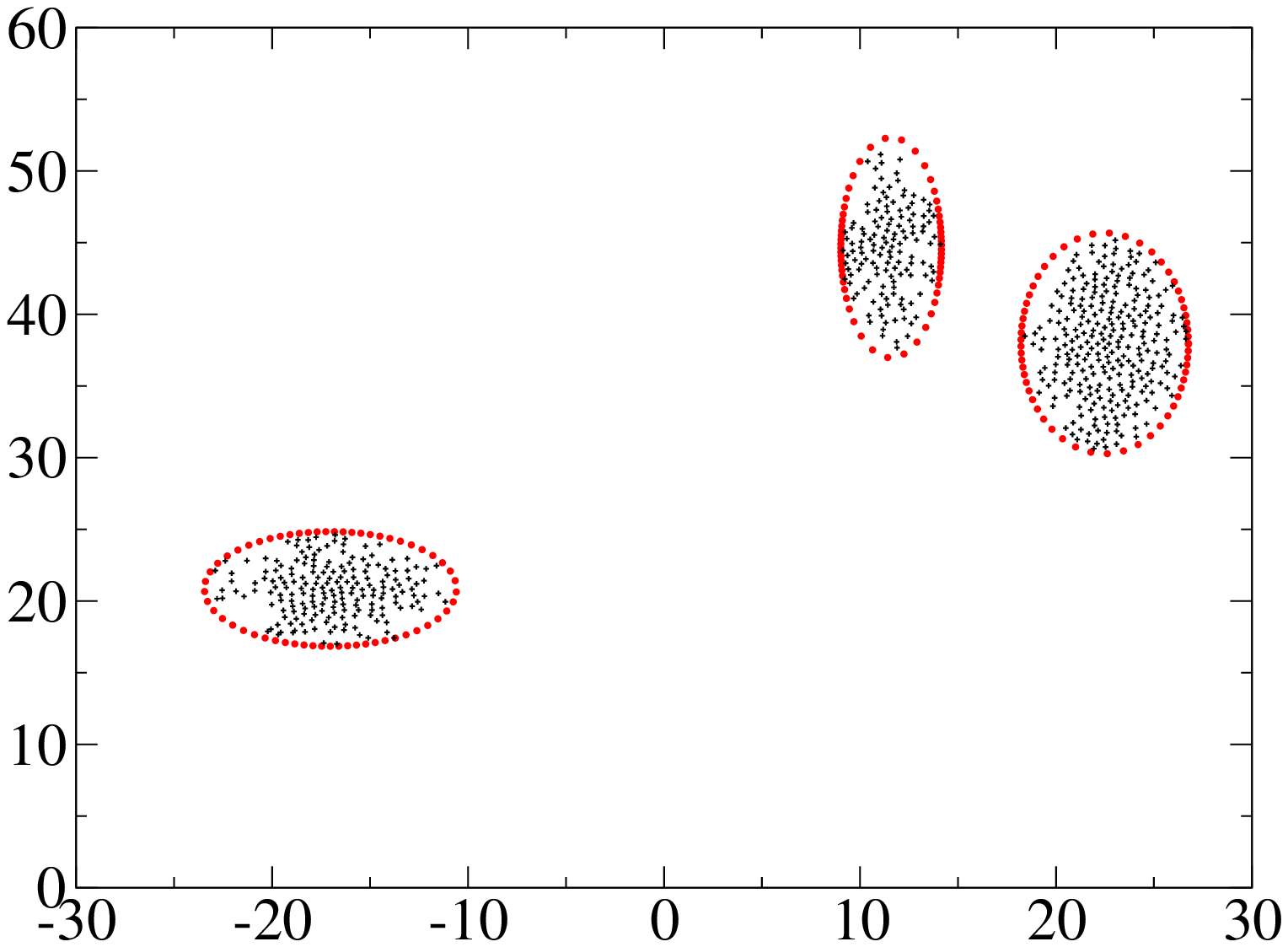}
}
\subfigure[]{
\includegraphics[scale=0.15]{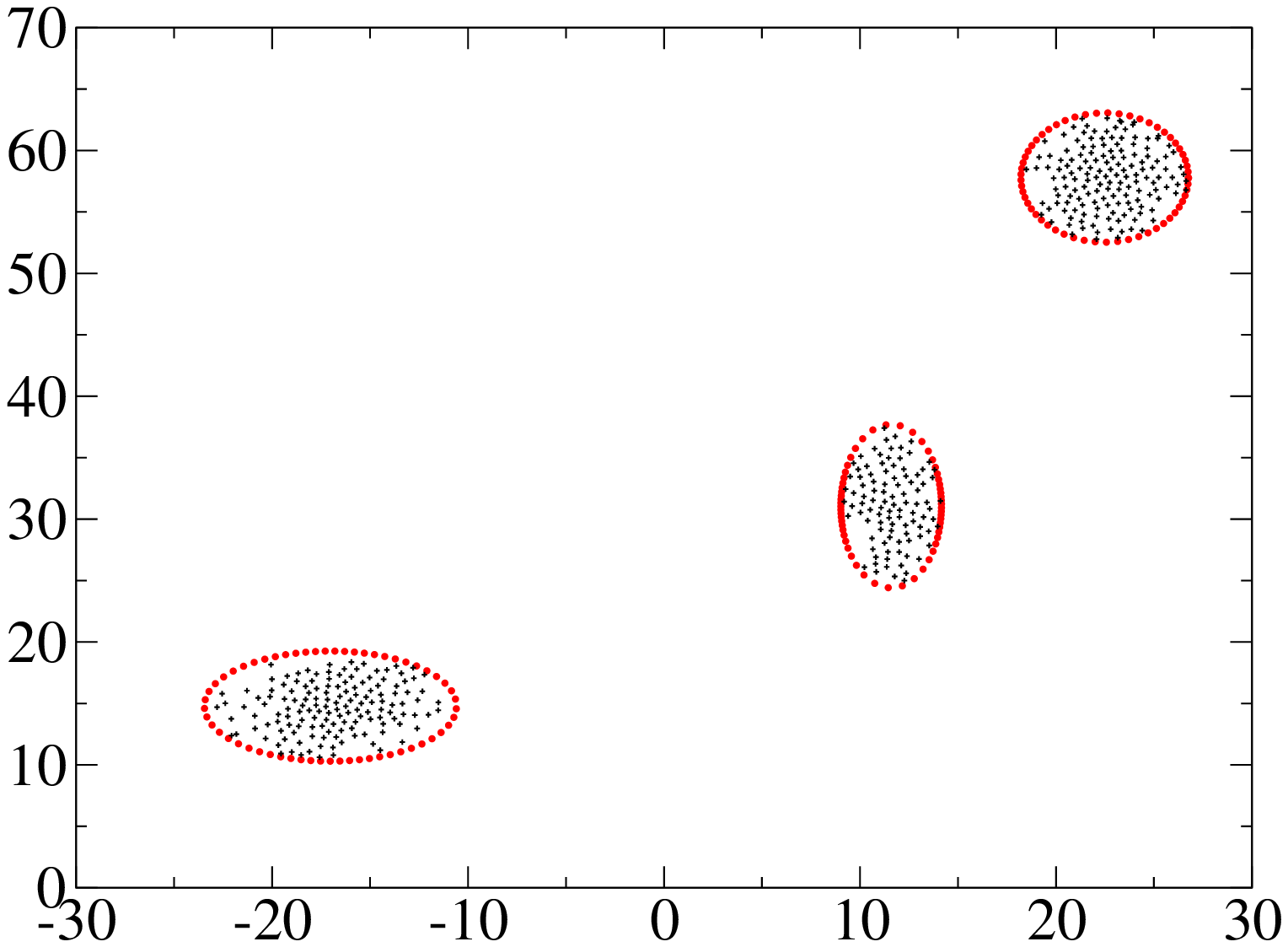}
}
\subfigure[]{
\includegraphics[scale=0.15]{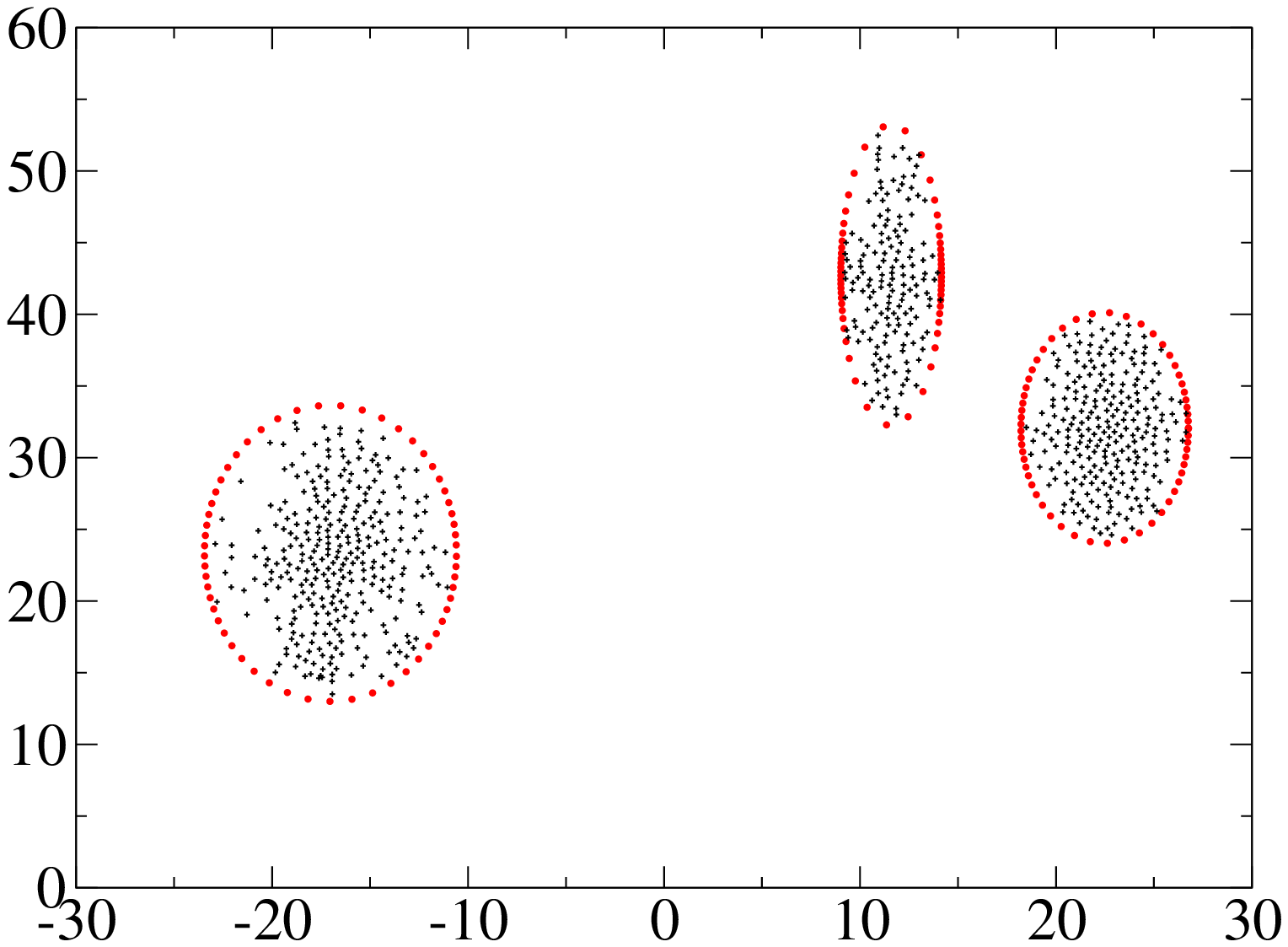}
}
\subfigure[]{
\includegraphics[scale=0.15]{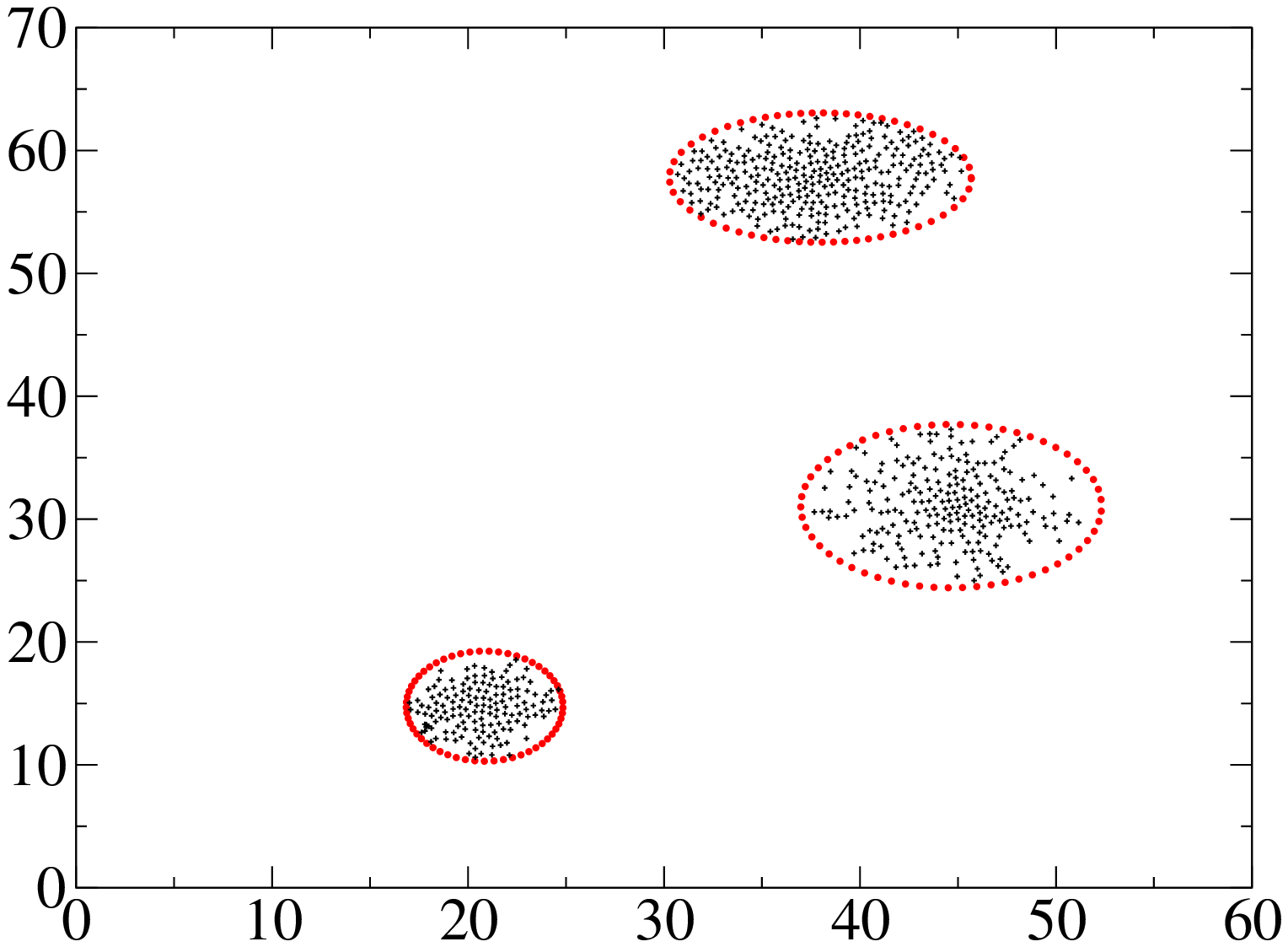}
}
\subfigure[]{
\includegraphics[scale=0.15]{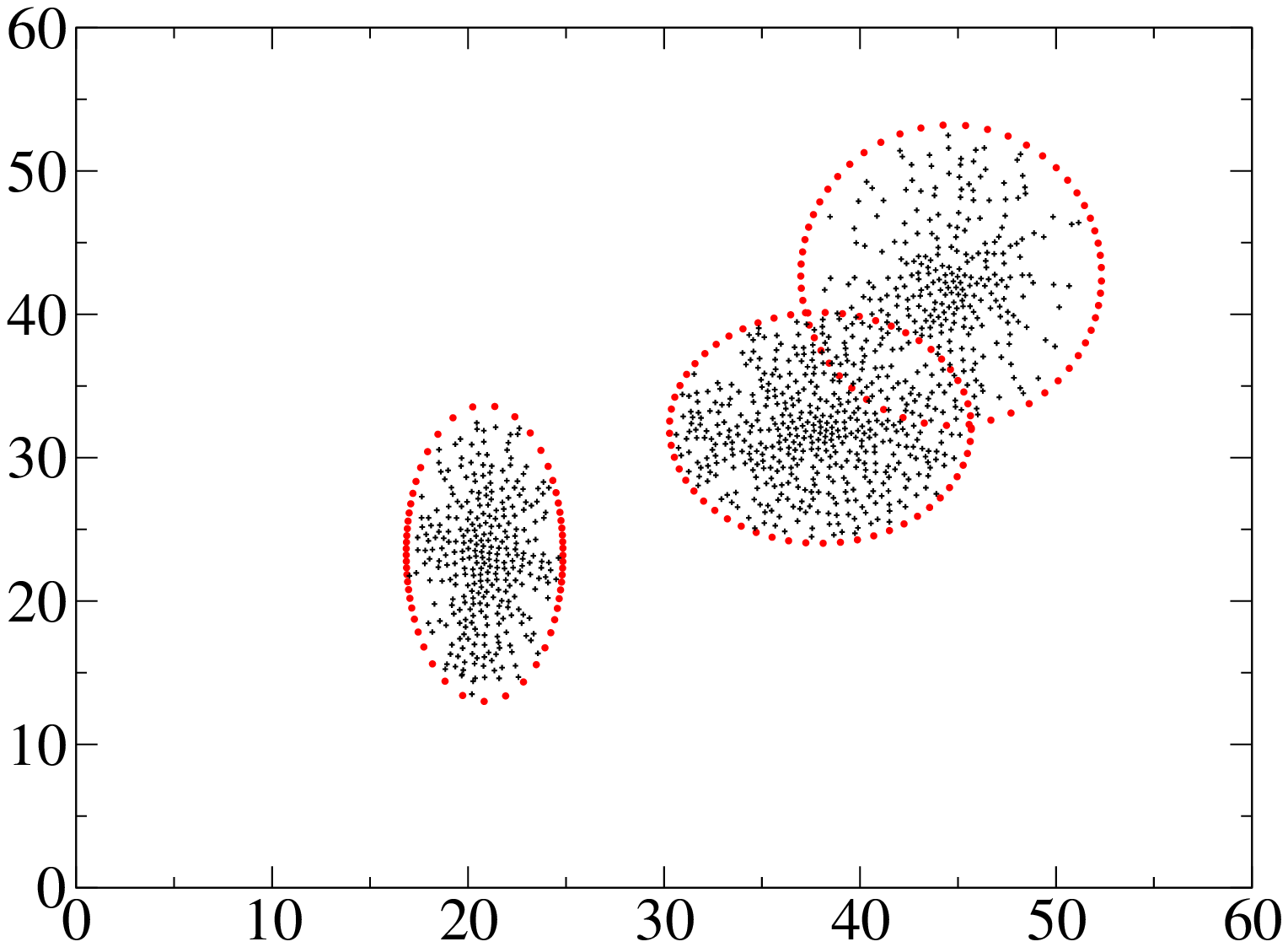}
}
\subfigure[]{
\includegraphics[scale=0.15]{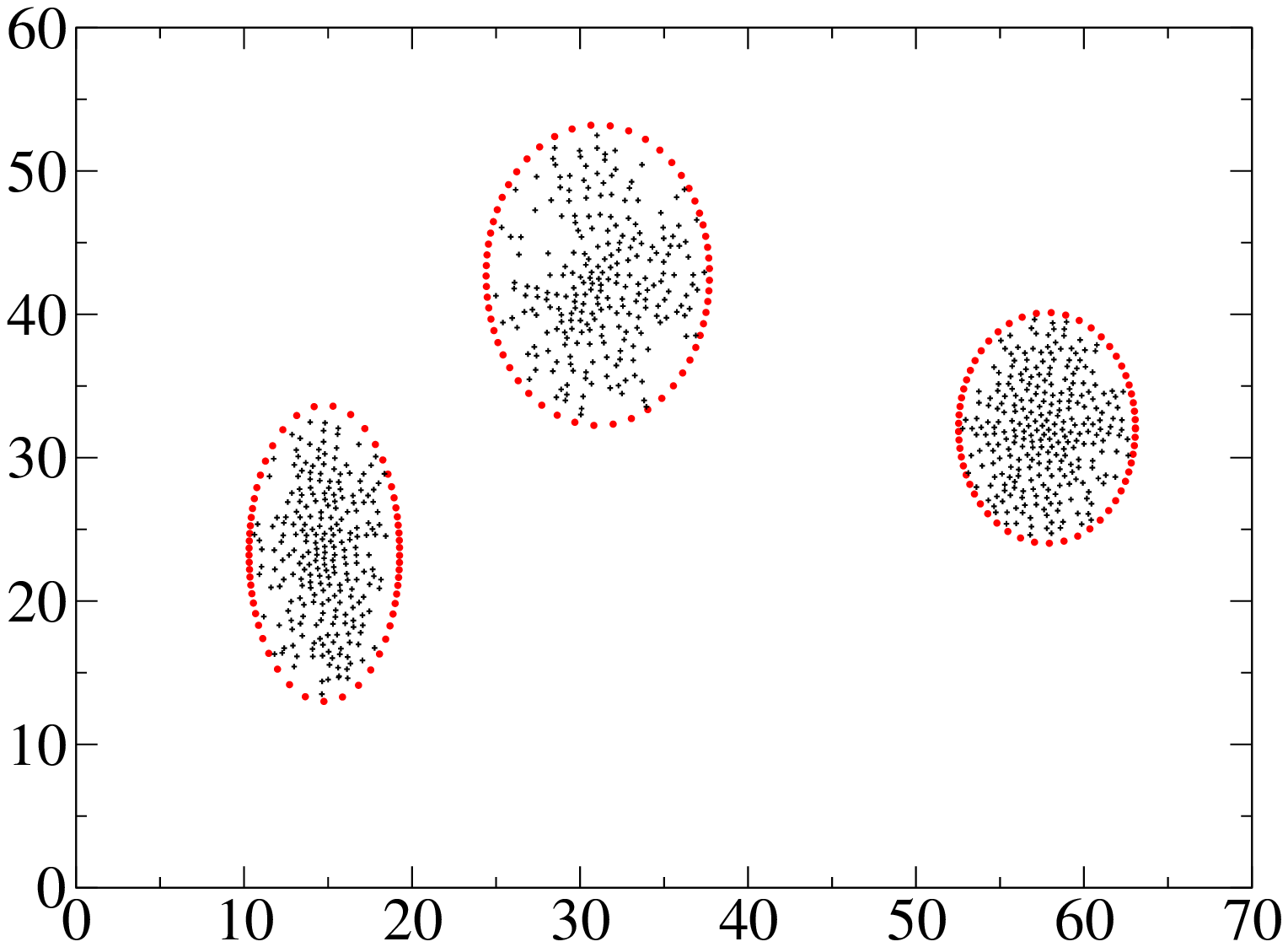}
}
\caption{
This plot shows all 10 2-dimensional sub-spaces of our 5-dimensional toy likelihood
function with 3 minima in $\chi^2$.  The black points are $\chi^2\le\chi^2_\text{lim}$
found by APS after 10,000 samplings.  The red contours are the known $\chi^2=\chi^2_\text{lim}$
contours of the function.
}
\label{fig:toyFreq3}
\end{figure*}

\begin{figure*}
\subfigure[]{
\includegraphics[scale=0.15]{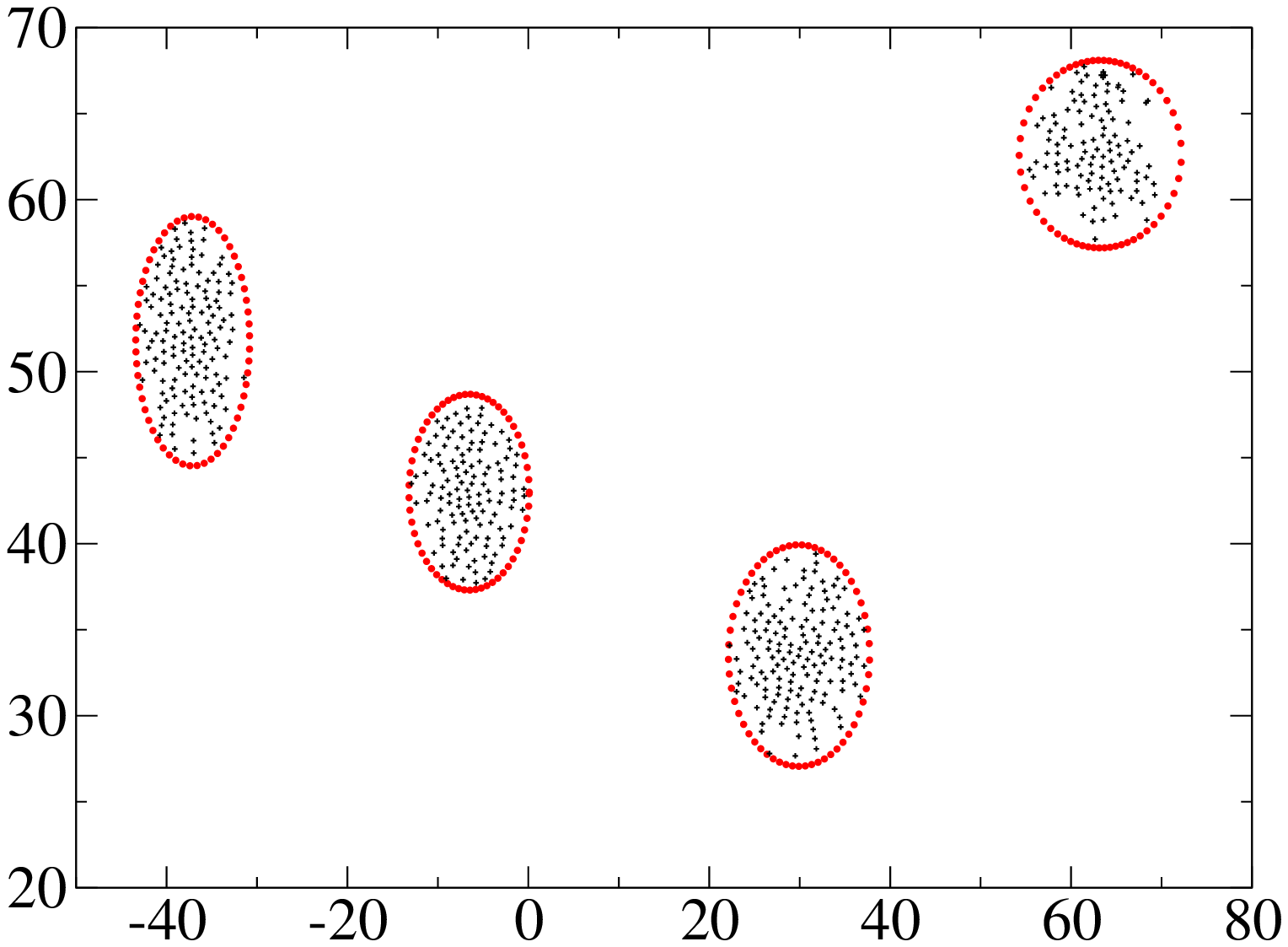}
}
\subfigure[]{
\includegraphics[scale=0.15]{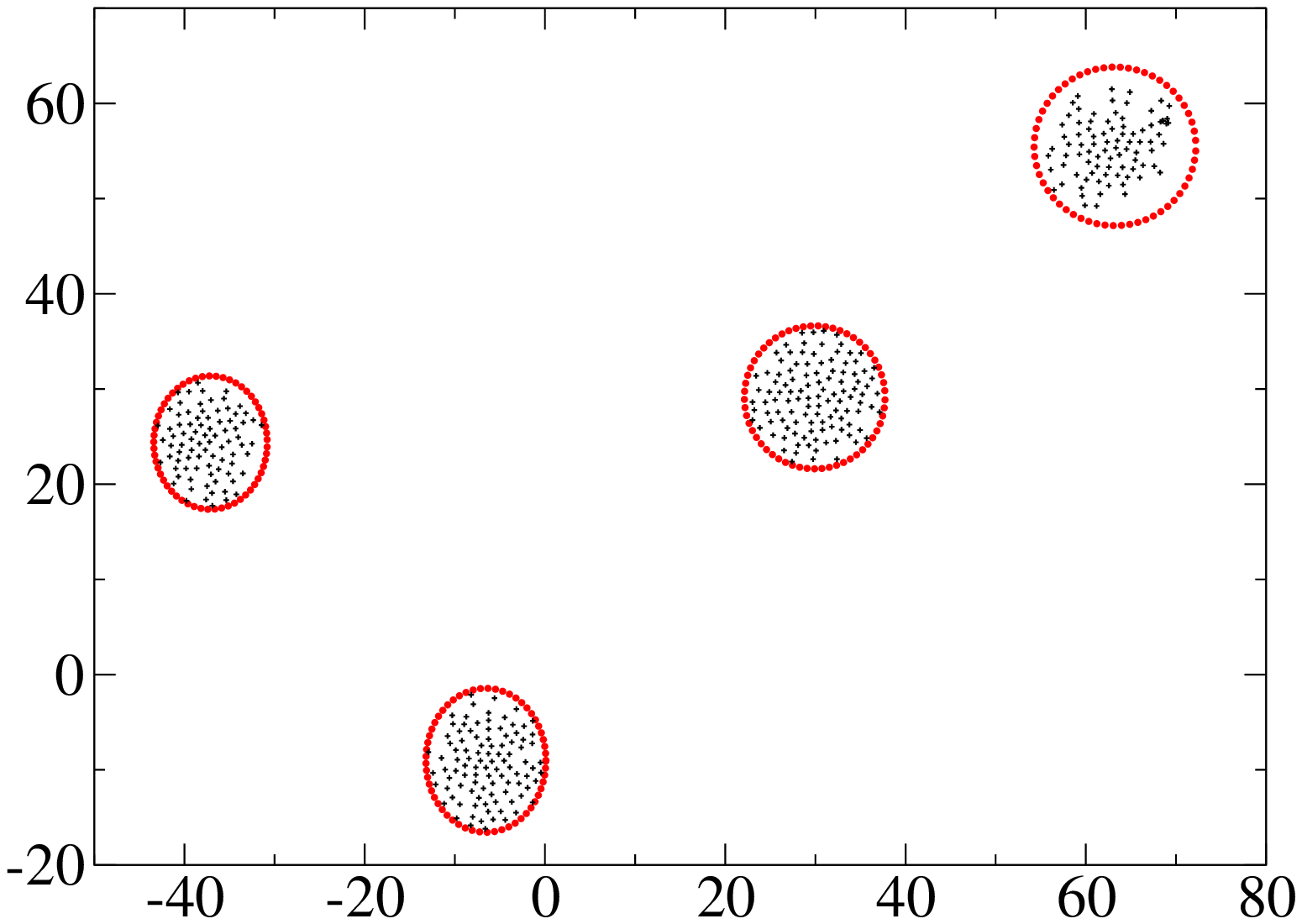}
}
\subfigure[]{
\includegraphics[scale=0.15]{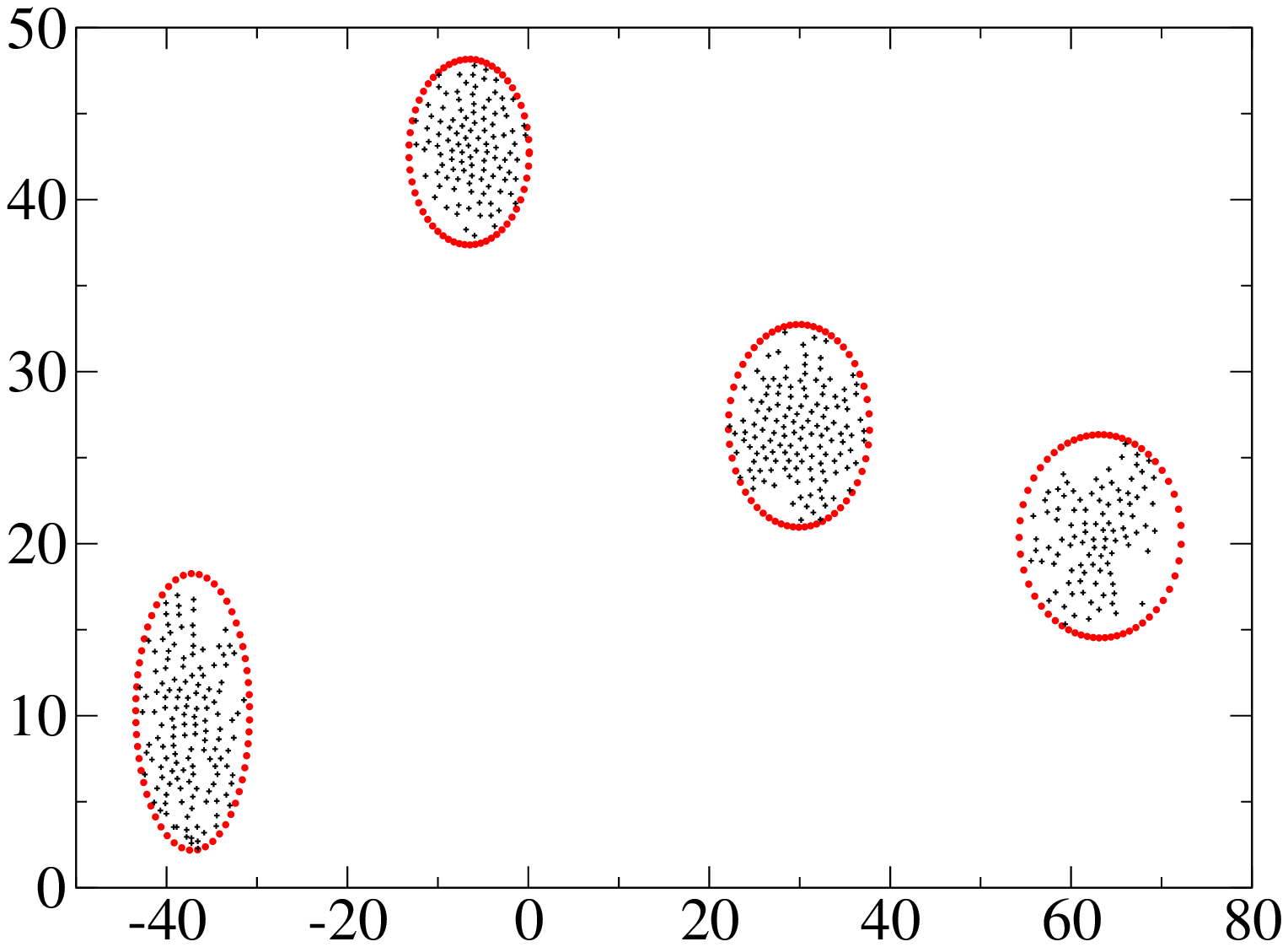}
}
\subfigure[]{
\includegraphics[scale=0.15]{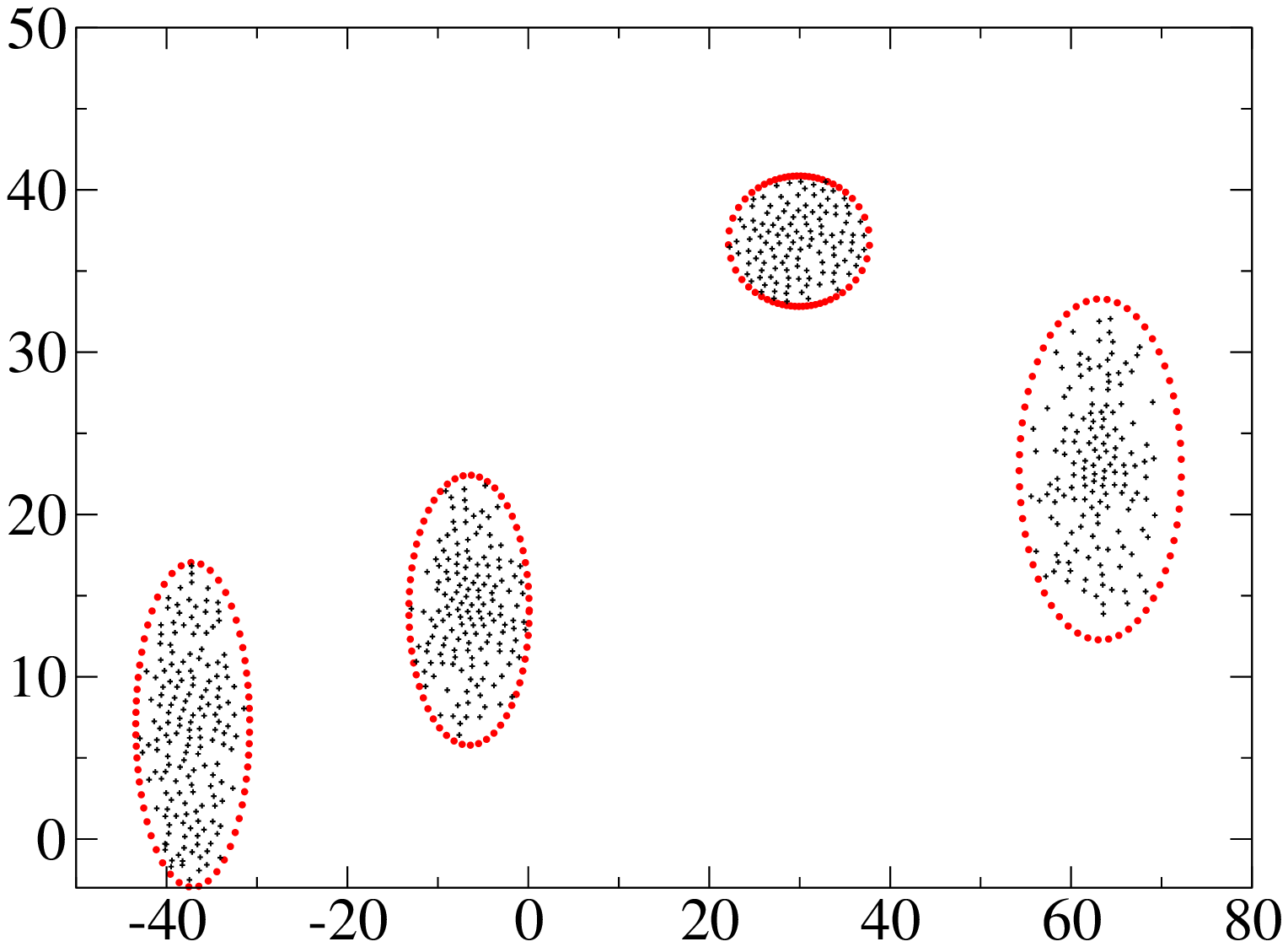}
}
\subfigure[]{
\includegraphics[scale=0.15]{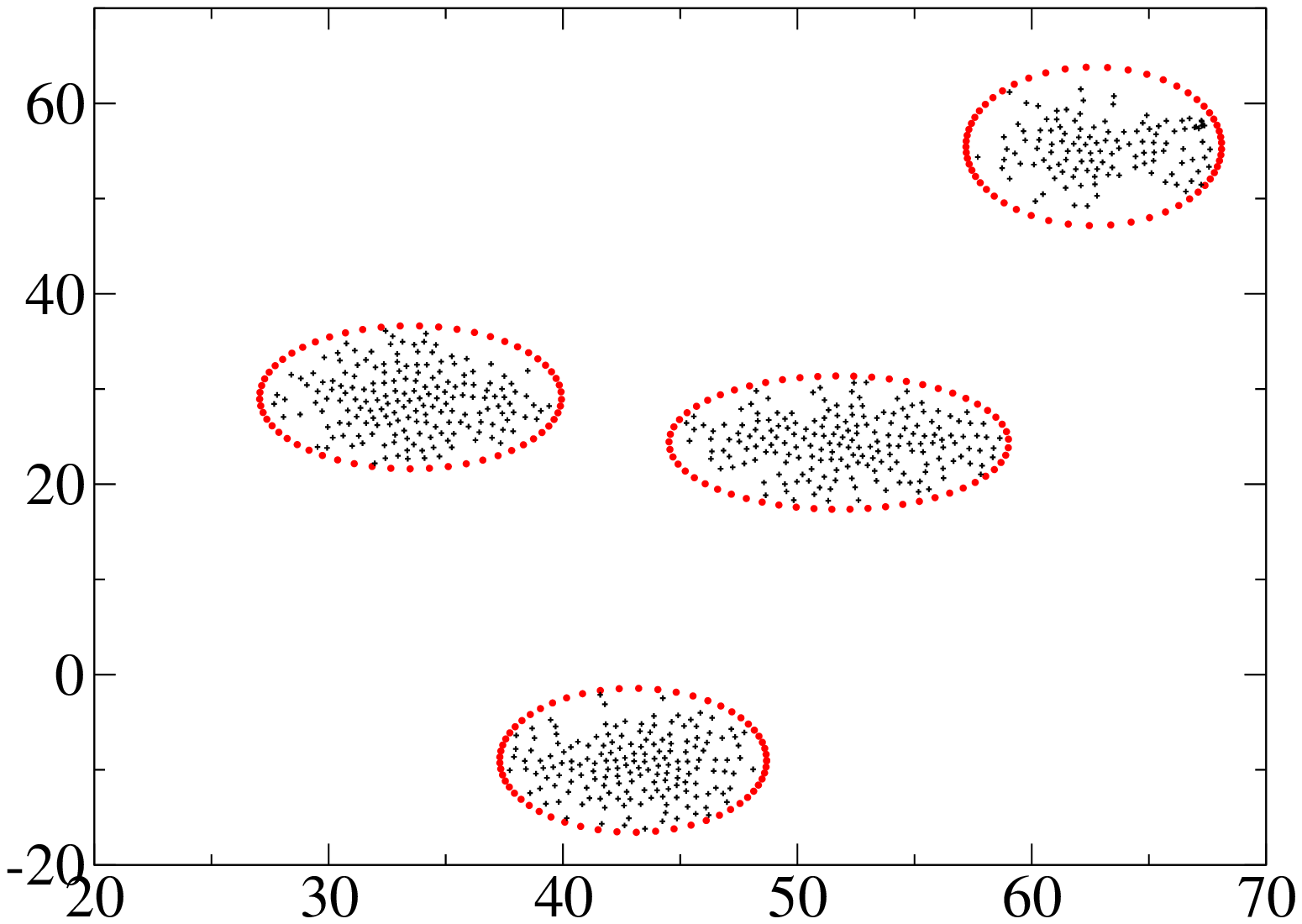}
}
\subfigure[]{
\includegraphics[scale=0.15]{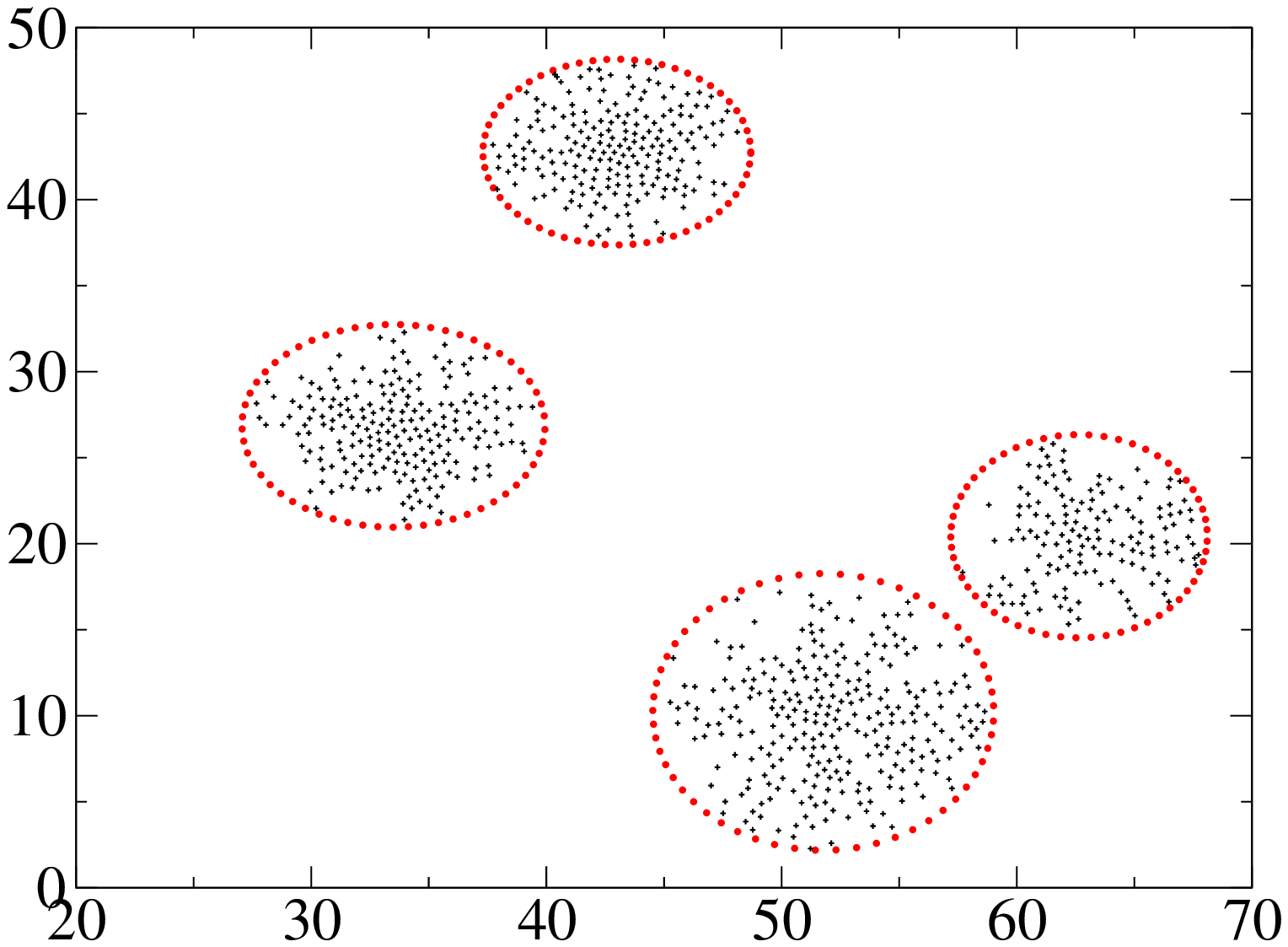}
}
\subfigure[]{
\includegraphics[scale=0.15]{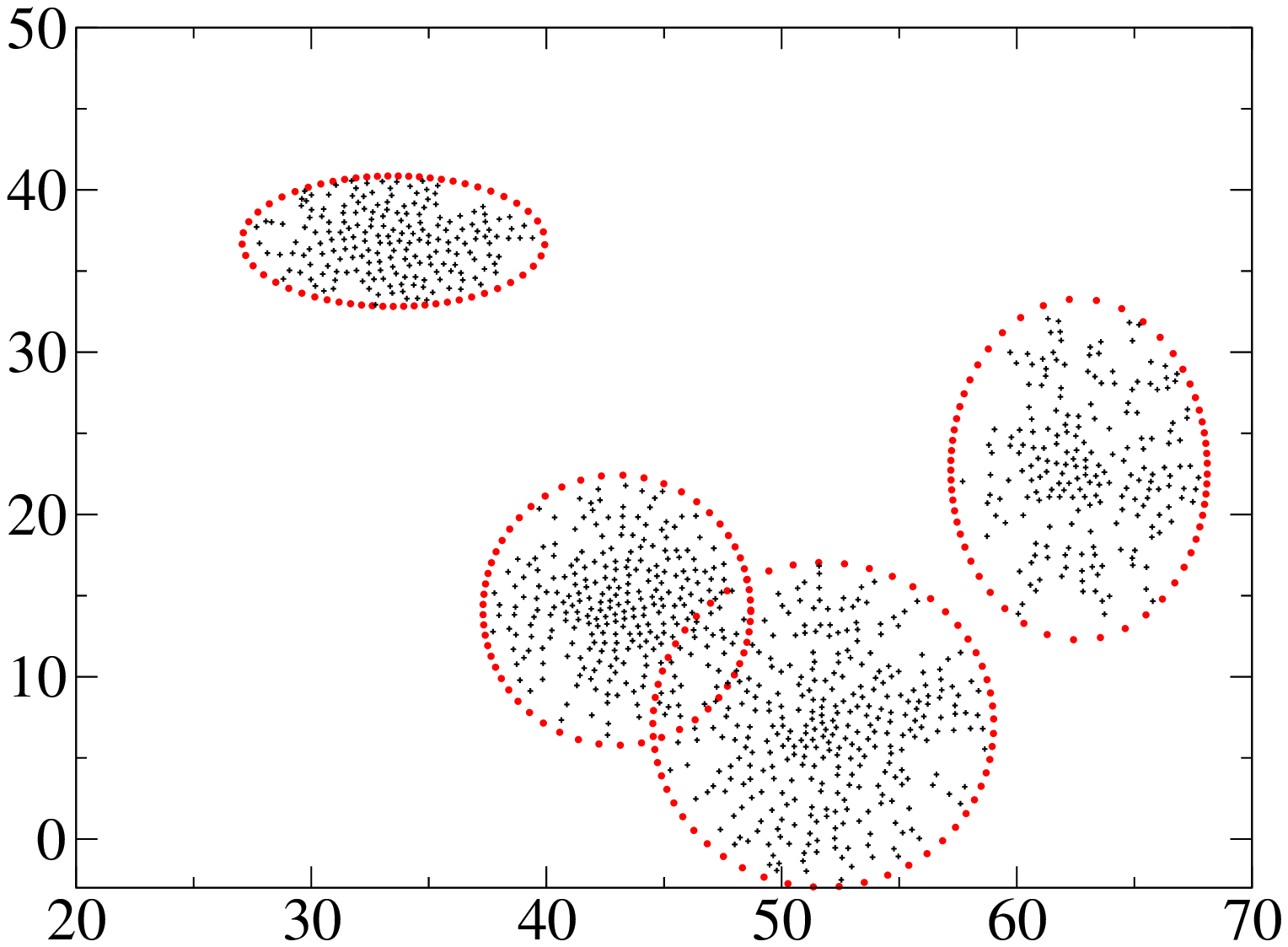}
}
\subfigure[]{
\includegraphics[scale=0.15]{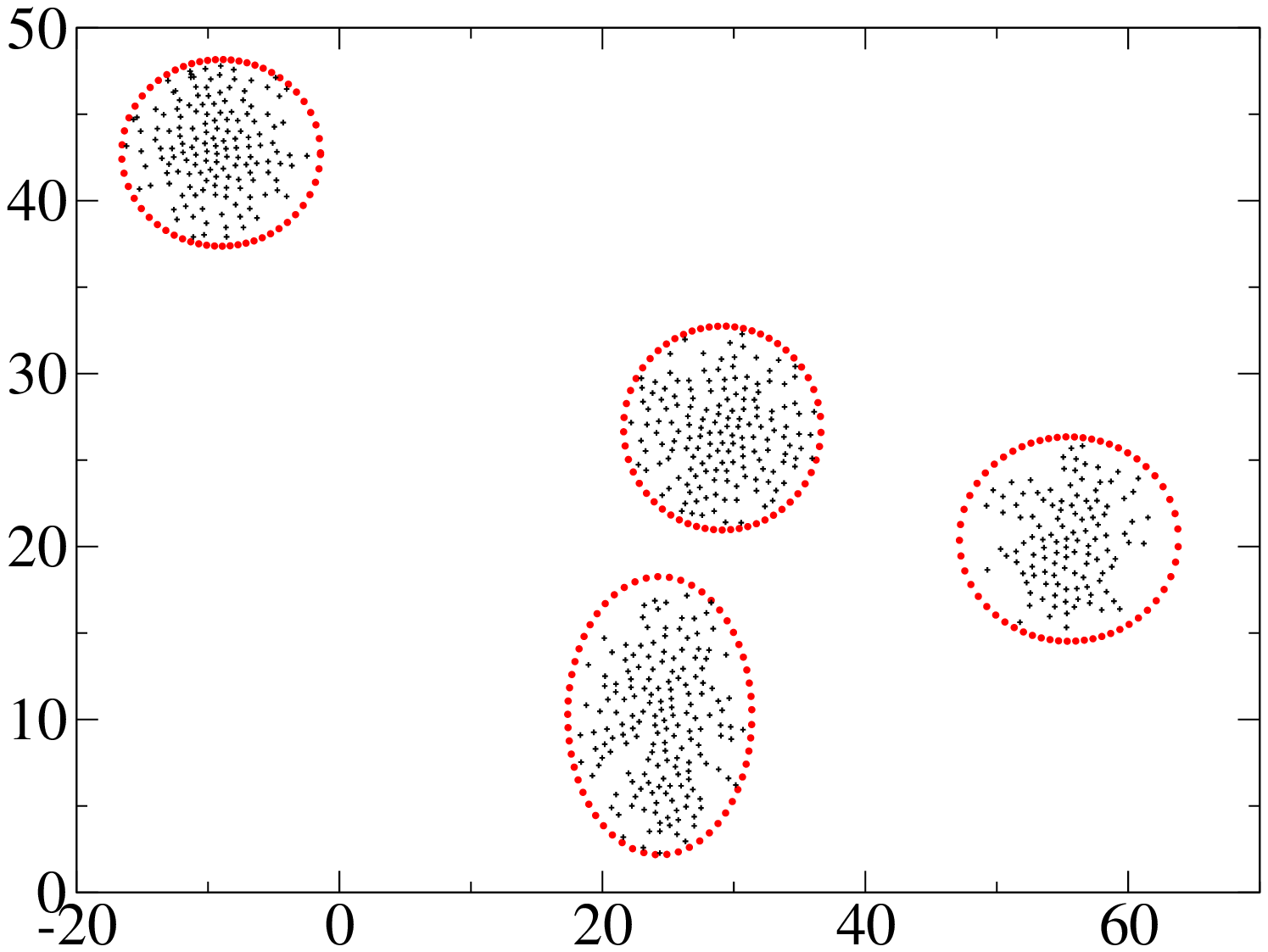}
}
\subfigure[]{
\includegraphics[scale=0.15]{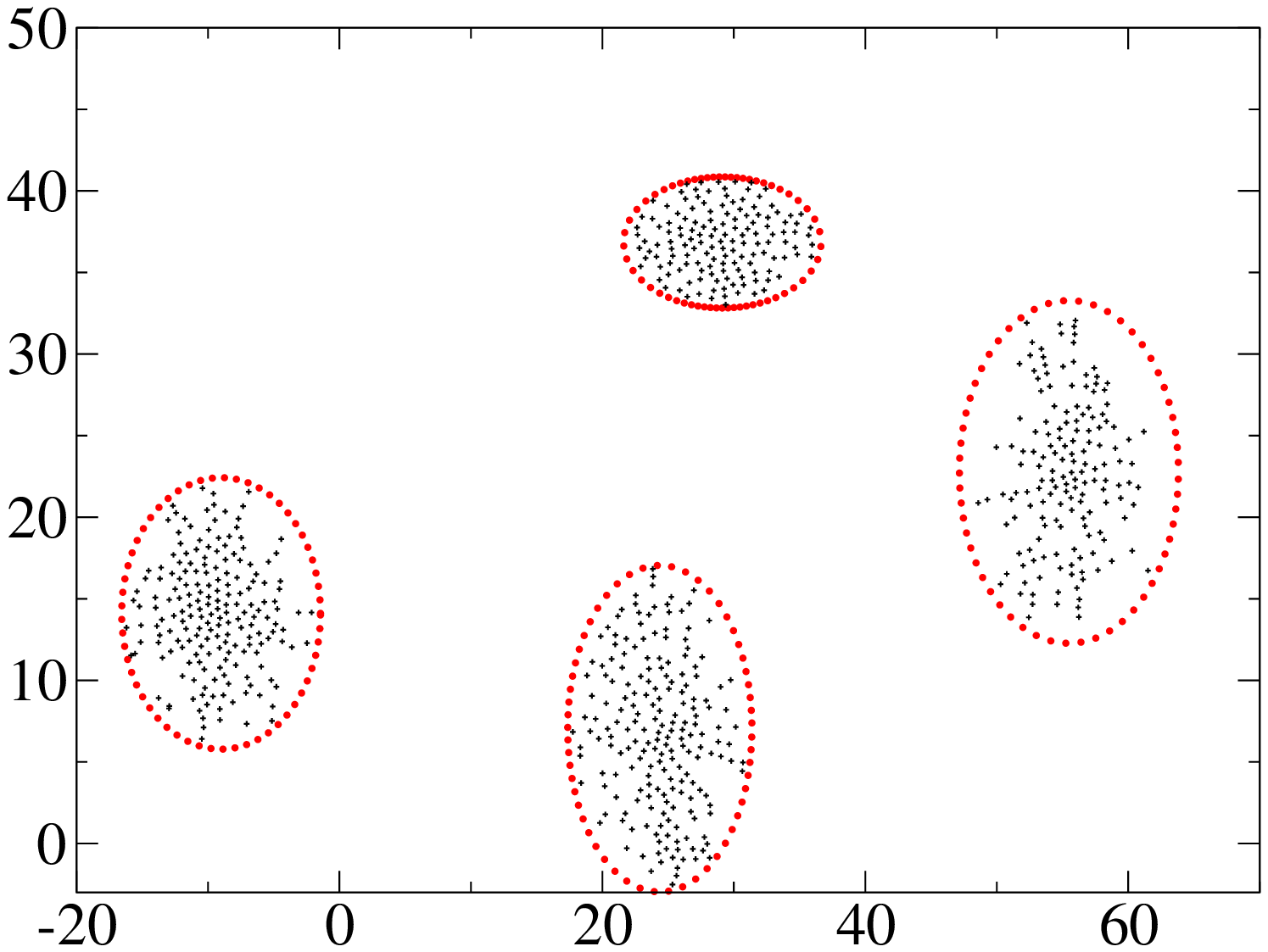}
}
\subfigure[]{
\includegraphics[scale=0.15]{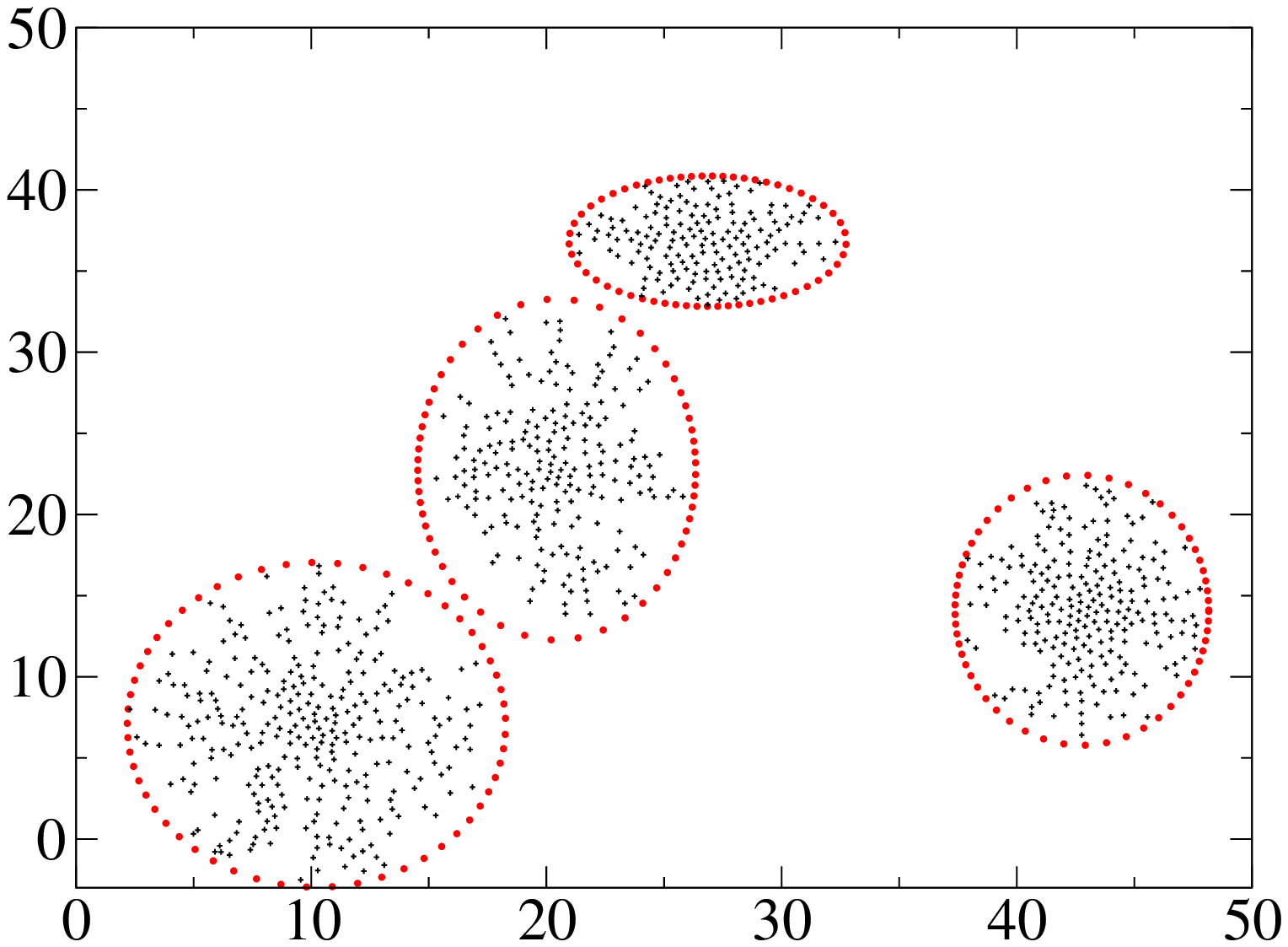}
}
\caption{
This plot shows all 10 2-dimensional sub-spaces of our 5-dimensional toy likelihood
function with 4 minima in $\chi^2$.  The black points are $\chi^2\le\chi^2_\text{lim}$
found by APS after 10,000 samplings.  The red contours are the known $\chi^2=\chi^2_\text{lim}$
contours of the function.
}
\label{fig:toyFreq4}
\end{figure*}

To test the performance of APS, we ran APS on versions of this
$\vec{\theta}\rightarrow\chi^2$ function with 2, 3, and 4 discrete minima in $\chi^2$.  We
allowed APS to sample a total of 10,000 points in the 5-dimensional parameter space.  
We had APS adaptively set $\chi^2_\text{lim}=\chi^2_\text{min}+\Delta\chi^2$, 
$\Delta\chi^2=11.0$ being the 95\%
confidence limit bound for $\chi^2$ probability distribution with 5 degrees of freedom.
We used a Mat\`ern covagiogram with $\nu=3/2$ for our Gaussian process (Rasmussen and Williams equation
4.17). Figures \ref{fig:toyFreq2}, \ref{fig:toyFreq3}, and \ref{fig:toyFreq4} plot the
$\chi^2\le\chi^2_\text{lim}$ points found by APS (black points) against the
known $\Delta\chi^2=11.0$ contours of the function (red contours).  Obviously, APS did a very
good job of exploring all of the available low-$\chi^2$ regions.

\begin{figure*}
\subfigure[]{
\includegraphics[scale=0.15]{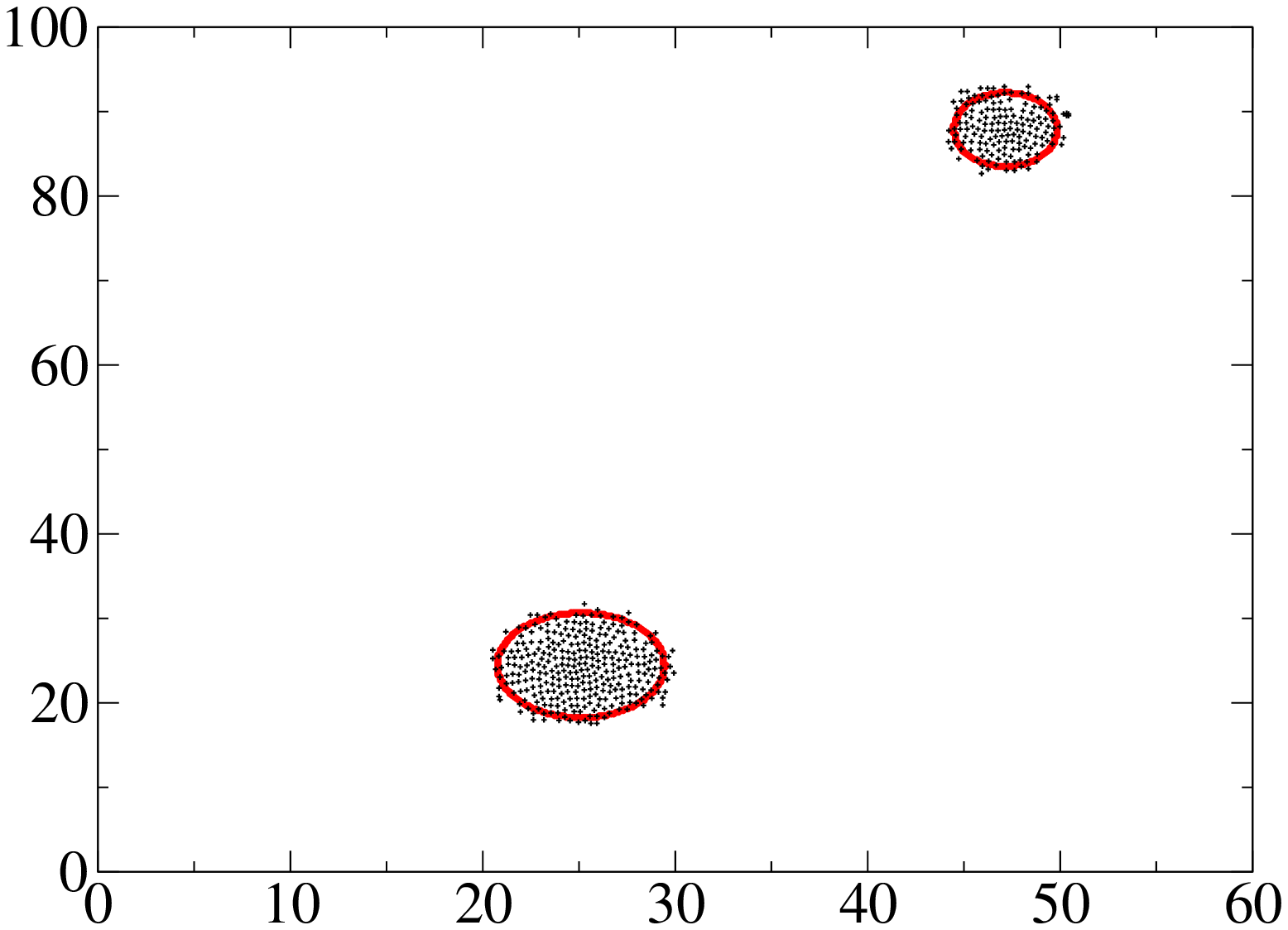}
}
\subfigure[]{
\includegraphics[scale=0.15]{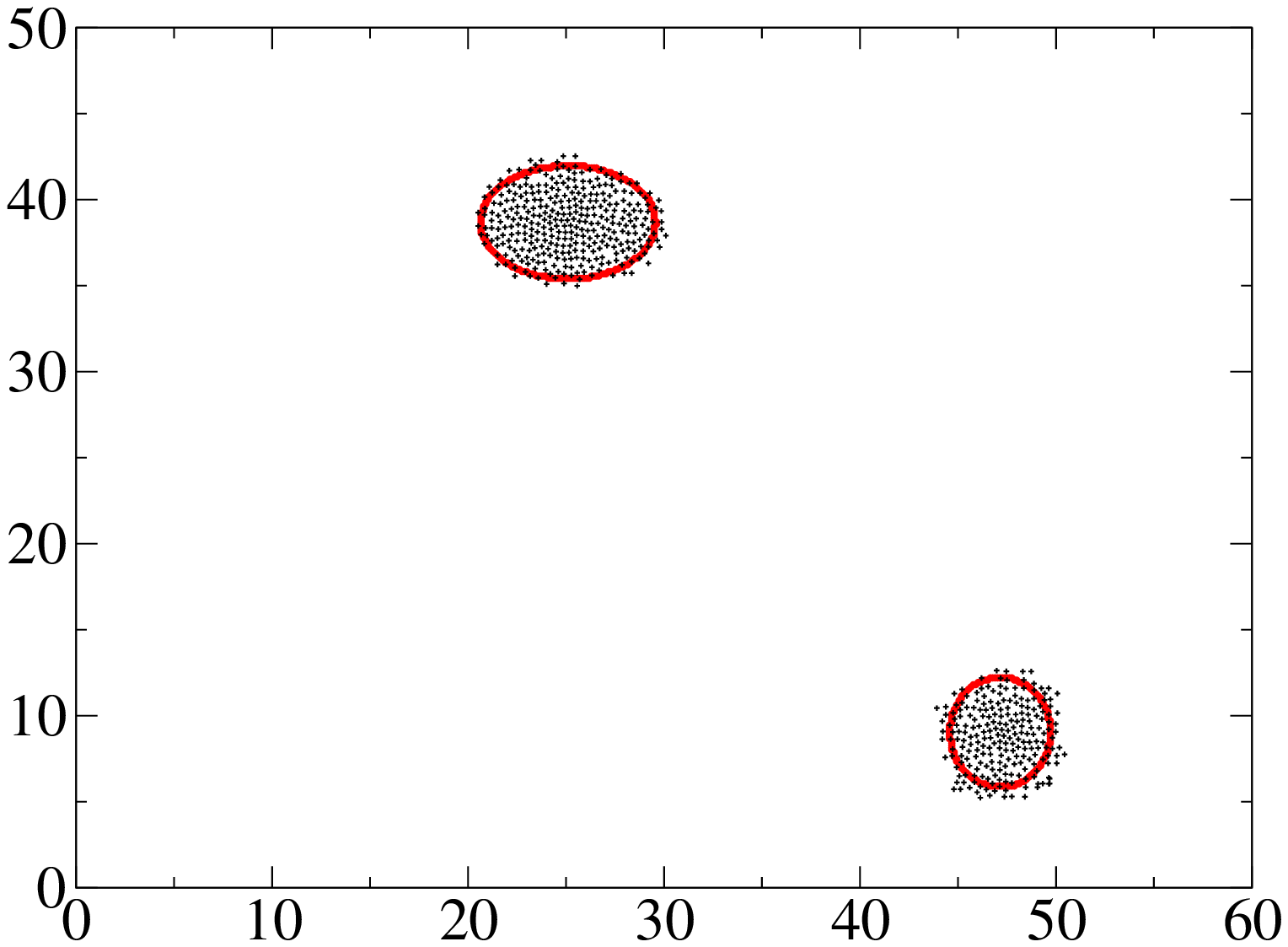}
}
\subfigure[]{
\includegraphics[scale=0.15]{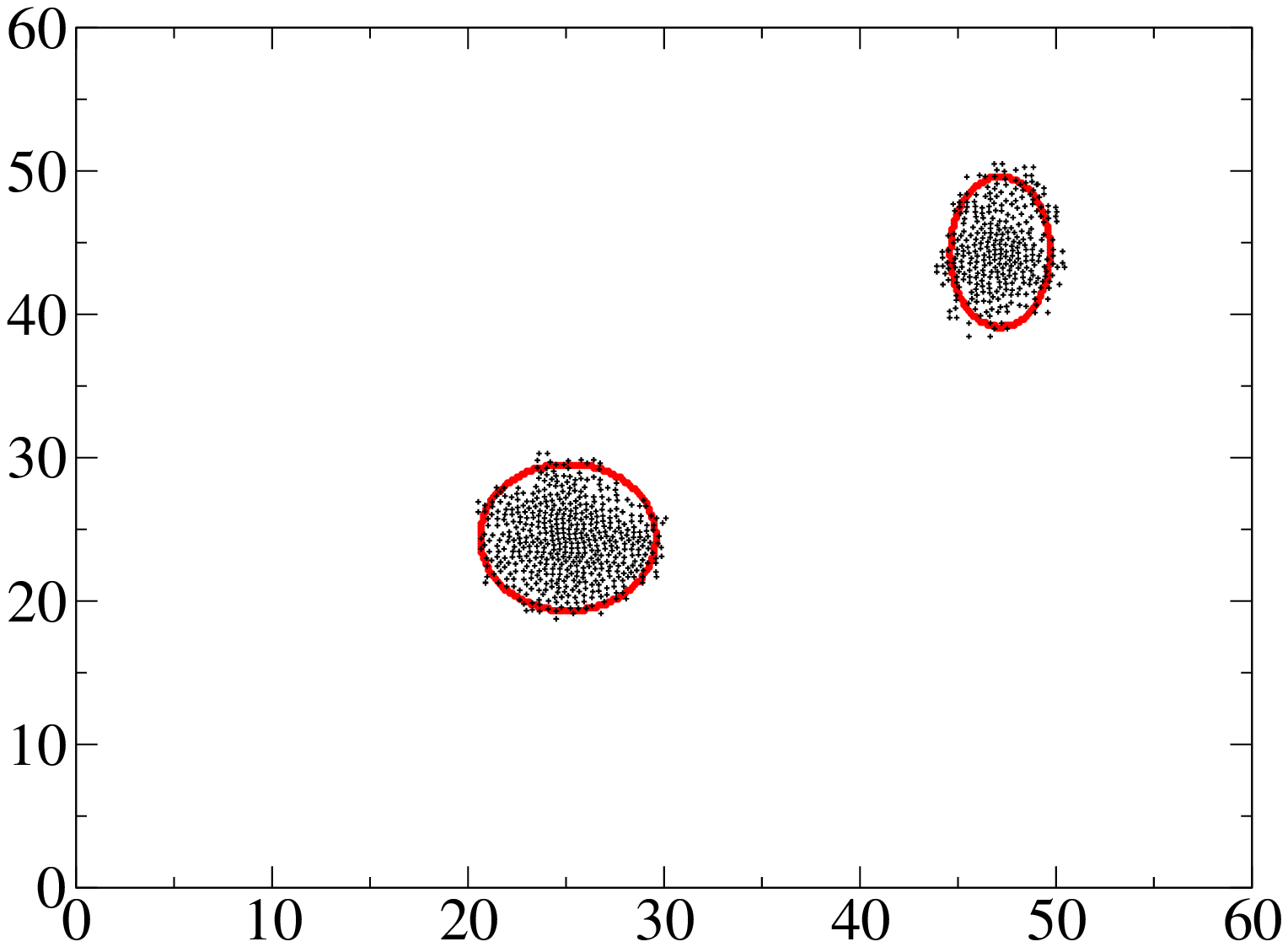}
}
\subfigure[]{
\includegraphics[scale=0.15]{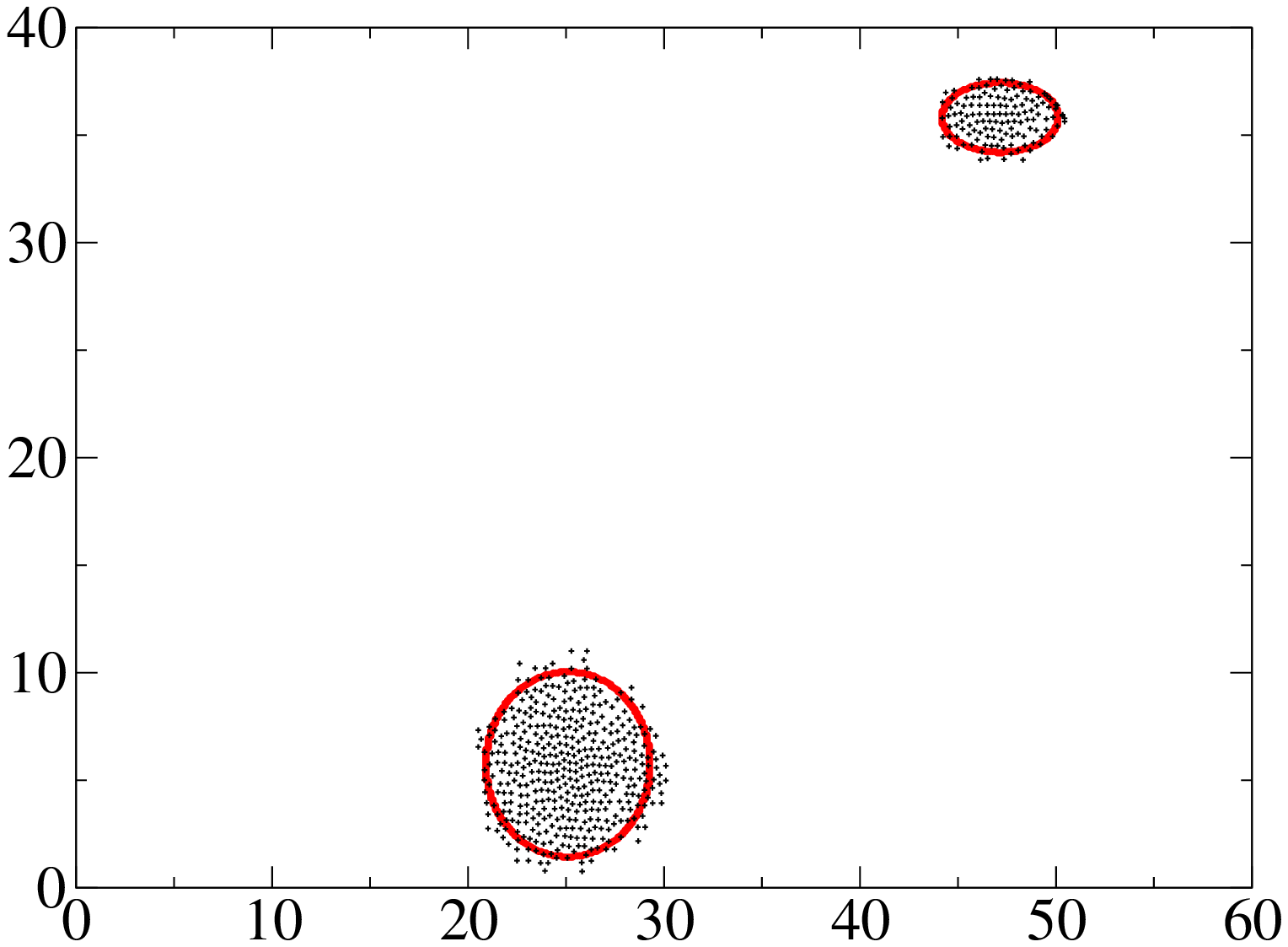}
}
\subfigure[]{
\includegraphics[scale=0.15]{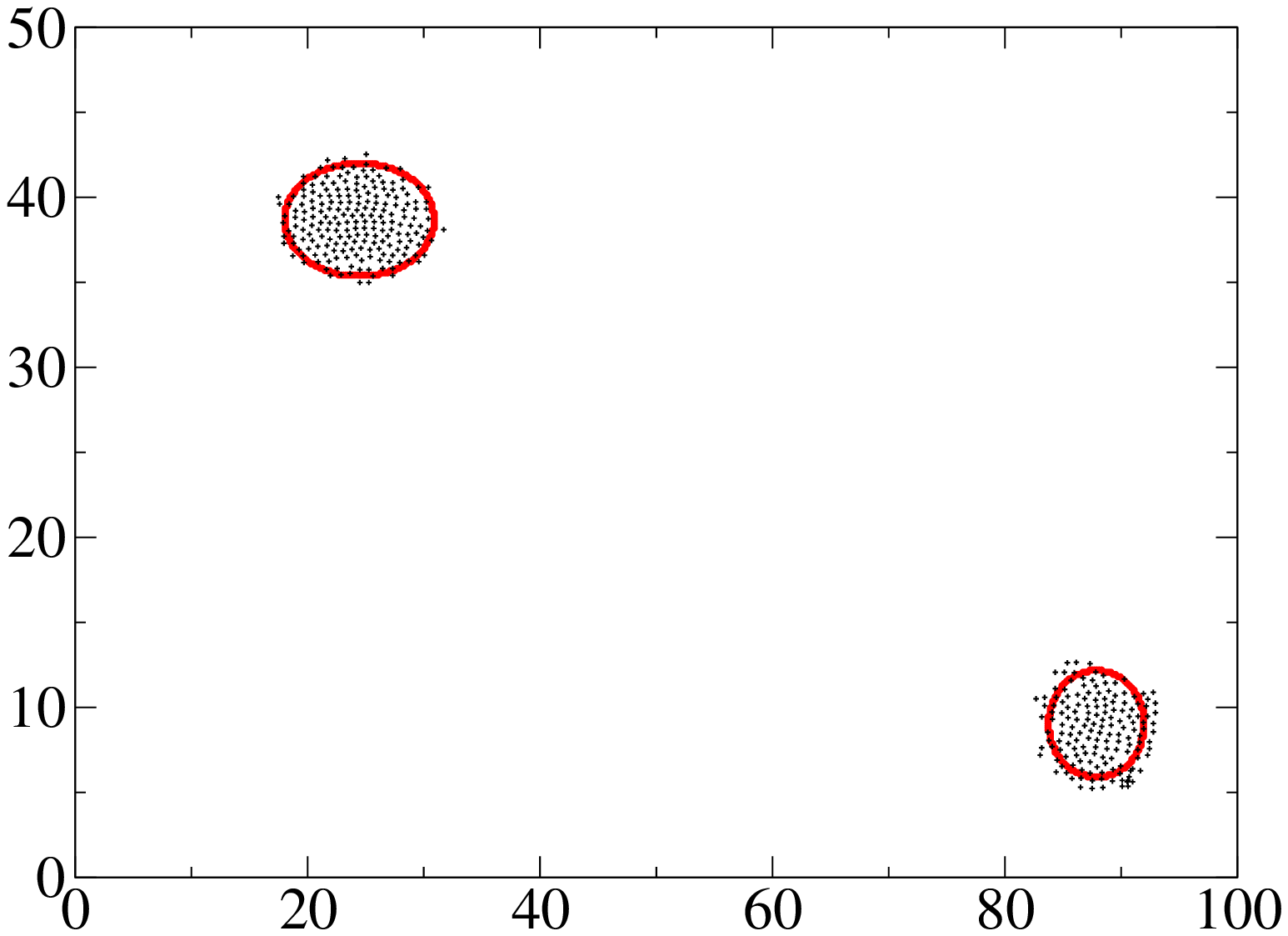}
}
\subfigure[]{
\includegraphics[scale=0.15]{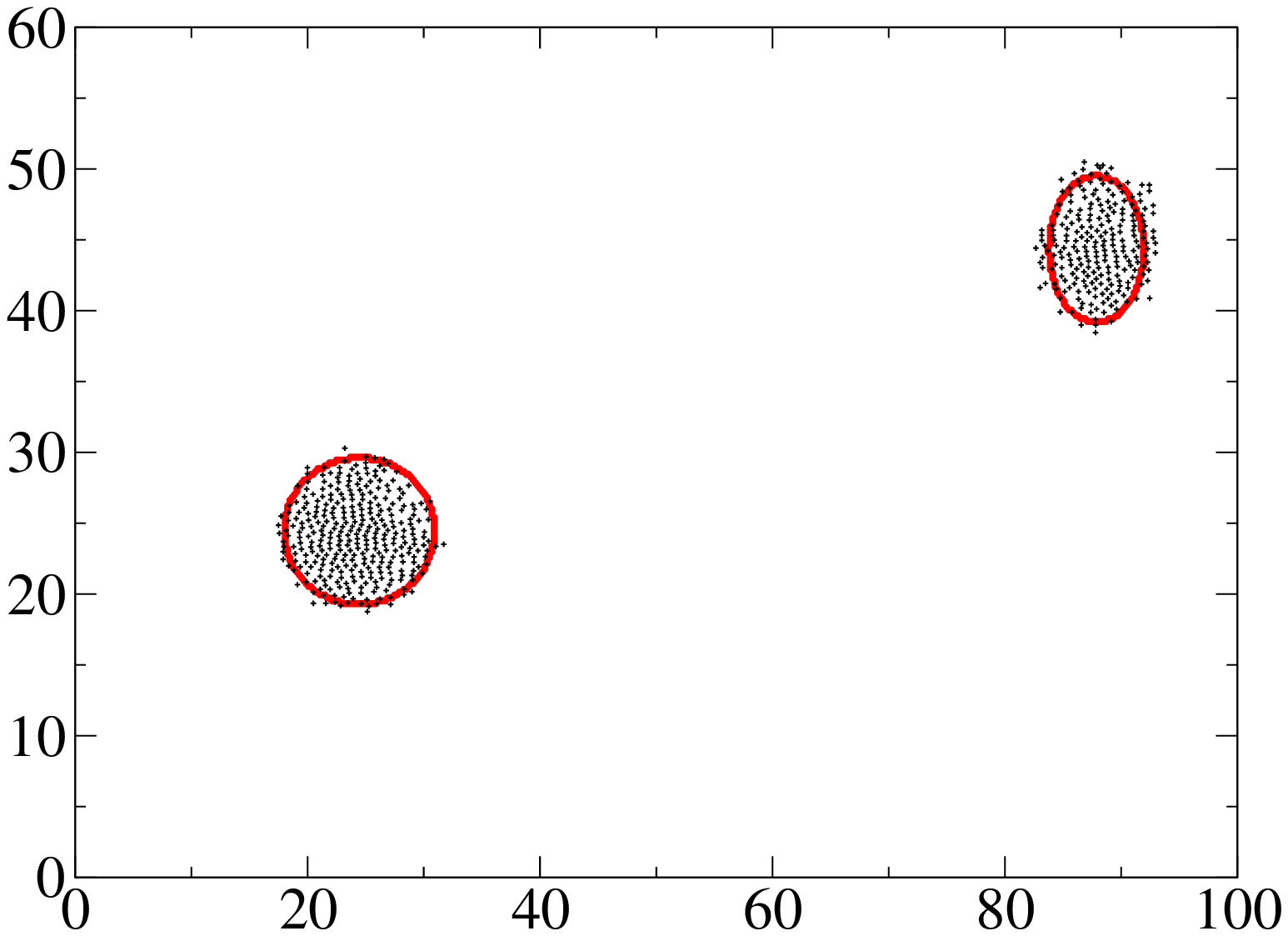}
}
\subfigure[]{
\includegraphics[scale=0.15]{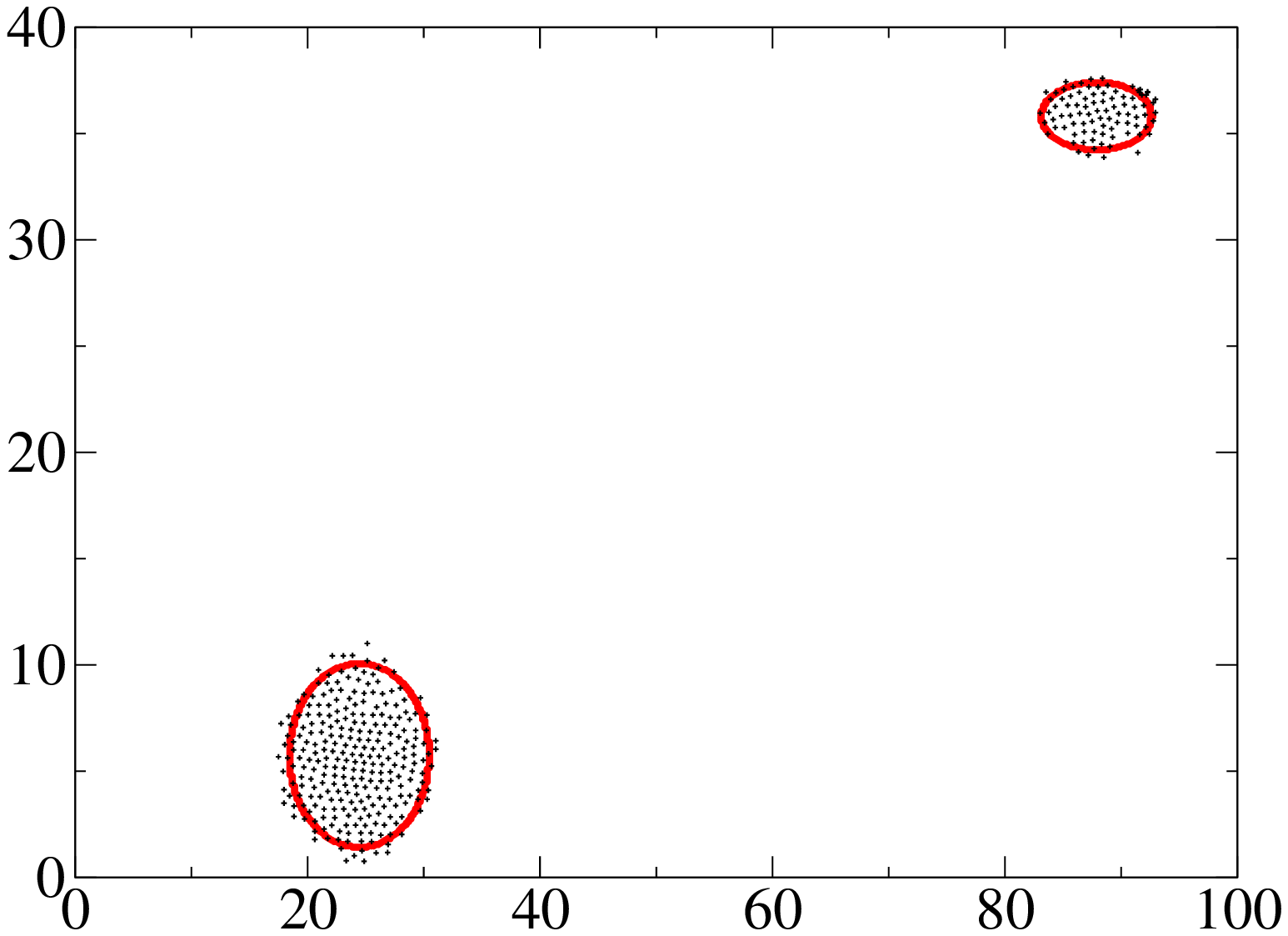}
}
\subfigure[]{
\includegraphics[scale=0.15]{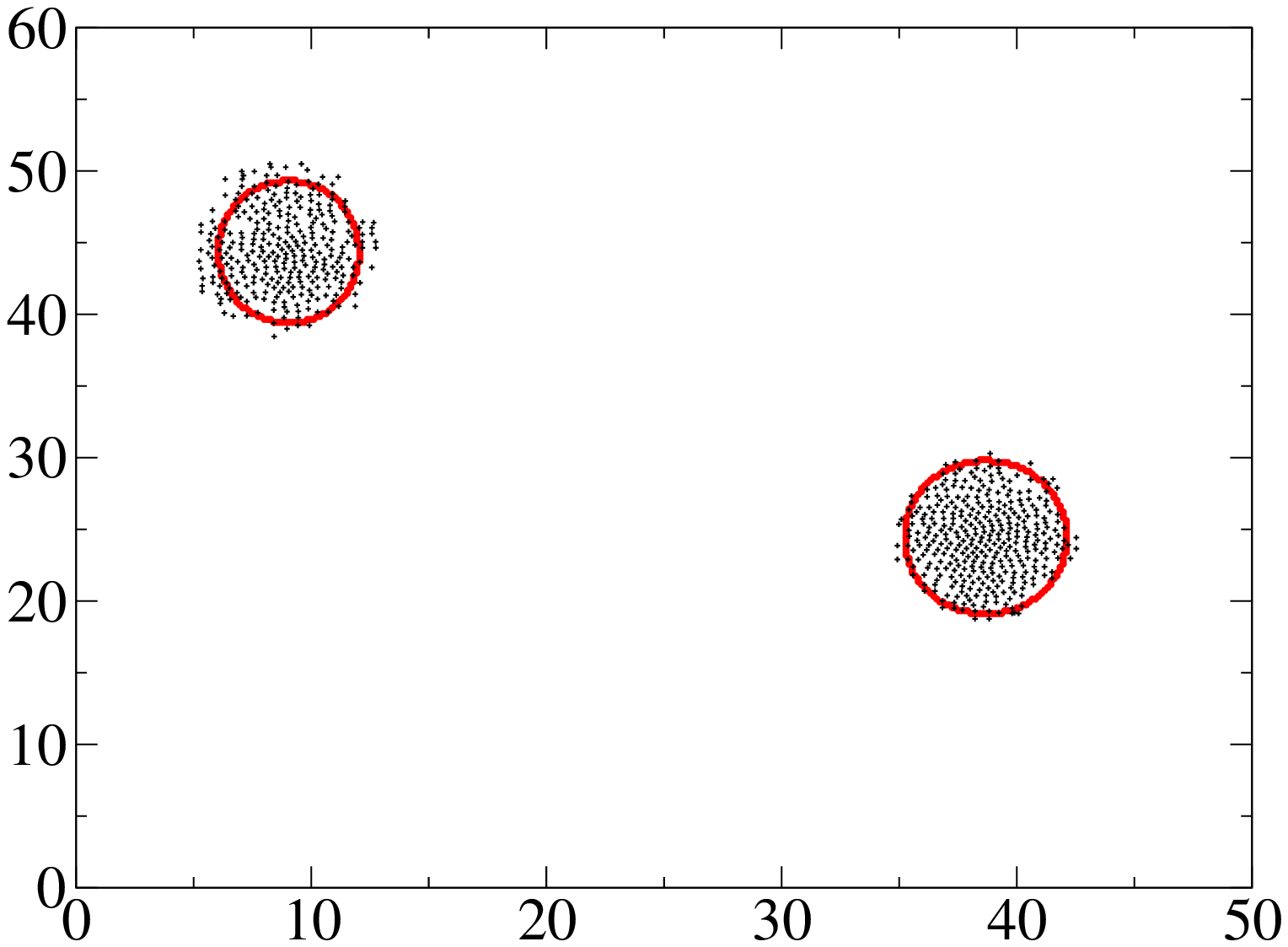}
}
\subfigure[]{
\includegraphics[scale=0.15]{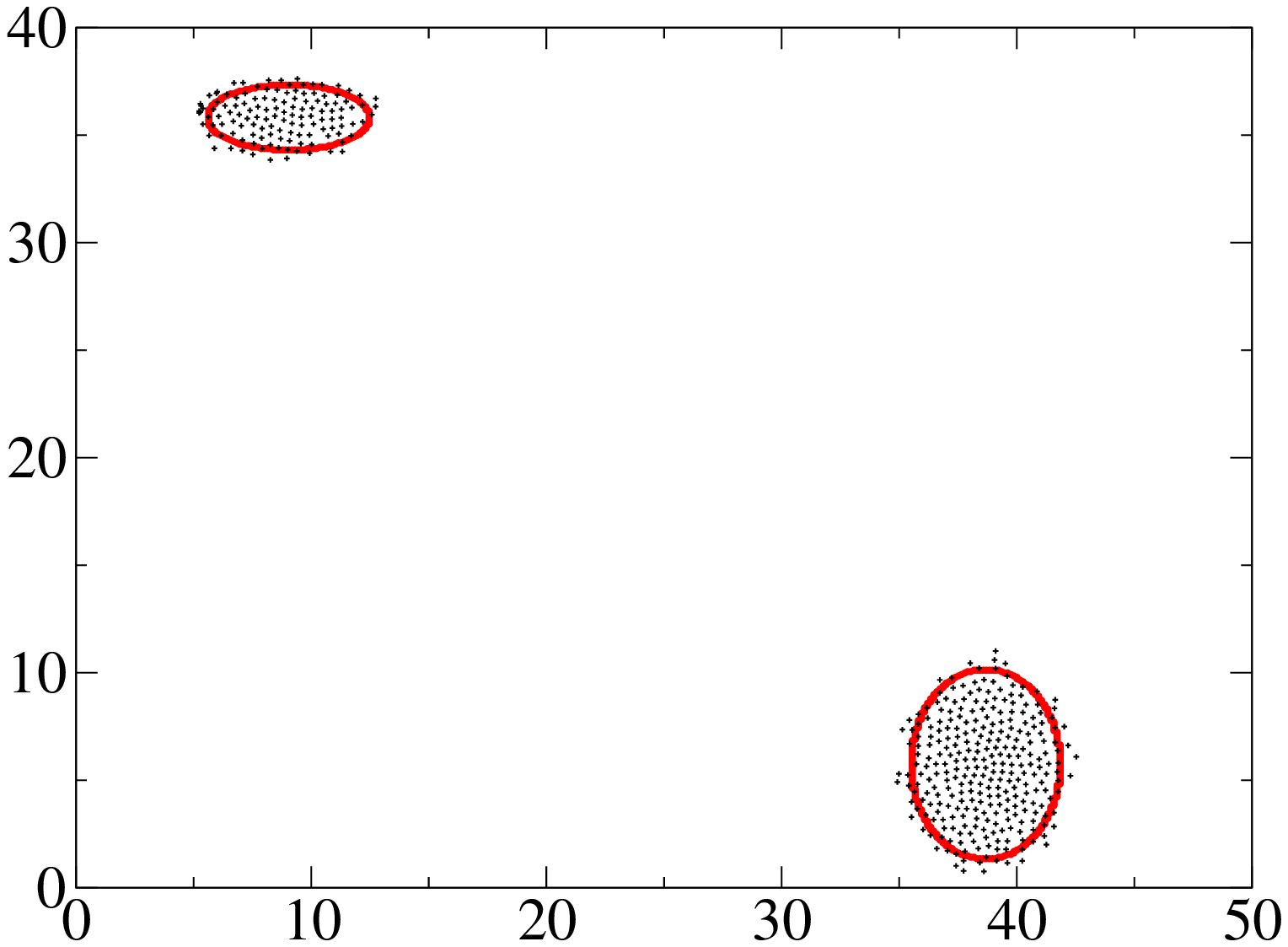}
}
\subfigure[]{
\includegraphics[scale=0.15]{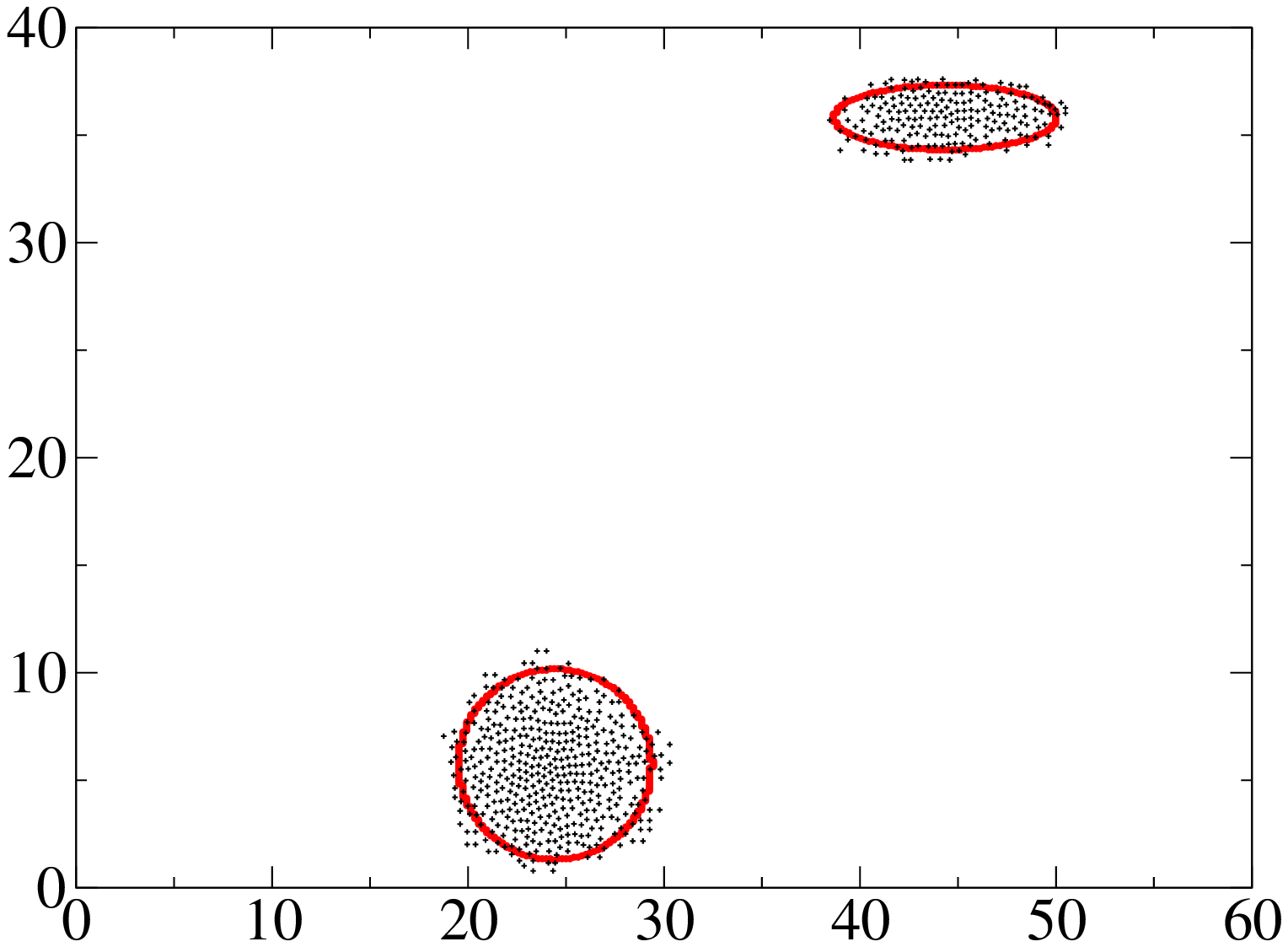}
}
\caption{
This plot shows all 10 2-dimensional sub-spaces of our 5-dimensional toy likelihood
function with 2 minima in $\chi^2$.  The black points represent the $95\%$ Bayesian credible
limit found by APS (and determined as described in Section \ref{sec:bayes}) after 10,000
samplings.  The red contours represent the known $95\%$ Bayesian credible limit of the toy
function.
}
\label{fig:toyBayes2}
\end{figure*}

\begin{figure*}
\subfigure[]{
\includegraphics[scale=0.15]{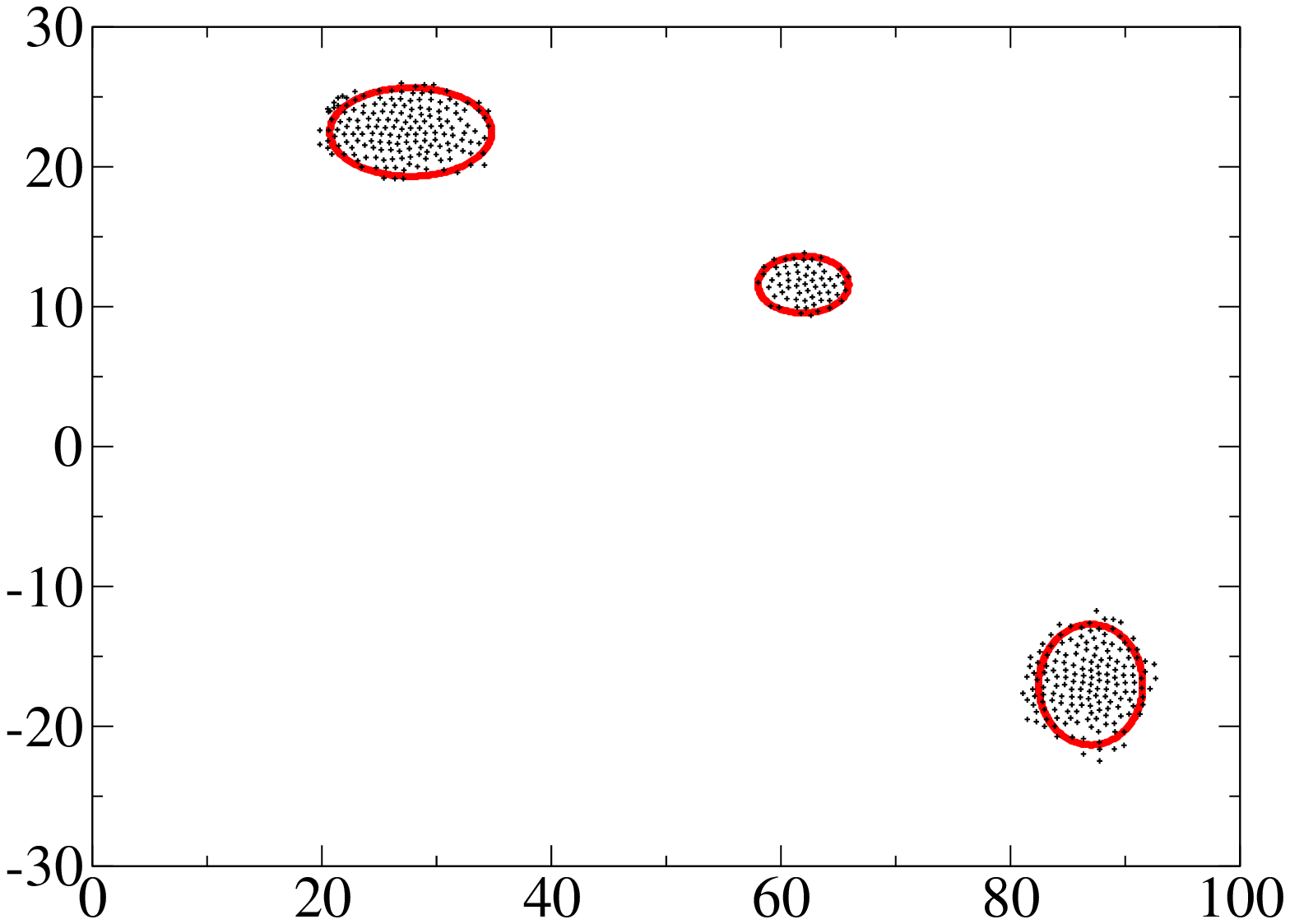}
}
\subfigure[]{
\includegraphics[scale=0.15]{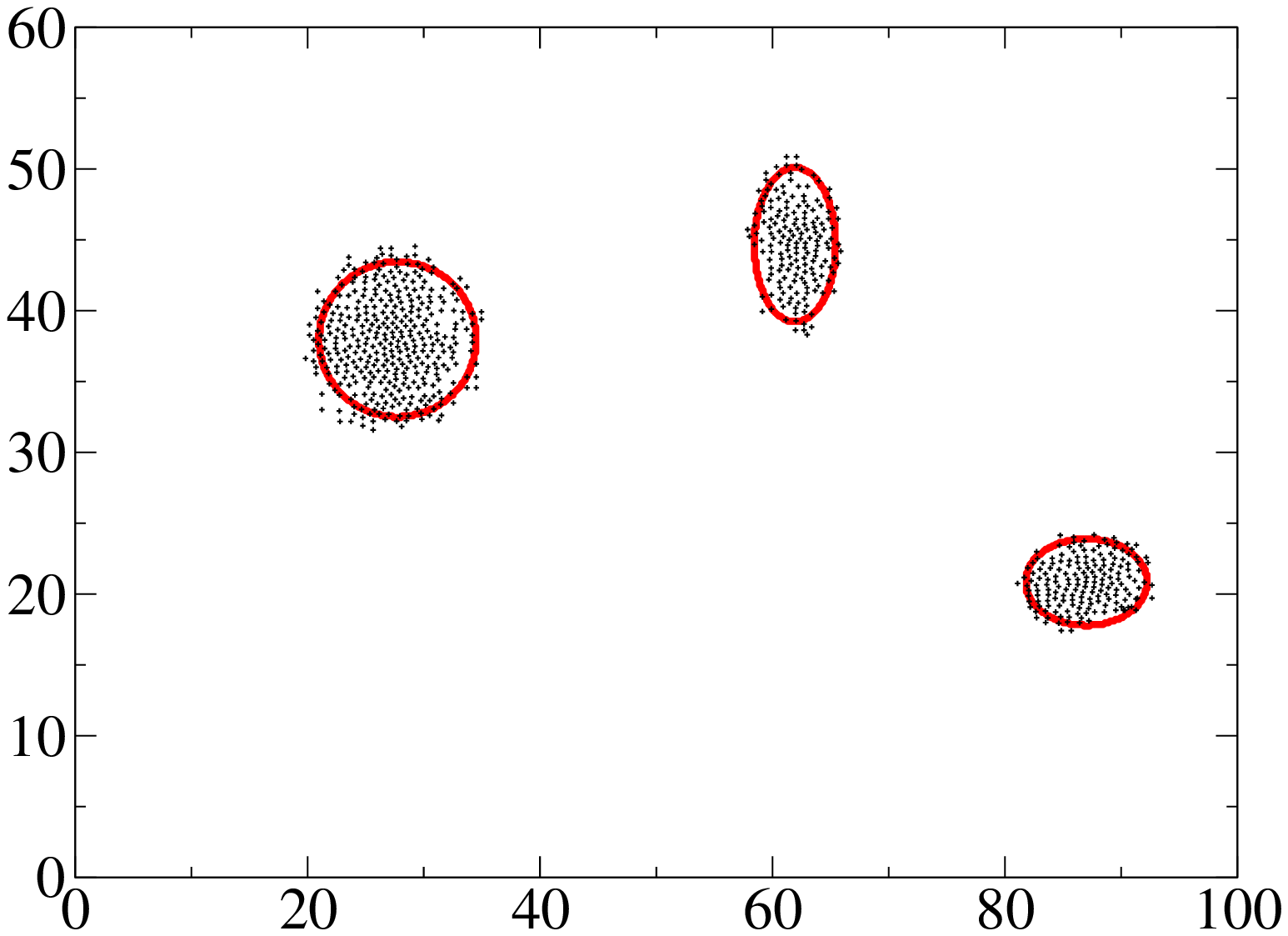}
}
\subfigure[]{
\includegraphics[scale=0.15]{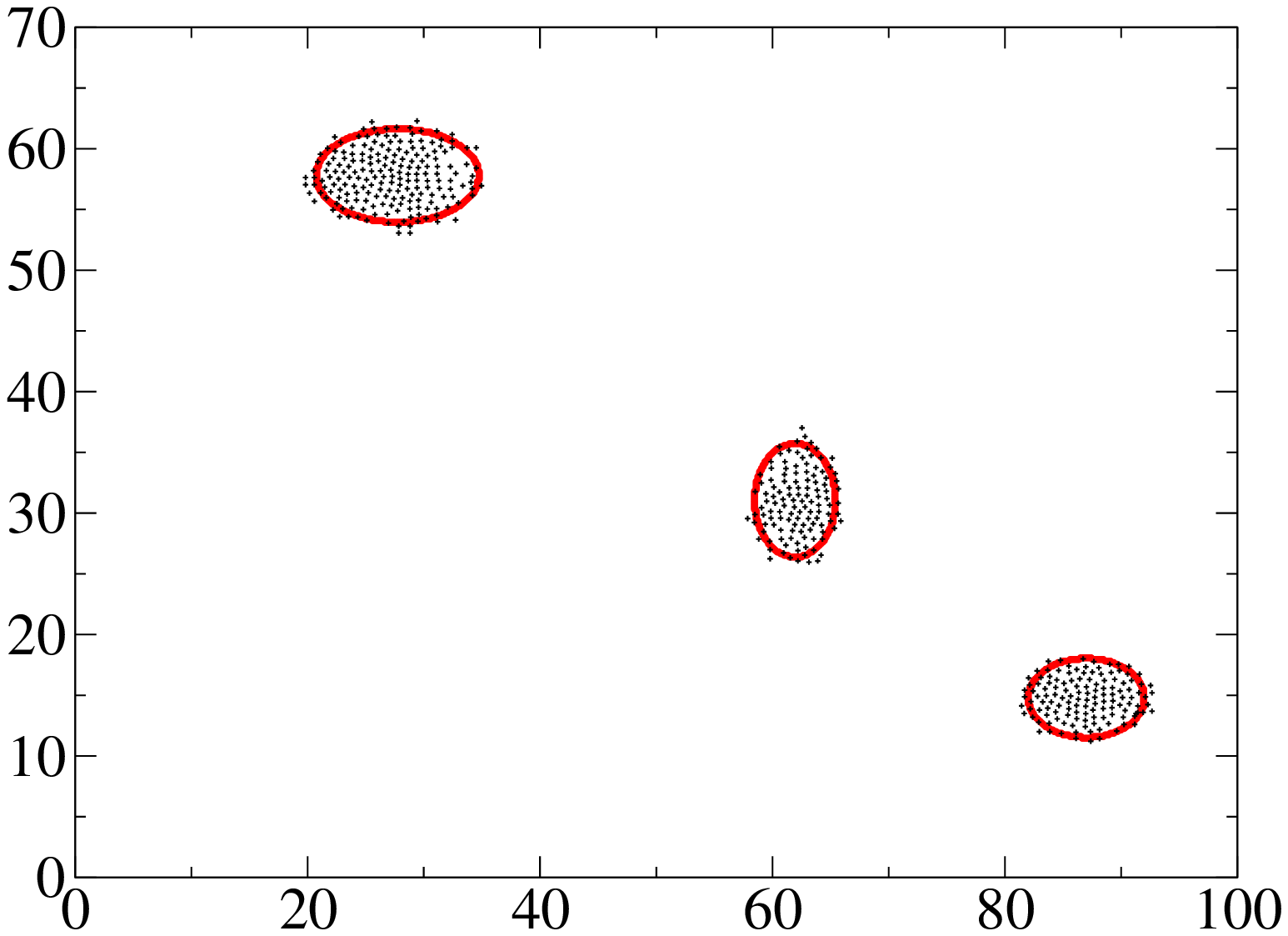}
}
\subfigure[]{
\includegraphics[scale=0.15]{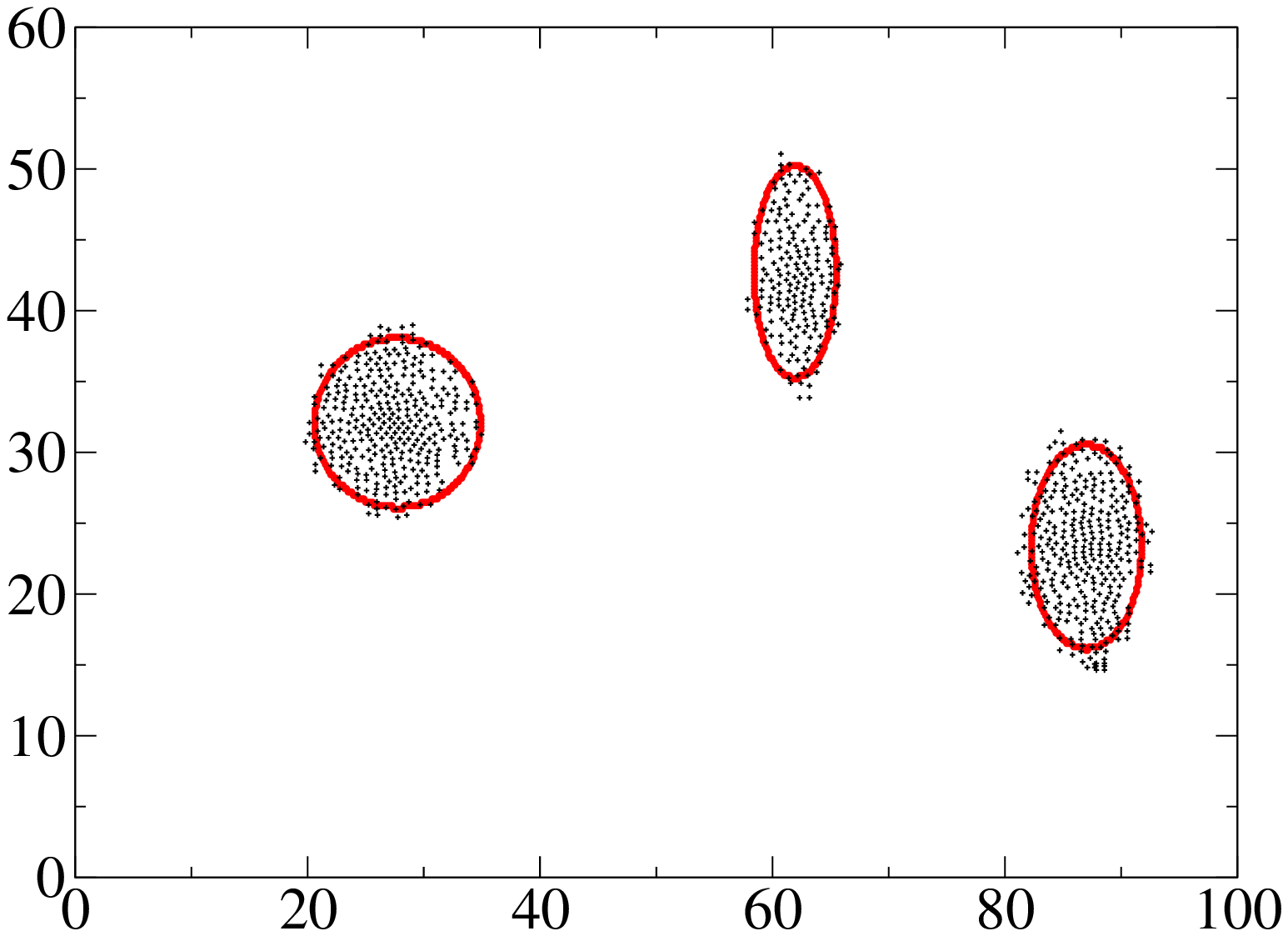}
}
\subfigure[]{
\includegraphics[scale=0.15]{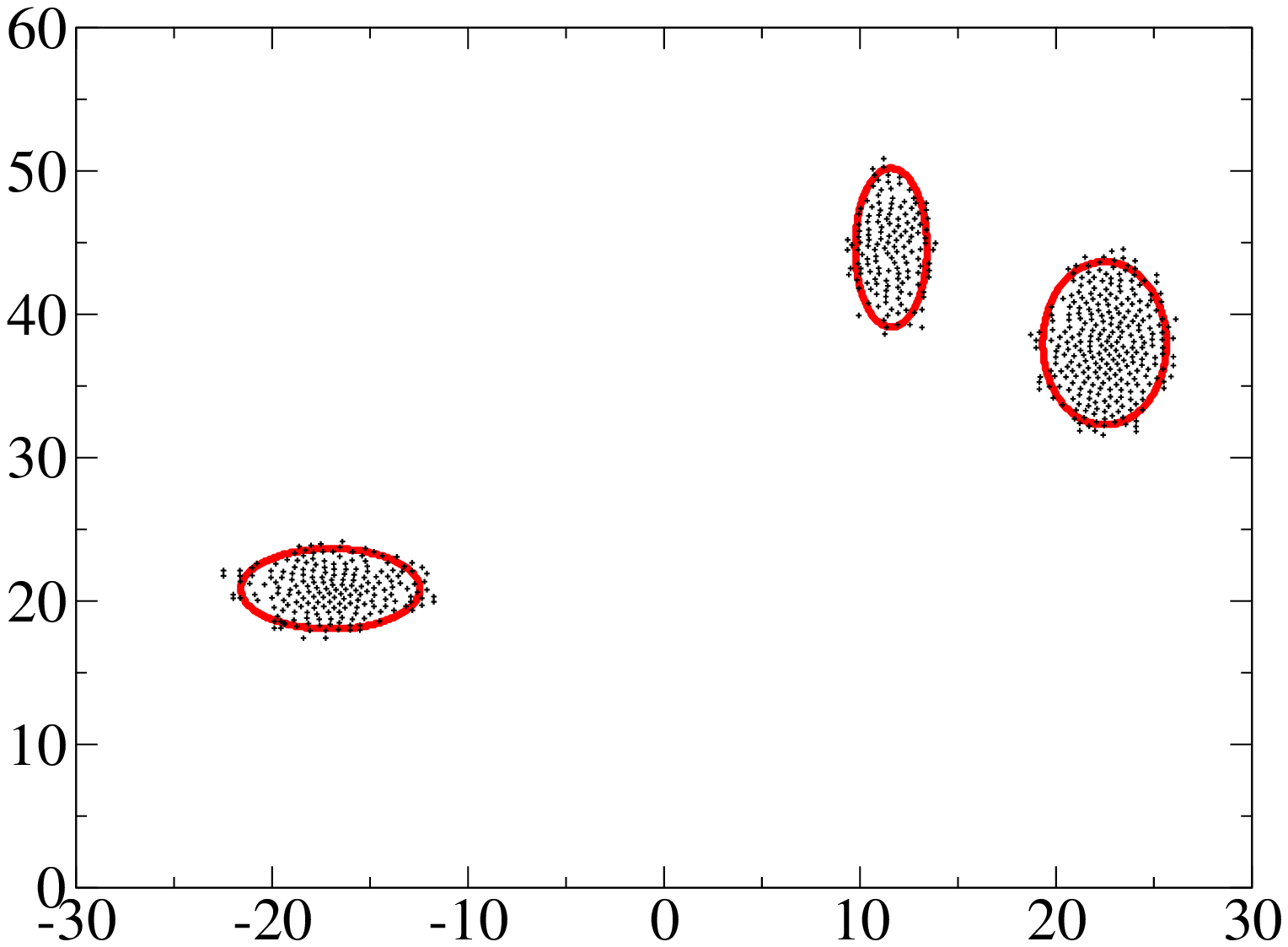}
}
\subfigure[]{
\includegraphics[scale=0.15]{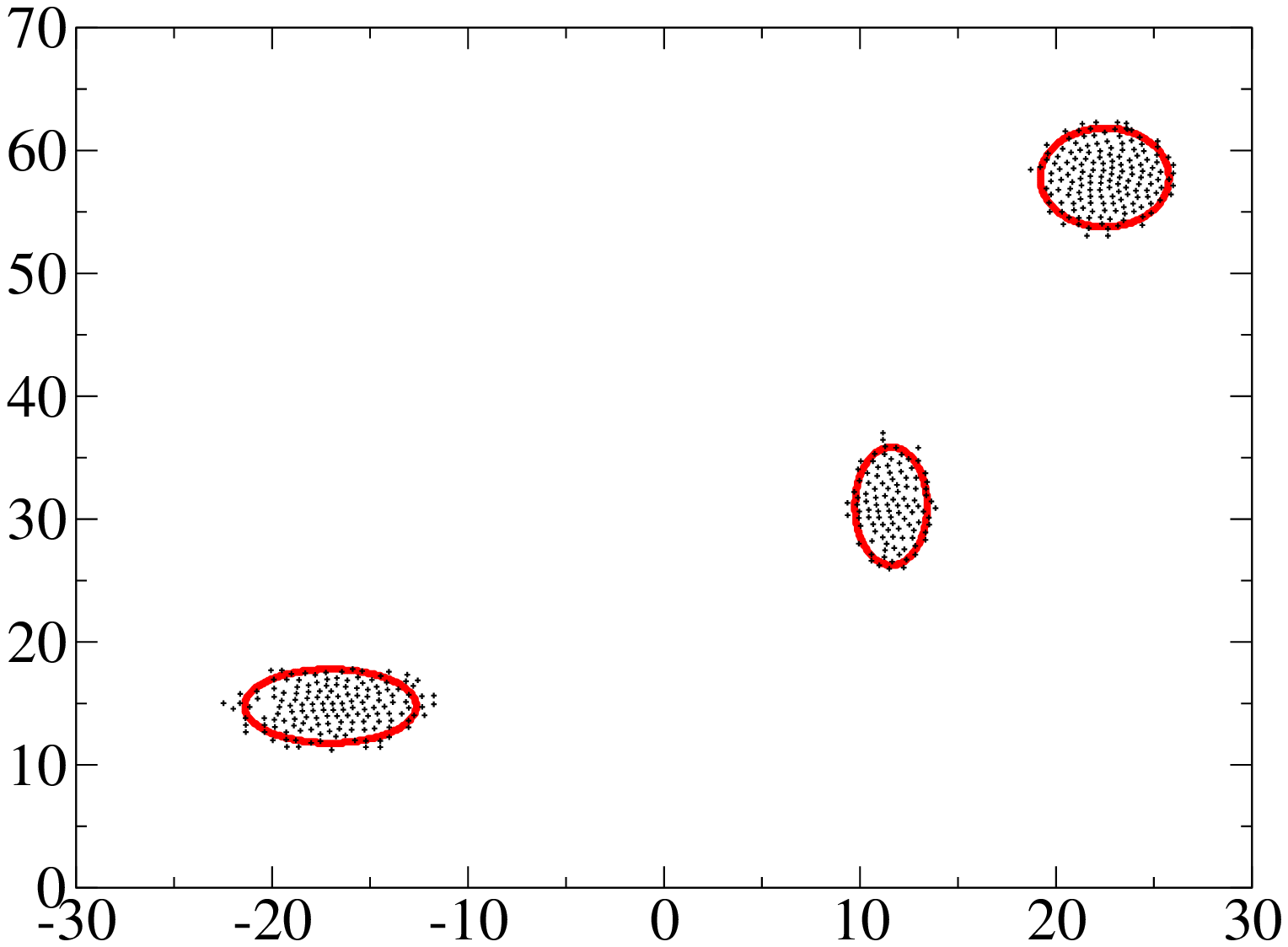}
}
\subfigure[]{
\includegraphics[scale=0.15]{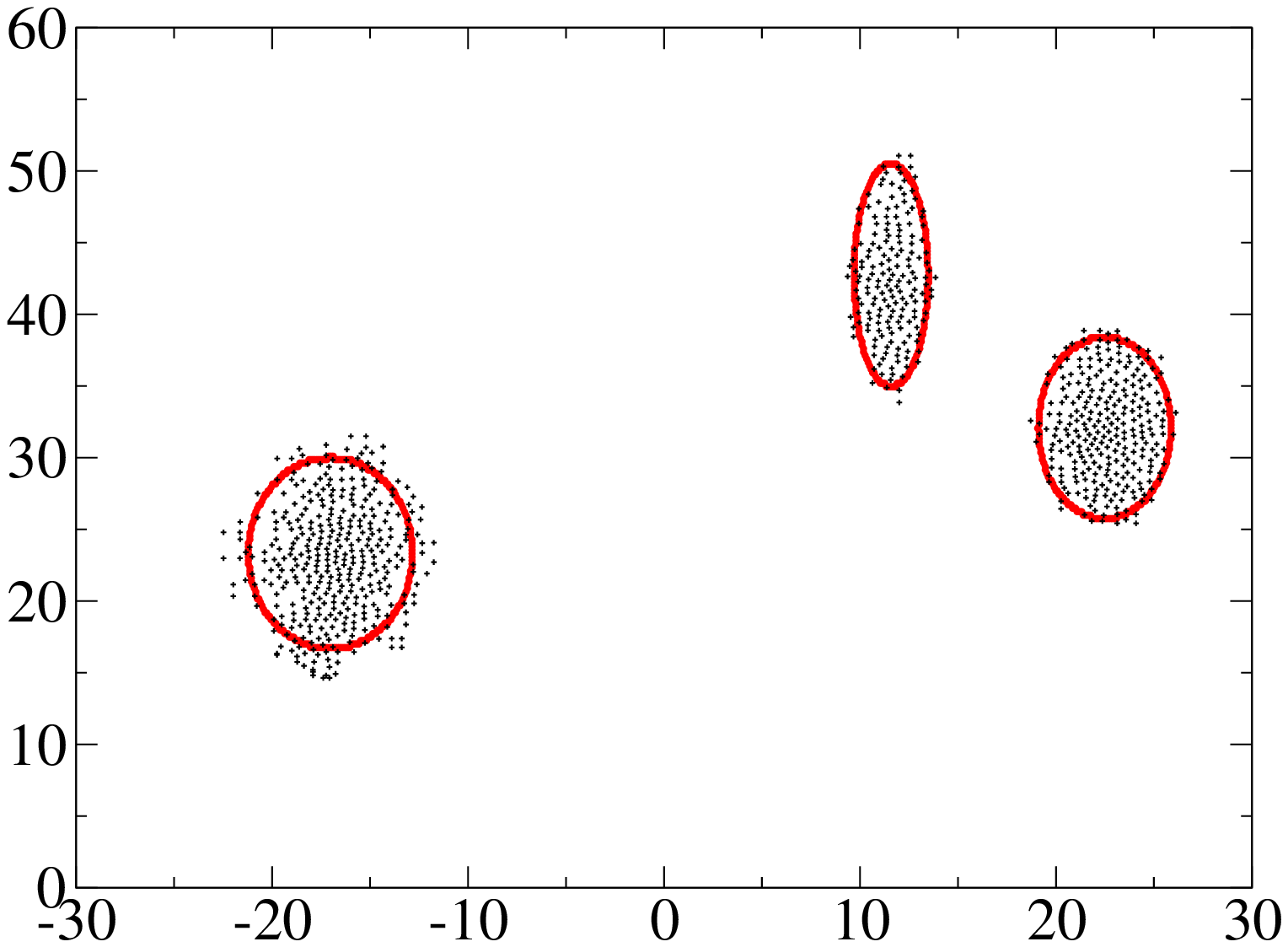}
}
\subfigure[]{
\includegraphics[scale=0.15]{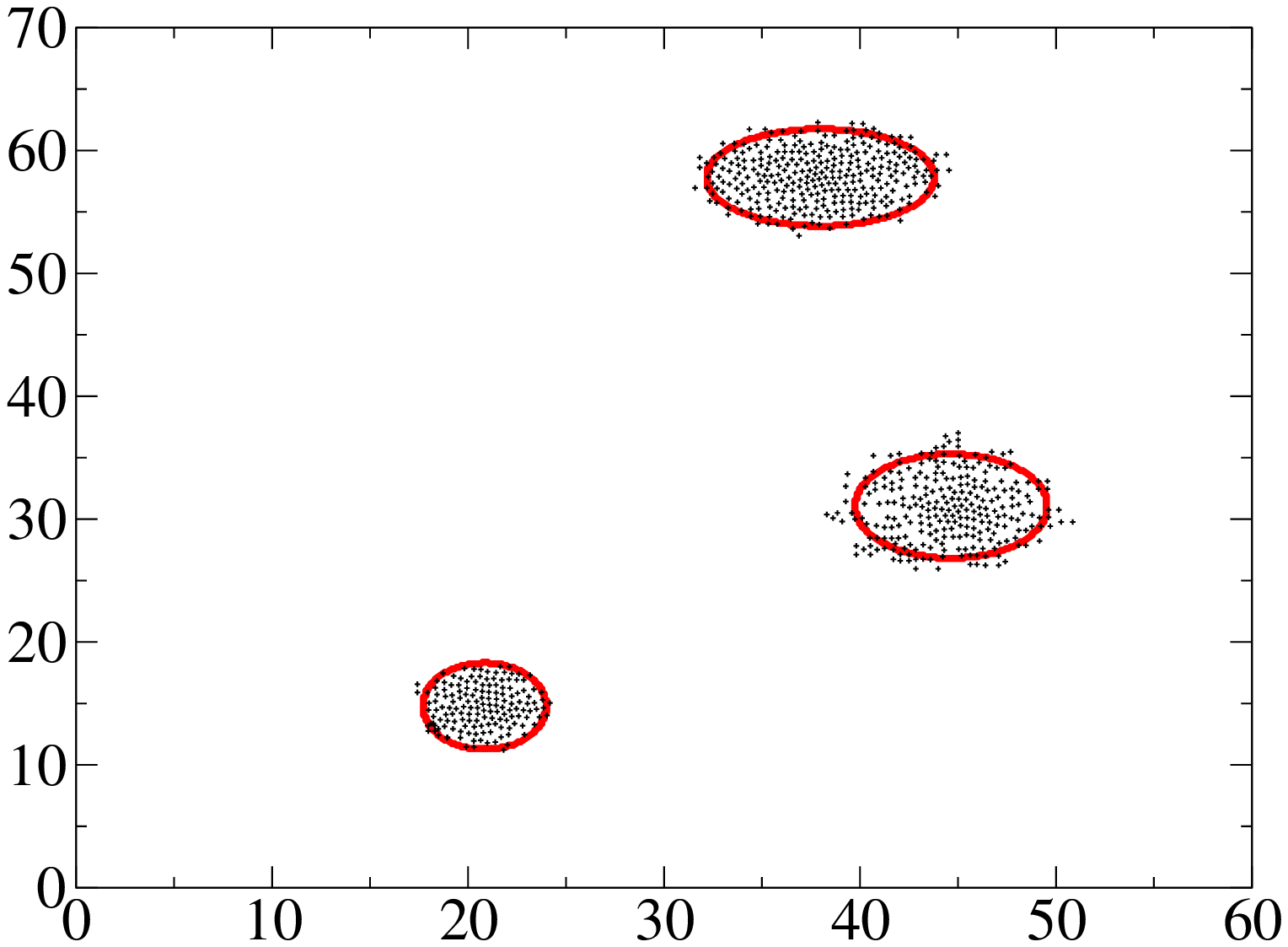}
}
\subfigure[]{
\includegraphics[scale=0.15]{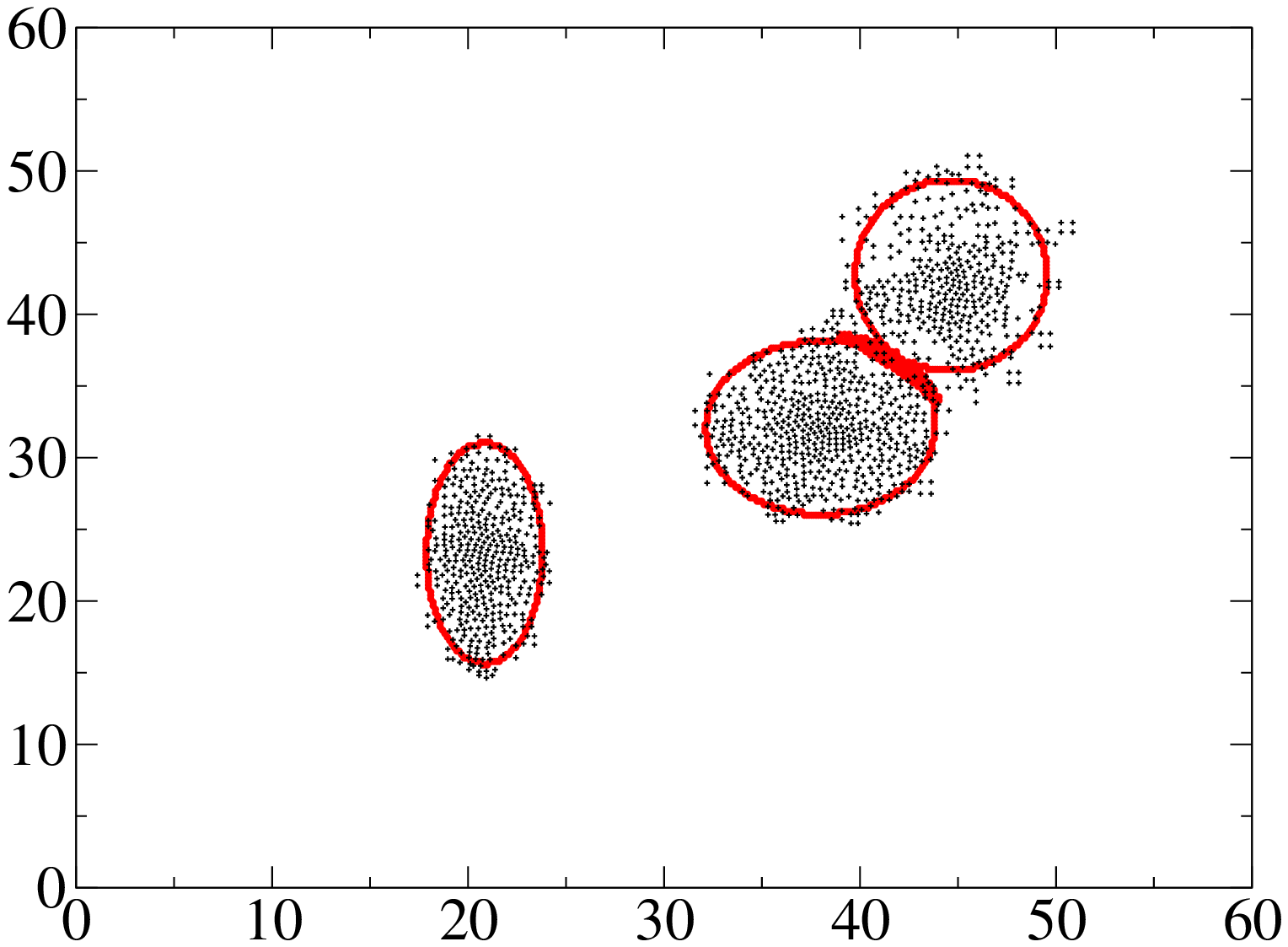}
}
\subfigure[]{
\includegraphics[scale=0.15]{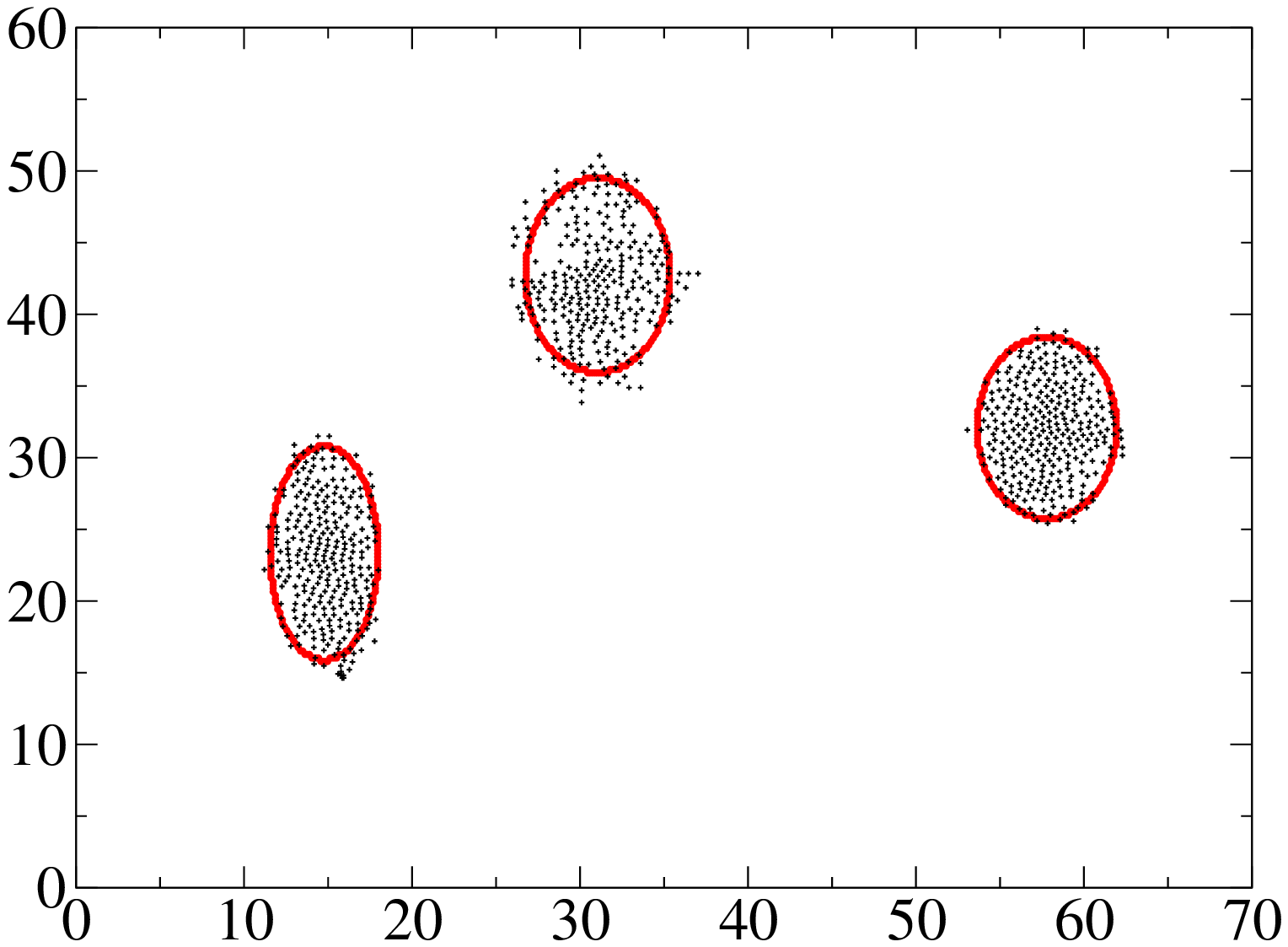}
}
\caption{
This plot shows all 10 2-dimensional sub-spaces of our 5-dimensional toy likelihood
function with 3 minima in $\chi^2$.  The black points represent the $95\%$ Bayesian credible
limit found by APS (and determined as described in Section \ref{sec:bayes}) after 10,000
samplings.  The red contours represent the known $95\%$ Bayesian credible limit of the toy
function.
}
\label{fig:toyBayes3}
\end{figure*}

\begin{figure*}
\subfigure[]{
\includegraphics[scale=0.15]{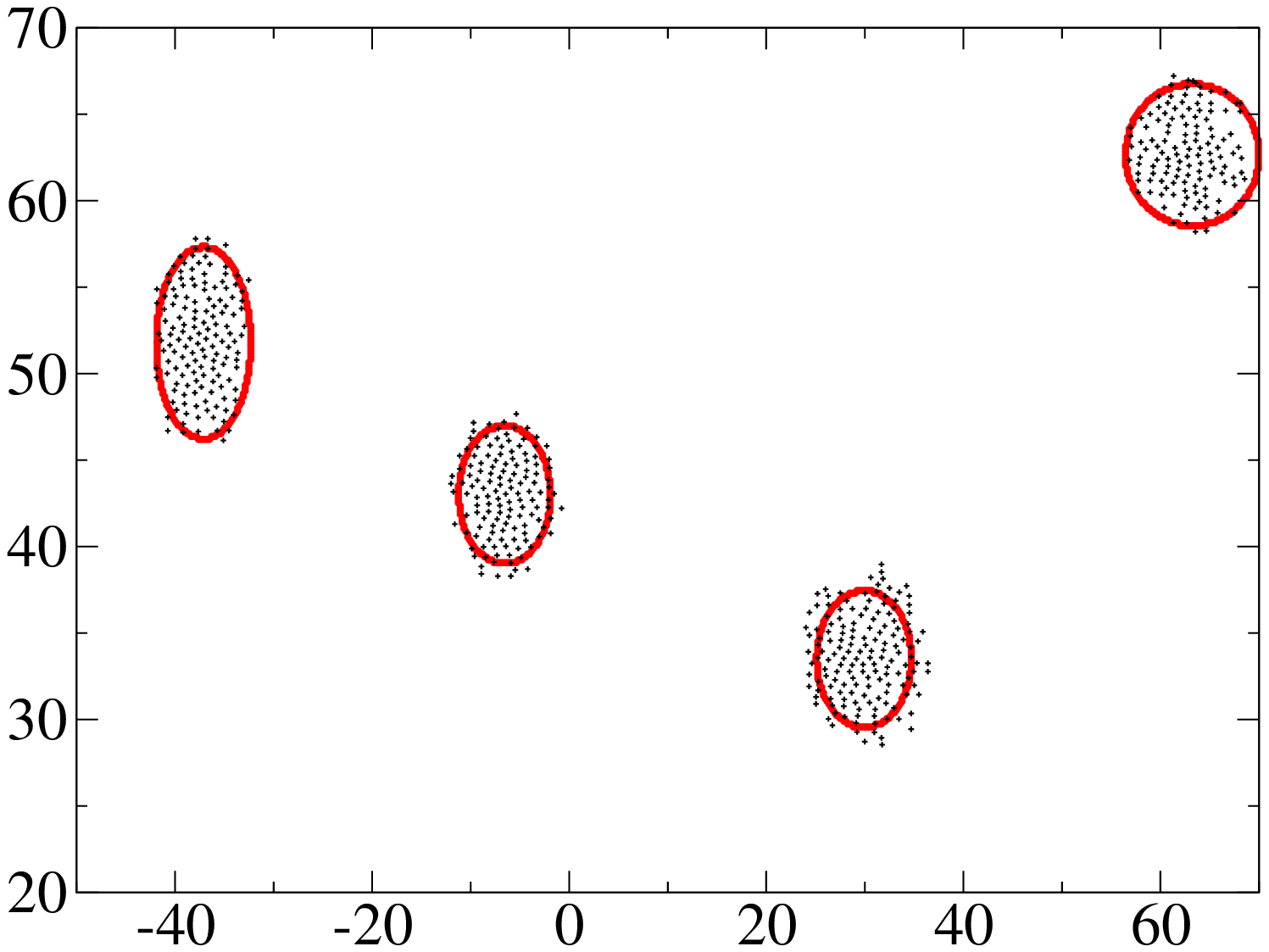}
}
\subfigure[]{
\includegraphics[scale=0.15]{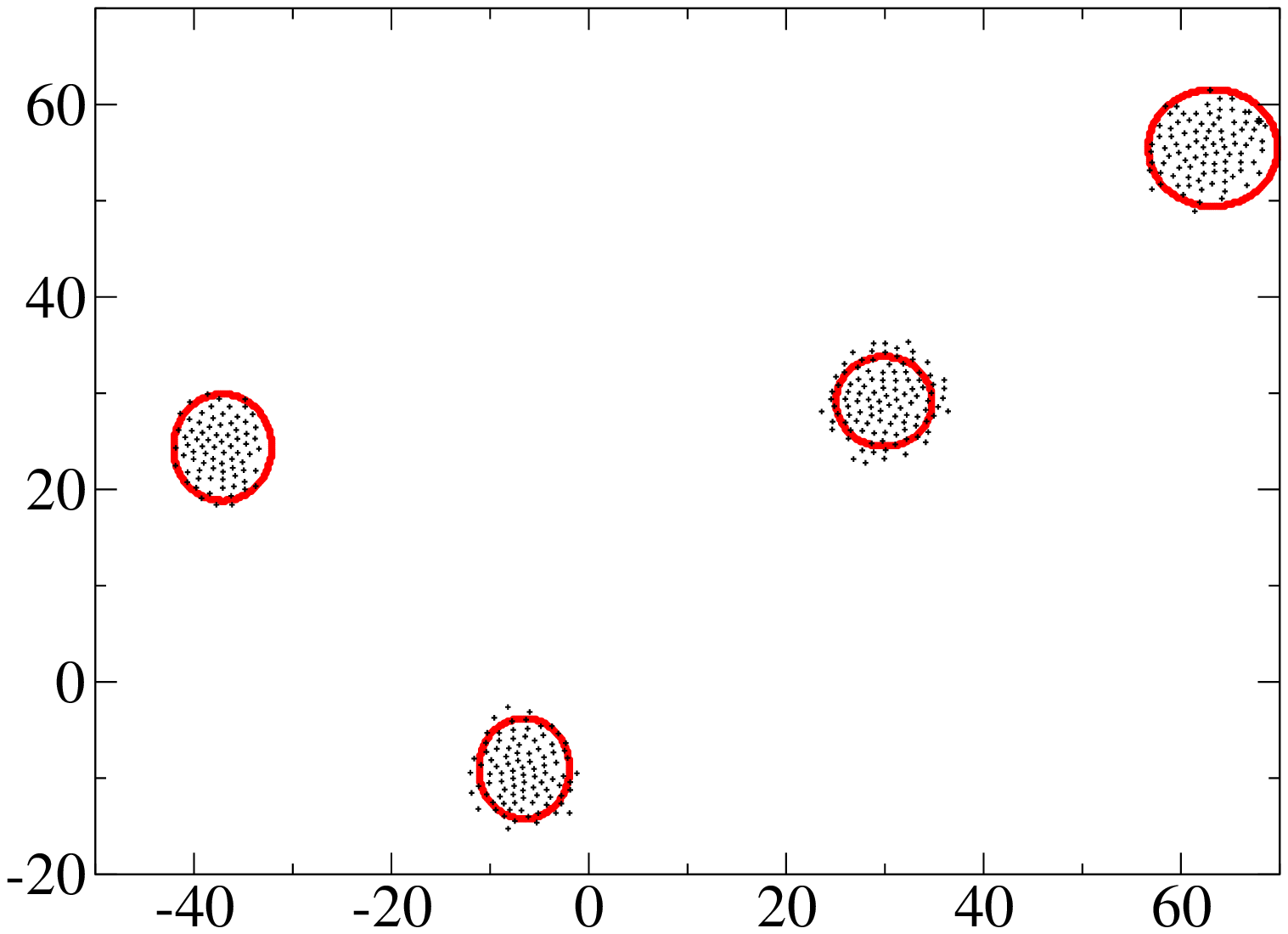}
}
\subfigure[]{
\includegraphics[scale=0.15]{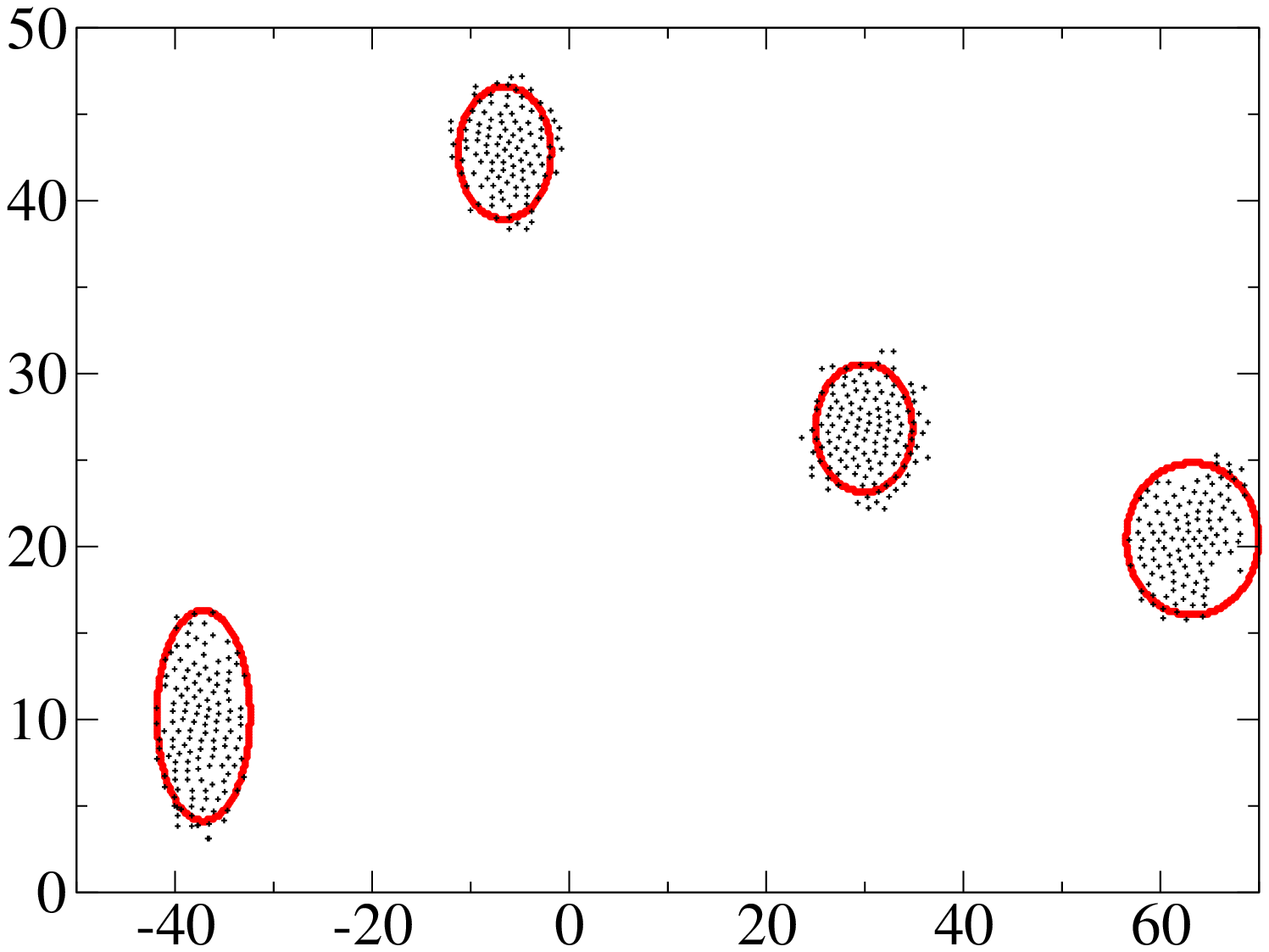}
}
\subfigure[]{
\includegraphics[scale=0.15]{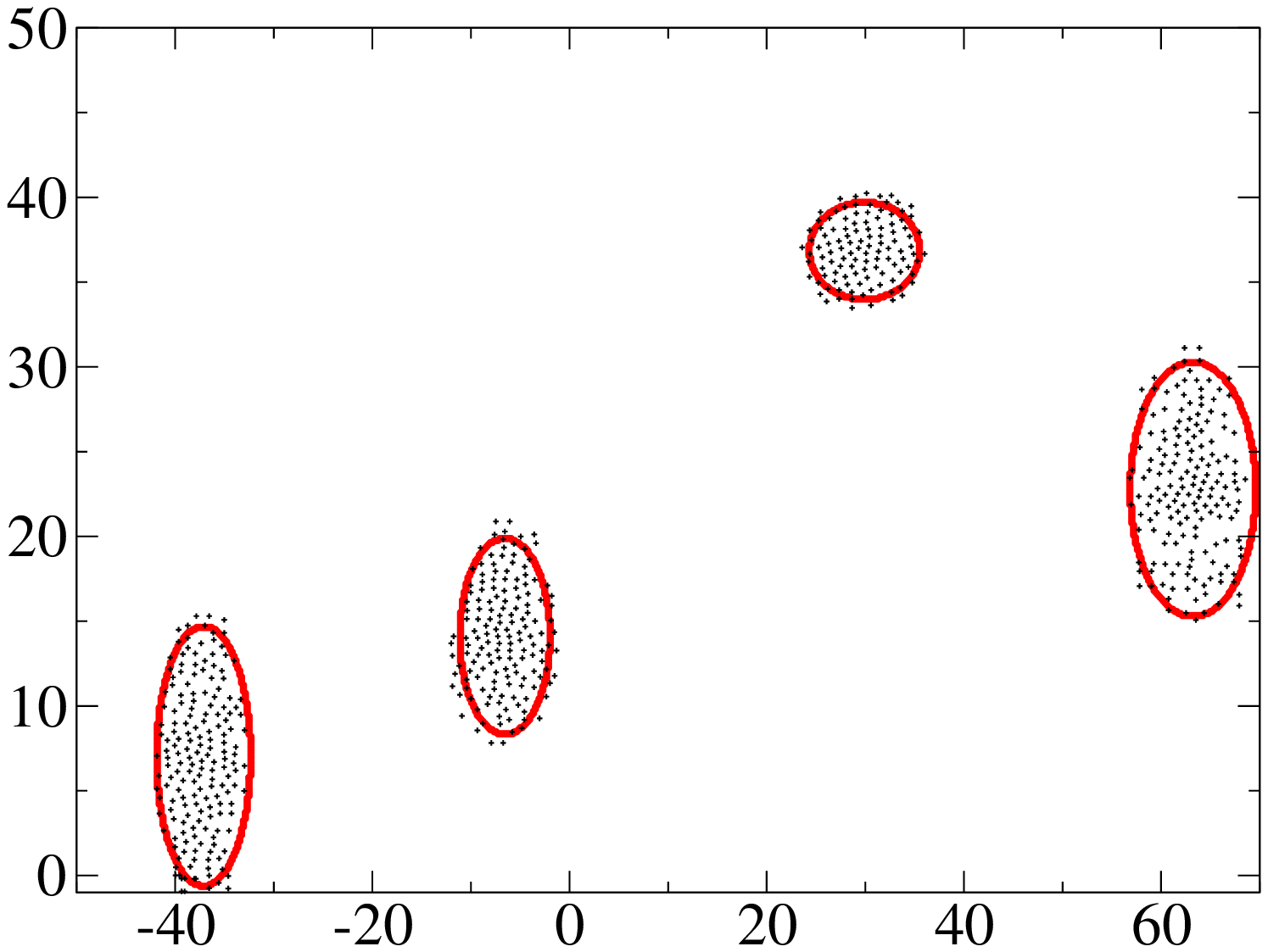}
}
\subfigure[]{
\includegraphics[scale=0.15]{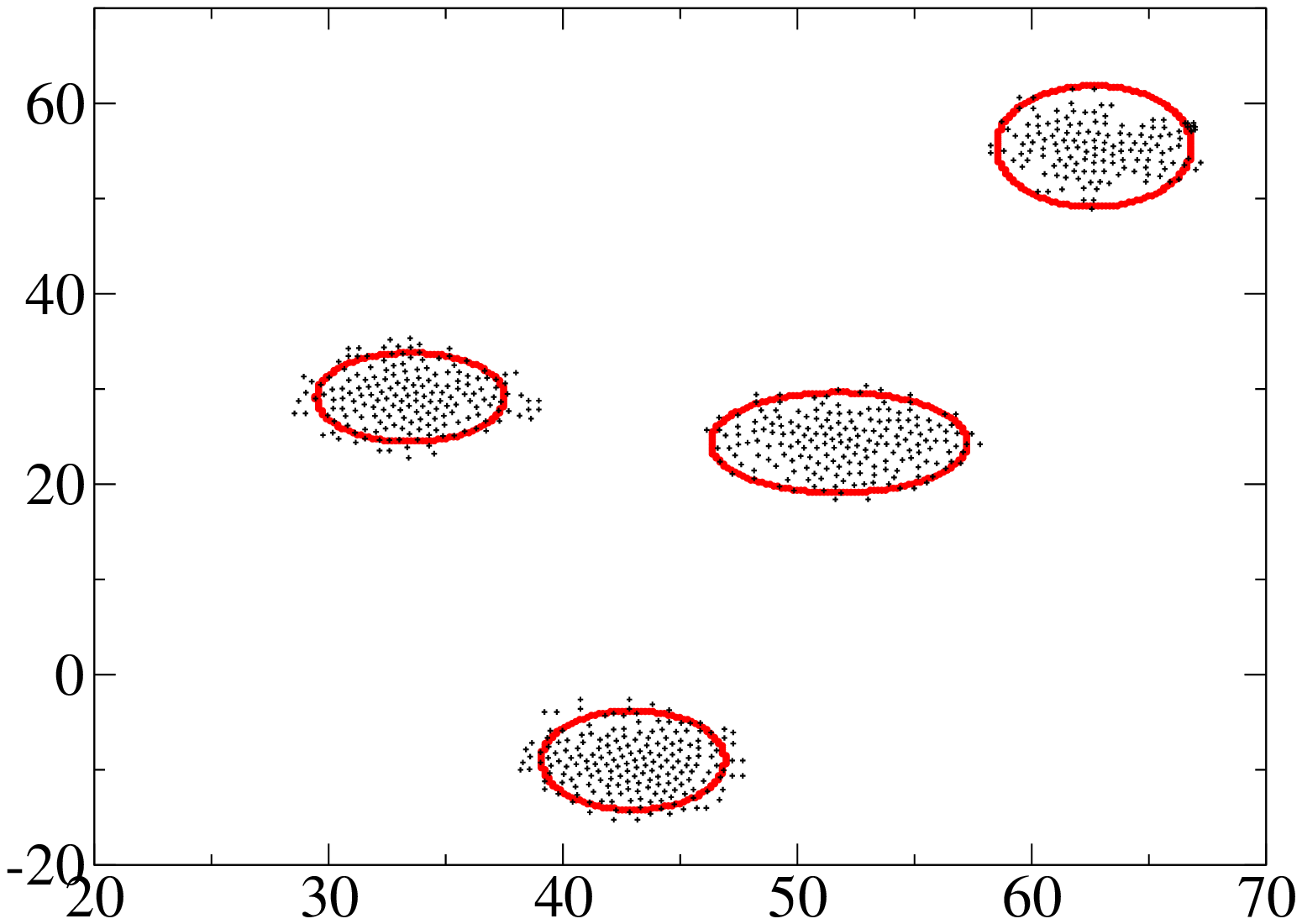}
}
\subfigure[]{
\includegraphics[scale=0.15]{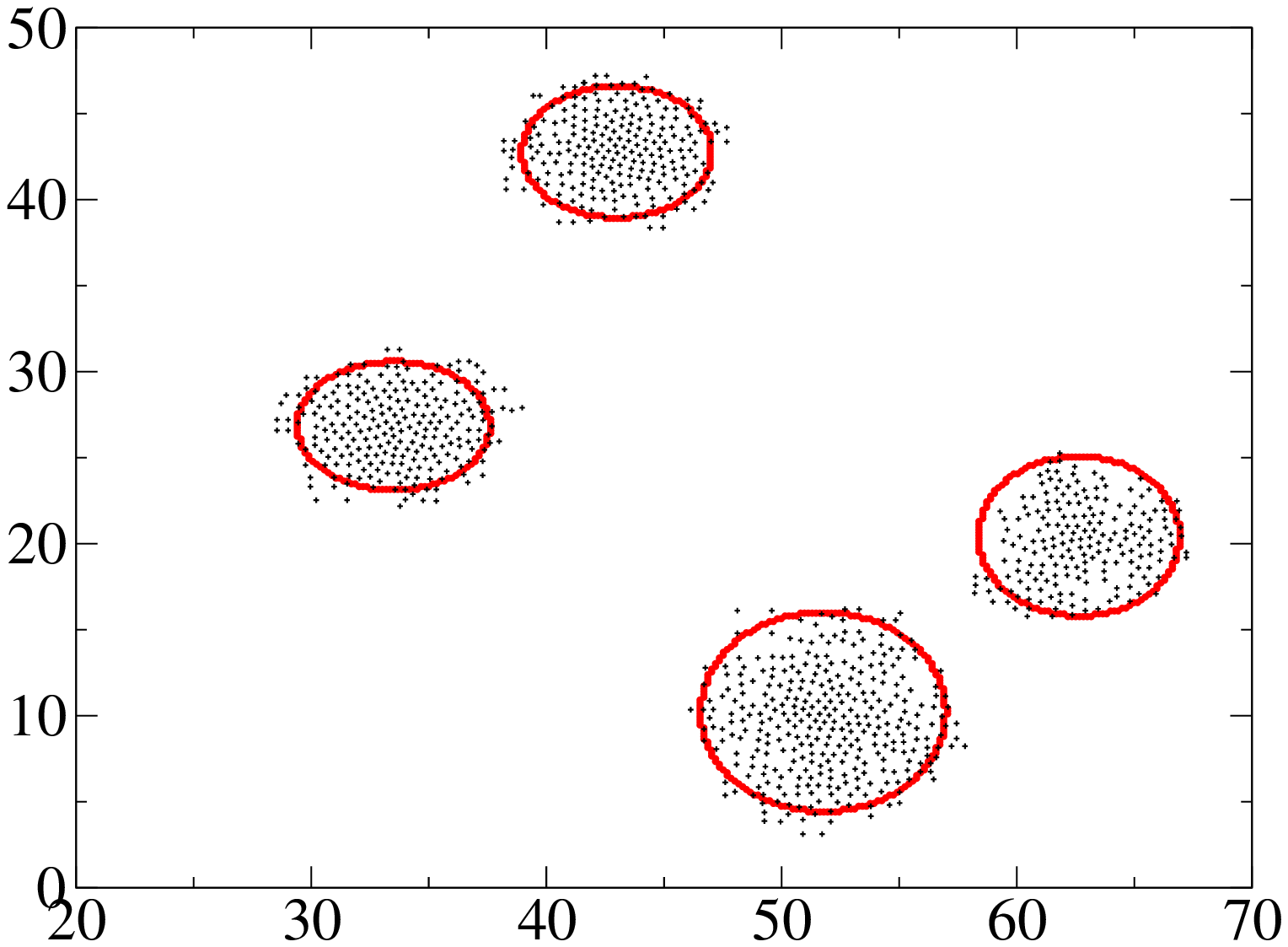}
}
\subfigure[]{
\includegraphics[scale=0.15]{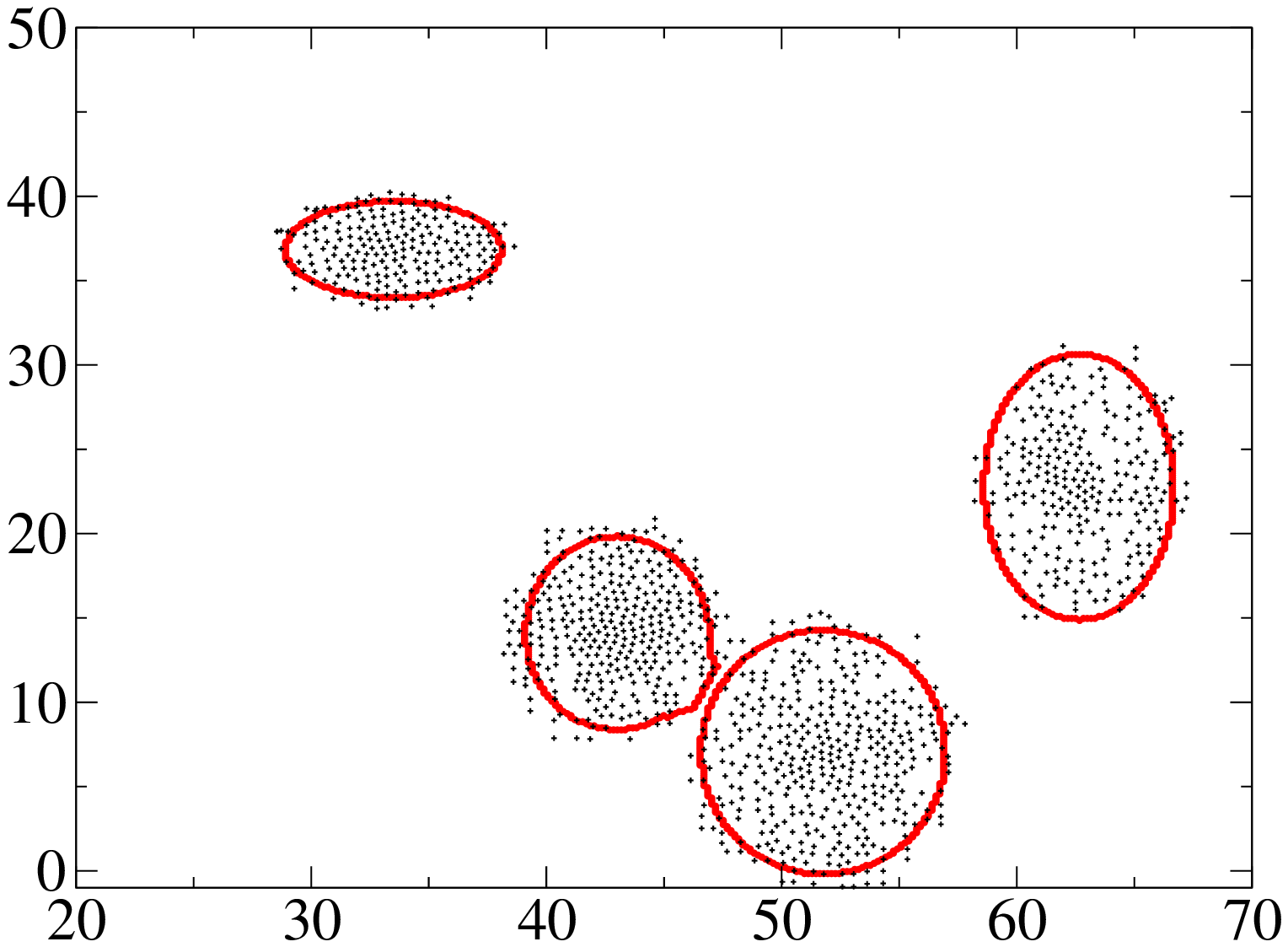}
}
\subfigure[]{
\includegraphics[scale=0.15]{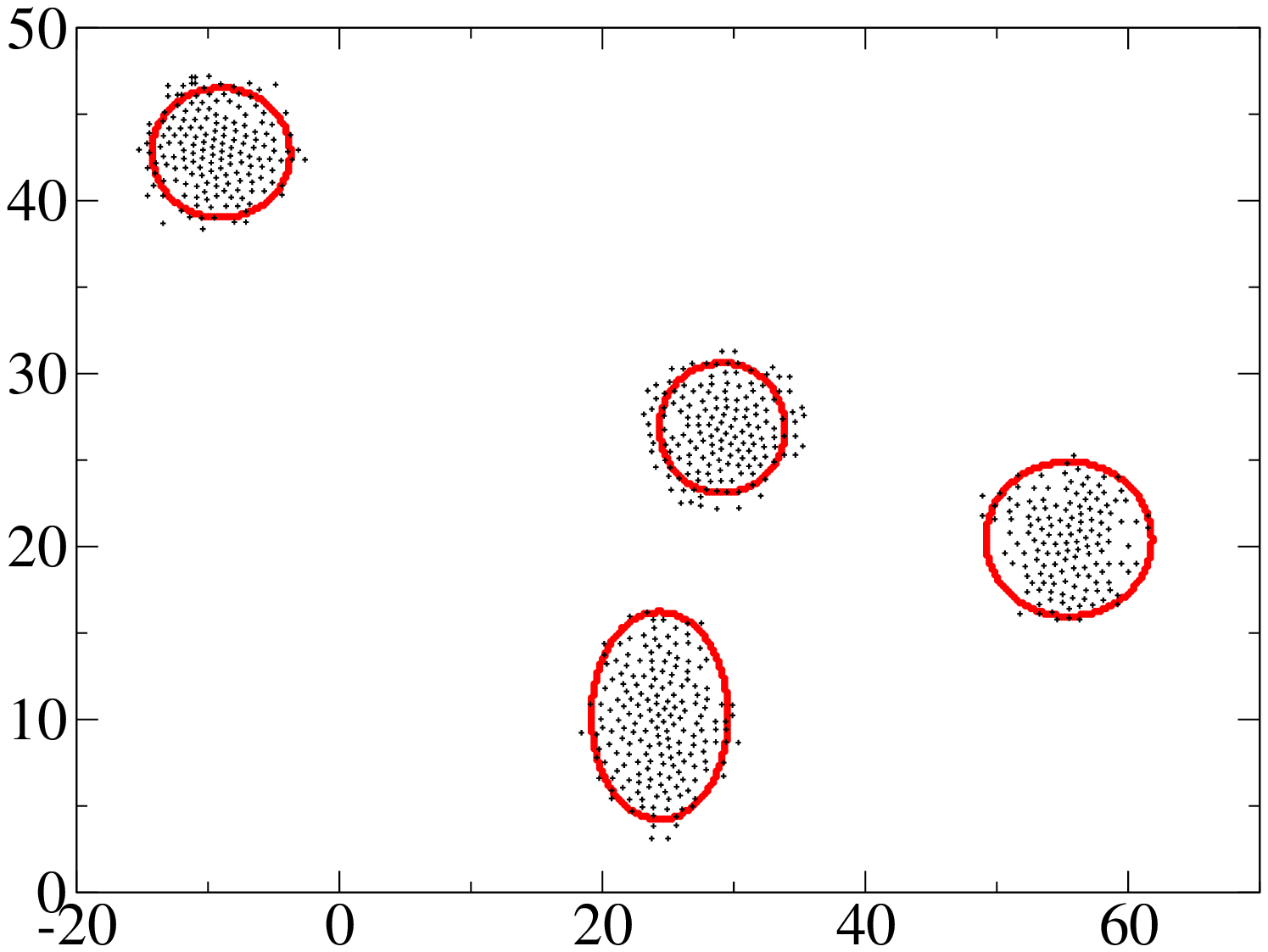}
}
\subfigure[]{
\includegraphics[scale=0.15]{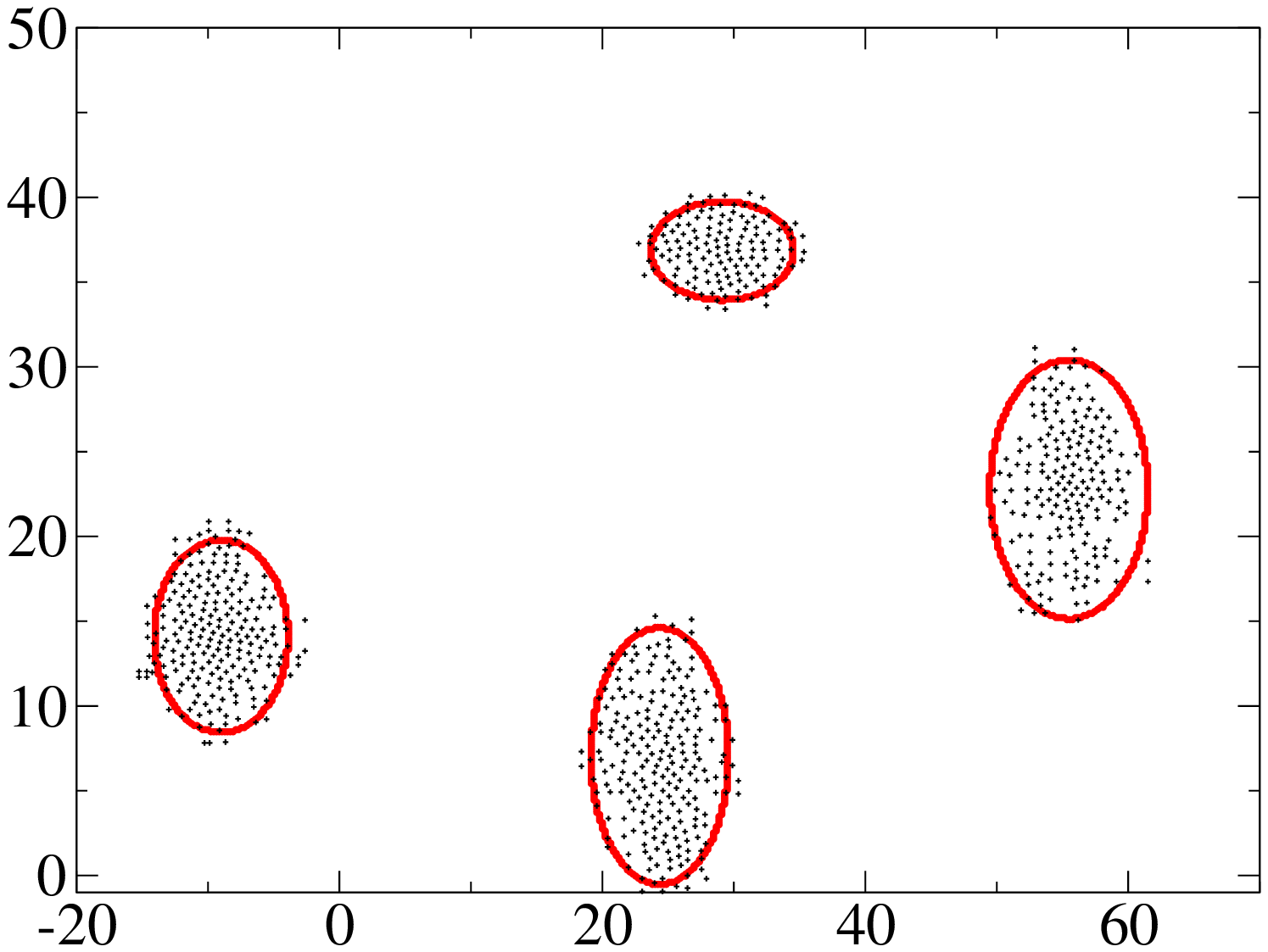}
}
\subfigure[]{
\includegraphics[scale=0.15]{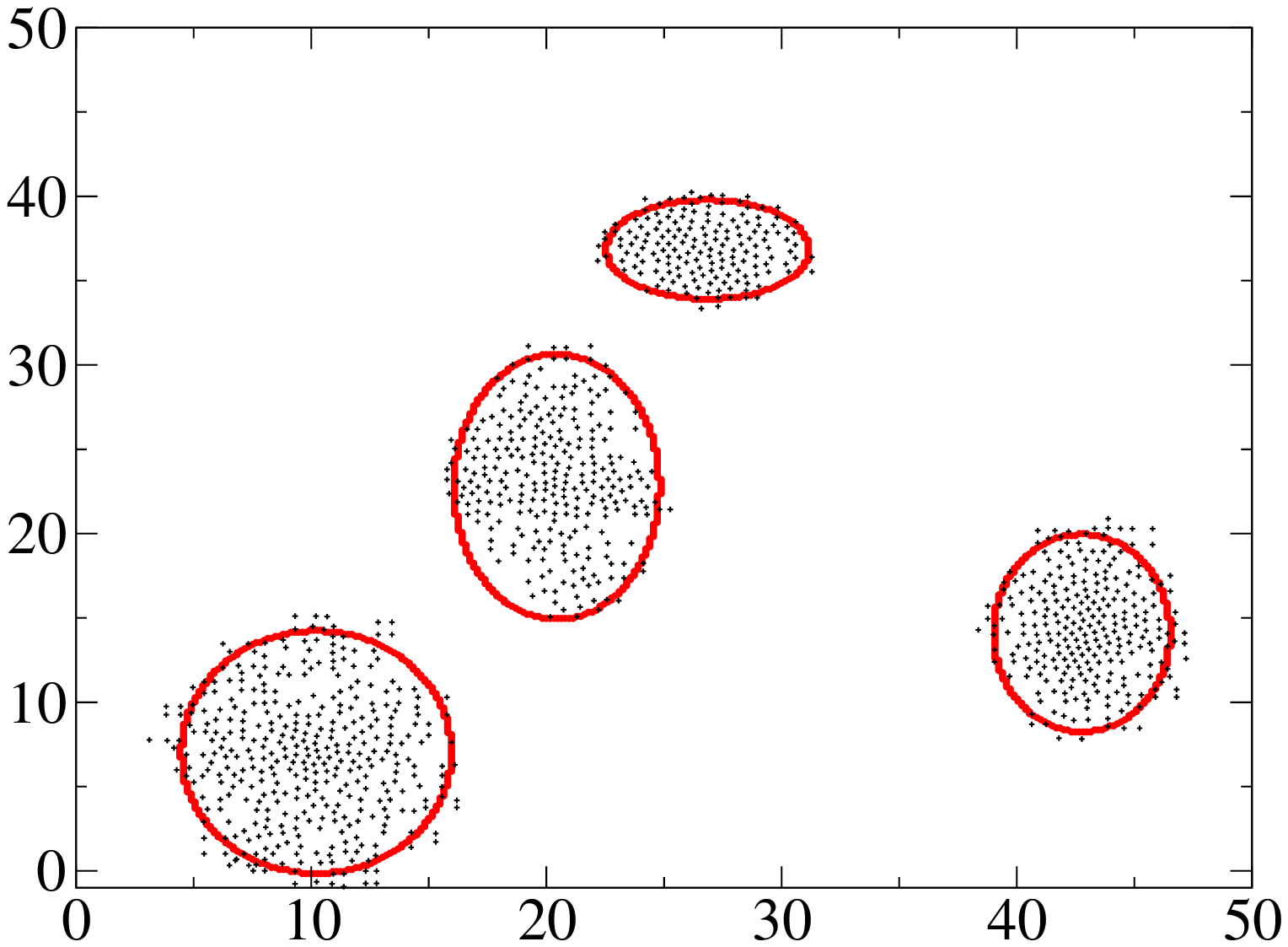}
}
\caption{
This plot shows all 10 2-dimensional sub-spaces of our 5-dimensional toy likelihood
function with 4 minima in $\chi^2$.  The black points represent the $95\%$ Bayesian credible
limit found by APS (and determined as described in Section \ref{sec:bayes}) after 10,000
samplings.  The red contours represent the known $95\%$ Bayesian credible limit of the toy
function.
}
\label{fig:toyBayes4}
\end{figure*}

As mentioned above, because our toy $\chi^2$ function is constructed without any
parameter degeneracies, we are able to directly integrate the posterior and
determine the true Bayesian credible limits in our parameter space.  Figures
\ref{fig:toyBayes2}, \ref{fig:toyBayes3}, and \ref{fig:toyBayes4} plot the true
95\% credible limits in all 10 2-dimensional sub-spaces of our 5-dimensional
parameter space (red contours).  They also show the APS-determined credible
limits as described in Section \ref{sec:bayes} (black points).  The
APS-determined credible limits correspond well with the true credible limits,
giving us confidence in APS's ability to map out Bayesian credible limits as well
as Frequentist confidence limits.

The tests above demonstrate APS's ability to correctly determine credible and
confidence limits.  They do not, however, address the question of APS's chief
advantage over Markov Chain Monte Carlo methods: its ability to successfully
locate all of the low-$\chi^2$ regions in a given parameter space.  To
demonstrate this, we ran APS 100 times on each of the likelihood functions
plotted in Figures \ref{fig:toyFreq2}-\ref{fig:toyFreq4}.  Each time, the
pseudo-random number generator driving APS was given a different seed.  In 
nearly all
cases, APS successfully located all of the local minima in $\chi^2$, usually in
fewer than 10,000 evaluations of $\chi^2$ (one case of the 2-mode $\chi^2$
function required 10,352 evaluations to find both low-$\chi^2$ regions; one
required 12,549 evaluations).  For comparison, we also ran 100 Markov Chain
Monte Carlo searches on these $\chi^2$ functions, each run consisting of five
independent chains.  We allowed the MCMCs to run
until they had sampled 100,000 points in parameter space.  
These searches were not nearly as successful as APS at
locating all of the low-$\chi^2$ regions.  In the case of the 2-mode function,
50 of the 100 MCMC searches failed to identify both modes.  In the case of the
3-mode function, 56 of the 100 MCMC searches failed to identify all of the
modes.  In the case of the 4-mode function, 87 of the 100 MCM searches failed to
identify all of the modes.  This is not to say that APS is perfect at
identifying all of the modes of a likelihood function.  If the
characteristic widths $w_{j,i}$ in equation (\ref{eqn:toychi}) are too small, 
we enter a regime
in which neither APS nor MCMC is capable of reliably
identifying all of the modes.  However, this test does indicate that APS is more
robust than MCMC against multi-modal likelihood functions, as was first
demonstrated by Bryan {\it et al}. (2007).

This should not be a surprising statement.  In order to speed convergence,
MCMC attempts to learn the size and shape of any high likelihood region it discovers.
It uses what it learns to propose new random steps in parameter space so that it can
efficiently converge to a state in which it is sampling the Bayesian posterior.  If
an MCMC chain discovers one of several high likelihood regions, it will learn a proposal
density that conforms to that high likelihood region, disregarding any other that may
exist.  It will, in other words, become trapped in that high likelihood region.
MCMC can overcome this difficulty by running several chains in parallel and hoping that,
if multiple high likelihood regions exist, each chain will become trapped in different
region and the whole MCMC will thus learn all of the modes in the parameter space.
This, however, requires initializing a number of chains that is large relative to both
the dimensionality of the parameter space and the number of discrete high likelihood
regions it contains so that the odds of discovering all of the high likelihood regions
are large.  If the user does not know how many high likelihood regions exist in
the parameter space, this could be problematic.  Because steps (2\~A-4\~A) keep APS
exploring widely through parameter space, even after a high likelihood region has been
identified, APS can discover the presence of
multiple high likelihood regions without requiring the user's prior knowledge of them.

\section{A Real Example: WMAP}
\label{sec:wmap}

Section \ref{sec:toy} demonstrates the robustness of APS against multi-modal
$\chi^2$ functions.  However, it does have some short-comings as a test of
APS's abilities.  As was already noted, there are no interesting parameter
degeneracies in equation (\ref{eqn:toychi}).  Furthermore,
because this function was constructed artificially, there is no noise.  At
any point in parameter space, $\chi^2$ falls smoothly and monotonically towards
the associated local minimum of $\chi^2$.  To really have confidence in APS
abilities, we must test it on actual physical data drawn from the universe,
with all of the noisy and ill-behaved properties that entails.
To that end, we will now consider for our $\vec{\theta}\rightarrow\chi^2$ 
function the likelihood function on the
7-year data release of the WMAP CMB satellite \cite{wmap7,wmap7likelihood}.
For simplicity, we consider only the anisotropy
power spectrum of the temperature-temperature correlation function.

The parameter space we explore is the six-dimensional space of the spatially
flat concordance $\Lambda$CDM cosmology:
\begin{itemize}
\item$\Omega_bh^2$ -- the density of baryons in the Universe
\\
\item$\Omega_\text{CDM}h^2$ -- the density of cold dark matter in the Universe
\\
\item$\tau$ -- the optical depth to the surface of last scattering.  $\tau$ cannot be
constrained without polarization data, which we do not consider here.  However, it functions
as a nuisance parameter to be marginalized over.  The fact that APS can handle it as well as
MCMC is an important test of APS's feasibility.
\\
\item$h$ -- the Hubble parameter in units of
$100~\text{km}~\text{s}^{-1}~\text{Mpc}^{-1}$
\\
\item$n_s$ -- the spectral index controlling the distribution of scalar density
perturbations in the Universe
\\
\item$\ln[10^{10}A_s]$ -- the amplitude of the initial scalar density
perturbations in the Universe
\end{itemize}
These are the parameters taken by the publicly available MCMC code CosmoMC,
which we use to run our baseline MCMC chains \cite{cosmomc}.  The
$\vec{\theta}\rightarrow\chi^2$ function in this case involves converting these
parameters into the power spectrum of anisotropy $C_\ell$s on the sky and
comparing those $C_\ell$s to the actual measurements registered by the WMAP
satellite.  Using the (very fast) Boltzmann code CAMB to calculate the $C_\ell$s
\cite{camb} takes 1.3 seconds on a 2.5 GHz processor.  
Evaluating the $\chi^2$ using the likelihood code provided by the WMAP team
\cite{wmap7likelihood} takes an additional 2 seconds (this can, of course, be sped up by using a faster
machine and multiple cores).  Hence the need
for an algorithm like APS or MCMC to accelerate the search.
We will consider both equations (\ref{eqn:GaussianCovar}) and (\ref{eqn:NNCovar})
as the covariogram for our Gaussian process, 
though we find that the choice has little effect on our
results.  We will also consider setting $\chi^2_\text{lim}=\chi^2+\Delta\chi^2$ 
as wel as setting $\chi^2_\text{lim}$ by hand.

\begin{figure}
\includegraphics[scale=0.3]{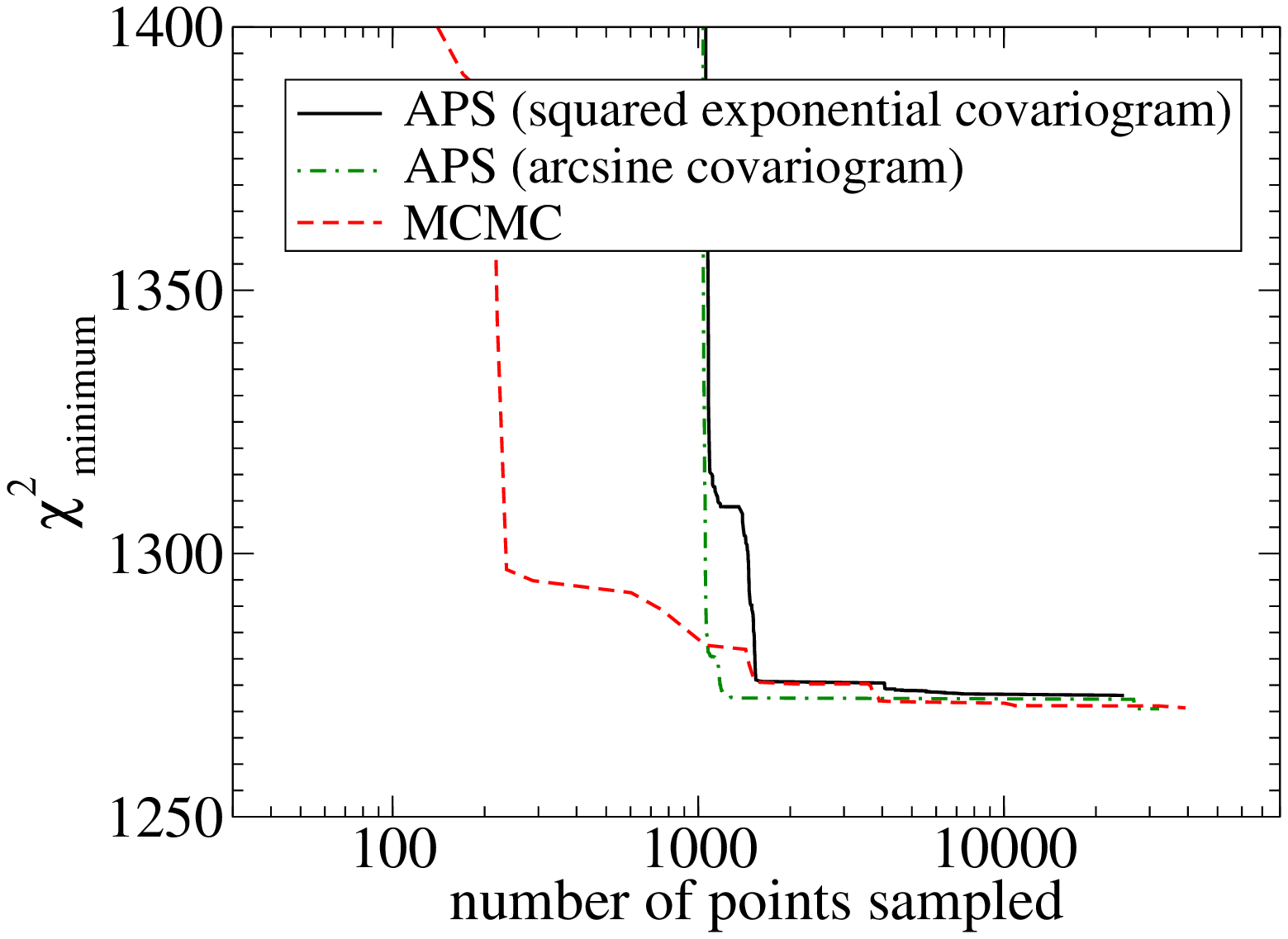}
\caption{
The minimum value of $\chi^2$ found by APS and MCMC on the WMAP 7 problem
as a function of 
steps sampled by each.
}
\label{fig:chimin}
\end{figure}

To gauge the appropriateness of applying APS to this problem, we consider
the two following tests.  In Figure \ref{fig:chimin} we plot the minimum discovered value
of $\chi^2$ yielded by APS and MCMC as a function of the number of samples drawn
by each.  If APS failed to find a suitable value of $\chi^2_\text{min}$, the use of
$\Delta\chi^2$  to find $\chi^2_\text{lim}$ would be questionable.  However, as we can see,
the $\chi^2_\text{min}$ yielded by APS ($1273.3$) 
is within a few of the $\chi^2_\text{min}$ yielded
by MCMC ($1270.7$).

\begin{figure*}
\subfigure[]{
\includegraphics[scale=0.15]{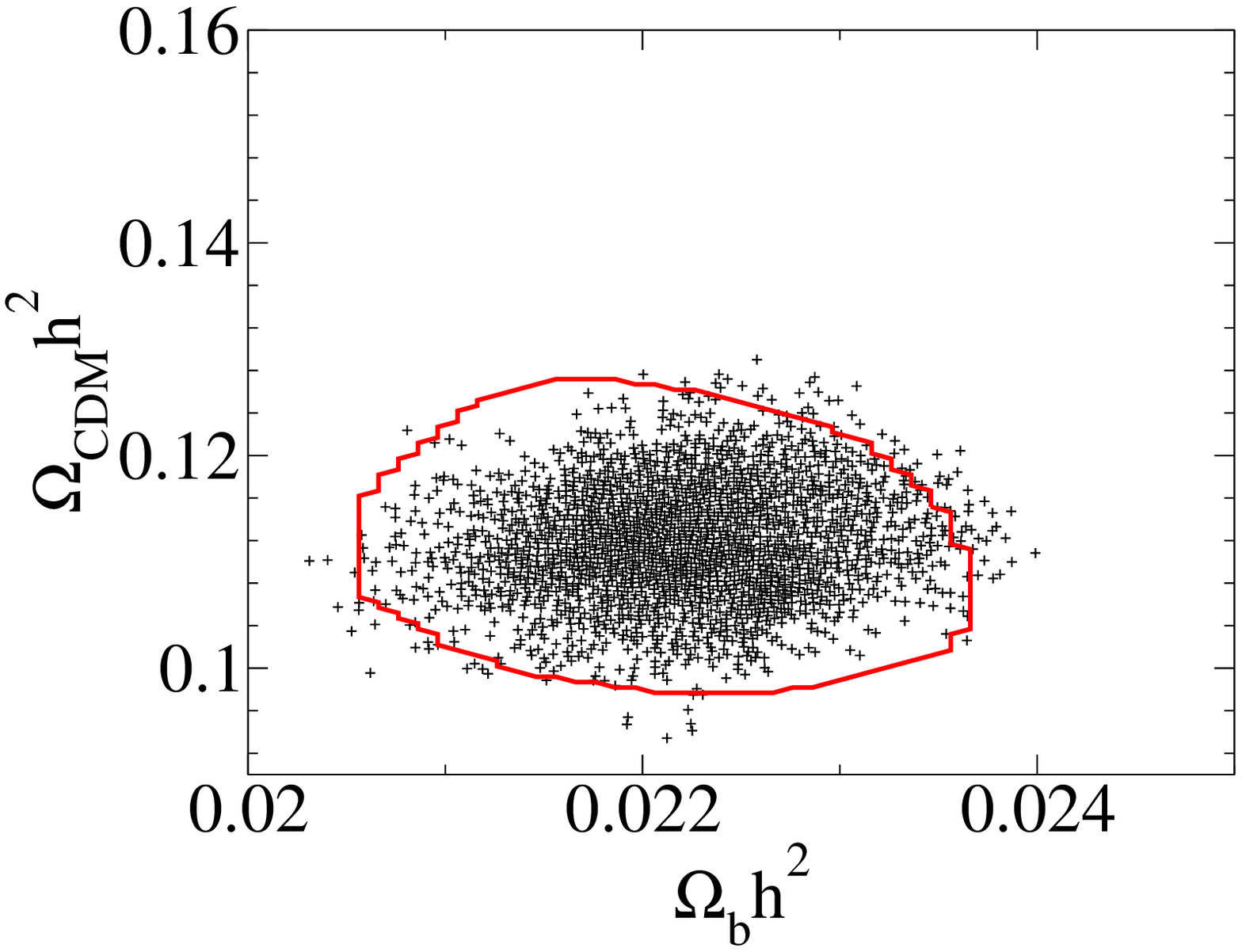}
}
\subfigure[]{
\includegraphics[scale=0.15]{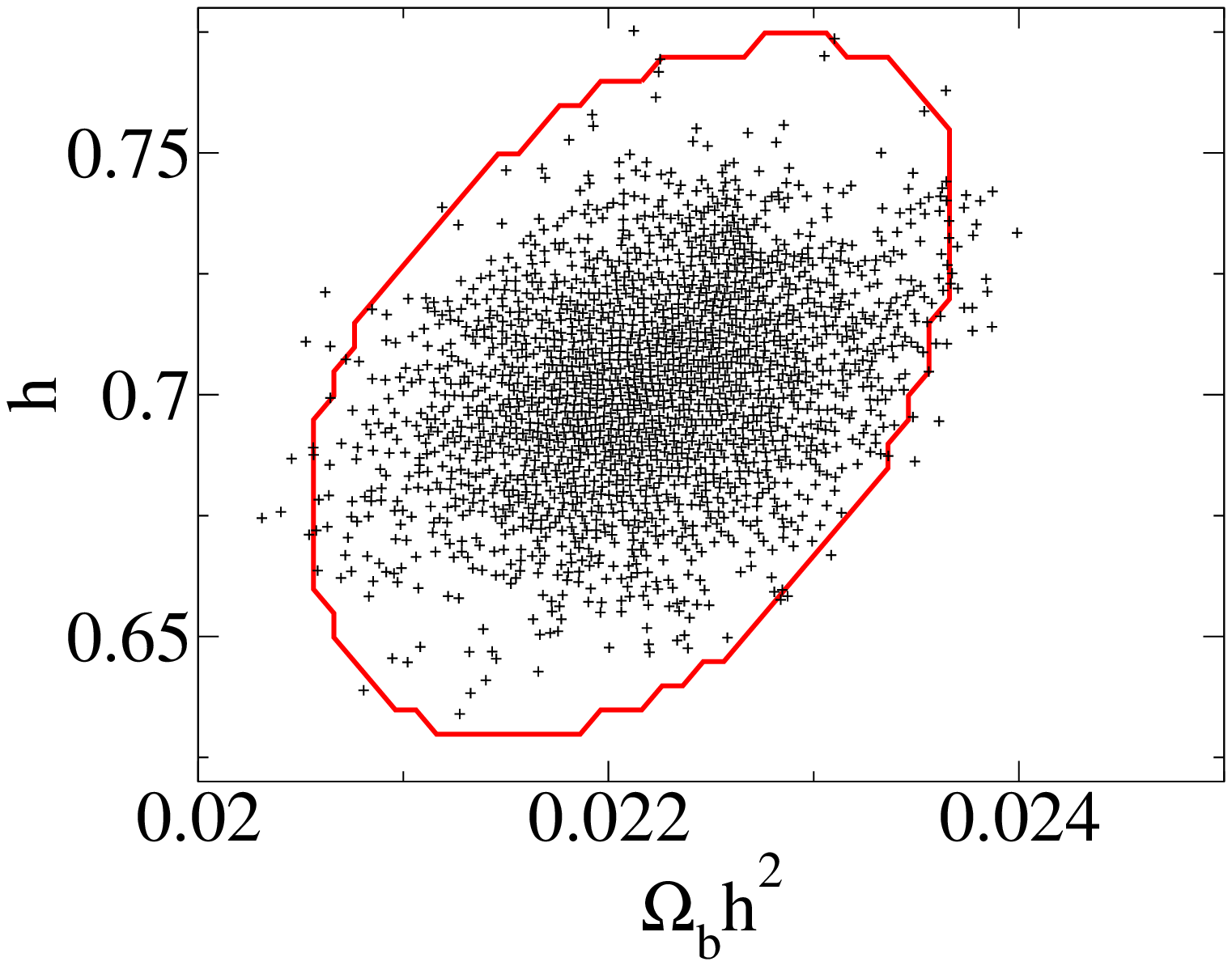}
}
\subfigure[]{
\includegraphics[scale=0.15]{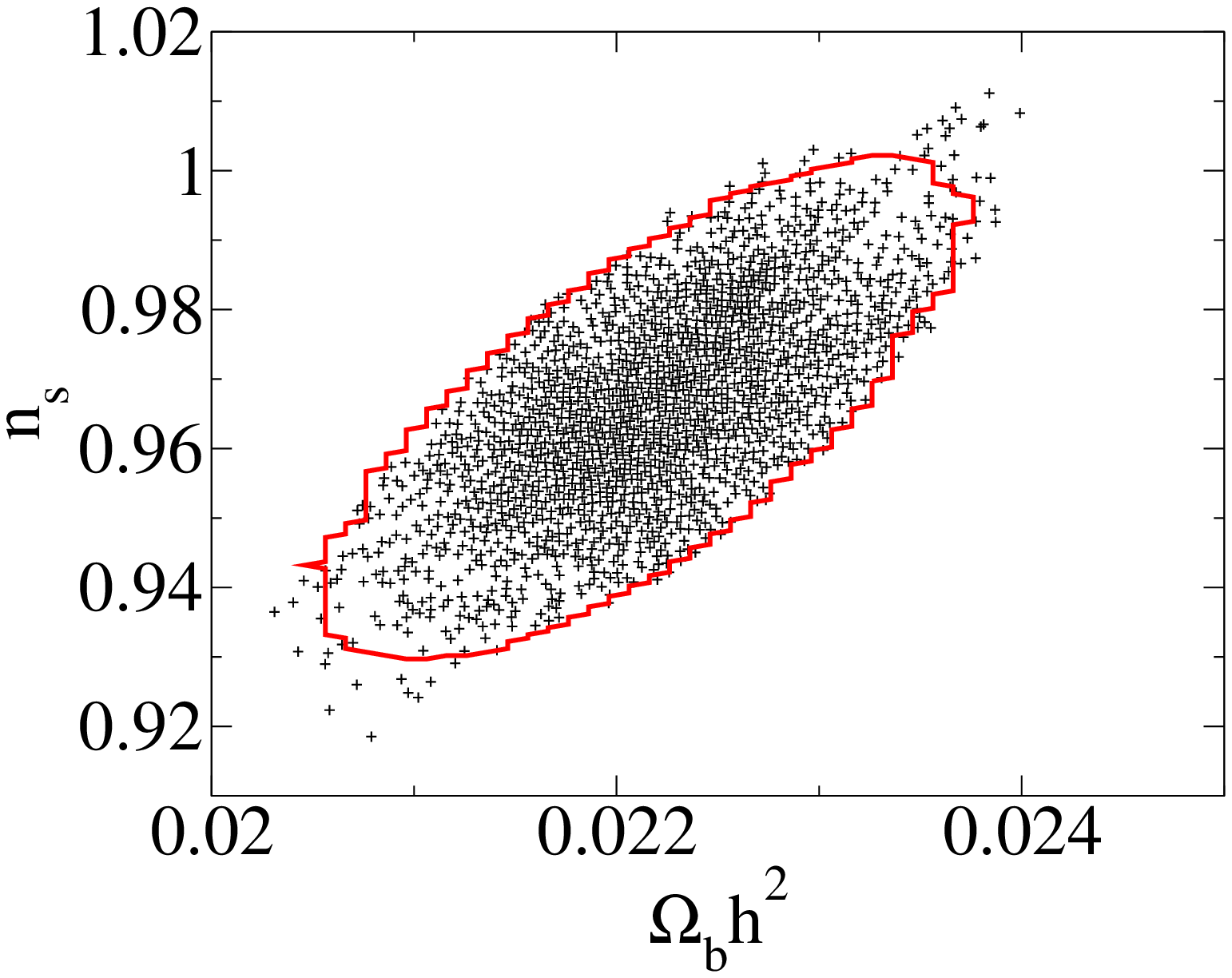}
}
\subfigure[]{
\includegraphics[scale=0.15]{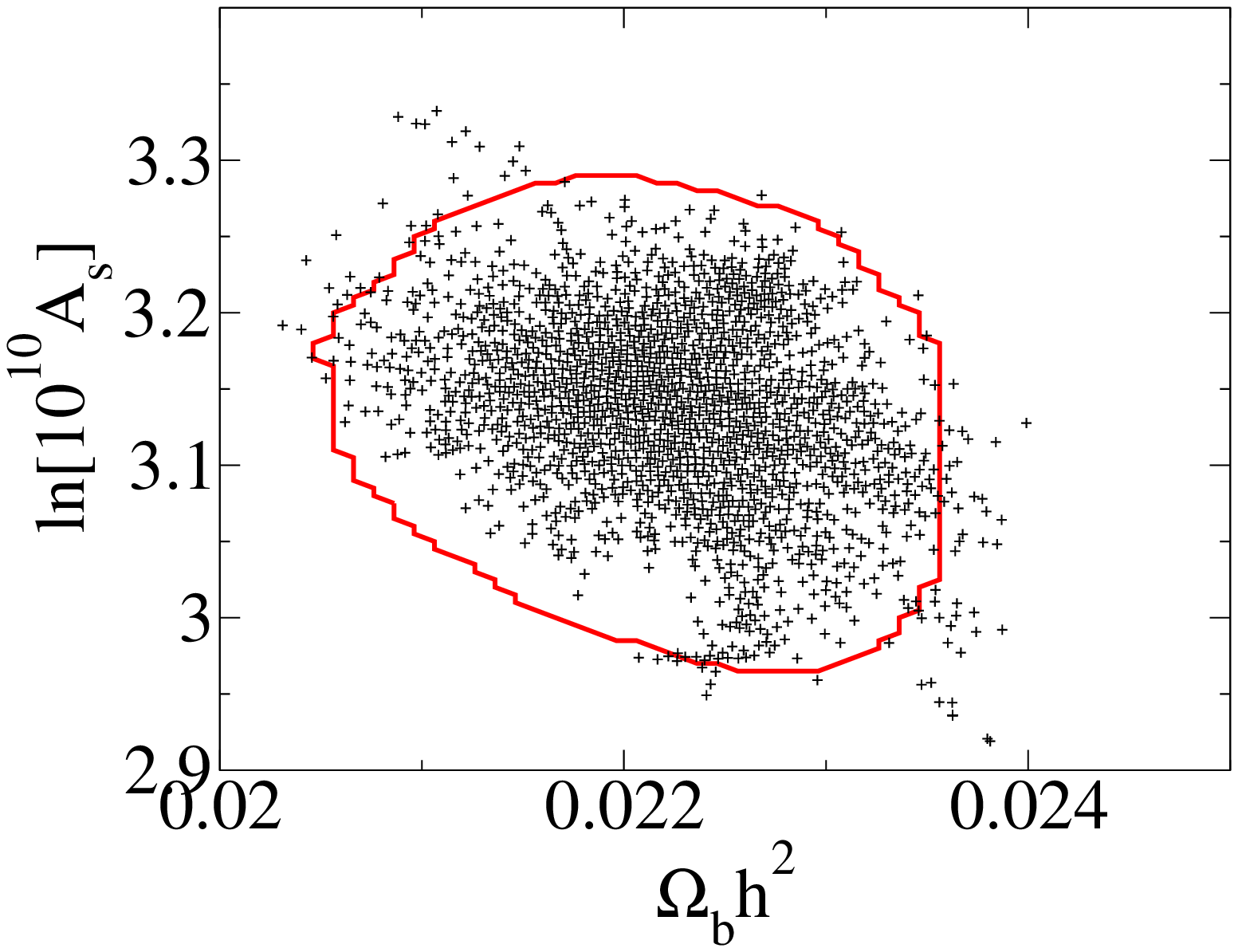}
}
\subfigure[]{
\includegraphics[scale=0.15]{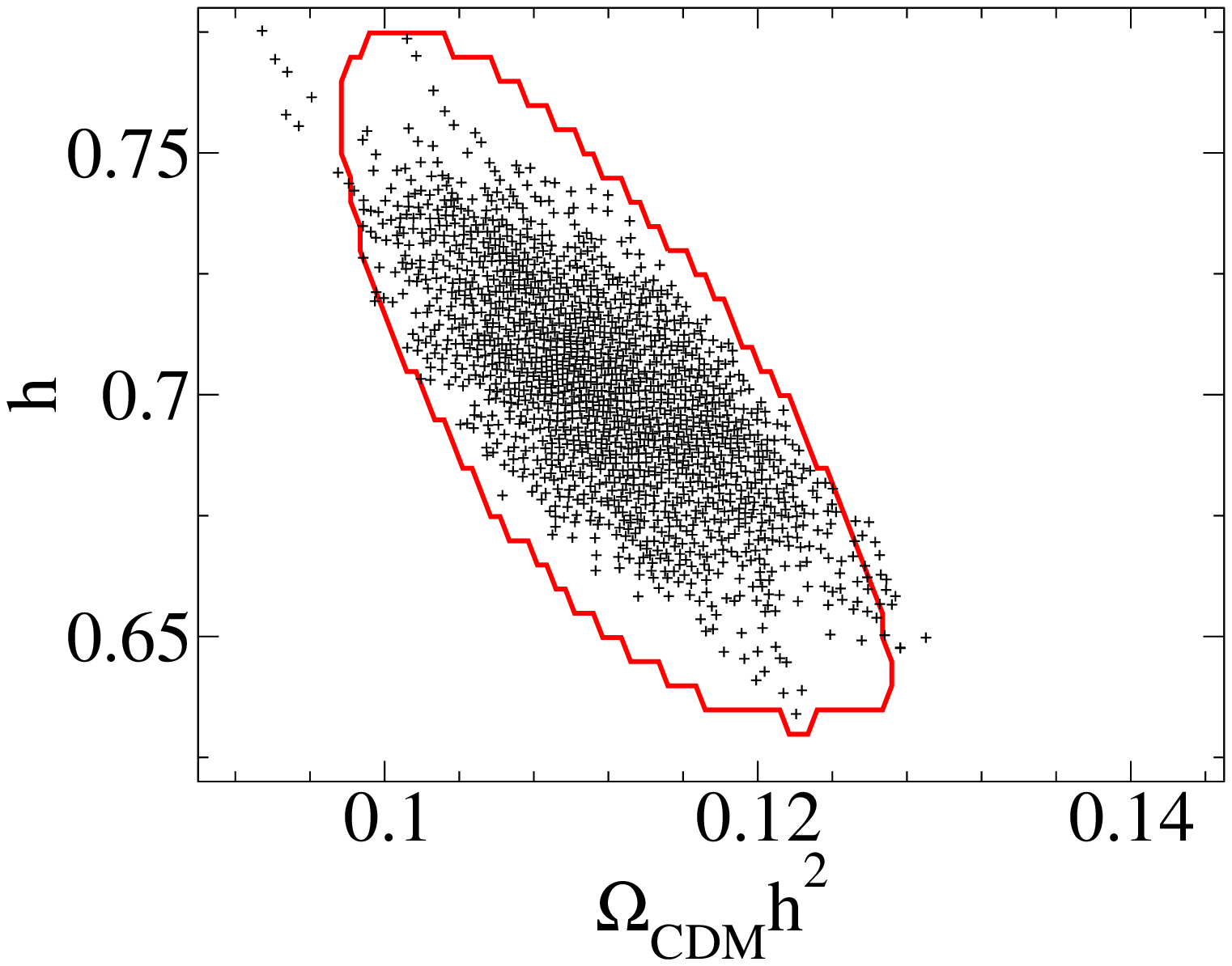}
}
\subfigure[]{
\includegraphics[scale=0.15]{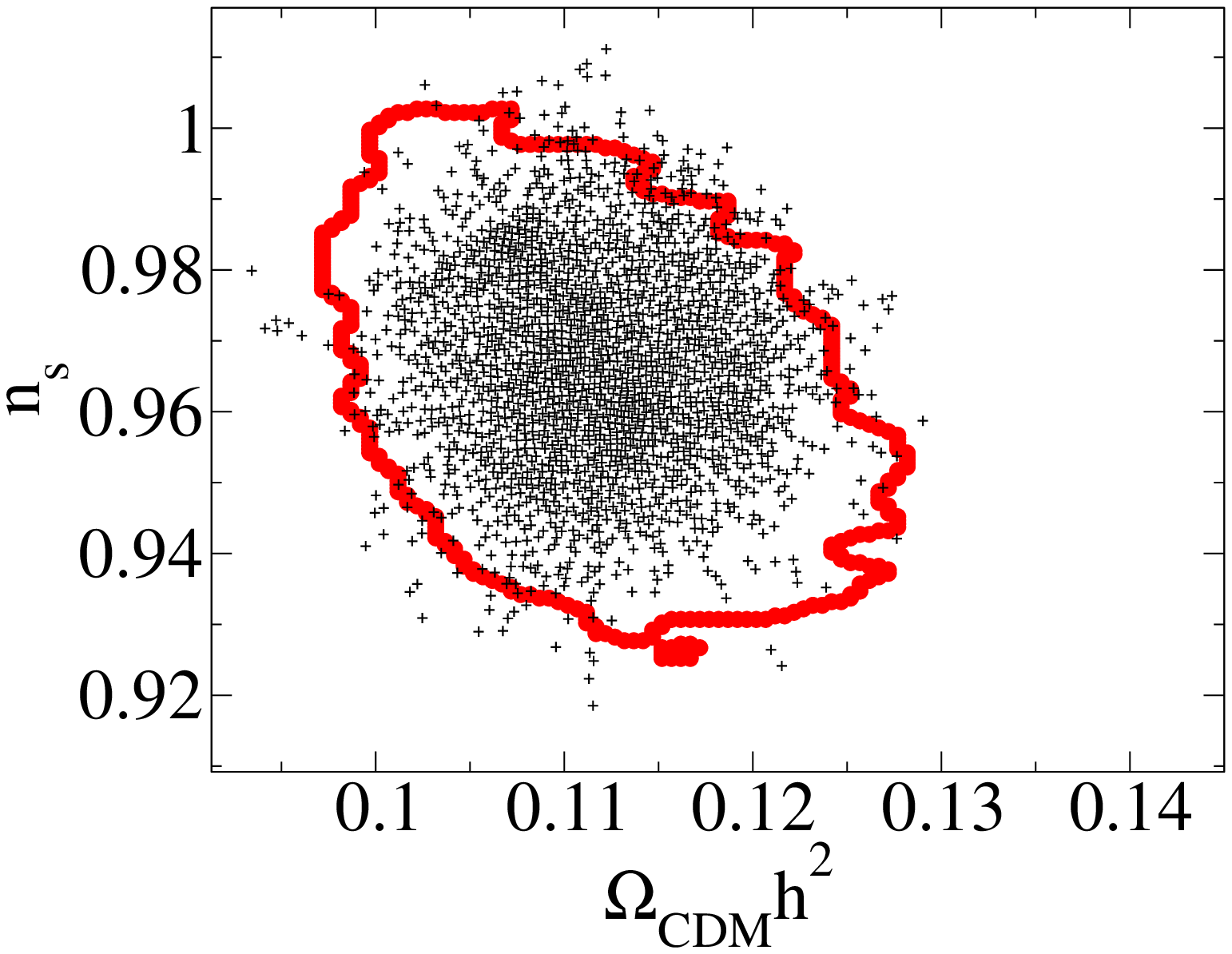}
}
\subfigure[]{
\includegraphics[scale=0.15]{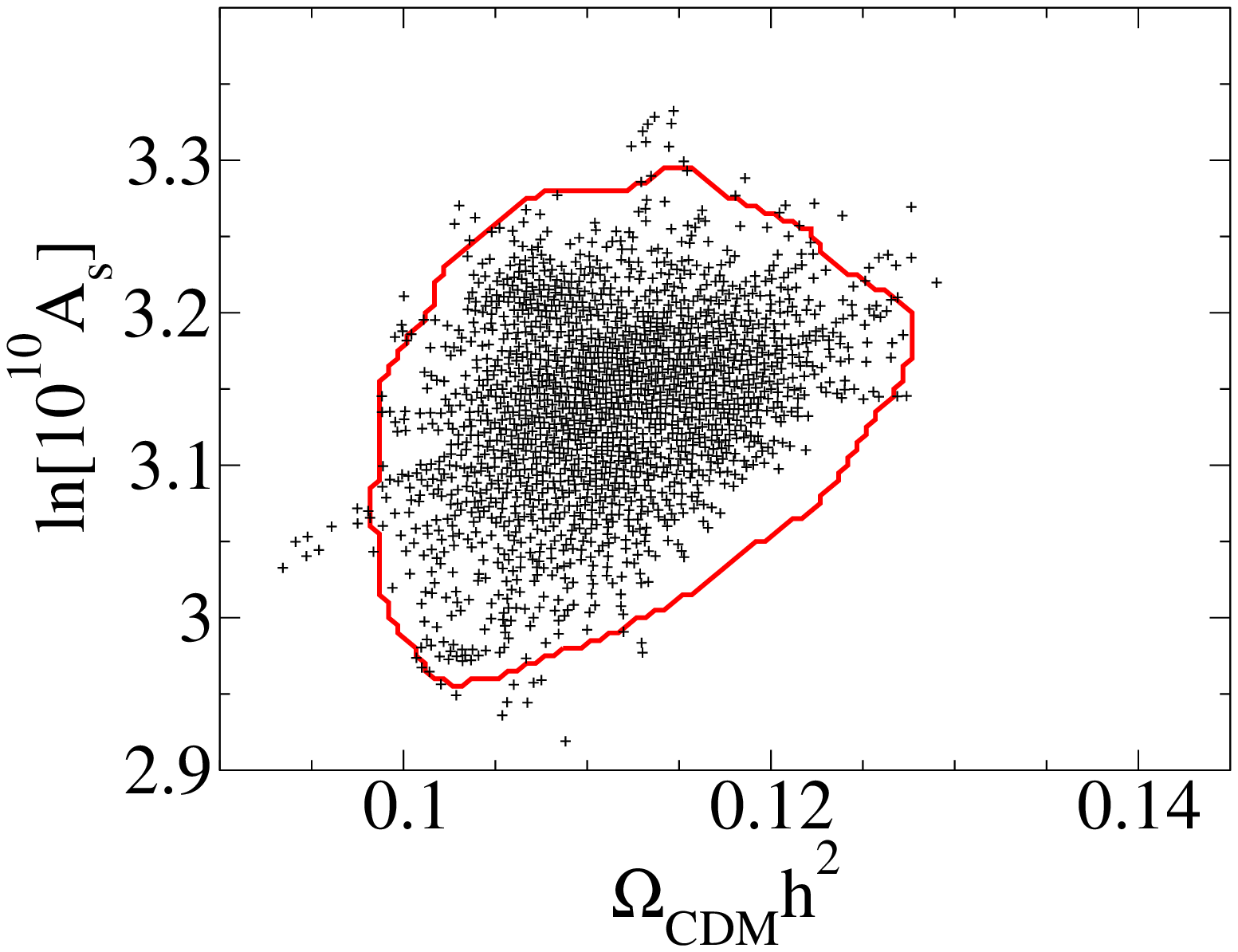}
}
\subfigure[]{
\includegraphics[scale=0.15]{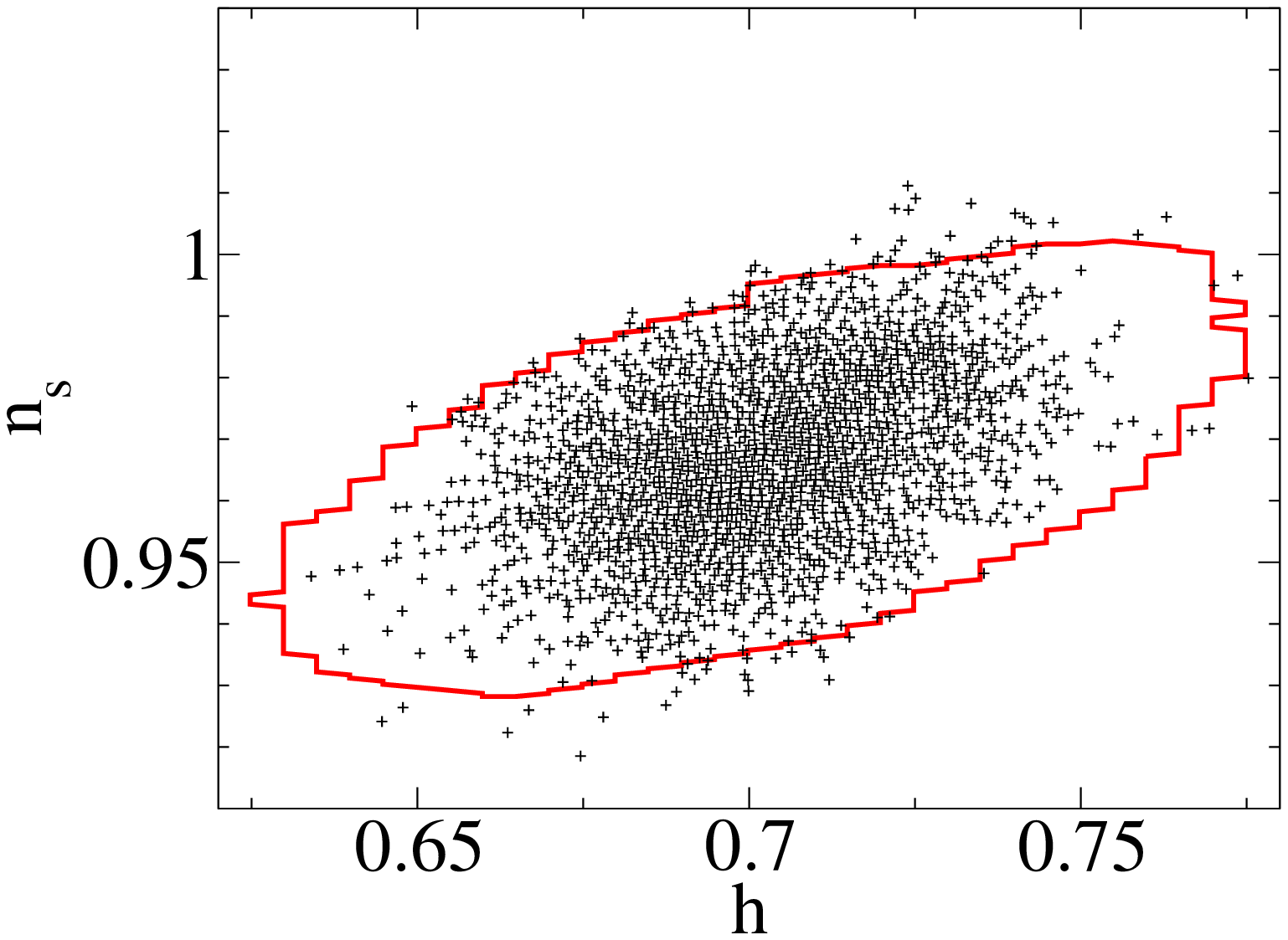}
}
\subfigure[]{
\includegraphics[scale=0.15]{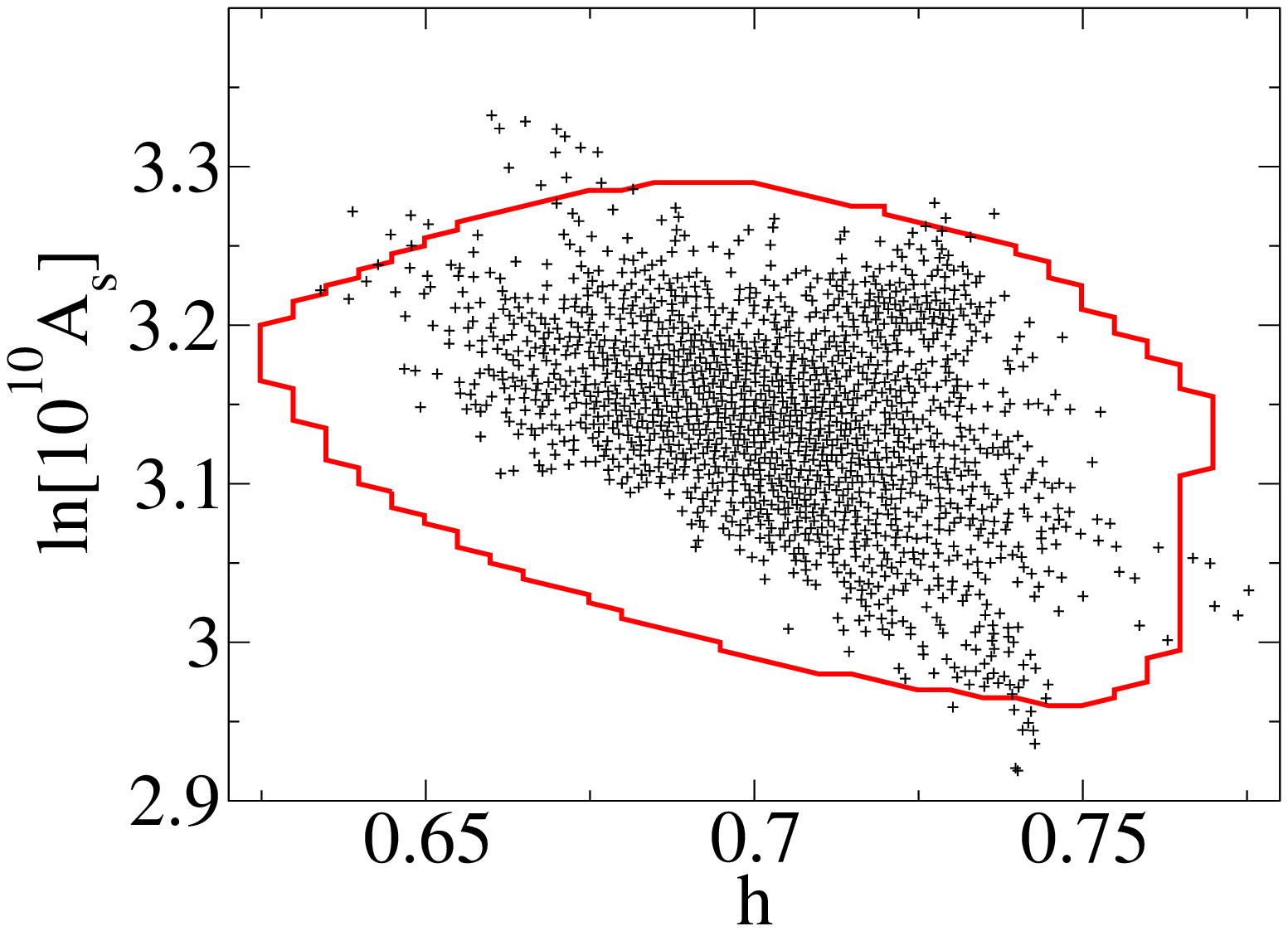}
}
\subfigure[]{
\includegraphics[scale=0.15]{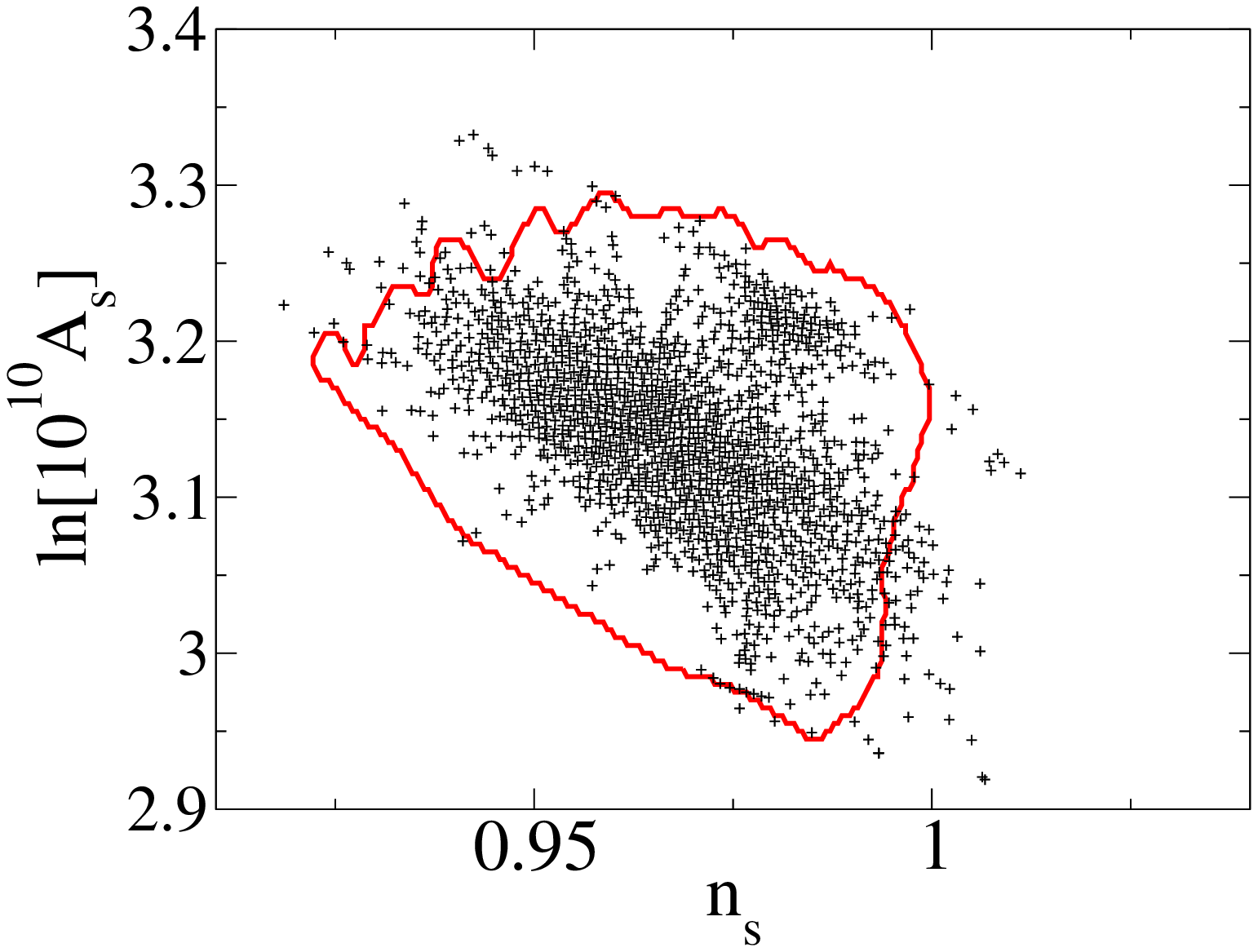}
}
\caption{
The 10 2-dimensional sub-spaces of our 6-dimensional parameter space
($\tau$ is ignored because no real constraint can be gleaned from this data).
The thick red contours are the 95\% Bayesian credible limits determined by MCMC after
460,000 calls to $\chi^2$.  The black points are the 95\% Frequentist confidence limits
determined by APS after 50,000 calls to $\chi^2$.  $\chi^2_\text{lim}=\chi^2_\text{min}+12.6$.
Equation (\ref{eqn:GaussianCovar}) is used for the Gaussian process covariogram.
}
\label{fig:contours_gauss}
\end{figure*}

\begin{figure*}
\subfigure[]{
\includegraphics[scale=0.15]{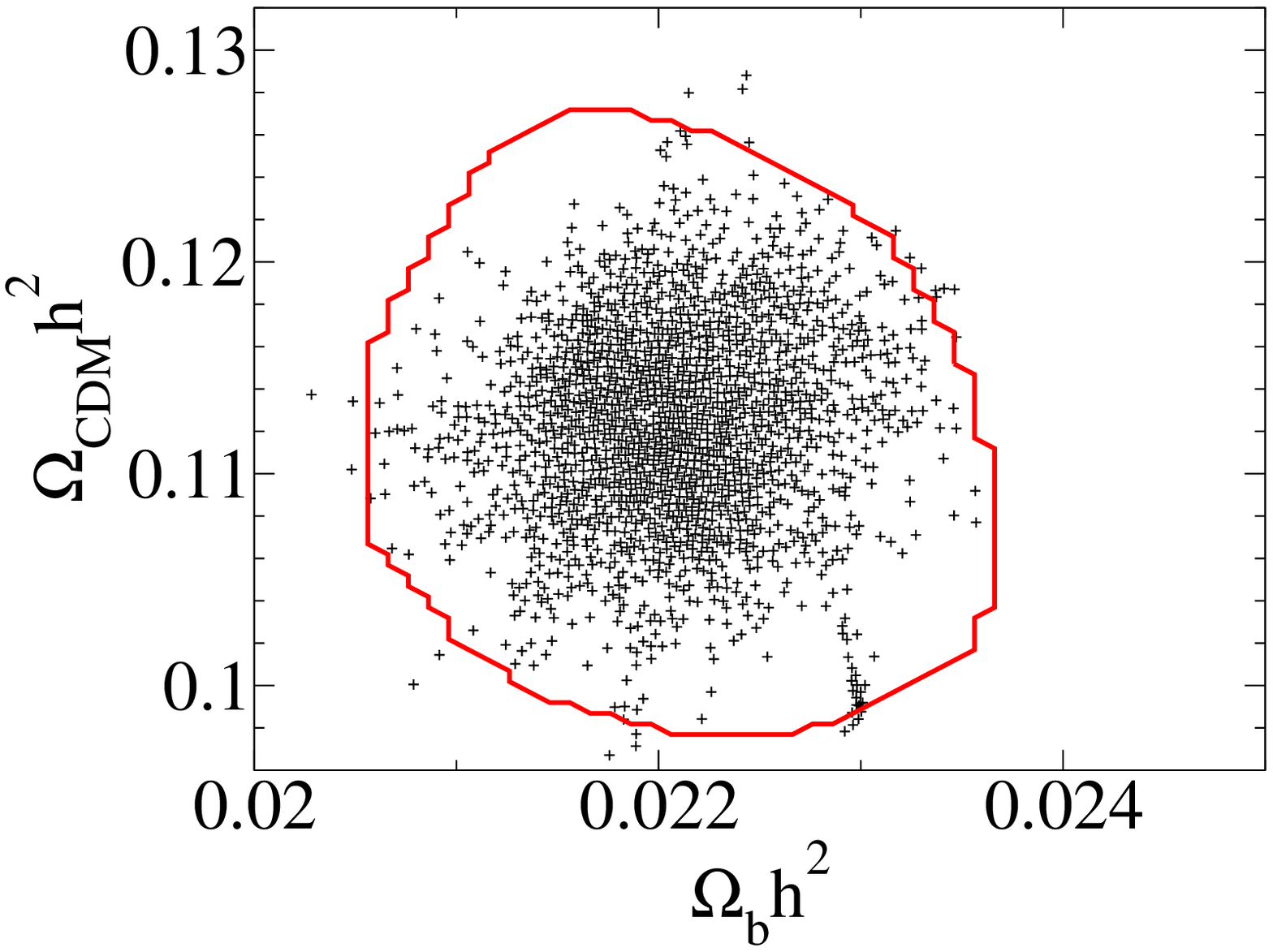}
}
\subfigure[]{
\includegraphics[scale=0.15]{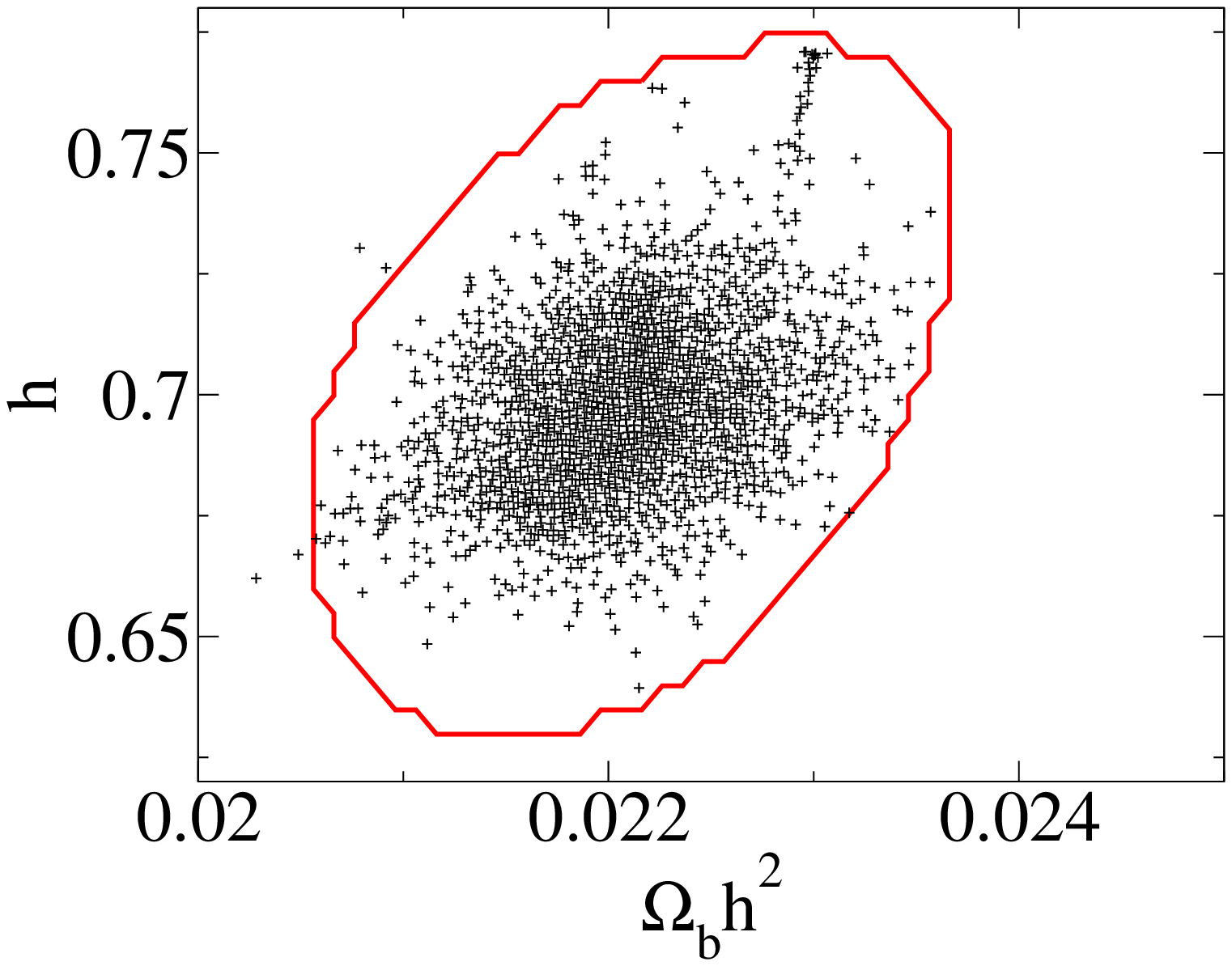}
}
\subfigure[]{
\includegraphics[scale=0.15]{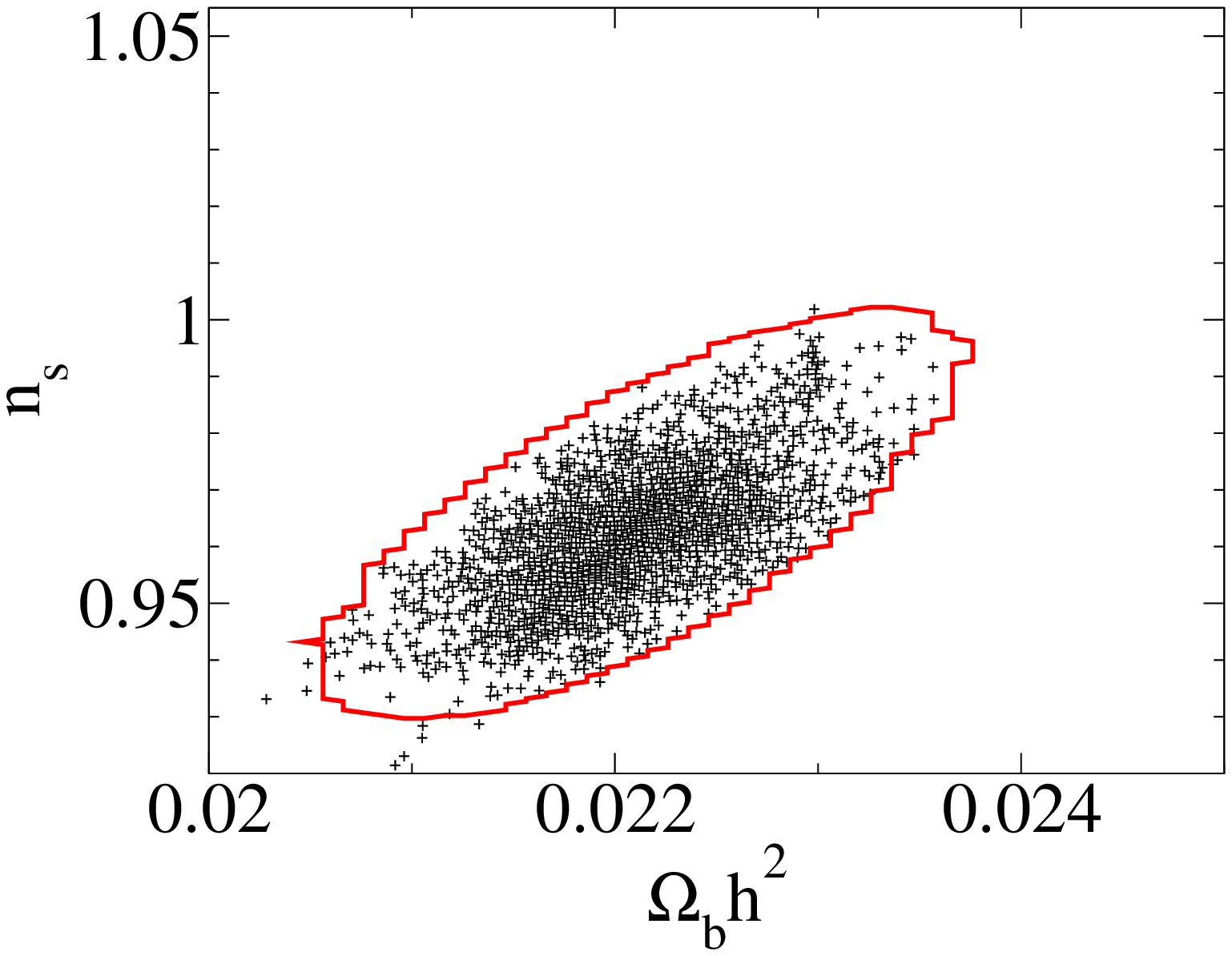}
}
\subfigure[]{
\includegraphics[scale=0.15]{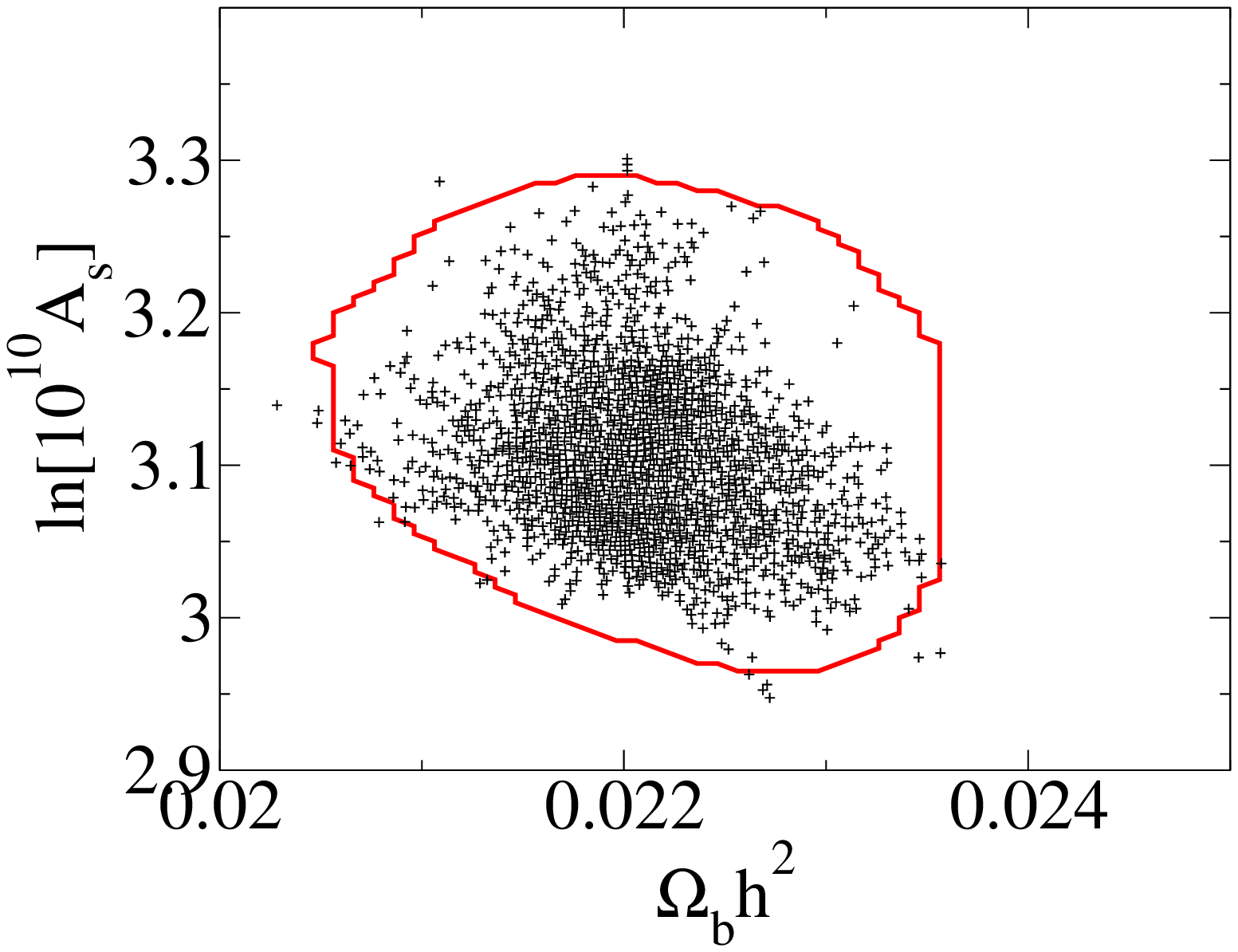}
}
\subfigure[]{
\includegraphics[scale=0.15]{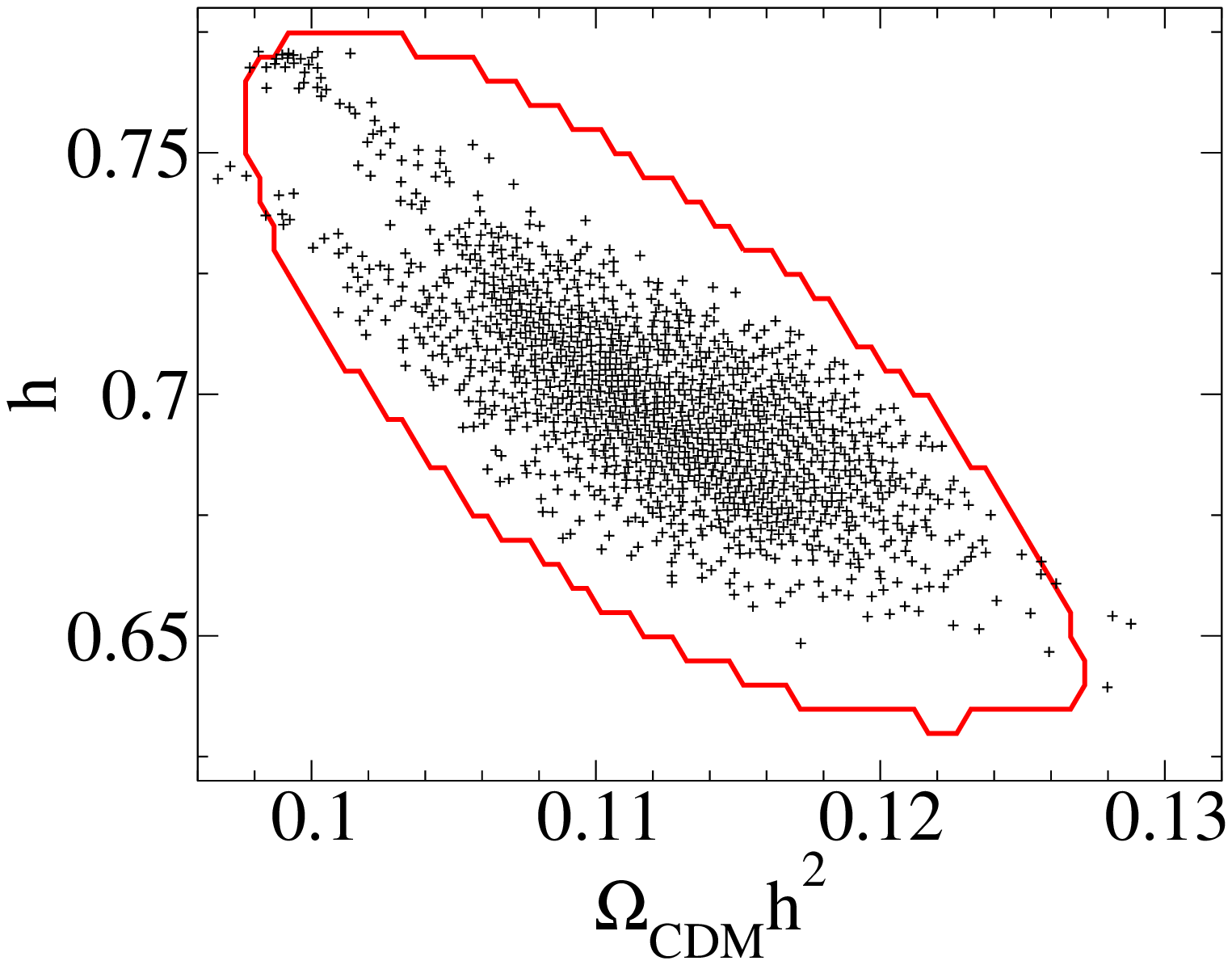}
}
\subfigure[]{
\includegraphics[scale=0.15]{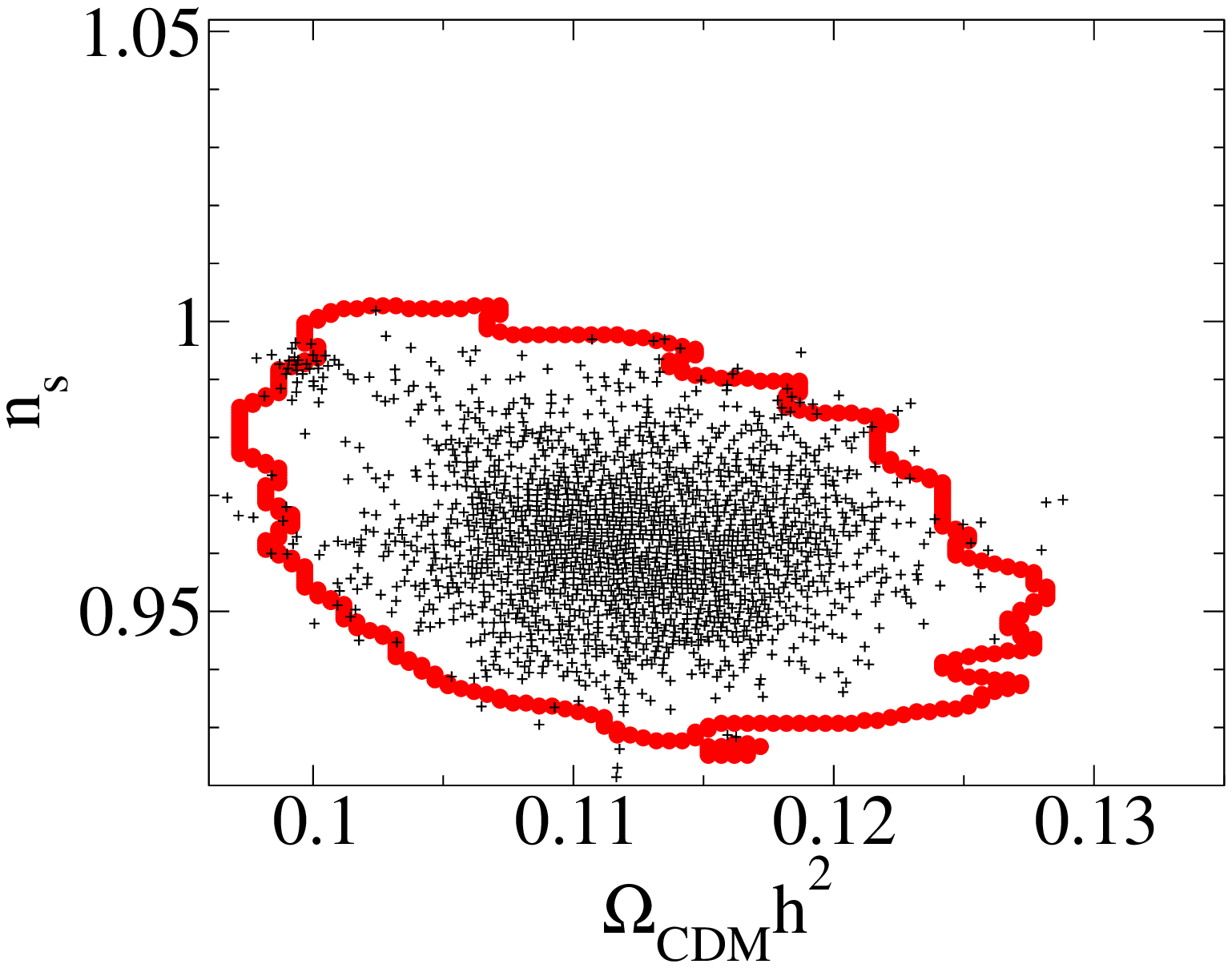}
}
\subfigure[]{
\includegraphics[scale=0.15]{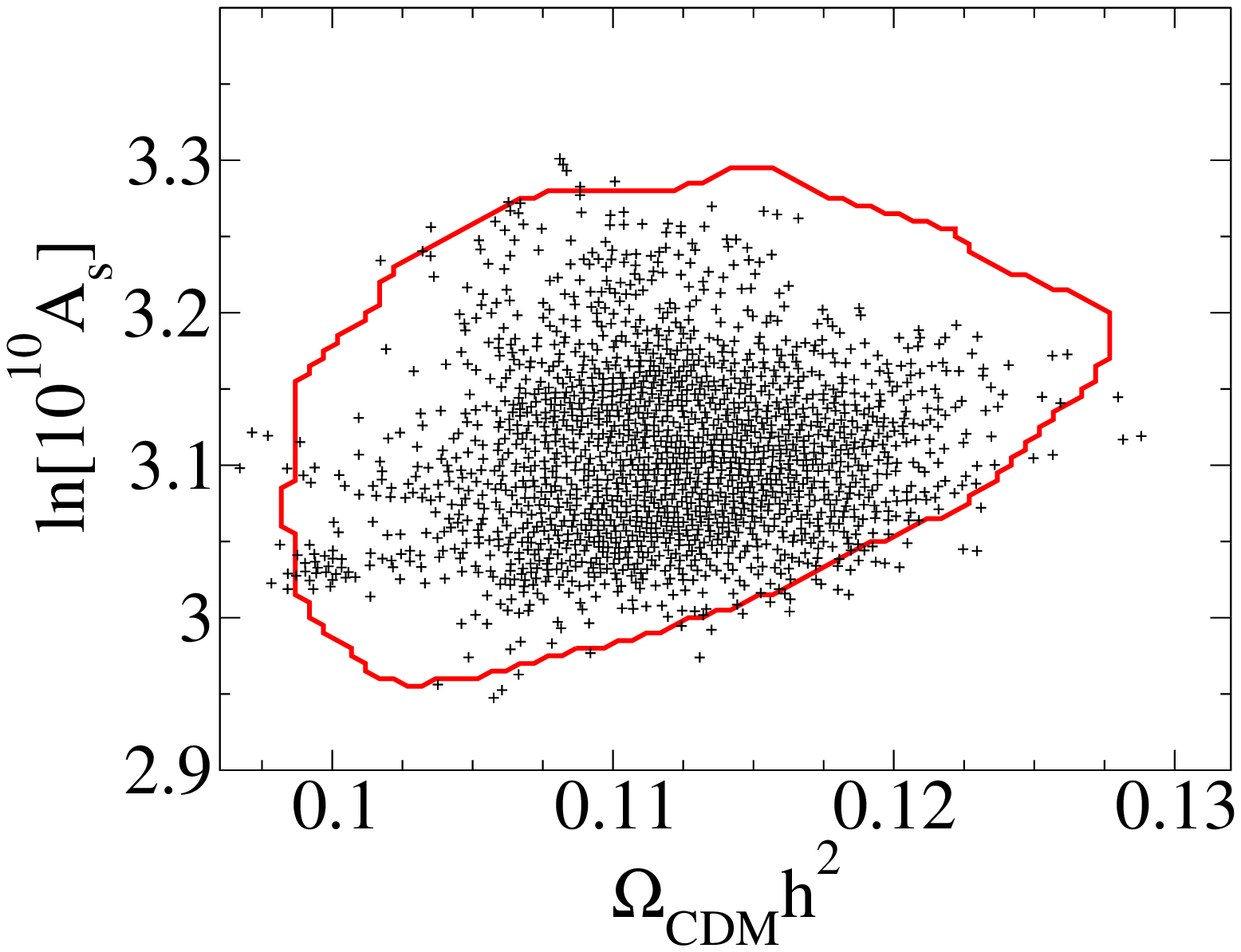}
}
\subfigure[]{
\includegraphics[scale=0.15]{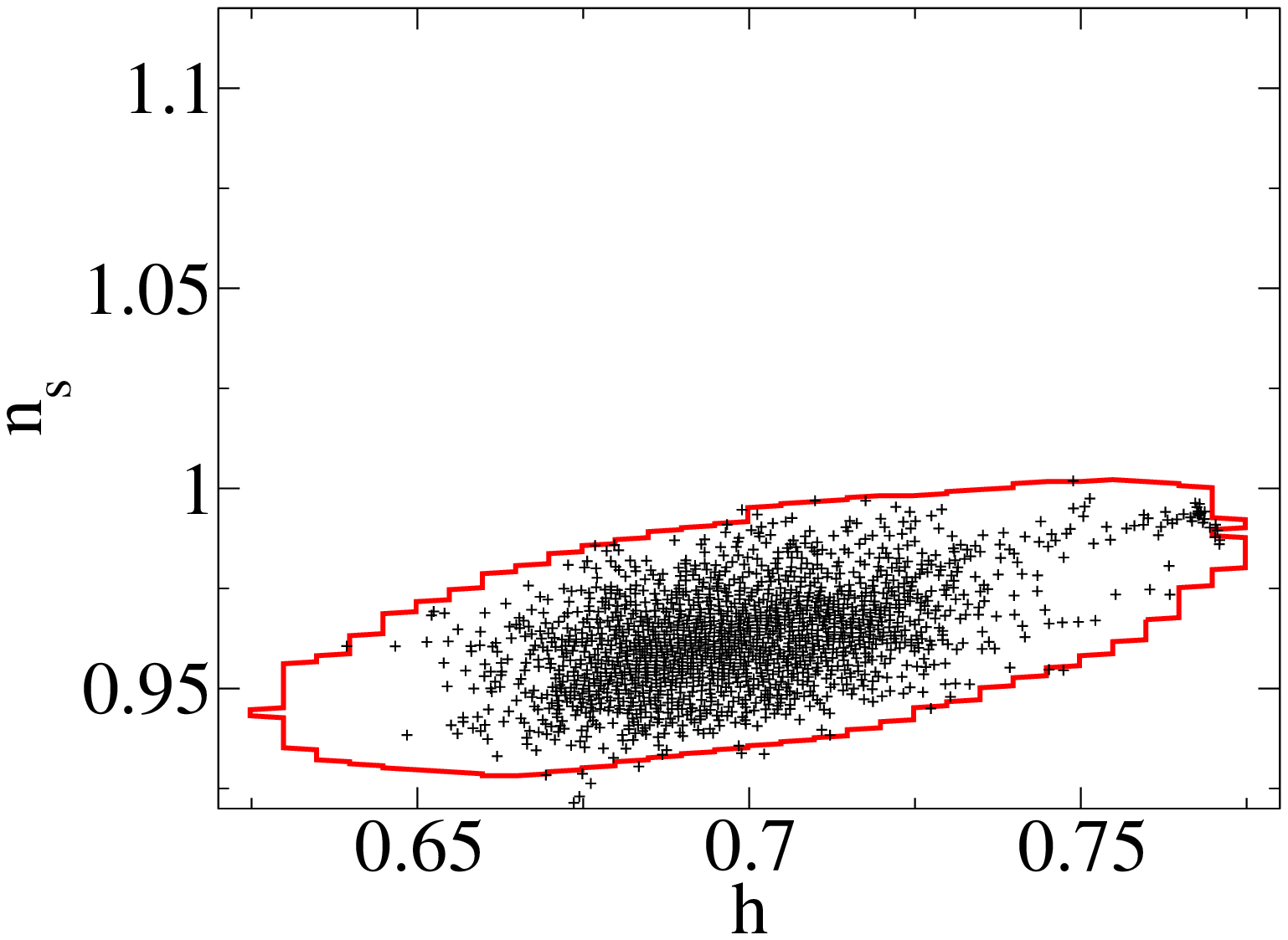}
}
\subfigure[]{
\includegraphics[scale=0.15]{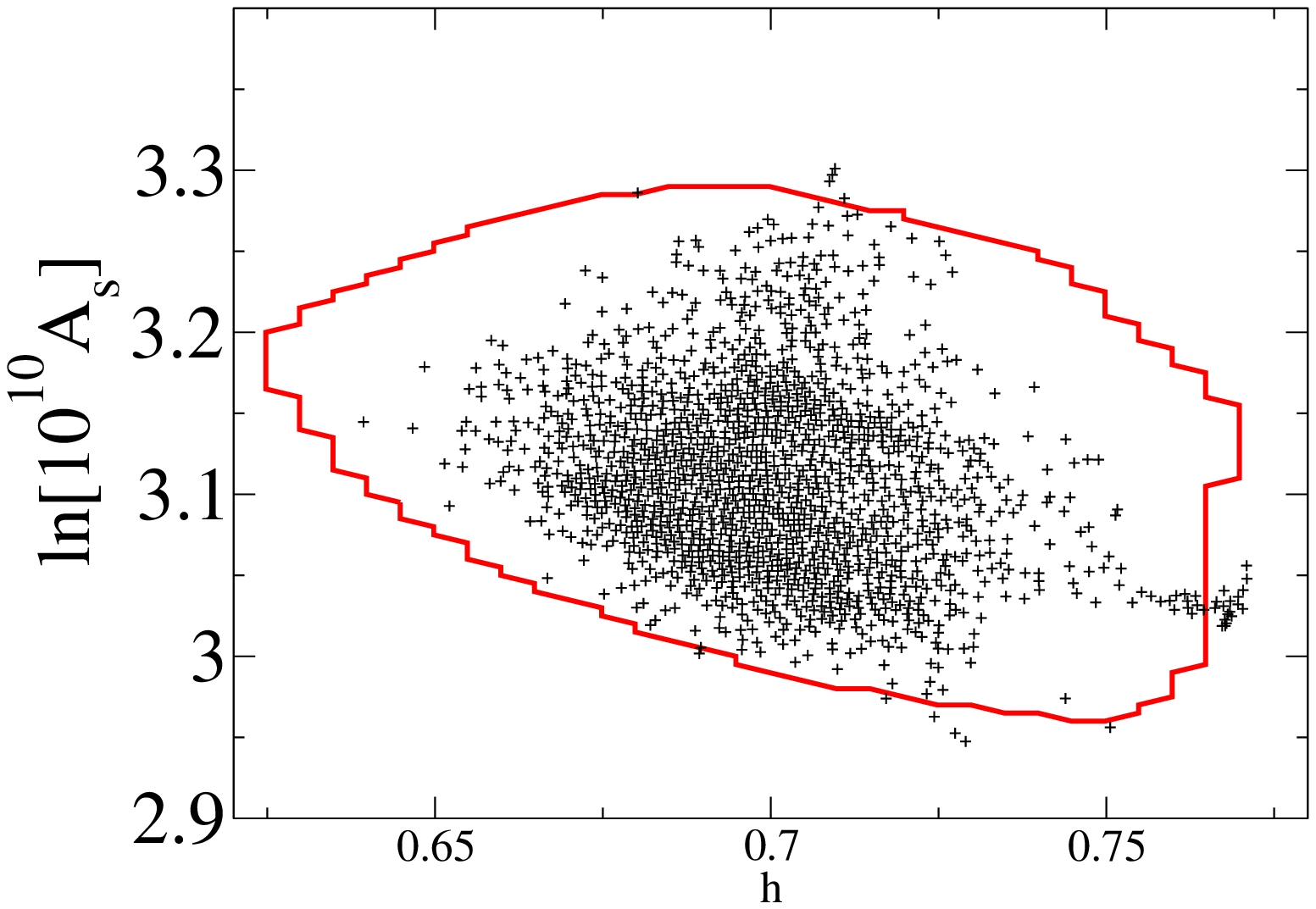}
}
\subfigure[]{
\includegraphics[scale=0.15]{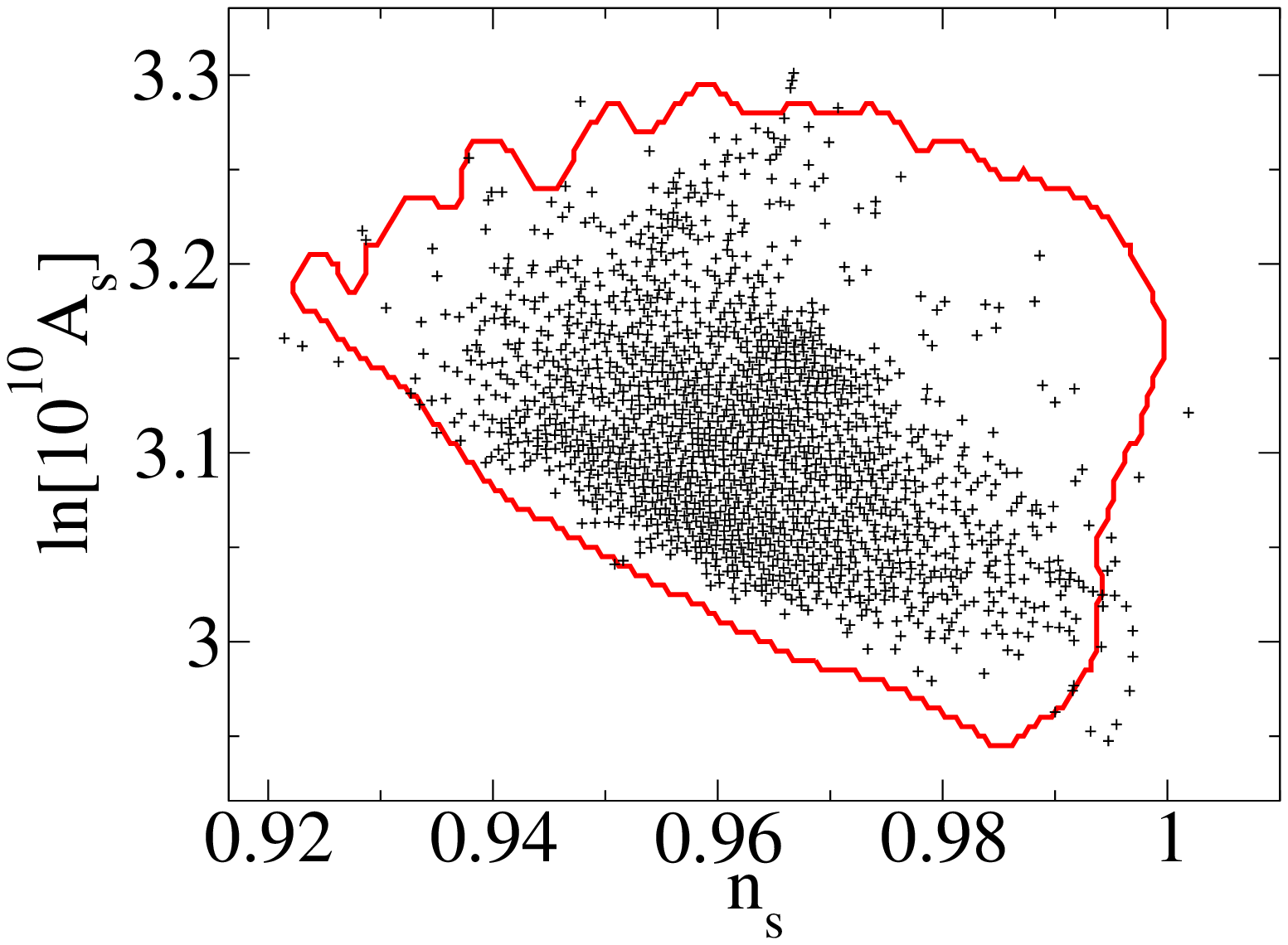}
}
\caption{
The 10 2-dimensional sub-spaces of our 6-dimensional parameter space
($\tau$ is ignored because no real constraint can be gleaned from this data).
The thick red contours are the 95\% Bayesian credible limits determined by MCMC after
460,000 calls to $\chi^2$.  The black points are the 95\% Frequentist confidence limits
determined by APS after 50,000 calls to $\chi^2$.  $\chi^2_\text{lim}=\chi^2_\text{min}+12.6$.
Equation (\ref{eqn:NNCovar}) is used for the Gaussian process covariogram.
}
\label{fig:contours_nn}
\end{figure*}

\begin{figure*}
\subfigure[]{
\includegraphics[scale=0.15]{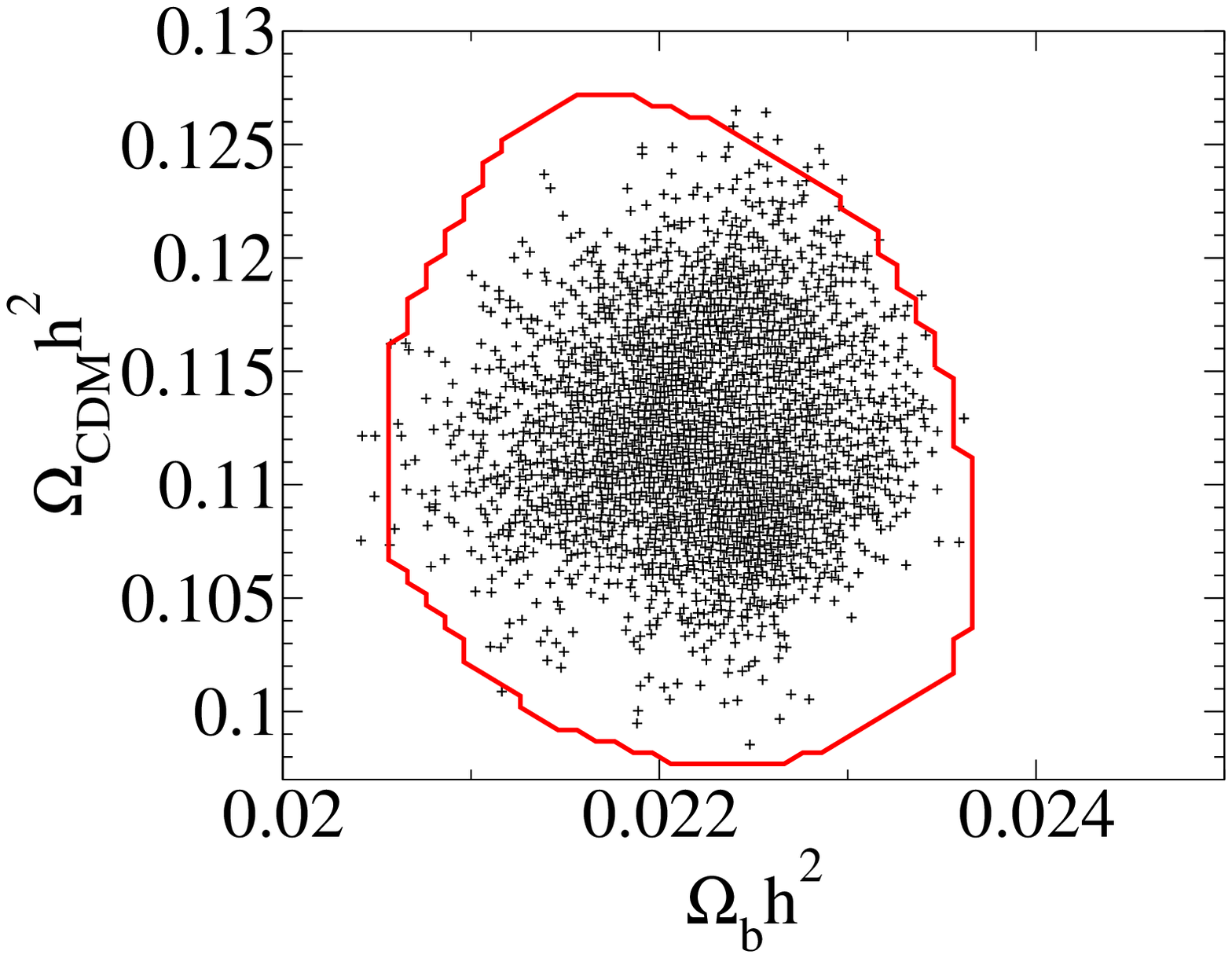}
}
\subfigure[]{
\includegraphics[scale=0.15]{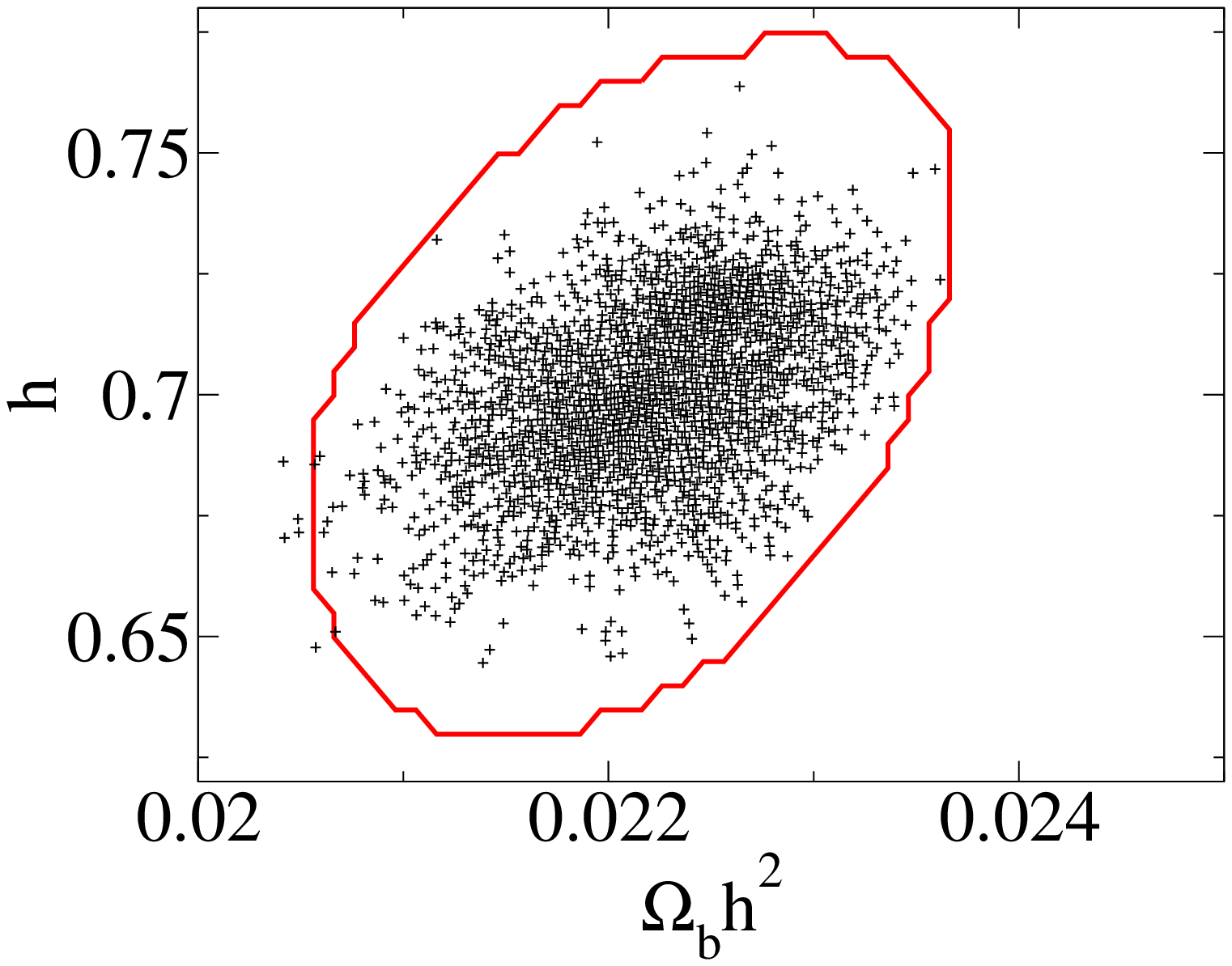}
}
\subfigure[]{
\includegraphics[scale=0.15]{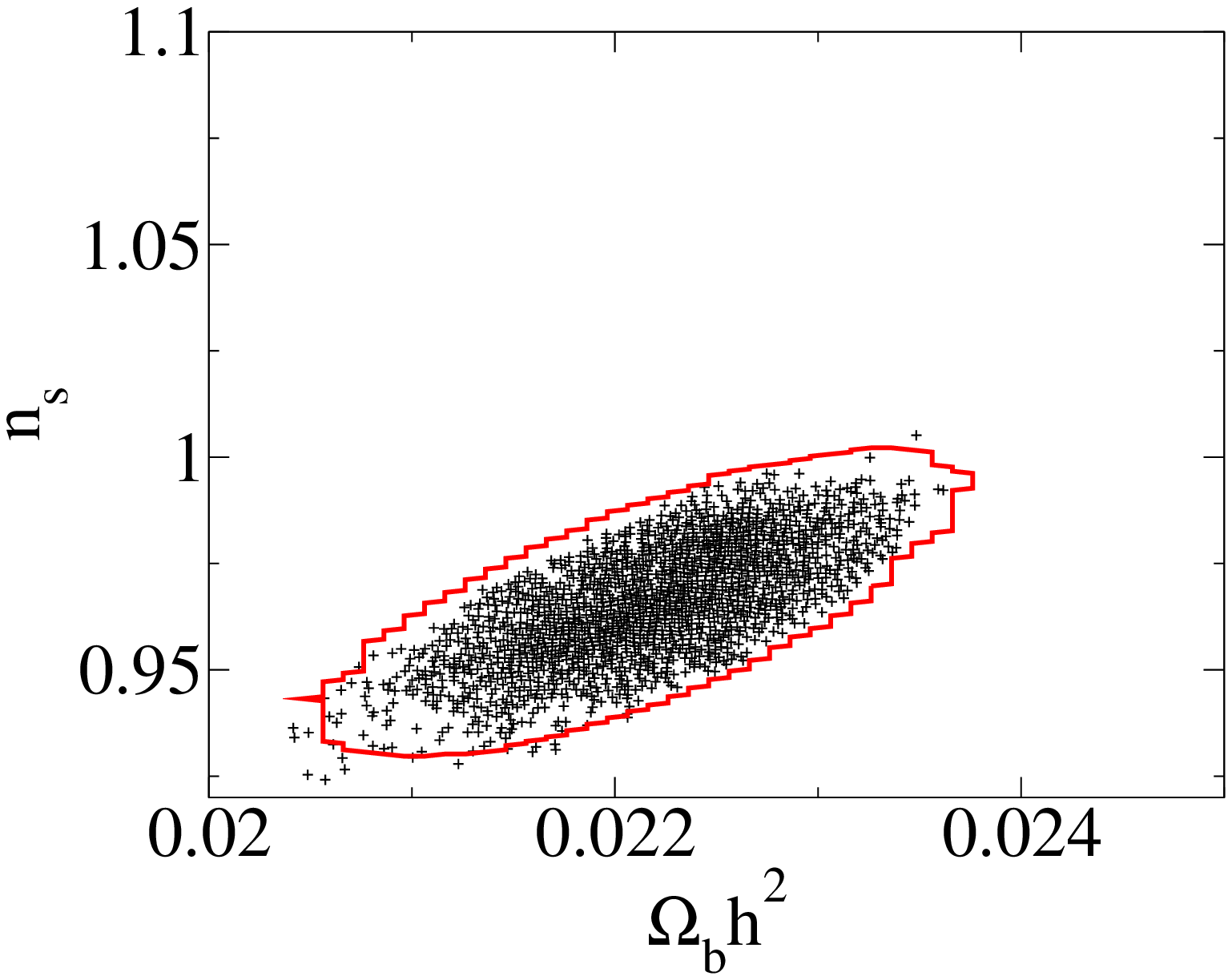}
}
\subfigure[]{
\includegraphics[scale=0.15]{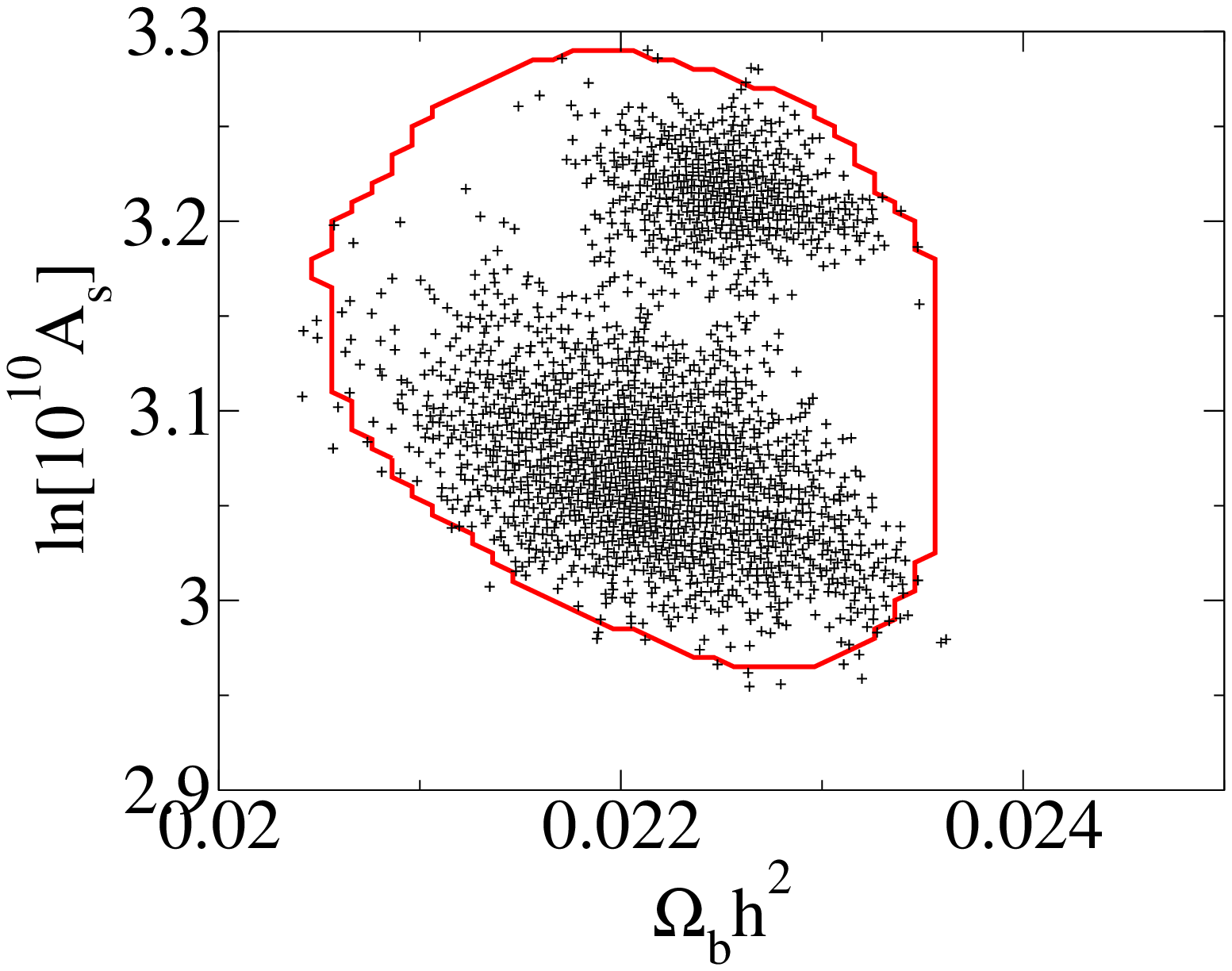}
}
\subfigure[]{
\includegraphics[scale=0.15]{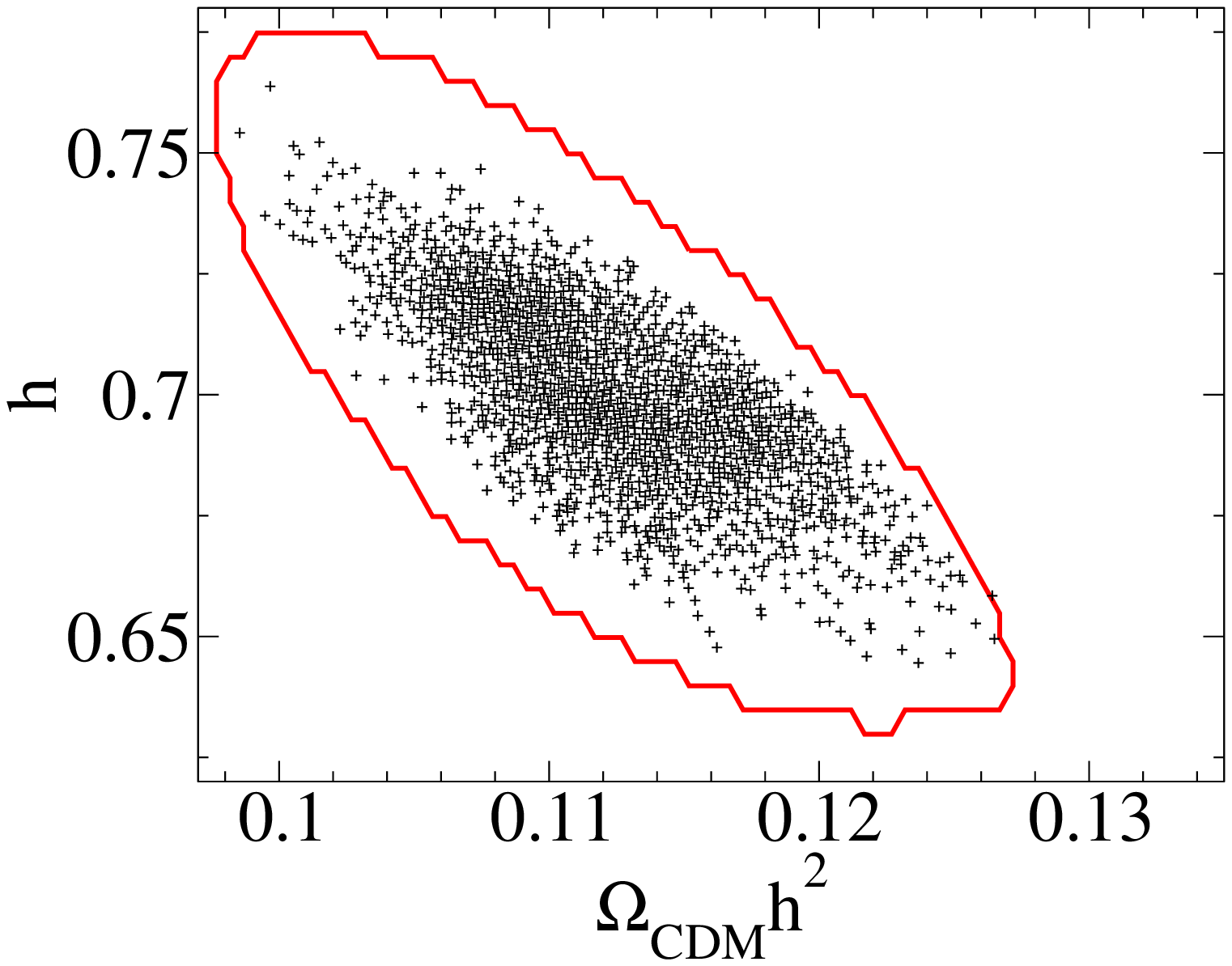}
}
\subfigure[]{
\includegraphics[scale=0.15]{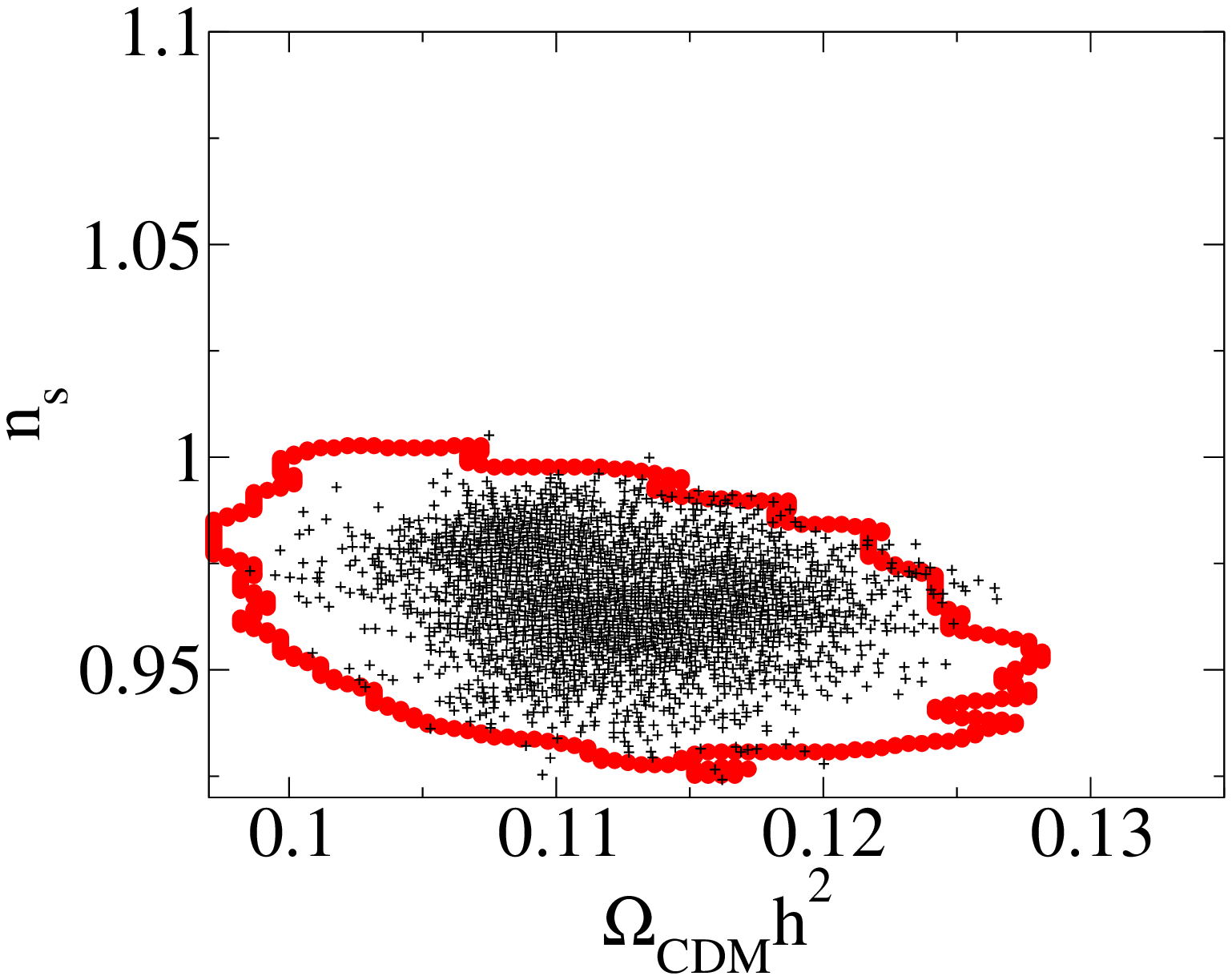}
}
\subfigure[]{
\includegraphics[scale=0.15]{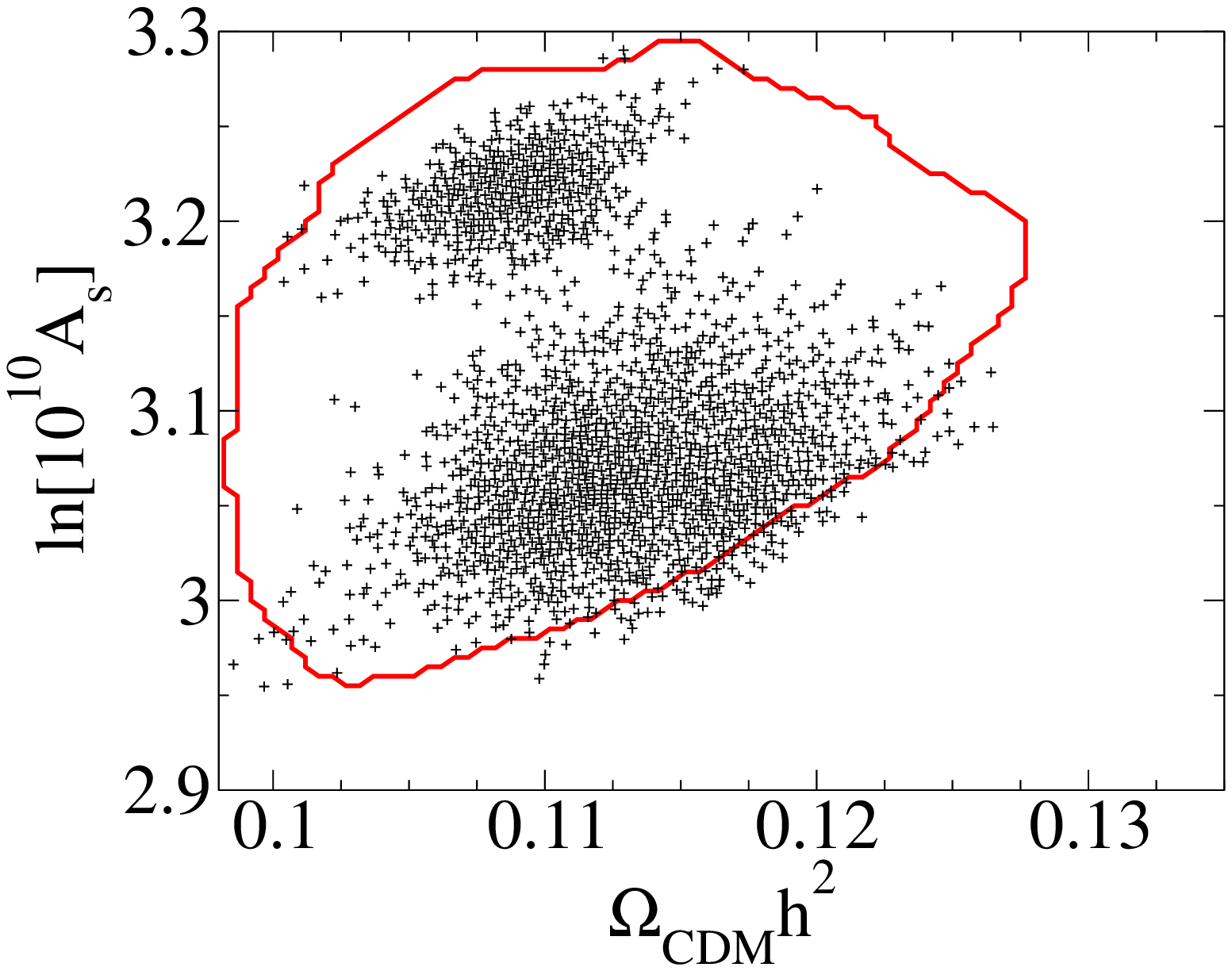}
}
\subfigure[]{
\includegraphics[scale=0.15]{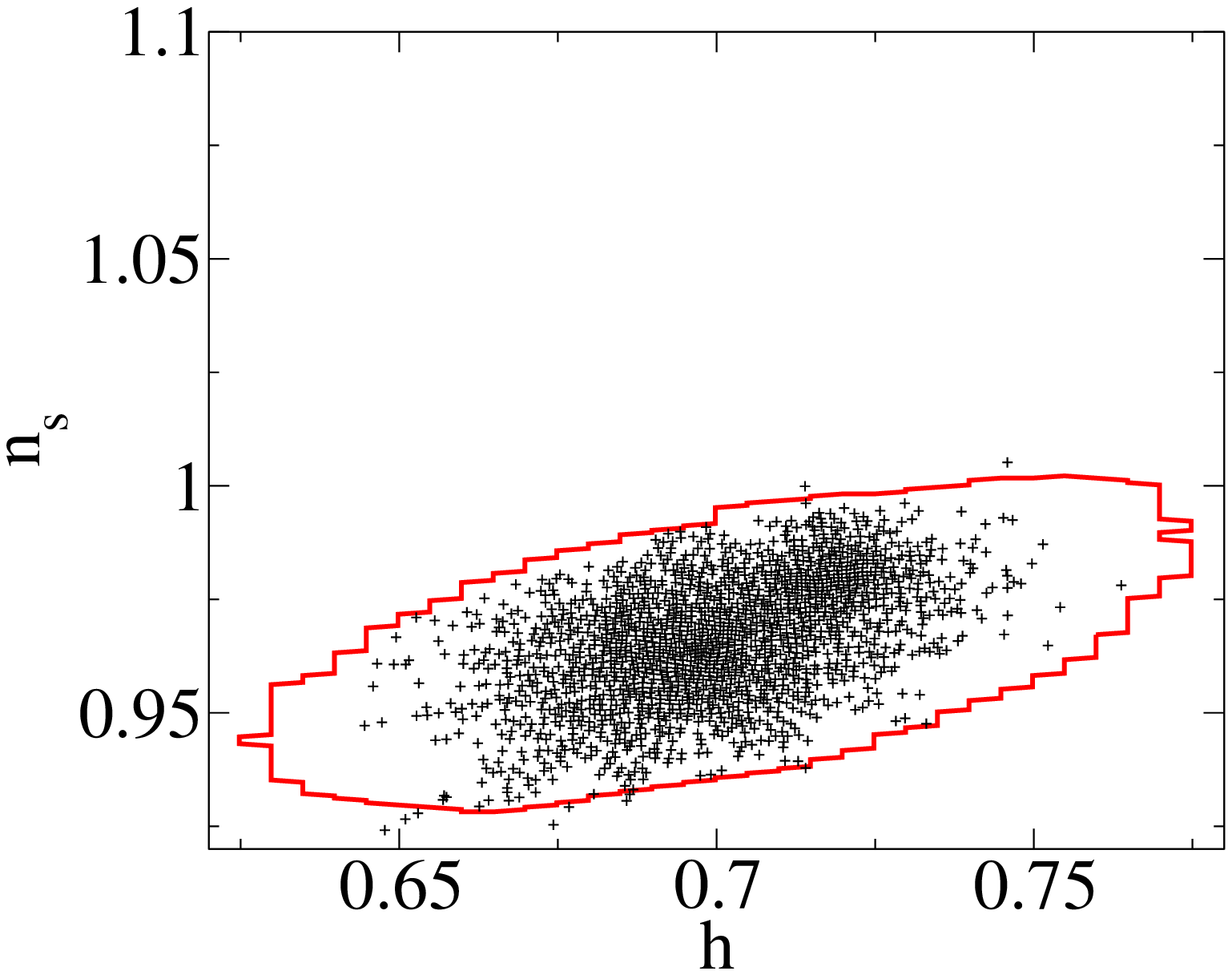}
}
\subfigure[]{
\includegraphics[scale=0.15]{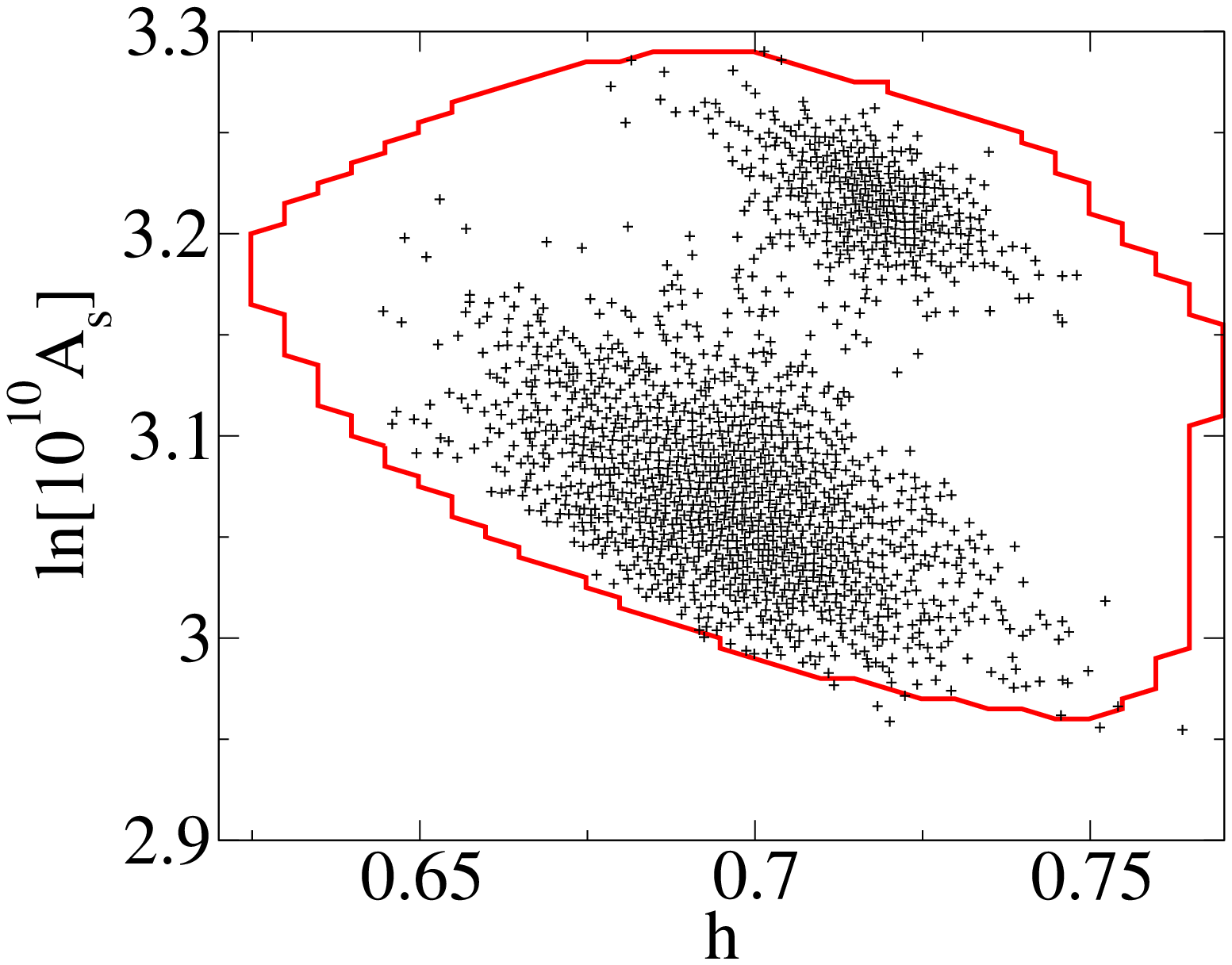}
}
\subfigure[]{
\includegraphics[scale=0.15]{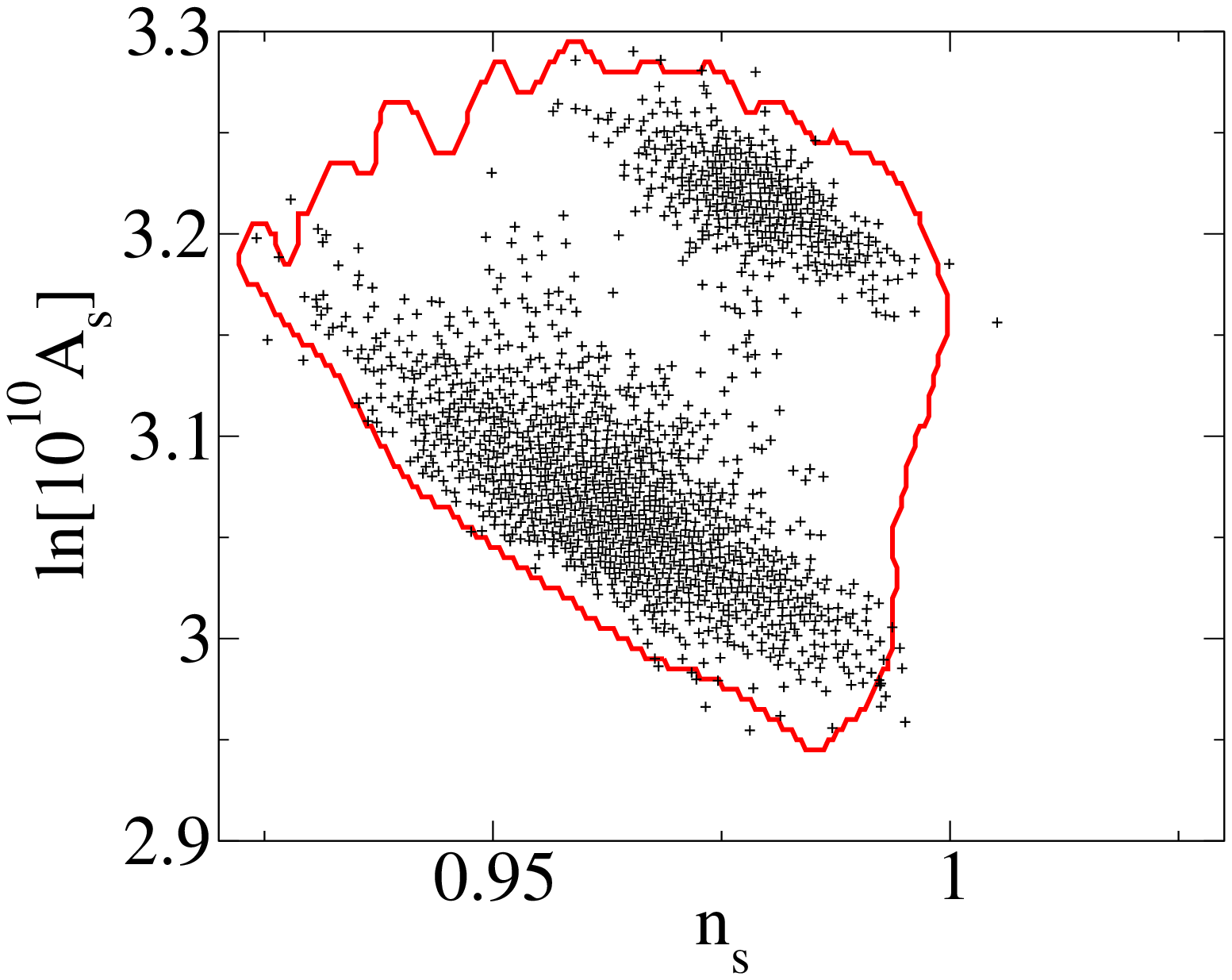}
}
\caption{
The 10 2-dimensional sub-spaces of our 6-dimensional parameter space
($\tau$ is ignored because no real constraint can be gleaned from this data).
The thick red contours are the 95\% Bayesian credible limits determined by MCMC after
460,000 calls to $\chi^2$.  The black points are the 95\% Frequentist confidence limits
determined by APS after 50,000 calls to $\chi^2$.  $\chi^2_\text{lim}=1280.7$.
Equation (\ref{eqn:GaussianCovar}) is used for the Gaussian process covariogram.
}
\label{fig:contours_br}
\end{figure*}

The reader will note that
MCMC discovers its $\chi^2_\text{min}$ much more rapidly than APS, however, 
we will see
in Figures \ref{fig:contours_gauss} through \ref{fig:contours_br} that this rapid
descent to $\chi^2_\text{min}$ does not translate into rapid convergence to
the final credible limit.  Because MCMC infers its limits based on the number
of samples drawn from the Bayesian posterior probability, MCMC still requires
many tens (or hundreds) of thousands of points to be sampled after $\chi^2_\text{min}$
has been discovered.  Because it is learning the shape of the function
directly, APS does not impose this requirement.
The reader should also consider Figure \ref{fig:contours_br}, which shows the Frequentist
confidence intervals discovered by APS when $\chi^2_\text{lim}$ is set by hand using
prior knowledge of the size of the WMAP 7-year data set.  In this case, APS does not need
to learn $\chi^2_\text{min}$ in order to proceed.  This is the original use case for which
APS was designed.  Here we see that the confidence intervals discovered agree very well with
the Bayesian credible intervals discovered by MCMC.

\begin{figure}
\includegraphics[scale=0.3]{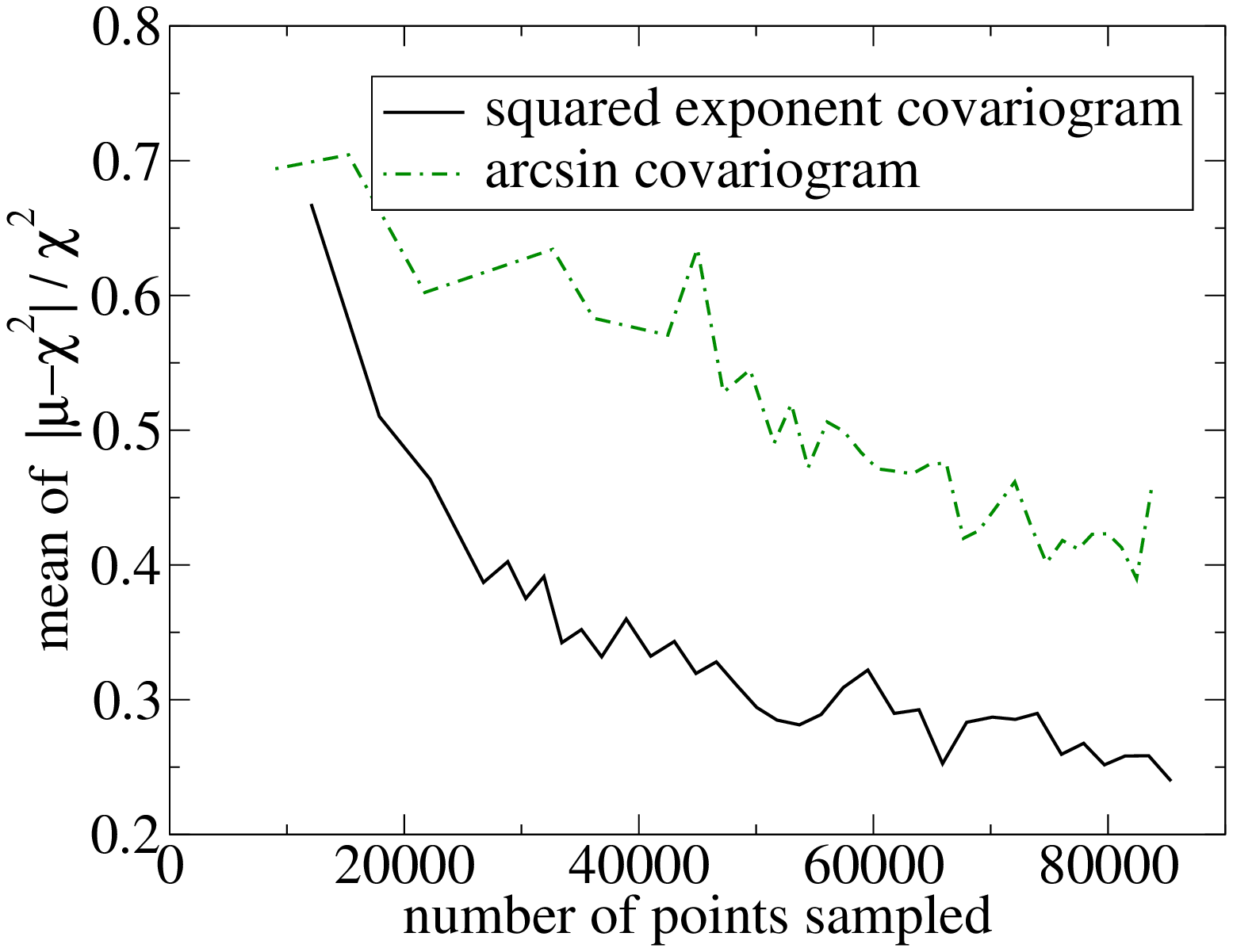}
\caption{
The average value of $|\mu-\chi^2|/\chi^2$ as a function of the number of points
sampled by APS in the WMAP 7 problem.  
This is shown as a check on our assumption that
$\vec{\theta}\rightarrow\chi^2$ can be approximated as a Gaussian process.
Though the squared exponential covariogram (\ref{eqn:GaussianCovar})
gives a better fit to the $\chi^2$ function than the arcsine covariogram
(\ref{eqn:NNCovar}),
we see in Figures \ref{fig:contours_gauss} and \ref{fig:contours_nn}
that this has little effect on the parameter constraints derived by APS.
}
\label{fig:goodness}
\end{figure}

The second question we consider is whether or not our assumption that
$\vec{\theta}\rightarrow\chi^2$ can be modeled by a Gaussian process is 
valid.  In Figure
\ref{fig:goodness} we plot the average fractional error in $\mu$ as a proxy for $\chi^2$ as a
function of the number of steps sampled (note: the vertical axis actually shows the average
fractional error over the preceding 500 steps so that the early, imprecise models do not
contaminate the later, precise models).  As APS samples more points, its Gaussian process
model becomes more accurate.  This effect is more pronounced for covariogram (\ref{eqn:GaussianCovar})
than for covariogram (\ref{eqn:NNCovar}).  However, we will see below that this difference
does not affect the parameter constraints determined by APS.

We now consider the ultimate results, namely the 95\% confidence limits on
our 6-dimensional cosmological parameter space yielded by APS.
To provide a baseline against which to test APS, we run 4 independent MCMC
chains using CosmoMC.  Like the MCMC code used in Section \ref{sec:toy},
CosmoMC periodically adjusts its proposal density by learning the covariance
matrix of the points it has already sampled.  At each step, it proposes
a new point in parameter space by randomly selecting an eigen vector
of that learned covariance matrix and stepping along it.
We integrate the posterior by taking the output chains, discarding the first
50\% of steps as a burn-in period, and thinning the remaining steps so that
we have a set of effectively independent samples drawn from the posterior.
Quantitatively, this is determined by keeping only every $L$th sample after
burn-in, where $L$ is set as the length such that the normalized correlation
\begin{equation}
\frac{\text{Cov}[\theta_n^{(i)},\theta_n^{(i+L)}]}
{\text{Var}[\theta_n]}\le 0.1
\end{equation}
where $n$ is the index over the $N_p$ parameters and $(i)$ is the index
over the samples drawn.

We find the 95\% credible
limits in two-dimensional slices of our parameter space using
a total of 460,000 MCMC steps.  These are assumed to be the
true credible limits on the parameter space and are plotted
as the thick red
contours in Figures \ref{fig:contours_gauss} through 
\ref{fig:contours_br} 
below.  Against these control contours, we plot the Frequentist
confidence limits (all points with $\chi^2\le\chi^2_\text{lim}$)
discovered by APS after only 50,000 evaluations of $\chi^2$.
We consider both an adaptive $\chi^2_\text{lim}=\chi^2_\text{min}+12.6$
(12.6 is the $\Delta\chi^2$ corresponding for 95\% confidence limit in the case
of 6 parameters)
and an absoulte $\chi^2_\text{lim}=1280.7$ (this is the 95\% confidence limit
for a $\chi^2$ probability distribution with the 1199 degrees of freedom corresponding
to the 1199 $C_\ell$s in the data set).
These are the black points in Figures \ref{fig:contours_gauss}, \ref{fig:contours_nn},
and \ref{fig:contours_br} respectively.  These confidence limits overlap very nicely
with the true credible limits discovered by MCMC,
comporting the the idea that, in the large data limit, Bayesian credible limits
and Frequentist confidence limits ought to align.

\begin{figure*}
\subfigure[]{
\includegraphics[scale=0.15]{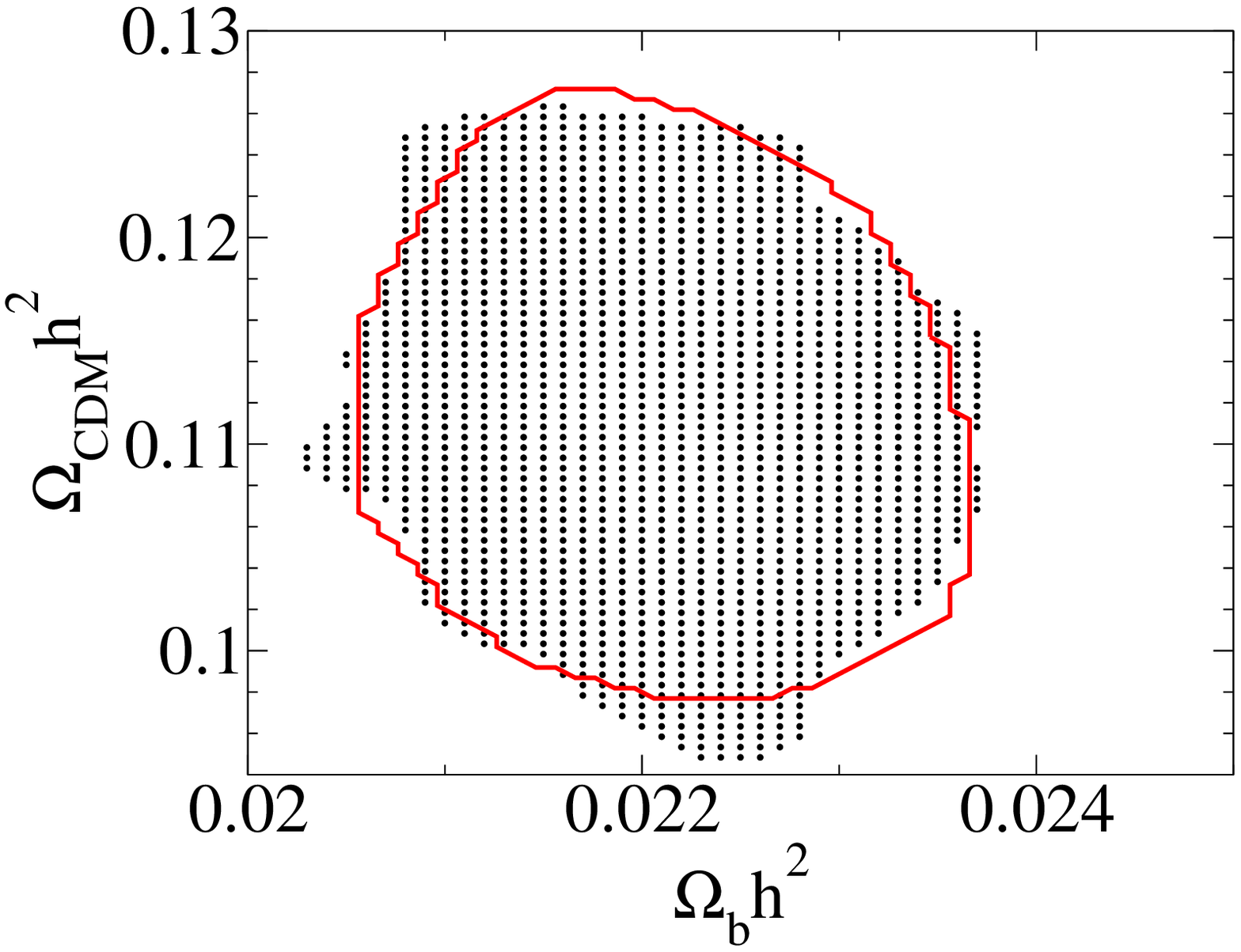}
\label{fig:mcmc_0_1}
}
\subfigure[]{
\includegraphics[scale=0.15]{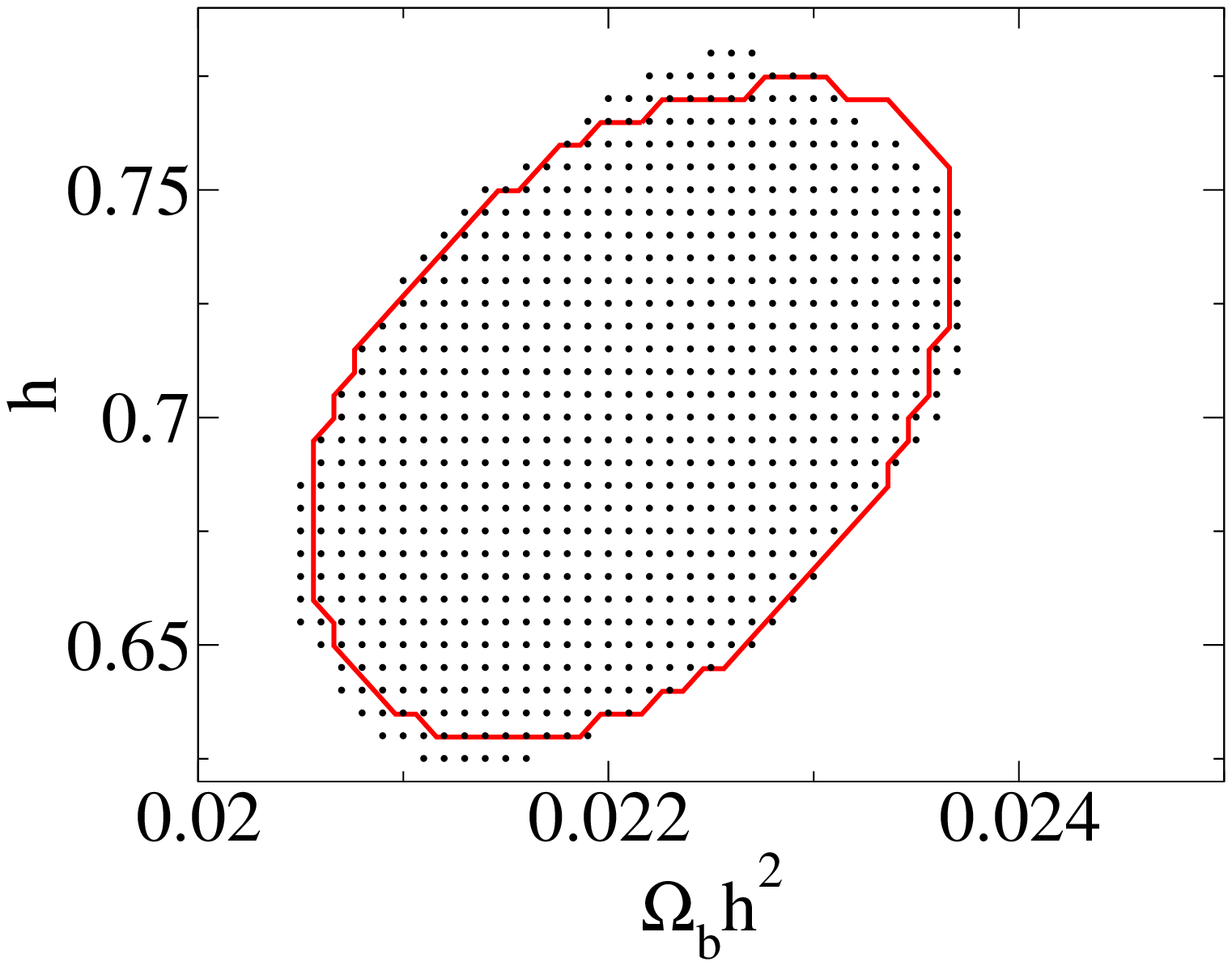}
\label{fig:mcmc_0_2}
}
\subfigure[]{
\includegraphics[scale=0.15]{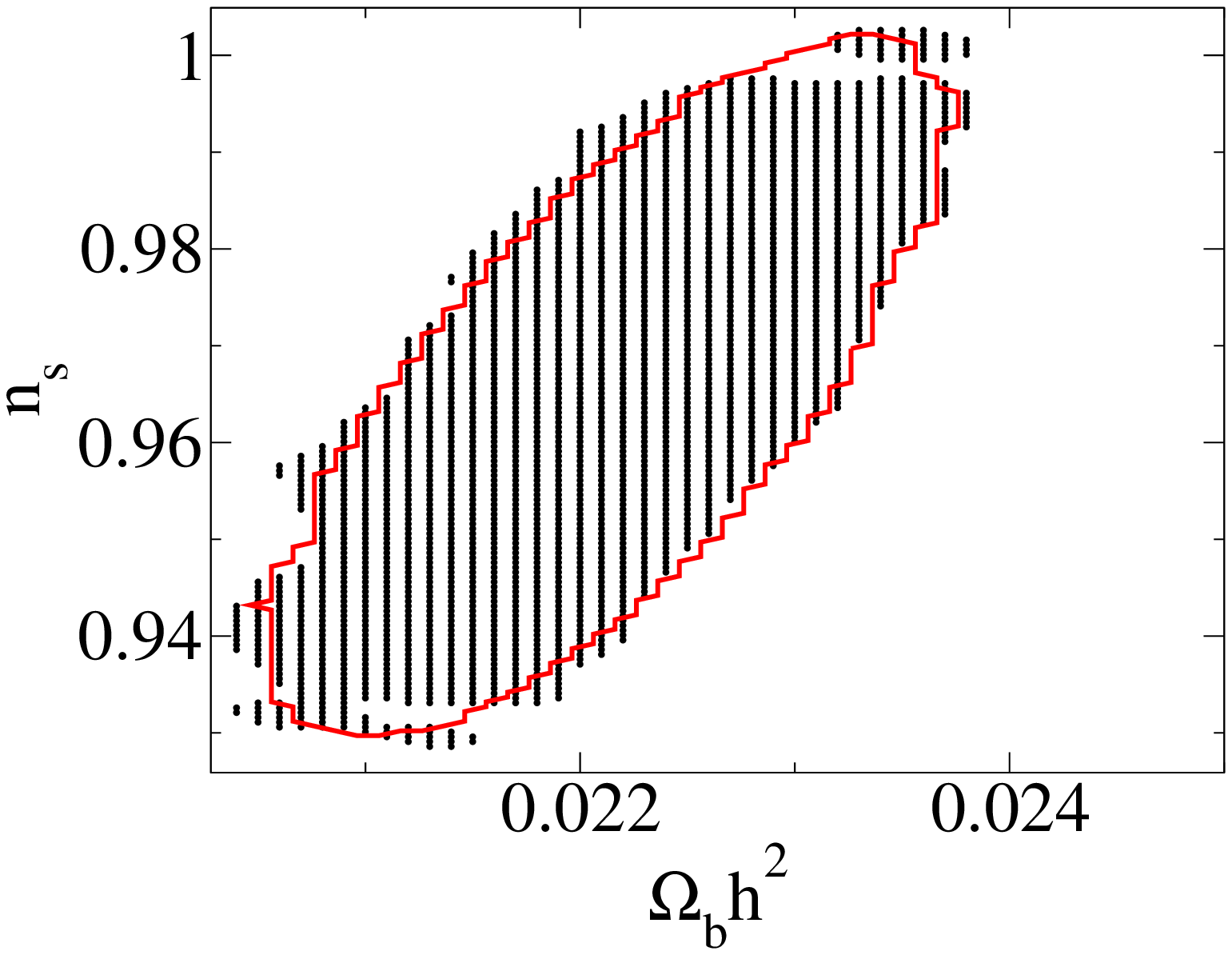}
\label{fig:mcmc_0_4}
}
\subfigure[]{
\includegraphics[scale=0.15]{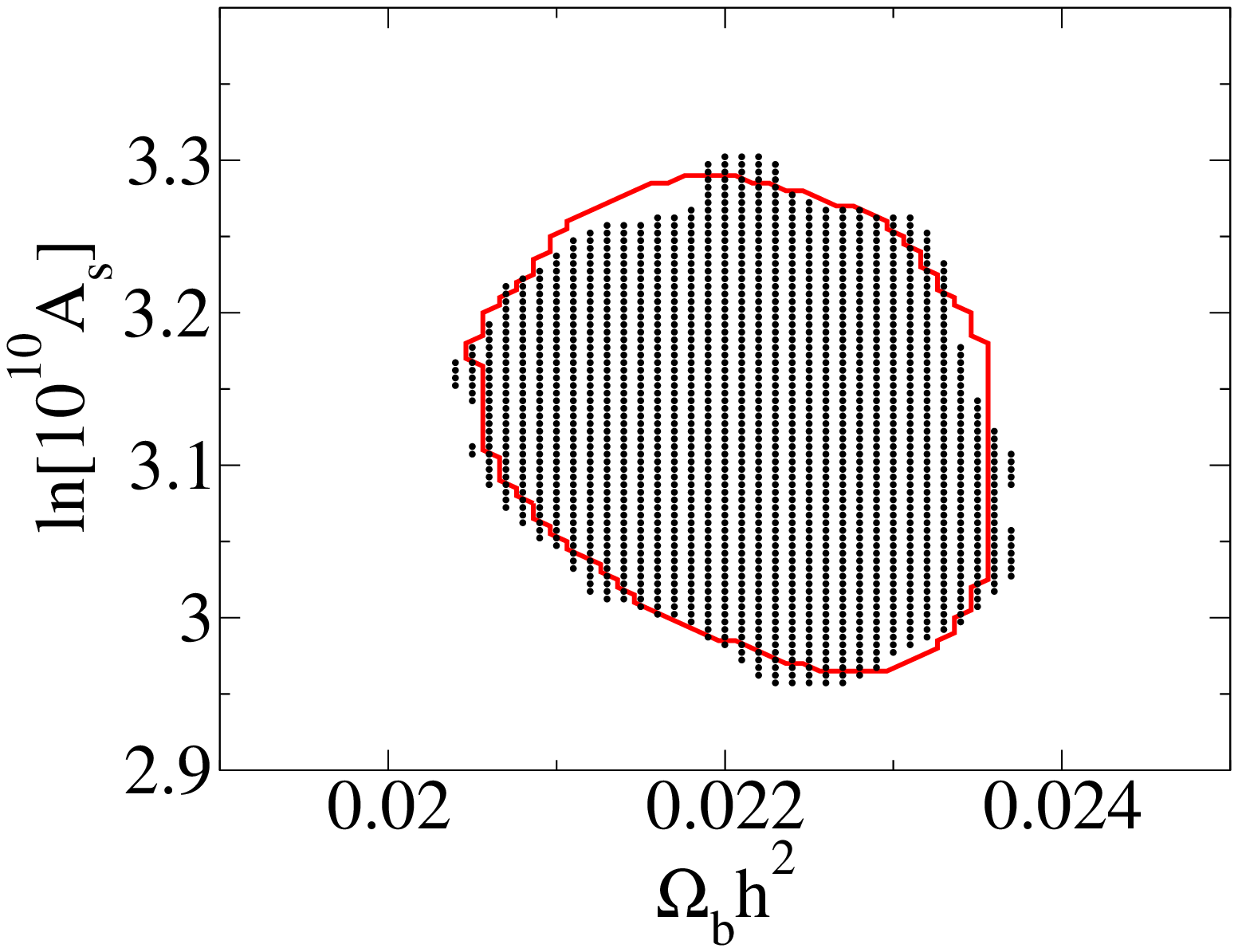}
\label{fig:mcmc_0_5}
}
\subfigure[]{
\includegraphics[scale=0.15]{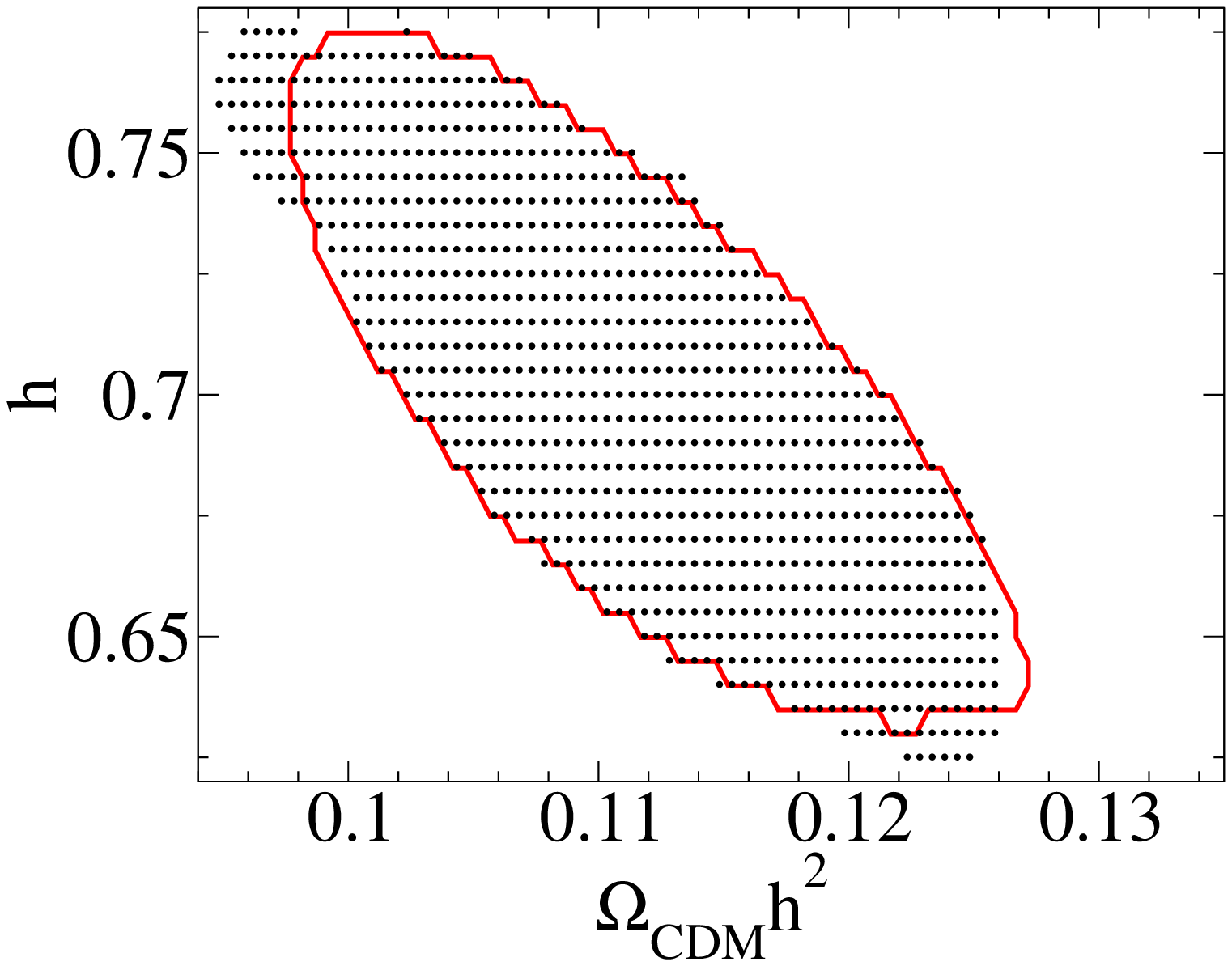}
\label{fig:mcmc_1_2}
}
\subfigure[]{
\includegraphics[scale=0.15]{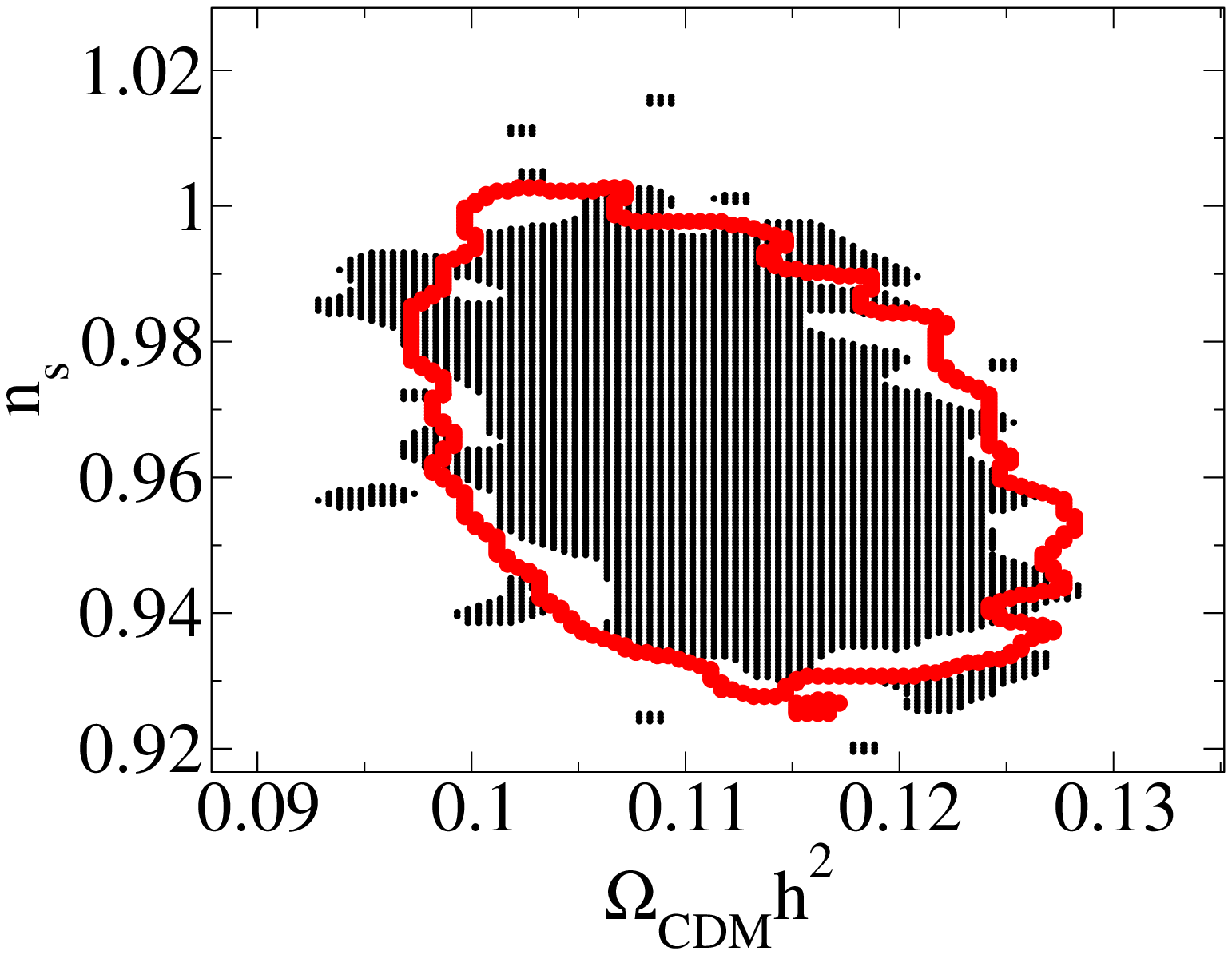}
\label{fig:mcmc_1_4}
}
\subfigure[]{
\includegraphics[scale=0.15]{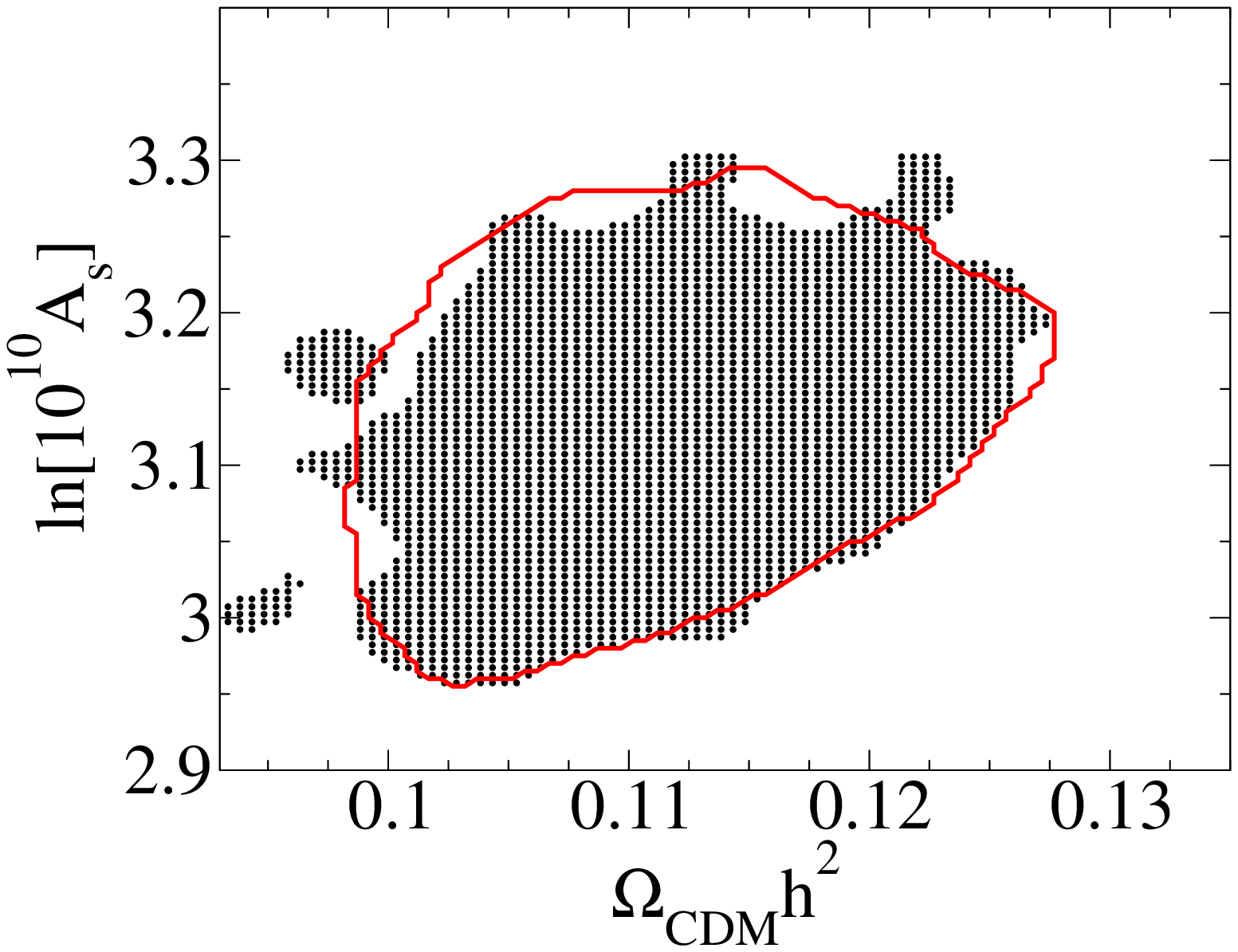}
\label{fig:mcmc_1_5}
}
\subfigure[]{
\includegraphics[scale=0.15]{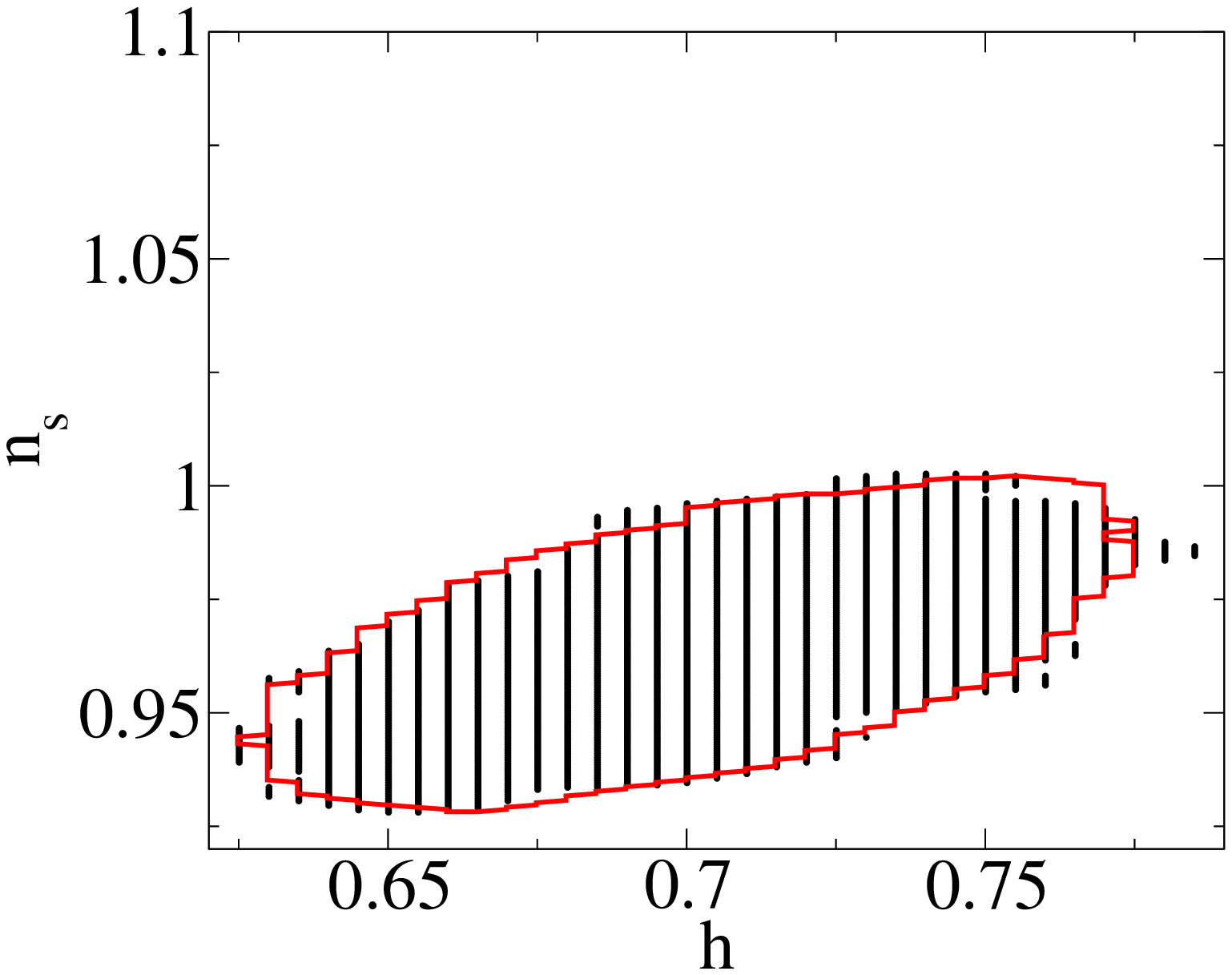}
\label{fig:mcmc_2_4}
}
\subfigure[]{
\includegraphics[scale=0.15]{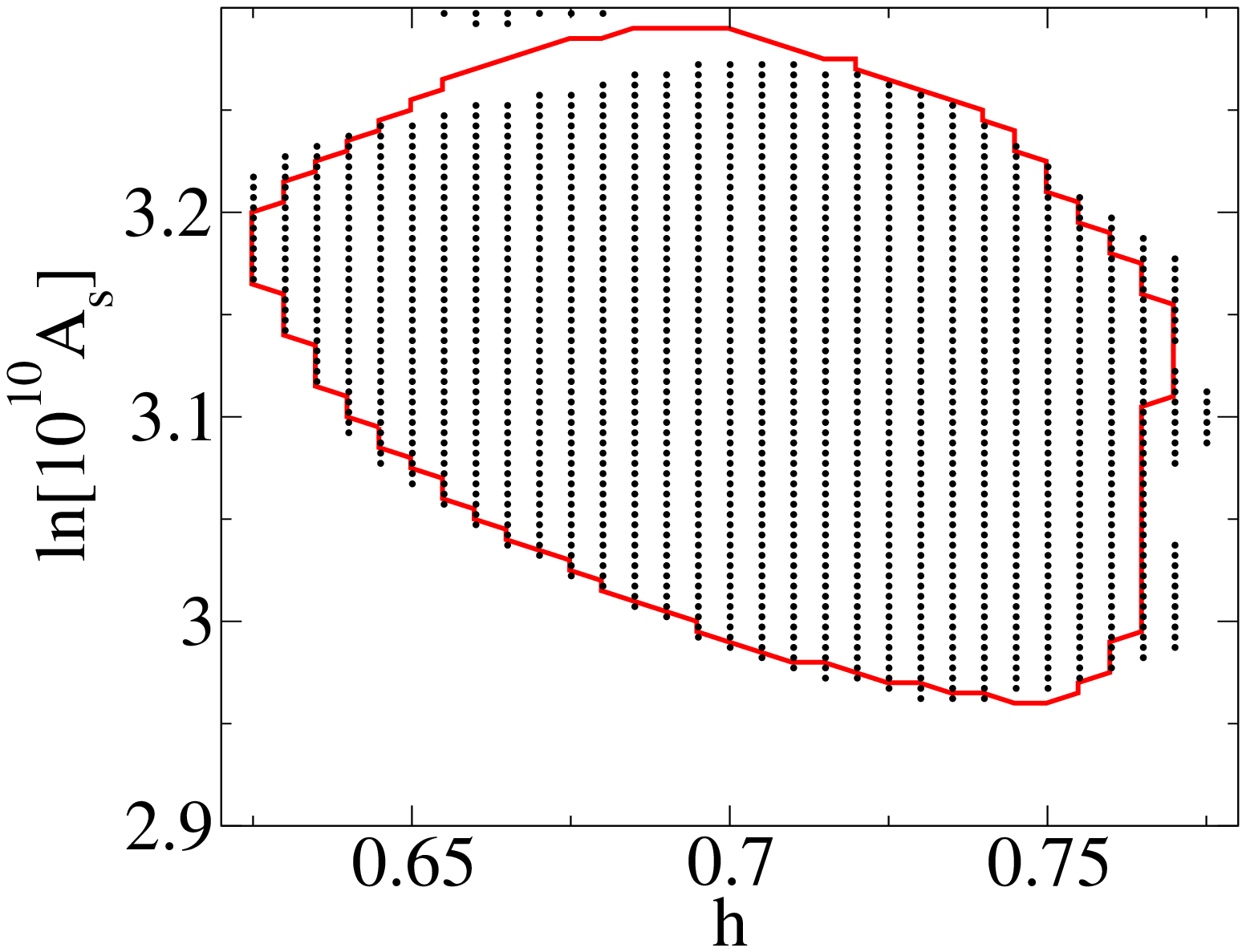}
\label{fig:mcmc_2_5}
}
\subfigure[]{
\includegraphics[scale=0.15]{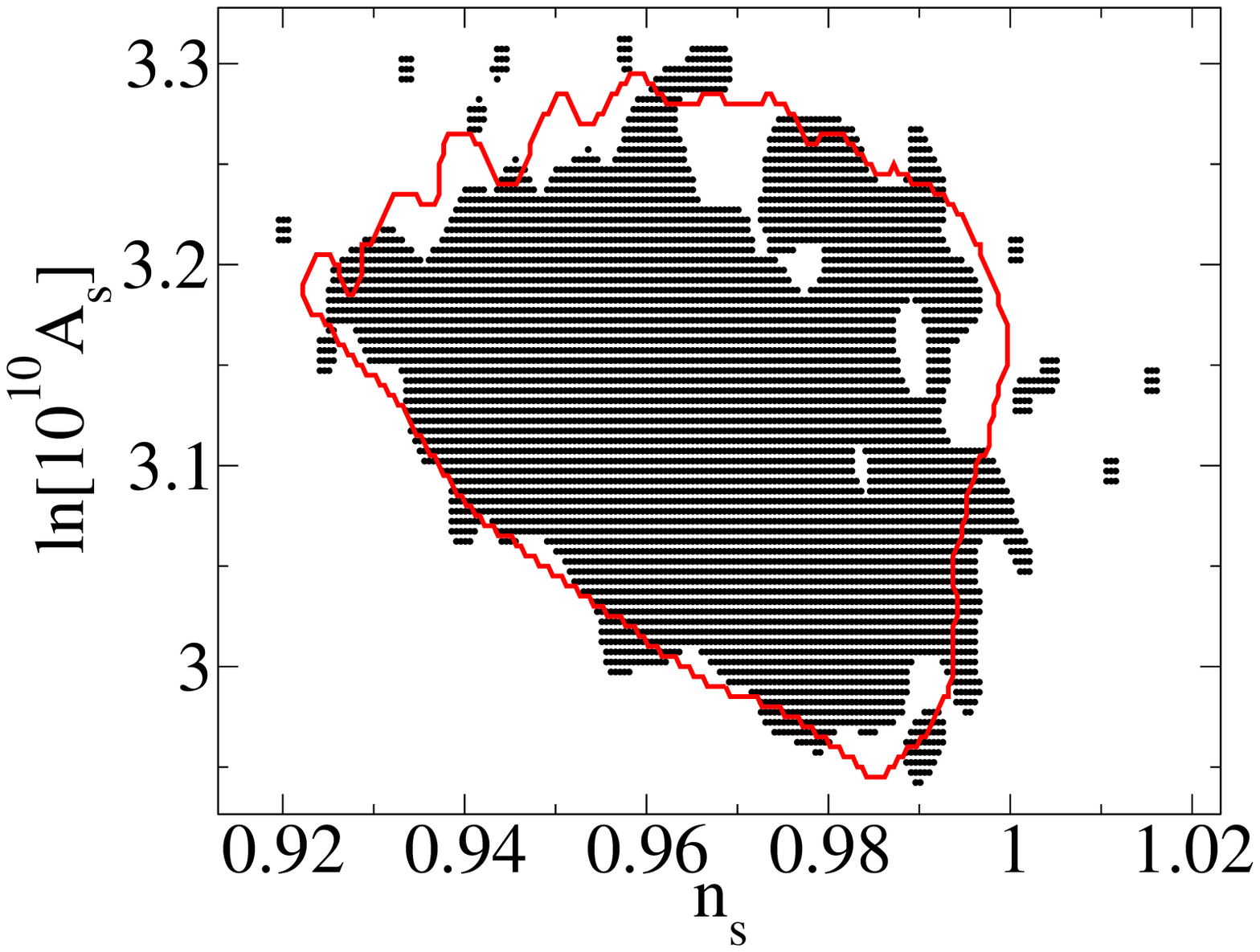}
\label{fig:mcmc_4_5}
}
\caption{
The 10 2-dimensional sub-spaces of our 6-dimensional parameter space
($\tau$ is ignored because no real constraint can be gleaned from this data).
The thick red contours are the 95\% Bayesian credible limits determined by MCMC after
460,000 calls to $\chi^2$.  The black regions are the 95\% Bayesian credible limits
determined by MCMC after 50,000 calls to $\chi^2$.
}
\label{fig:mcmc_comparison}
\end{figure*}

For comparison of convergence properties, we plot the Bayesian credible limits discovered
by MCMC after only 50,000 $\chi^2$ evaluations.  These are the black regions in
Figure \ref{fig:mcmc_comparison}.  
While MCMC has clearly converged to the final
confidence limits in some of the 2-dimensional sub-spaces shown,
others (Figures \ref{fig:mcmc_1_4}, \ref{fig:mcmc_1_5}, and \ref{fig:mcmc_4_5})
evince the presence of significant noise in the MCMC chains.
Indeed, even the ``final'' credible limit contour (the red contour) in Figure
\ref{fig:mcmc_1_4} shows significantly more noise along the edges than, for instance,
Figure \ref{fig:mcmc_0_1}.  This is meant, not to disparage the convergence
properties of MCMC, but to show that any convergence advantage MCMC may have over APS is
not absolute.  This should be compared with the clear advantage APS demonstrated over 
MCMC in identifying multi-modal likelihood functions in Section \ref{sec:toy} above.

\begin{figure*}
\subfigure[]{
\includegraphics[scale=0.15]{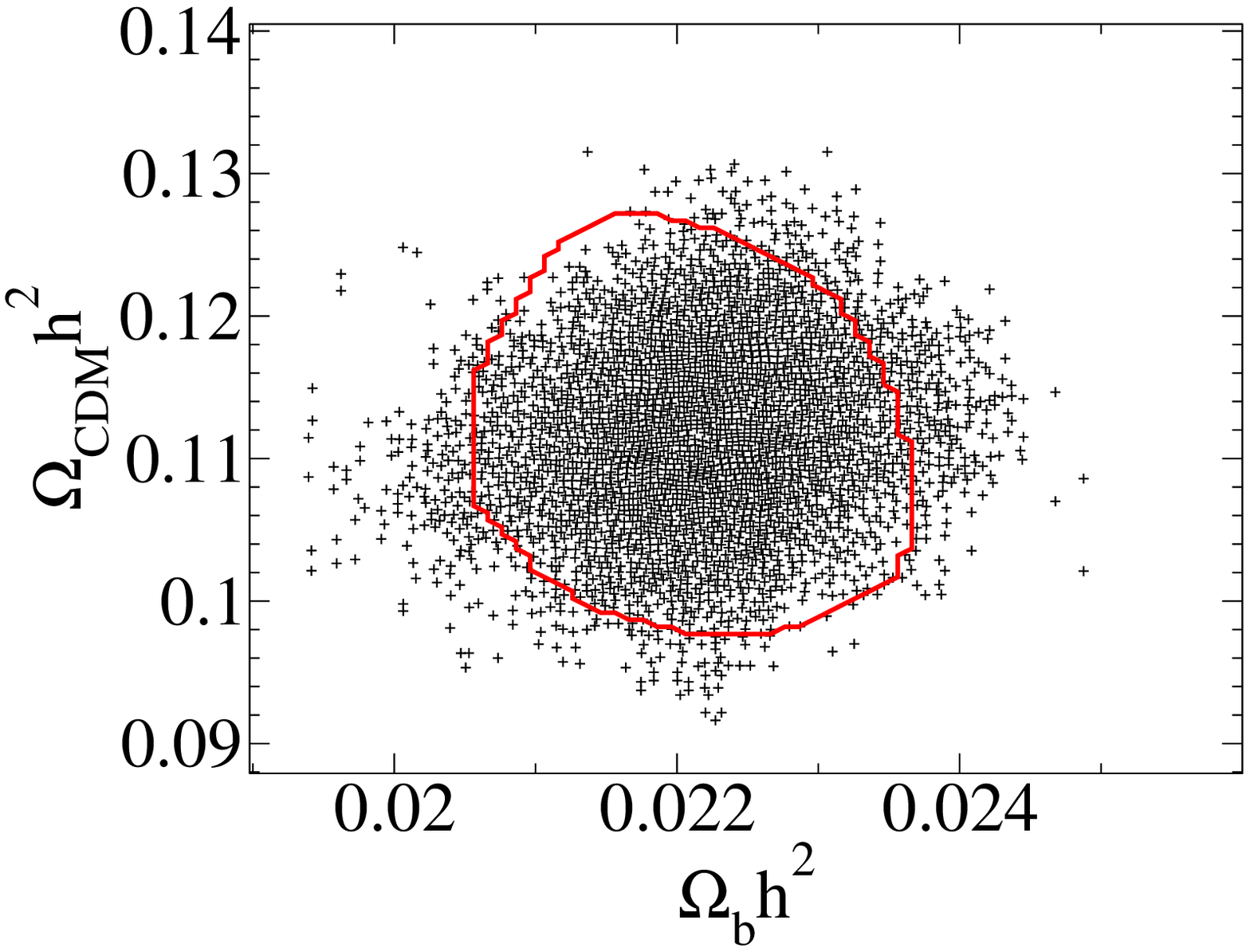}
}
\subfigure[]{
\includegraphics[scale=0.15]{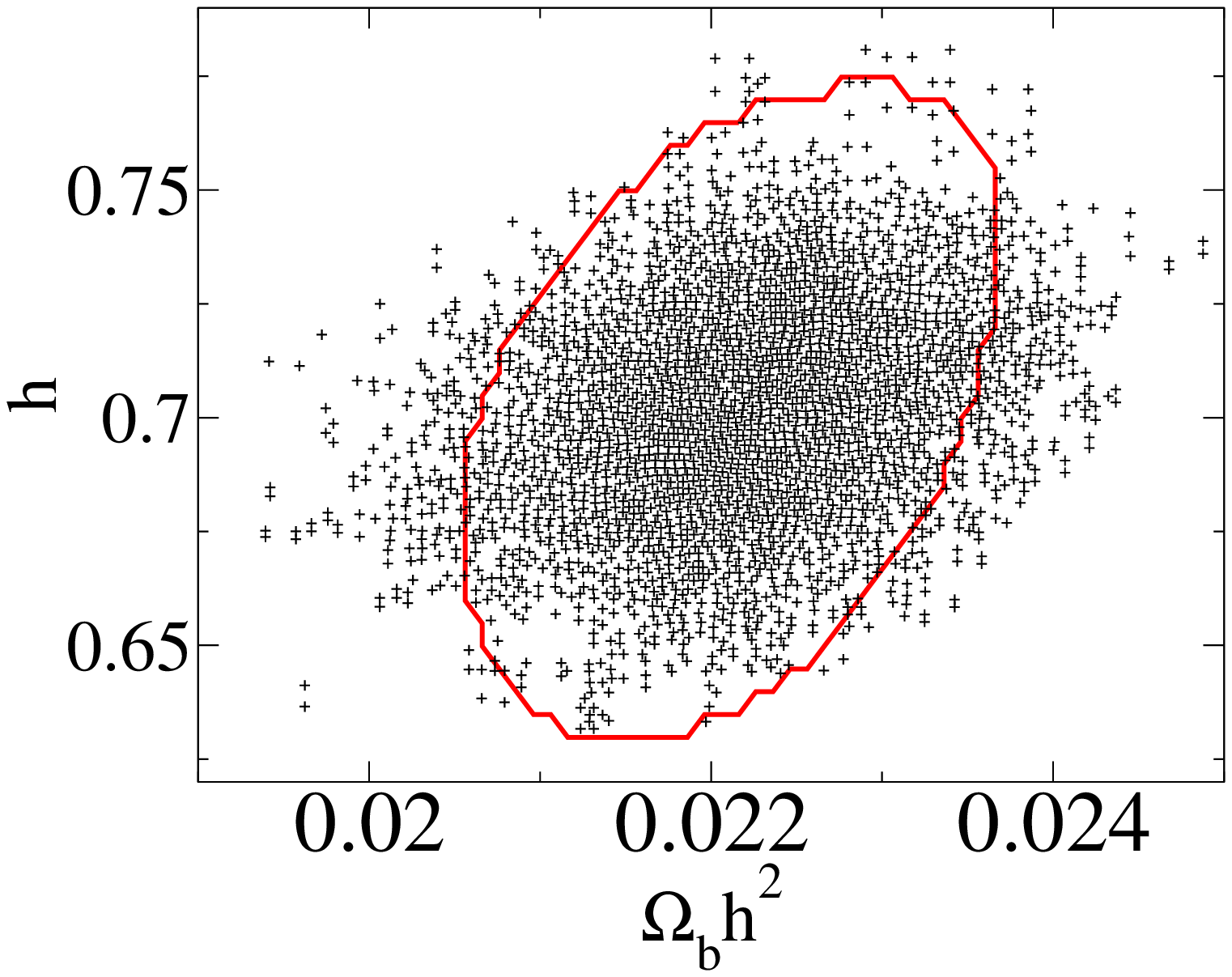}
}
\subfigure[]{
\includegraphics[scale=0.15]{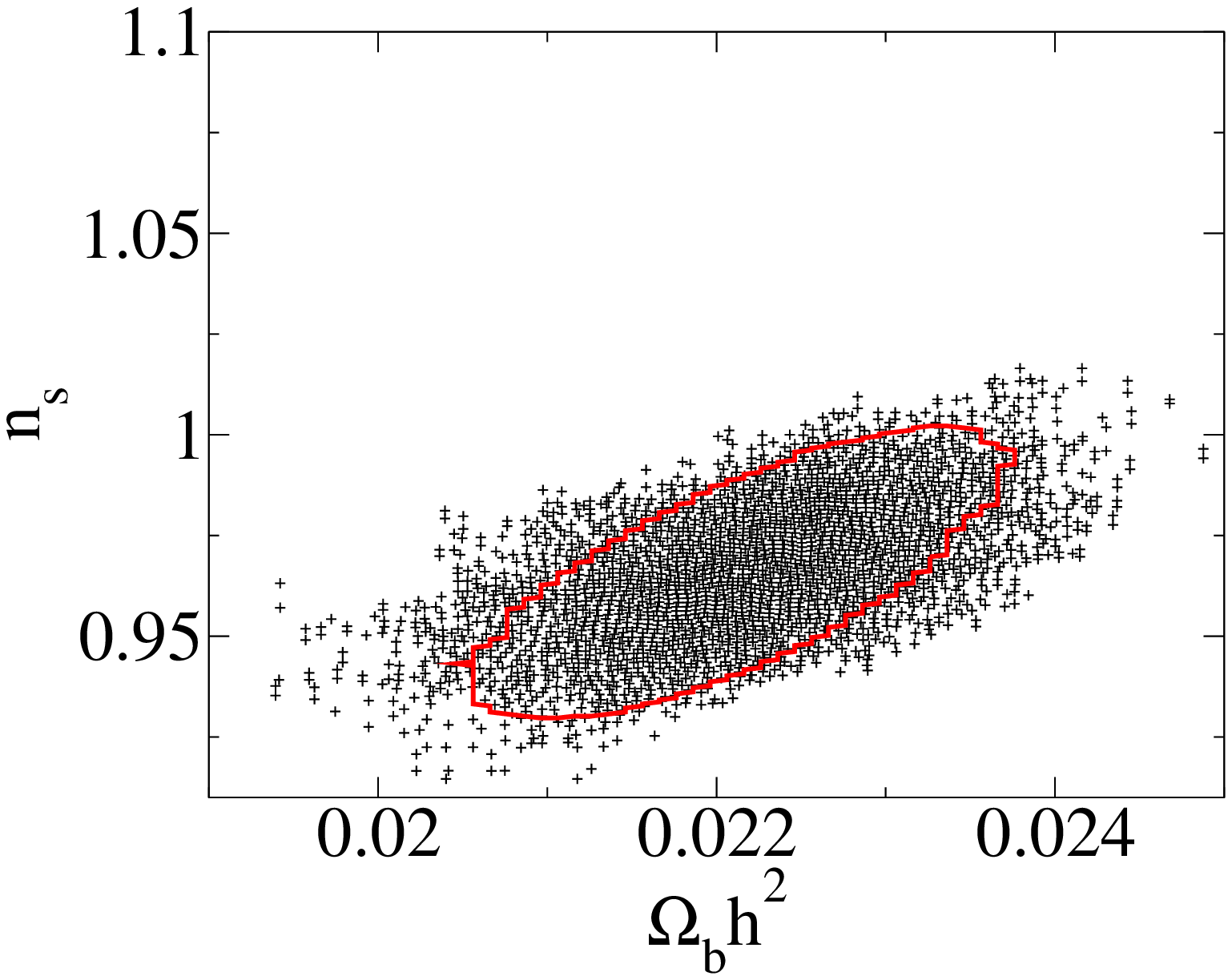}
}
\subfigure[]{
\includegraphics[scale=0.15]{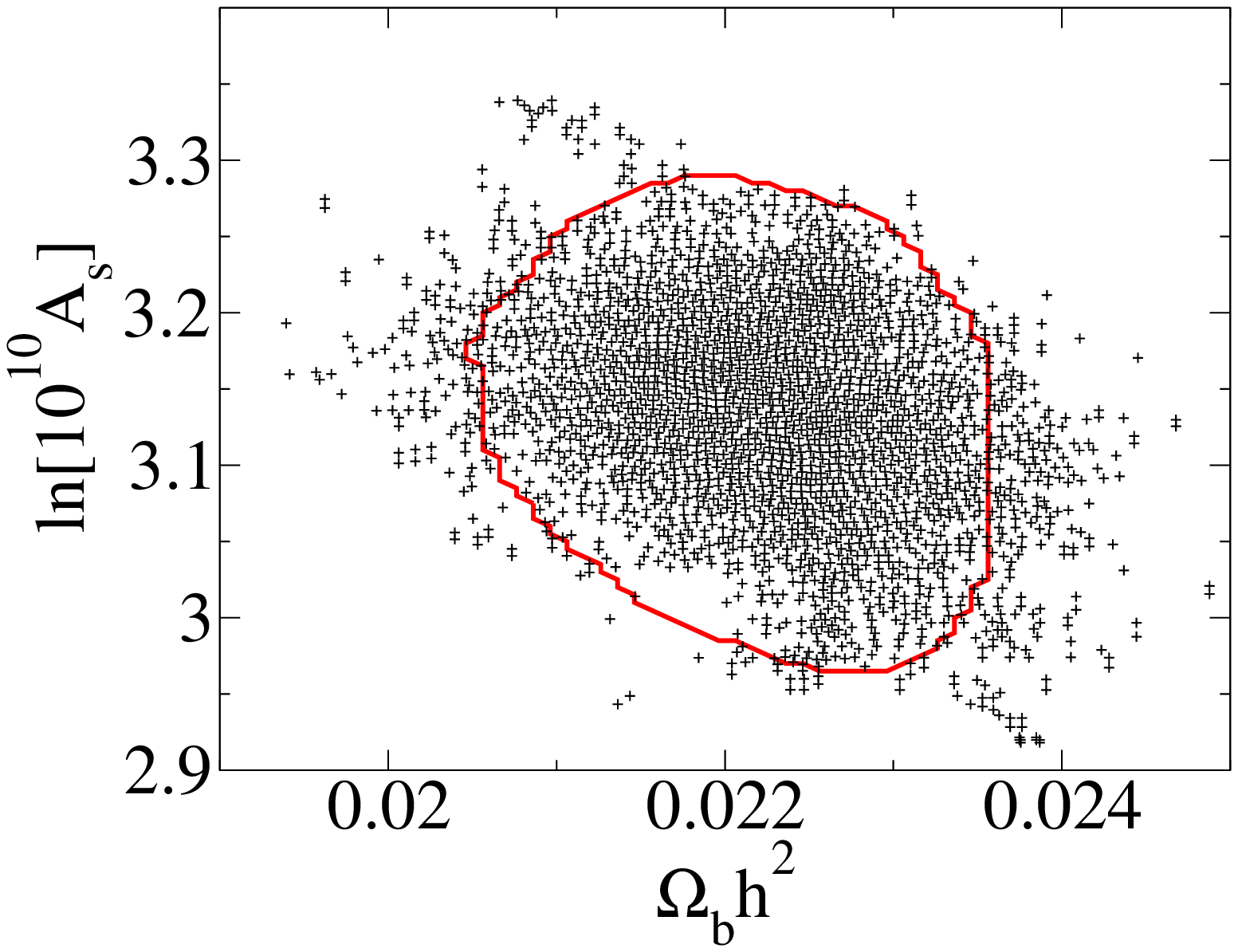}
}
\subfigure[]{
\includegraphics[scale=0.15]{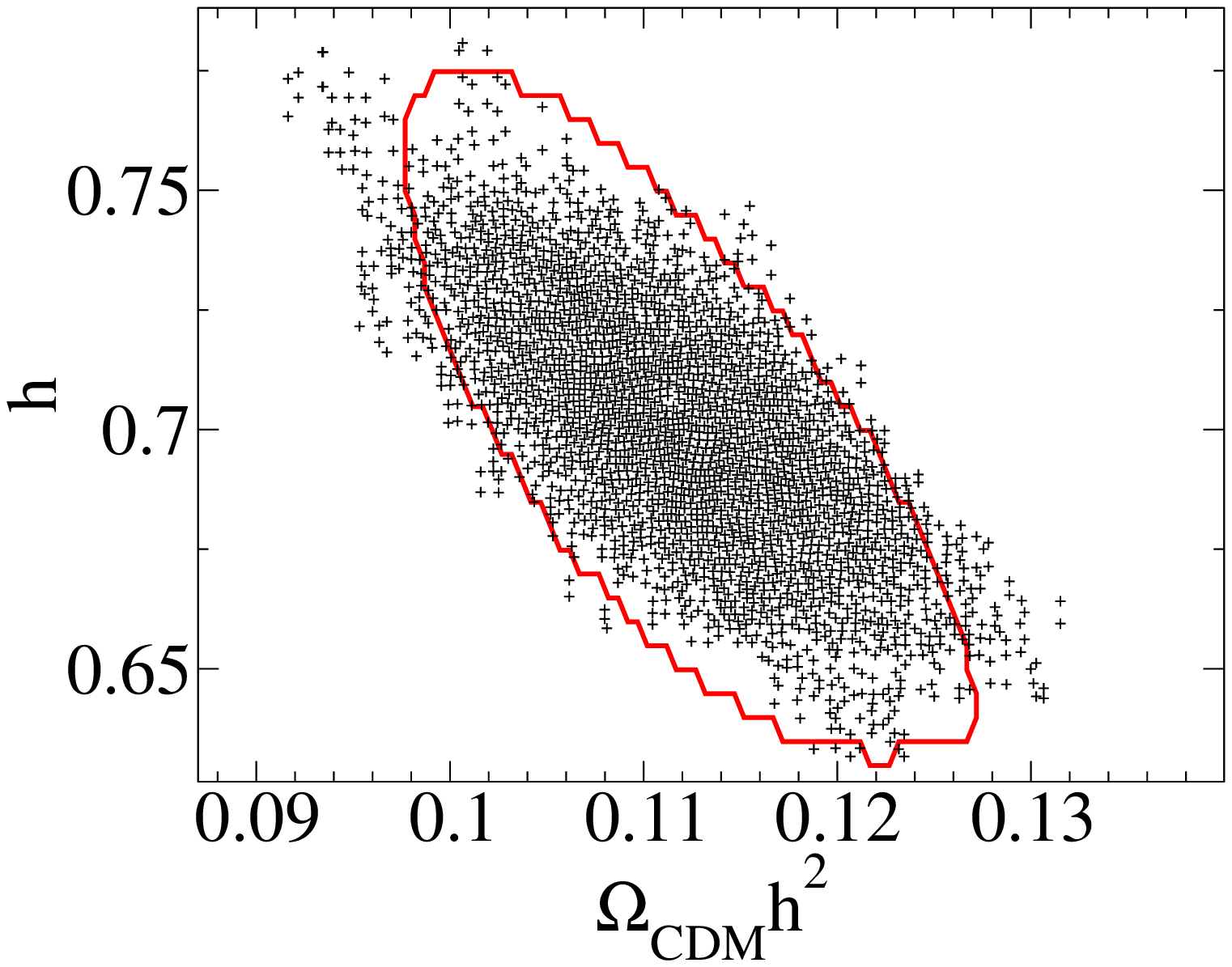}
}
\subfigure[]{
\includegraphics[scale=0.15]{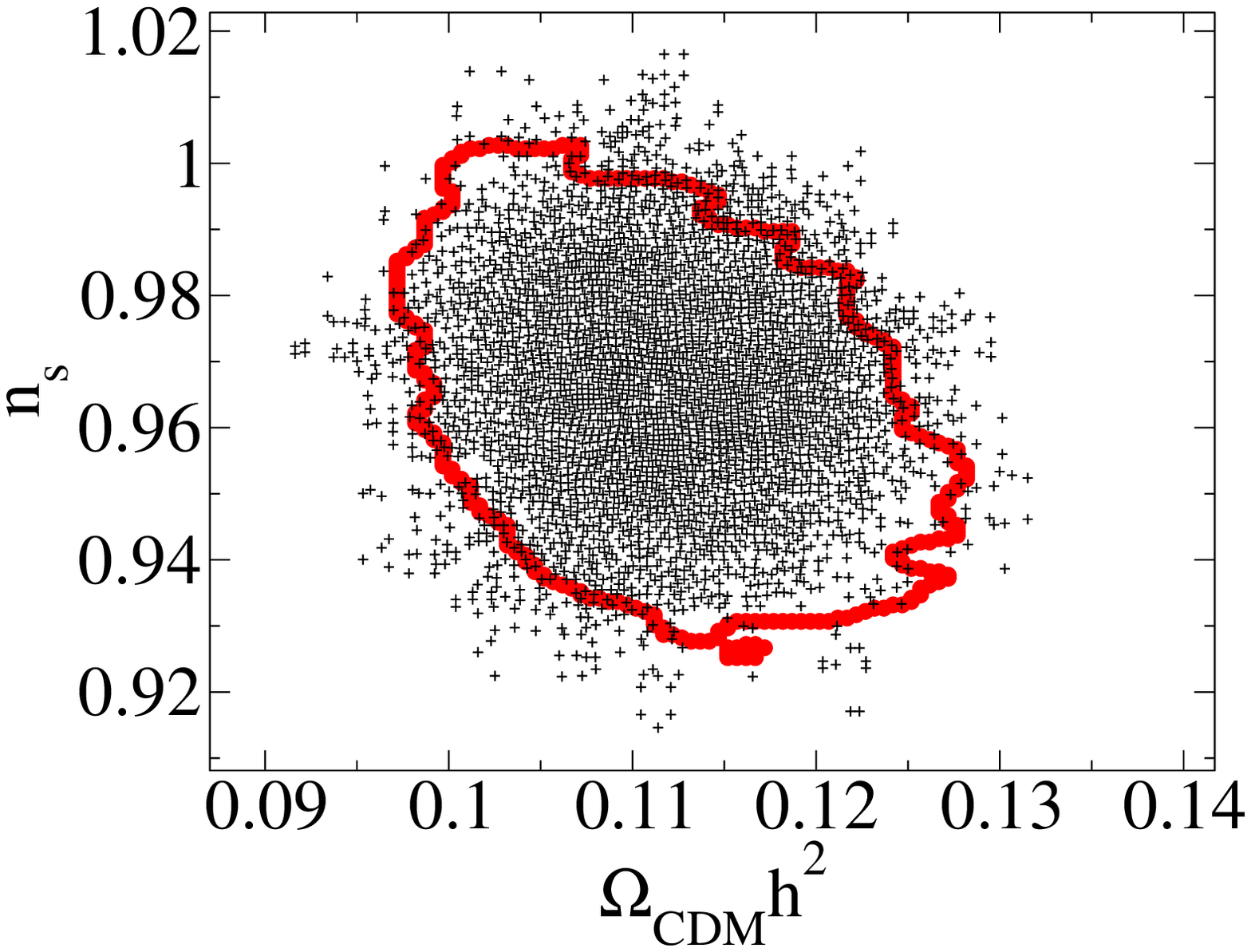}
}
\subfigure[]{
\includegraphics[scale=0.15]{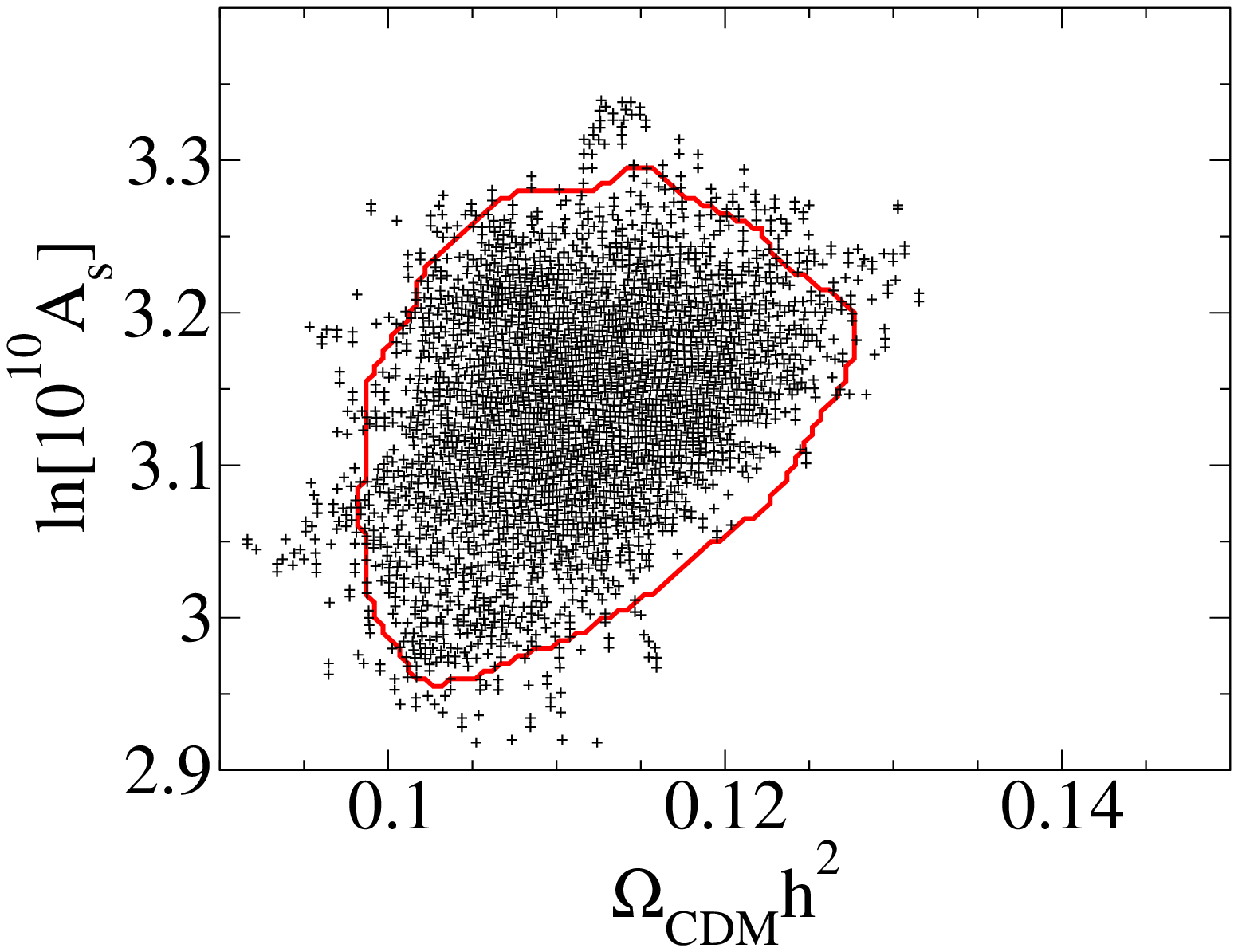}
}
\subfigure[]{
\includegraphics[scale=0.15]{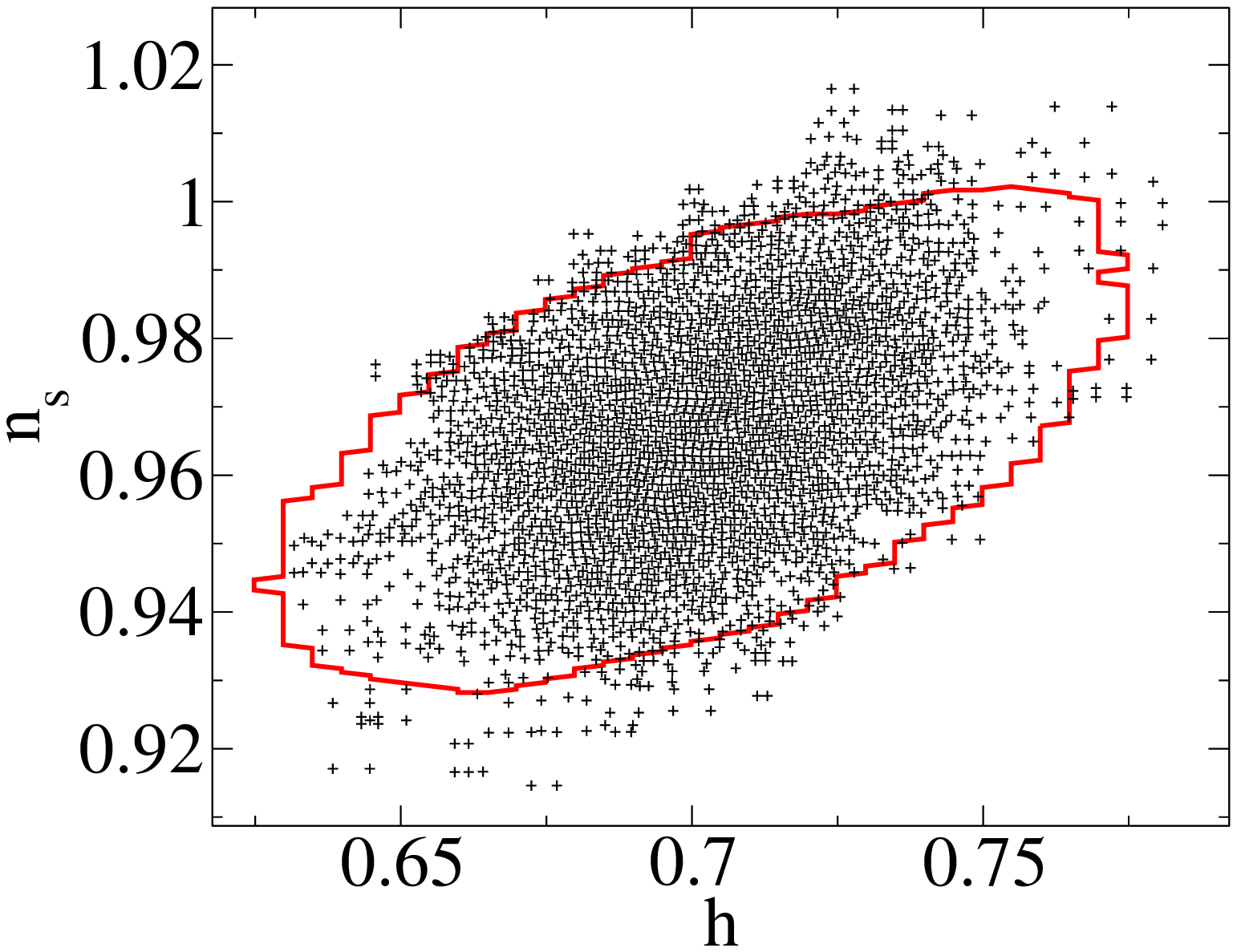}
}
\subfigure[]{
\includegraphics[scale=0.15]{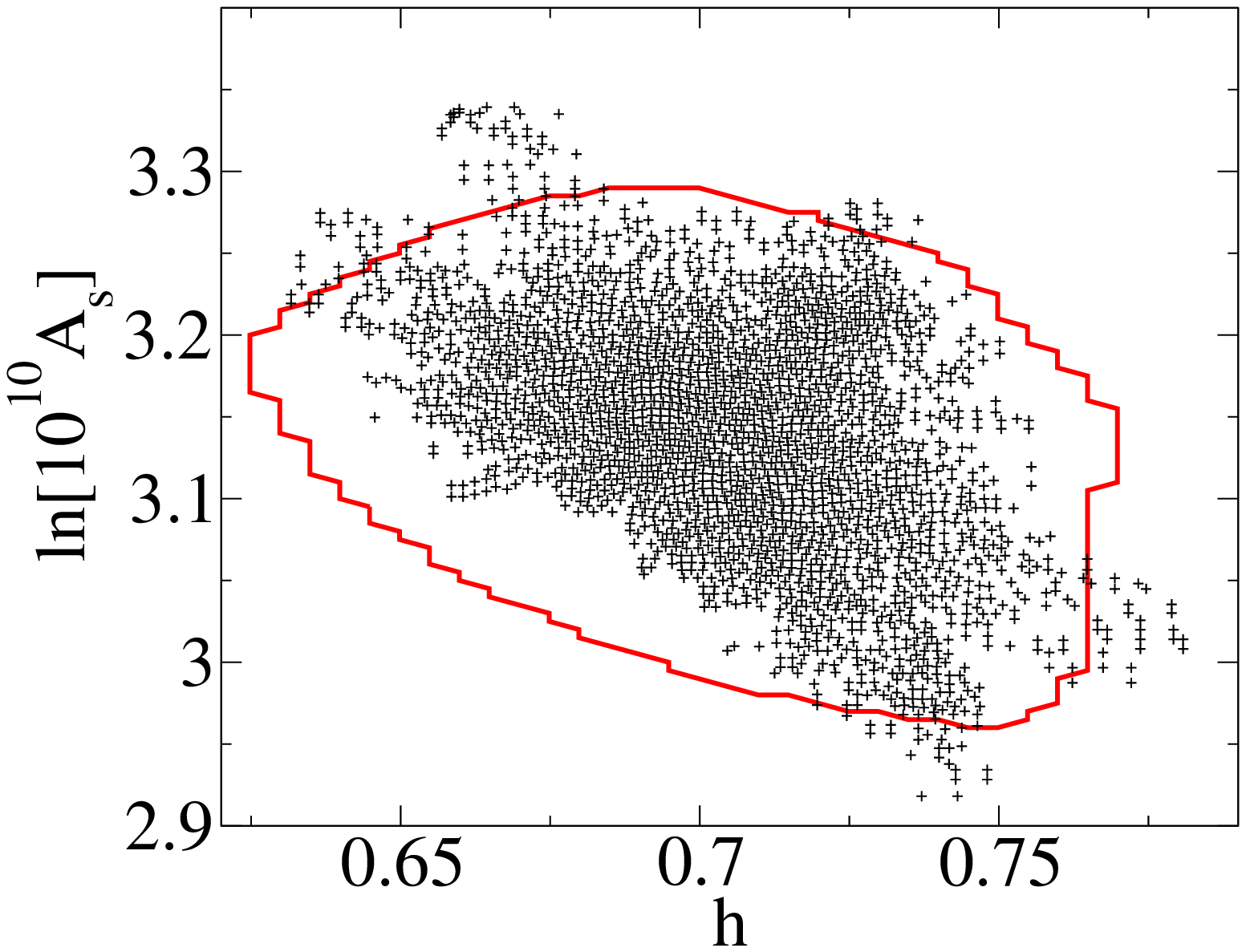}
}
\subfigure[]{
\includegraphics[scale=0.15]{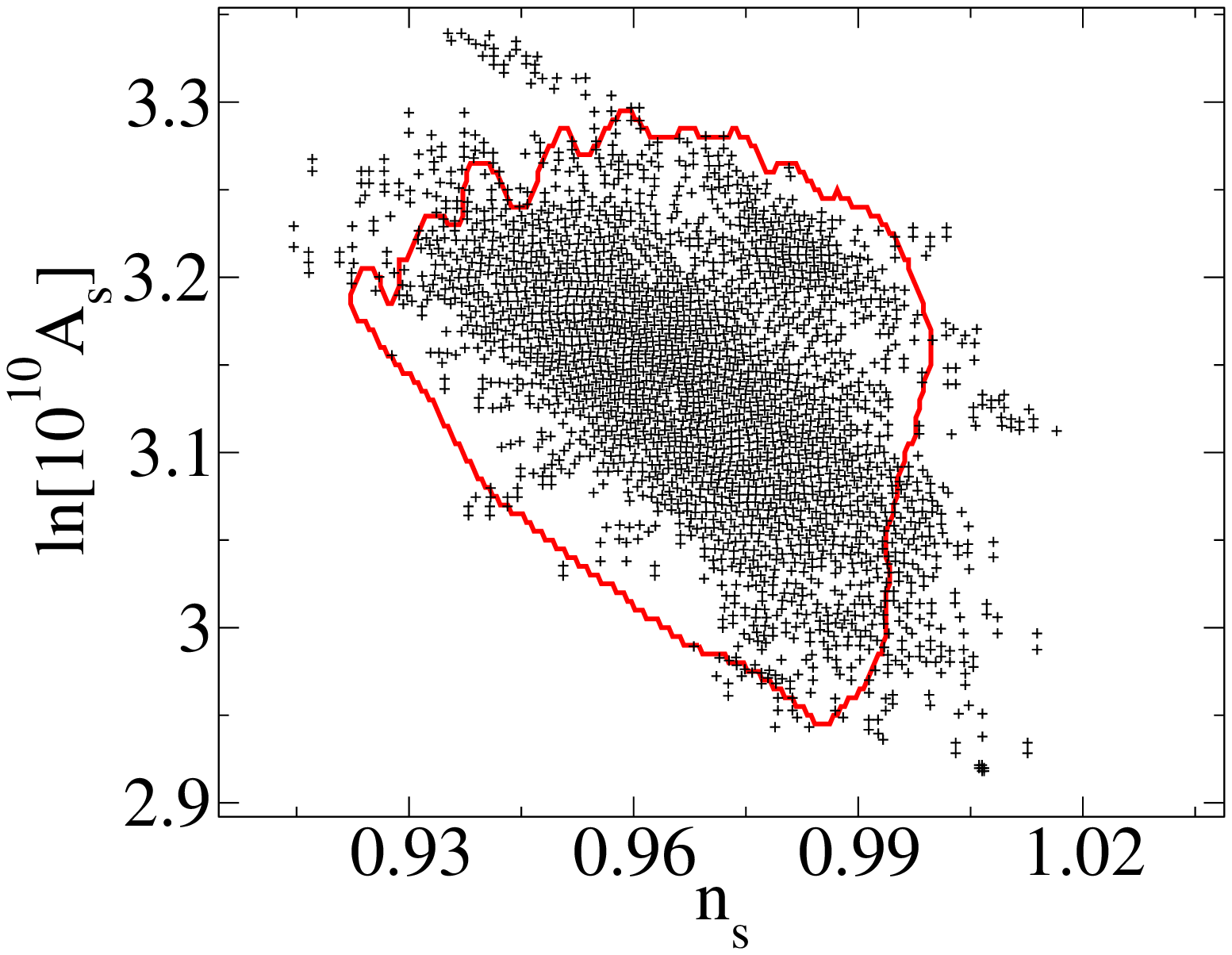}
}
\caption{
The 10 2-dimensional sub-spaces of our 6-dimensional parameter space
($\tau$ is ignored because no real constraint can be gleaned from this data).
The thick red contours are the 95\% Bayesian credible limits determined by MCMC after
460,000 calls to $\chi^2$.  The black points are the 95\% Bayesian credible limits determined
by APS according to the method outlined in Section \ref{sec:bayes}.  While the APS credible limit
covers the entire MCMC limit, it also covers regions of parameter space excluded by MCMC.
}
\label{fig:contours_bayes}
\end{figure*}

In Figure \ref{fig:contours_bayes}, we plot the 95\% Bayesian credible limit discovered
by APS after 50,000 evalutations of $\chi^2$ and determined by the method outlined in
Section \ref{sec:bayes}.  Unfortunately, we do not see here the convergence to the true
credible limits that we saw in Figures \ref{fig:toyBayes2}-\ref{fig:toyBayes4}.  While
it does appear that APS has correctly identified the locations and general shapes of the
true credible limits, it has, in most cases, drawn credible limits that are larger than
those found by MCMC.  This should not be surprising.  APS was designed with Frequentist
confidence limits in mind, and Figures \ref{fig:contours_gauss}-\ref{fig:contours_br}
seem much better converged than Figure \ref{fig:contours_bayes}.  However, it does seem
that APS might be useful in cases where little prior knowledge is 
known about $\vec{\theta}\rightarrow\chi^2$
and users need to be confident that they have discovered all of the low-$\chi^2$
regions in a particular parameter space.  Once APS has been used to identify rough
Bayesian credible limits, more detailed algorithms (like MCMC) can be run using the prior
knowledge gained by APS to refine the credible limits.

As a final note, we do not plot any confidence or 
credible limits for $\tau$.
This is because $\tau$ needs CMB polarization data for meaningful constraints.  For the sake
of simplicity, we did not consider CMB polarization data here.  Thus, $\tau$ has become a
``nuisance parameter'': it affects the fit of our theoretical models, but cannot itself be
meaningfully constrained.  MCMC handles such parameters by effectively integrating over them
when delivering the two-dimensional contours of Figures
\ref{fig:contours_gauss} through \ref{fig:contours_br}.  
APS handles such nuisance parameters by
simply allowing them to vary and returning any combinations of useful-plus-nuisance parameters
that satisfy the $\chi^2\le\chi^2_\text{lim}$ criterion.  The fact that this nuisance
paramter did not significantly impede the convergence of APS relative to MCMC speaks 
well of
APS' ability to handle actual physical problems which 
often include parameters that have
little physical interest but are required to, e.g., 
calibrate noise distributions or
systematics.

We reiterate that the purpose of this section was to demonstrate that the presence of
real-world noise in the data and likelihood function do not significantly impede APS's
ability to determine confidence and credible limits.  The problem of drawing parameter
constraints from the WMAP likelihood function is, by now, well-understood, and shrewdly
drawn priors and proposal densities are readily available.  
Indeed, prior knowledge of early WMAP results allowed us to use the following 
initial proposal densities for the MCMC runs displayed in this section:
\begin{eqnarray}
\Omega_\text{b}h^2&=&0.03\pm0.005\label{eqn:mcmcProposal}\\
\Omega_\text{CDM}h^2&=&0.1\pm0.05\nonumber\\
\theta&=&1.0\pm0.01\nonumber\\
\tau&=&0.07\pm0.03\nonumber\\
n_s&=&0.95\pm0.05\nonumber\\
\ln\left[10^{10}A_s\right]&=&3.0\pm0.2\nonumber
\end{eqnarray}
where $\theta$ is 100 times the ratio of the sound horizon to the angular diameter
distance and recombination (this is the input parameter CosmoMC uses in place of $h$).
These distributions are fairly similar to the final, 
marginalized, one-dimensional distributions of
those same parameters
\begin{eqnarray}
\Omega_\text{b}h^2&=&0.0221\pm0.0006\nonumber\\
\Omega_\text{CDM}h^2&=&0.112\pm0.006\nonumber\\
\theta&=&1.039\pm0.003\nonumber\\
\tau&=&0.05\pm0.03\nonumber\\
n_s&=&0.96\pm0.02\nonumber\\
\ln\left[10^{10}A_s\right]&=&3.12\pm0.07\nonumber\\
h&=&0.7\pm0.03\nonumber
\end{eqnarray}
Conversely, for the purposes of our APS runs, we initialize the algorithm only with the
gross assumptions that
\begin{eqnarray}
0.01\le&\Omega_\text{b}h^2&\le0.04\label{eqn:apsProposal}\\
0.01\le&\Omega_\text{CDM}h^2&\le0.3\nonumber\\
0.4\le&h&\le1.0\nonumber\\
0.005\le&\tau&\le0.15\nonumber\\
0.7\le&n_s&\le1.3\nonumber\\
2.0\le&\ln\left[10^{10}A_s\right]&\le4.0\nonumber
\end{eqnarray}
In less well-studied problems, users may not understand their parameter spaces
well enough to be able to supply shrewd proposal densities in the vein of equations 
(\ref{eqn:mcmcProposal}).  In such a situation, the
exploratory behavior of APS demonstrated in Section \ref{sec:toy} 
may prove more important
than the theoretical rigor offered by MCMC's sampling algorithm.

\section{Conclusions}
\label{sec:conclusions}

In Section \ref{sec:toy} we indicated that, consistent with the findings
of Bryan {\it et al}. (2007), APS is more robust
than MCMC against multi-modal likelihood functions (at least on the idealized
cartoon $\vec{\theta}\rightarrow\chi^2$ functions considered).
In Section \ref{sec:wmap}, we showed that the presence of realistic noise
does not interfere with APS's ability to derive Frequentist confidence limits.
However, the Bayesian credible
limits yielded by APS remain noisy and imprecise.  We believe that this performance
can still be of use to the community, especially when trying to evaluate proof-of-concept
parameter constraints (perhaps for forecasting parameter constraints
from future experiments)
with very expensive $\vec{\theta}\rightarrow\chi^2$ functions on poorly-understood
parameter spaces.  Certainly, for users interested in plotting Frequentist confidence 
limits, APS is a viable option.

\begin{figure}
\includegraphics[scale=0.3]{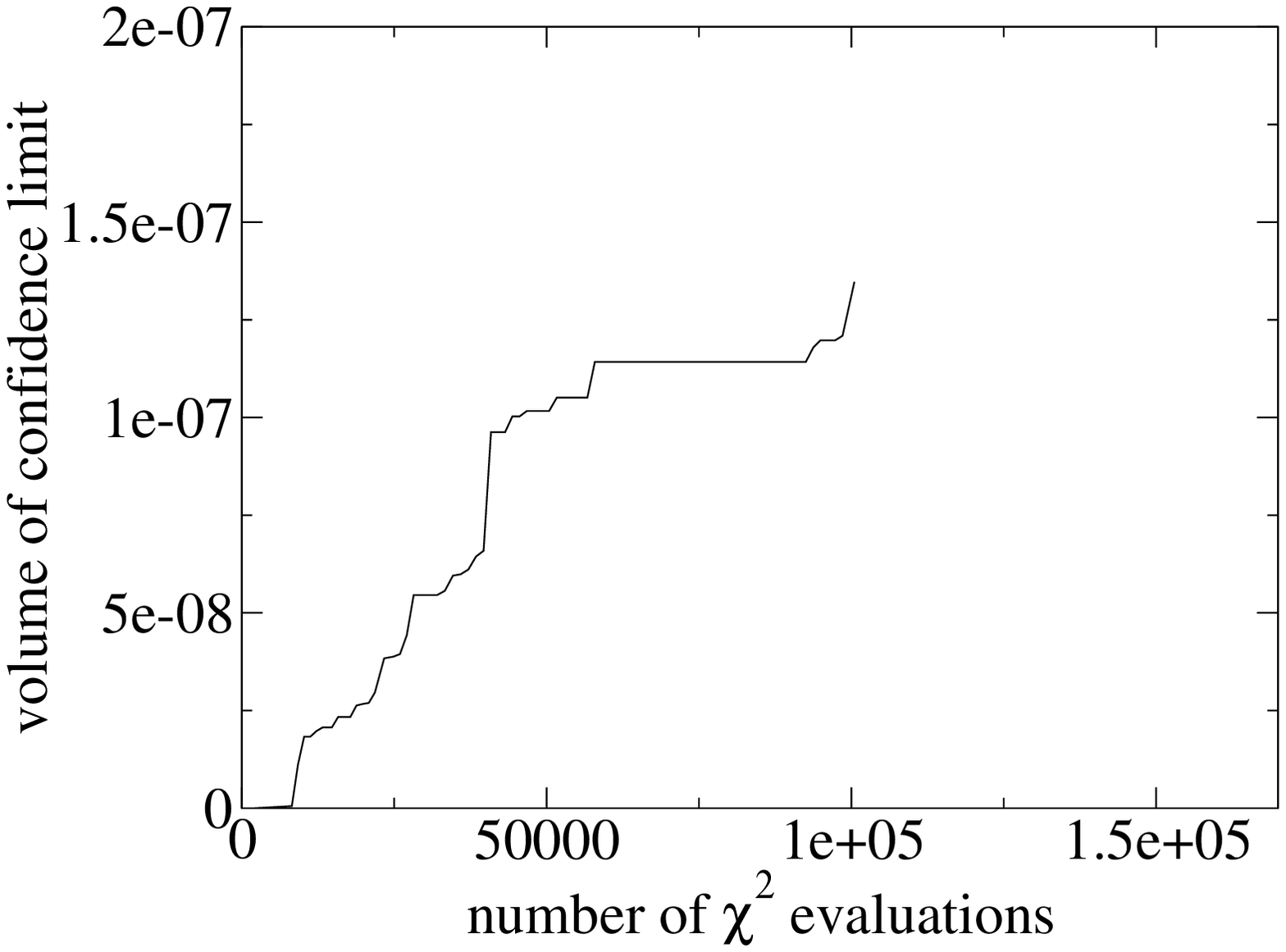}
\caption{
The parameter space volume of the Frequentist confidence limit
in figure \ref{fig:contours_gauss} as a function of the number of
$\vec{\theta}\rightarrow\chi^2$ evaluations.
}
\label{fig:convergence}
\end{figure}

There remain, however, several concerns to be addressed in the use of APS.  Unlike MCMC, there is
no obvious criterion for convergence of the APS algorithm.  One possibility is to interpret the Frequentist
confidence limit as a hyperbox (i.e. keep track of the allowable maximum and minimum values of each parameter
as $\chi^2=\chi^2_\text{lim}$ is explored) and watch the growth of the volume of that hyperbox as a function
of the number of $\vec{\theta}\rightarrow\chi^2$ evaluations.  Figure \ref{fig:convergence} plots this
convergence metric for the APS run in Figure \ref{fig:contours_gauss}.  While it appears that APS may have
converged after $\sim$50,000 evaluations, the volume of the Frequentist confidence limit continues to grow,
though Figure \ref{fig:contours_gauss} already shows good coverage of the control Bayesian credible limit. 
It is, of course, possible that we are seeing the effect of the slight difference 
between the Bayesian
credible limit and the Frequentist confidence limt, 
even in the large-data case.

\begin{figure}
\includegraphics[scale=0.3]{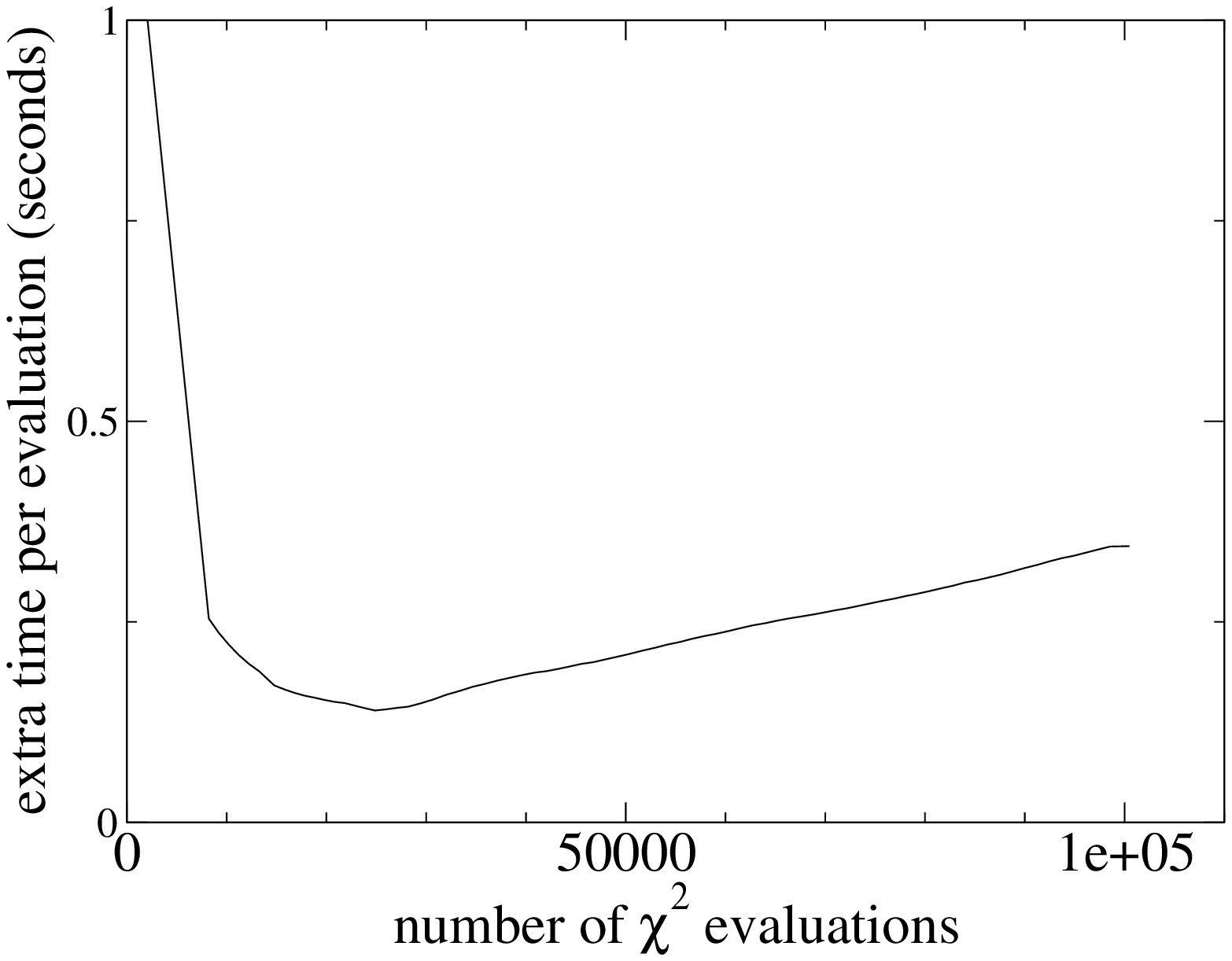}
\caption{
The extra time added by APS to a $\vec{\theta}\rightarrow\chi^2$ 
as a function of the number of
$\vec{\theta}\rightarrow\chi^2$ evaluations.
}
\label{fig:time}
\end{figure}

Another concern is that, while APS was explicitly designed to rapidly converge
to the Frequentist confidence limit (where rapidity is measured in the number of
$\vec{\theta}\rightarrow\chi^2$ evaluations made), the searches in Section \ref{sec:algorithm}
are complicated enough that they will add extra clock time to each $\vec{\theta}\rightarrow\chi^2$
evaluation.  We have attempted to balance APS so that it does not 
rely too heavily on expensive calculations.
However, some overhead is inevitably required.  Figure \ref{fig:time} shows the average overhead
time in seconds added by APS to each $\vec{\theta}\rightarrow\chi^2$ evaluation as a function
of the number of evaluations.  As the number of evaluations grow and the number of points APS must search
when constructing its Gaussian processes also grows, so does this extra time (the initial spike is due to the
first expensive call to the hyper-parameter optimization in Section \ref{sec:hyperparams}).
Though the extra time stays within the bounds of a few 0.1 seconds per $\vec{\theta}\rightarrow\chi^2$
evaluation, this may still be too expensive in the case of functions for whom a direct
$\vec{\theta}\rightarrow\chi^2$ takes a vanishingly small amount of time.

Finally, one advantage of MCMC that APS cannot overcome is its comparative ease to
implement.  To this end,
we make our code available at \verb|https://github.com/uwssg/APS/|.  
The code is presented as a series of C++ classes with directions
indicating where the user can interface with 
the desired $\vec{\theta}\rightarrow\chi^2$ function.  
Those with questions about the code or the algorithm should not
hesitate to contact the authors.


\section*{Acknowledgments}
The authors would like to thank the referee for comments and suggestions
which we hope have made this paper more useful to astrophysical audiences.
They would also like to thank Christopher Genovese for useful discussions
regarding statistial theory and interpretation.
SFD would like to thank Eric Linder, Arman Shafieloo, and Jacob Vanderplas 
for useful conversations
about Gaussian processes.  SFD would also like to acknowledge the hospitality of
the Institute for the Early Universe at Ewha Womans University in 
Seoul, Korea, who
hosted him while some of this work was done.
We acknowledge support from DOE award grant number DESC0002607 and NSF grant
IIS-0844580.

\begin{appendix}
\section{Frequentist Confidence Intervals}
\label{sec:frequentism}

Frequentist confidence limits are not nearly as common in the astrophysical
literature as their Bayesian counterparts.  For this reason, we define them below.

Suppose we are going to conduct some experiment resulting in a random data set
$\vec{d}$.  Suppose also that we have some theory by which the distribution of
$\vec{d}$ is controlled by a set of parameters $\vec{\theta}$.  We would like to
use $\vec{d}$ to make some statement about the true values of $\vec{\theta}^{(0)}$
i.e., the values of $\vec{\theta}$ actually realized by the universe.
Rather than adopt the Bayesian perspective 
and talk about the peak and width of the
the posterior
``probability of $\vec{\theta}$ given $\vec{d}$,'' Frequentists attempt to
construct sets of estimators
$\vec{\theta}^{(\text{CL})}(\vec{d})$ such that the
probability that 
$\vec{\theta}^{(0)}\in\vec{\theta}^{(\text{CL})}(\vec{d})$
is $(1-\alpha)\%$, i.e. if the experiment were repeated many times, yielding
many different sets $\vec{d}$, and the confidence limit set
$\vec{\theta}^{(\text{CL})}$
was calculated for each of those data sets, the true
value of $\vec{\theta}^{(0)}$ would fall within $\vec{\theta}^{(\text{CL})}$ 
$(1-\alpha)\%$ of the time.
This is the original definition of confidence intervals posited by Neyman (1937;
see his equation 20 and the attendant discussion).  It also underlies the
confidence belt construction discussed 
in Chapter 9 of Eadie {\it et al}. (1971), Chapter 20 of Stuart and Ord (1991)
and section II B of Feldman and Cousins (1998).

The specific example we consider in Section \ref{sec:wmap}
of this paper is the WMAP 7 year data
release measuring the anisotropy in the Cosmic Microwave Background.  This 
data set takes the form of a
power spectrum composed of 1199 $C_\ell$s ($2\le\ell\le1200$) 
characterizing the
anisotropy power on different angular scales.  
These 1199 measurements represent
independent Gaussian random variables. 
The probability density on $C_\ell$ space associated with
measuring a particular set of $C_\ell$s given a particular candidate set of
$\vec{\theta}^{(c)}$ is
$$P\propto\exp
\big[-\frac{1}{2}
\sum_{\ell,\ell^\prime}(C_\ell(\vec{d})-C_\ell(\vec{\theta}^{(c)})) 
\text{Cov}^{-1}_{\ell\ell^\prime}
(C_{\ell^\prime}(\vec{d})-C_{\ell^\prime}(\vec{\theta}^{(c)}))\big]$$
where 
$C_\ell(\vec{d})$ are the 1199 measured values of $C_\ell$,
$C_\ell(\vec{\theta}^{(c)})$ are the 1199 values of $C_\ell$ predicted
by the candidate theory,
and $\text{Cov}^{-1}_{\ell\ell^\prime}$ is the inverse of the covariance matrix
relating the measured $C_\ell$s.
A change of variables to
$$\chi^2\equiv \sum_{\ell\ell^\prime}
(C_\ell(\vec{d})-\vec{C}_\ell(\vec{\theta}^{(c)}))
\text{Cov}^{-1}_{\ell\ell^\prime}
(C_{\ell^\prime}(\vec{d})-C_{\ell^\prime}(\vec{\theta}^{(c)}))$$
gives the well-known result that the 1199 independent, Gaussian-distributed
$C_\ell$s give a $\chi^2$ statistic distributed according to the eponymous
$\chi^2$ distribution with 1199 degrees of freedom.  The $\chi^2$ distribution
is well-understood.  In the case of 1199 degrees of freedom, 95\% of the
probability is enclosed by $\chi^2\le1280.7$.  
We may use this fact to construct a
95\% confidence limit.

Recall the definition of Frequentist confidence limits.  The true
value of the parameters
$\vec{\theta}^{(0)}$ is fixed, but $\vec{d}$ (in the specific example above, the
set of 1199 $C_\ell$s) is random and will change each time we conduct the
experiment.  In this case, ``conducting the experiment'' means creating a new Universe
with a new Cosmic Microwave Background, randomly generated from the same
$\vec{\theta}^{(0)}$, and measuring it with WMAP;
while this is impossible, we ask the readers to
suspend their disbelief for the sake of this illustration.  
Because of what we have just noted about $\chi^2$, in
95\% of our repeated WMAP experiments $\chi^2$ calculated relative to the true
parameter set $\vec{\theta}^{(0)}$ will be less than or equal to 1280.7.  If, for
each set of $C_\ell$s we find all of the combinations of $\vec{\theta}$ that
yield $\chi^2\le1280.7$, that set of $\vec{\theta}$ will contain $\vec{\theta}^{(0)}$
95\% of the time.
Taking our one realization of $C_\ell$ and finding all of the values of
$\vec{\theta}^{(\text{CL})}$ which give $\chi^2\le1280.7$,
we can be confident that we have contained $\vec{\theta}^{(0)}$ with 95\%
probability.  This is what is meant by a 95\% Frequentist confidence limit.

In the large data limit, Frequentist confidence limits are expected to
give comparable results to Bayesian inference (as we see in Figure
\ref{fig:contours_gauss}).  In the limit of small data, Frequentist confidence limits
may differ from Bayesian limits.  If statistical fluctuations dominate the
signal, it is possible for Frequentist methods to return an empty confidence
limit (no points in parameter space fit the data).  While this may seem an
unpalatable outcome, it is useful to know that one's data set is noisy so that
one can interpret the explored likelihood surface with all due caution.
Lyons (2002 and 2008) discusses this distinction, as well as the
comparative strengths and weaknesses of other statistical perspectives, at
greater length than this work.

The Frequentist approach to confidence limits described herein
is actually a conservative version of
the non-parametric confidence ball approach popular among academic statisticians
(see the introduction to Baraud 2004).  Statisticians typically are concerned
with using data to constrain the form of a function in arbitrary function space.
They use the data to derive an estimator of the function 
from which it was drawn (in our example, the smooth theoretical $C_\ell$ function)
and then use
theoretical considerations to draw a hypersphere in function space about that
fit constituting the $(1-\alpha)\%$ confidence limit
\cite{bd,li,baraud,cai,davies}.
Note that these hyperspheres (referred to in the literature as ``confidence
balls'') are calculated from the data alone without reference to any
underlying physical model or set of (in Wilks' 1938 words) ``admissible
hypotheses.''  In this way, they function identically to the $\chi^2$ test
adopted in the present work.  

Astrophysicists can utilize the
``confidence ball'' formalism by demanding that their theoretical
models give predicted functions that fall within 
the hypersphere drawn in function space.  This is how
Bryan {\it et al}. (2007) originally drew their constraints on the CMB.  It is
also how Genovese {\it et al}. (2004) confirmed the statistical significance of
the harmonic peaks in the WMAP 1-year CMB data.  Because the confidence
hypersphere is drawn a-priori, without any consideration for the admissible
hypotheses within the space of physical parameters, this method would be
straightforward to interface with \APS (indeed, it was included in the Bryan
{\it et al}. draft of the code).  
We leave such an implementation to future work.

The more familiar Likelihood Ratio Test detailed
by Wilks (1938)  and Neyman and Pearson (1933) is an asymptotic simplification
of Frequentist confidence limits
for cases in which we know that the true model must exist
within some limited set of hypotheses.  
In this case, one assumes that the adopted $\vec{\theta}$ parametrization
is the only possible way of explaining that data.  In that case, the
$\chi^2_\text{min}$ on parameter space must, by definition, be the smallest
achievable $\chi^2$.  $\Delta\chi^2\equiv\chi^2-\chi^2_\text{min}$
is then distributed according to the $\chi^2$ distribution with $N_p$ 
degrees of freedom.  This is the assumption underlying our use of
$\Delta\chi^2$ in the test of Section \ref{sec:toy} and \ref{sec:wmap} above.

\end{appendix}


\label{lastpage}

\end{document}